\documentclass[reprint,twocolumn,longbibliography,
notitlepage,amsmath,amssymb,aps,prb,
]{revtex4-1}

\usepackage{array,makecell}
\usepackage[dvipsnames]{xcolor}
\usepackage{mathrsfs}
\usepackage{pst-node}
\usepackage{yfonts}

\DeclareFontFamily{U}{mathx}{}
\DeclareFontShape{U}{mathx}{m}{n}{<-> mathx10}{}
\DeclareSymbolFont{mathx}{U}{mathx}{m}{n}
\DeclareMathAccent{\widecheck}{0}{mathx}{"71}

\usepackage{amsthm,tikz,tikz-cd}

\usepackage{comment}
\usepackage{graphicx}

\usepackage[bbgreekl]{mathbbol}
\usepackage{amscd}

\usepackage{quiver}
\input{macros.sty}

\usepackage[colorlinks, breaklinks=true, linkcolor=red, citecolor=blue, linktocpage=true]{hyperref}

\usepackage{tikz}
	\usetikzlibrary{calc}
\usetikzlibrary{arrows,shapes, positioning, matrix}
\usetikzlibrary{decorations.markings}
\newcommand*{\halfway}{0.5*\pgfdecoratedpathlength+.5*4pt}\tikzstyle arrowstyle=[scale=1]
\tikzstyle arrowstyle=[scale=1]
\tikzstyle directed=[postaction={decorate,decoration={markings,
    mark=at position .15 with {\arrow[arrowstyle]{stealth}}}}]
\tikzstyle string=[thick,postaction={decorate},decoration={markings,
    mark=at position \halfway with {\arrow[arrowstyle]{stealth}}}]
\tikzstyle string2=[thick,postaction={decorate},decoration={markings,
    mark=at position .5 with {\arrow[arrowstyle]{stealth}}}]
\tikzstyle trivial=[dashed, thick]

\tikzstyle dw=[thick,postaction={decorate,decoration={markings,
    mark=at position 1 with {\arrow[arrowstyle]{stealth}}}}]
\tikzstyle group=[]

\newcommand{\comments}[1]{}

\newcommand{\mb}[1]{\mathbf{#1}}

\renewcommand{\cal}[1]{\mathcal{#1}}

\newcommand{\ag}[2]{#1_\mb{#2}}

\makeatletter
\newcommand{\vast}{\bBigg@{3}}
\newcommand{\Vast}{\bBigg@{4}}
\makeatother

\newcommand{\be}{\beta}
\newcommand{\om}{\omega}

\begin{document}

\title{Exactly solvable models for fermionic symmetry-enriched topological phases and fermionic 't Hooft anomaly}
\author{Jing-Ren Zhou}
\affiliation{Department of Physics, The Chinese University of Hong Kong, Shatin, New Territories, Hong Kong}
\author{Zheng-Cheng Gu}
\email{zcgu@phy.cuhk.edu.hk}
\affiliation{Department of Physics, The Chinese University of Hong Kong, Shatin, New Territories, Hong Kong}

\begin{abstract}
The interplay between symmetry and topological properties plays a very important role in modern physics. In the past decade, the concept of symmetry-enriched topological (SET) phases was proposed and their classifications have been systematically studied for bosonic systems. Very recently, the concept of SET phases has been generalized into fermionic systems and their corresponding classification schemes are also proposed. Nevertheless, how to realize all these fermionic SET (fSET) phases in lattice models remains to be a difficult open problem. In this paper, we first construct exactly solvable models for non-anomalous non-chiral 2+1D fSET phases, namely, the symmetry-enriched fermionic string-net models, which are described by commuting-projector Hamiltonians whose ground states are the fixed-point wavefunctions of each fSET phase. Mathematically, we provide a partial definition to $G$-graded super fusion category (when the total symmetry $G^f$ is a direct product of physical symmetry $G$ and fermion-parity symmetry $\mathbb{Z}_2^f$), which is the input data of a symmetry-enriched fermionic string-net model. Next, we construct exactly solvable models for non-chiral 2+1D fSET phases with 't Hooft anomaly, especially the $H^3(G,\mathbb{Z}_2)$ fermionic 't Hooft anomaly which is different from the well known bosonic $H^4(G,U(1)_T)$ anomaly. In our construction, this $H^3(G,\mathbb{Z}_2)$ fermionic 't Hooft anomaly is characterized by a violation of fermion-parity conservation in some of the surface $\mathcal{F}$-moves (a type of renormalization moves for the ground state wavefunctions of surface fSET phases), and also by a new fermionic obstruction $\Theta$ in the surface pentagon equation. We demonstrate this construction in a concrete example that the surface topological order is a $\mathbb{Z}_4$ gauge theory embedded into a fermion system and the total symmetry $G^f=\mathbb{Z}_2^f\times\mathbb{Z}_2\times\mathbb{Z}_4$. We further conjecture that the $H^2(G,\mathbb{Z}_2)$ fermionic 't Hooft anomaly is characterized by a possible violation of non-terminating q-type (called $\sigma$-type) string in the fusion rules of surface models. 
\end{abstract}

\maketitle
    
\tableofcontents

\section{Introduction}
\subsection{Overview}
The concept of symmetry has been introduced into modern physics for more than a century. 
In condensed matter physics, for long time people believed that the Landau symmetry-breaking theory could describe all possible phases and (continuous) phase transitions. Until the discovery of the fractional quantum Hall effect (FQHE) and high-temperature superconducting cuprates, it was pointed out that systems exhibiting anyon excitations with fractional charge are described by a new type of order beyond Landau paradigm, namely, the topological orders characterized by long-range entanglement. Thereafter, the interplay between symmetry and topological properties opened up a new chapter in modern physics. For example, topological insulators and superconductors were proposed and intensively studied in the past two decades. The concept of symmetry protected topological (SPT) order was proposed and their full classifications for interacting systems have also been achieved recently\cite{chen2012symmetry,chen2011complete,chen2011classification,gu2009tensor,cheng2018classification, Gu2014,Wang2018,Wang2020}. 

Symmetry-enriched topological (SET) phases are gapped quantum phases with intrinsic topological order and global symmetry, where the symmetry can interact with the topological order in a nontrivial way, thus enriching the classification of these phases. In general, symmetry can also act non-trivially on the anyon excitations (we will use anyon, quasiparticle excitation, anyon excitation interchangeably) in an SET phase, such that the anyons can carry fractionalized quantum numbers under global symmetry, called symmetry fractionalization\cite{chen2017symmetry,essin2013classifying}. For example, the gapped quantum spin liquids in frustrated magnets are SET phases, where the spin-$1/2$ spinon excitation transforms projectively under the $SO(3)$ spin-rotation symmetry\cite{Wen2002}. Aside from symmetry fractionalization, another feature of SET phases is that symmetry may permute the anyon type. For example, considering the toric code topological order with onsite $\mathbb{Z}_2$ symmetry, there are several possible SET phases, in one of which anyons $e$ and $m$ are permuted by the $\mathbb{Z}_2$ symmetry\cite{Cheng2017}. Anyons and symmetry defects in 2+1D SET phases are systematically classified in Ref.~\onlinecite{BSET} via $G$-crossed braided tensor categories for a given symmetry group $G$. While exactly solvable models for 2+1D non-chiral SET phases, called symmetry-enriched string-net models, are constructed in Refs.~\onlinecite{heinrich2016symmetry,Cheng2017}, which focus on discrete finite group case (the continuous group case is discussed in Ref.~\onlinecite{wang2022exactly}).

On the other hand, surface topological orders with 't Hooft anomaly have been studied in many previous works\cite{burnell2014exactly,fidkowski2013non,cheng2018microscopic,metlitski2014interaction,hong2017topological,wang2017anomaly,wang2014interacting,shirley2022three,delmastro2021global,cho2014conflicting,wang2016bulk,xu2013three,wang2013boson,vishwanath2013physics,bonderson2013time,metlitski2015symmetry,tata2023anomalies,cheng2016translational,barkeshli2022classification}. It was shown that 2+1D bosonic SET phases with 't Hooft anomaly are characterized by an obstruction that is a 4-cocycle in $H^4(G,U(1)_T)$\cite{chen2015anomalous,barkeshli2020relative}, 
and this obstruction can be cancelled by putting it on the surface of a 3D bulk SPT phase classified by a 4-cocycle.
If we study the anyon excitations of the surface SET phase, this 4-cocycle $o_4\in H^4(G,U(1)_T)$ is realized as the obstruction for a $G$-crossed extension for the unitary modular tensor category that describes the anyons\cite{BSET}. ($G$-crossed extension means that we consistently add deconfined symmetry defects into the system, such that the anyons and symmetry defects together form a larger category, called a $G$-crossed braided tensor category.) 
Gauging the symmetry is another commonly used tool in studying topological phases with symmetry\cite{levin2012braiding,wang2015topological,lan2023gauging}. It transforms a global symmetry into a local gauge symmetry, which brings new topologically ordered states into the system. 

Although fruitful results have been achieved for bosonic systems in the past decade, the interplay between symmetry and topological orders is much richer and more complicated in fermionic systems. Here we will focus on 2+1D fermionic SET (fSET) phases, which are SET phases whose underlying degrees of freedom are fermions (such as electrons). Denoting this physical fermion as $f$, fermion-parity is always conserved in any fermionic system, i.e., there is always a fermion-parity symmetry $\Z_2^f$. Therefore, the total symmetry group is always a central extension of a bosonic symmetry group $G$ by $\Z_2^f$, denoted by $G^f$. Another feature is that all anyon excitations in an fSET phase always have trivial double-braiding with this physical fermion $f$, whereby $f$ is also called the transparent fermion. The famous fractional quantum Hall (FQH) states are examples of chiral fSET phases\cite{wen1995topological}, where at filling factor $\nu=1/q$, anyon excitations carry fractional charge $e/q$ under the $U(1)$ charge conservation symmetry. An algebraic framework for the classification of anyons and how they behave under symmetry in fSET phases is studied in Refs.~\onlinecite{bruillard2019classification,bruillard2020classification}, and symmetry defects are further included in Ref.~\onlinecite{FSET} as a more complete description. Roughly speaking, $G^f$-crossed super modular categories describe anyons and symmetry defects in 2+1D fSET phases. 
On the other hand, if we gauge $\Z^f_2$, we obtain $G$-crossed spin modular categories (a spin modular category is one of the 16 possible $\Z^f_2$-modular extensions of a super modular category).

Similar to bosonic systems, surface topological orders with 't Hooft anomaly can also be generalized into fermion systems. In the context of quasiparticle excitations, 't Hooft anomalies in 2+1D fSET phases are distinguished into four layers in Refs.~\onlinecite{bulmash2022anomaly,FSET}, characterized by the following obstruction functions: a 1-cocycle $o_1\in H^1(G,\Z_T)$, a 2-cocycle $o_2\in H^2(G,\Z_2)$, a 3-cocycle $o_3\in H^3(G,\Z_2)$ and a 4-cocycle $o_4\in H^4(G,U(1)_T)$. The four layers of 't Hooft anomalies are in one-to-one correspondence (up to some equivalence relations) to the four layers in classifying 3+1D fermionic SPT (fSPT) phases: $p+ip$ superconductor layer, Kitaev chain layer, complex fermion layer and bosonic layer\cite{Gu2014,Wang2018,Wang2020} respectively. This implies the picture that any 2+1D fSET phase with 't Hooft anomaly is living on the surface of a certain bulk 3+1D fSPT phase. 
The $H^4(G,U(1)_T)$ anomaly corresponds to the obstruction to do $G$-crossed extension or gauging $G$ for the surface bosonic SET phase. While the other three anomalies, we call fermionic 't Hooft anomalies, all correspond to the obstruction to gauging $\Z^f_2$ in different levels.

Although a lot of progress has been made for understanding the intriguing properties of fSET phases in recent years, so far it is still unclear how to systematically realize all these fSET phases in lattice models. 
The aim of this paper is to construct exactly solvable models for 2+1D fSET phases, including both non-anomalous phases and phases with fermionic 't Hooft anomaly. In particular, we will propose a generic framework to construct fixed-point wavefunctions for non-chiral fSET phases based on equivalence class of generalized fermionic symmetric local unitary (gfSLU) transformations. Given a $G$-graded super fusion category $\cS_G$ (described in Subsection \ref{subsec.gsuperfusion}) as input, we can construct a fermionic symmetry-enriched string-net model that fully describes the fixed-point ground state wavefunctions and the parent commuting-projector Hamiltonian of a non-anomalous non-chiral 2+1D fSET phase. {We present a simple but illustrative example of the exactly solvable model for the toric code together with a physical fermion with $\Z_4^f\cong\Z_2^f\times_{\omega_2}\Z_2$ symmetry in Subsection \ref{subsec.ex}. 
We further present two more-involved examples: the fermionic symmetry-enriched string-net models with input $\Z_2$-graded super fusion category being the super Tambara-Yamagami category and ${^\svec \text{SU}(2)_6}$, both of which are exactly solvable models for non-chiral fSET phases with $\Z_2^f\times\Z_2$ symmetry.}

The concept of equivalence class of fermionic symmetric local unitary transformations can also be generalized into constructing anomalous fSET states by including the bulk 3+1D fSPT states. In our construction, we focus on the $H^3(G,\Z_2)$ fermionic 't Hooft anomaly characterized by a violation of fermion-parity conservation in some of the surface $\cF$-moves, and also by a new fermionic obstruction $\Theta$ in the surface pentagon equation defined in Eq.~(\ref{eq.omega}). We demonstrate this construction in a concrete example that the surface topological order is a $\Z_4$ gauge theory embedded into a fermion system and the total symmetry is $G^f=\Z_2^f\times\Z_2\times\Z_4$. We further conjecture that the $H^2(G,\Z_2)$ fermionic 't Hooft anomaly can only be supported by surface topological order with Ising $\sigma$-type strings (a  q-type string\cite{fc,fSN} is called $\sigma$-type if it is $\Z_2$-conserved), and this anomaly corresponds to a violation of the $\Z_2$-conservation of $\sigma$-type strings in some of the fusion rules, and the obstructed pentagon equation is in a similar structure as the $H^3(G,\Z_2)$ anomaly. 

Throughout the paper, we will use the following conventions, and we summarize important notations used in this paper in Table I.

\begin{convention}\label{conv.iso}
    We may use the set of isomorphism classes of simple objects in a fusion category to denote this category. For example, $\cZ_1(\textbf{Vec}_{\mathbb{Z}_2})=\{1,e,m,\psi\}$ (the notations are introduced in Table I above), where $1,e,m,\psi$ are four (representatives of) isomorphism classes of simple objects, which physically correspond to anyon types in the toric code topological order.
\end{convention}

\begin{convention}\label{conv.cochain}
    We may use the following abbreviation to denote any homogeneous $n$-cochain $f_n\in C^n(G,M)$ ($M$ is a $G$-module, which is an Abelian group):
    \begin{align}
        f_n(012...n)
        :=f_n(\Bg_0,\Bg_1,\Bg_2,...\Bg_n),
    \end{align}
    where $\Bg_0,\Bg_1,\Bg_2,...\Bg_n\in G$.
    
    In addition, let $f_n\in Z^n(C,M)$ be a $n$-cocycle. We may abuse the notation $f_n\in H^n(C,M)$ instead of the rigorous notation $[f_n]\in H^n(C,M)$, where $[f_n]$ denotes the cohomology class of $f_n$.
\end{convention}

\begin{center}
\begin{tabular}{ |c|c| } 
  \hline
  $G$ & Bosonic symmetry group \\
  \hline
  $\one$ & Trivial group element in $G$ \\
  \hline
  $\Z^f_2$ & fermion-parity symmetry group \\
  \hline
  $\omega_2$ & \makecell{The 2-cocycle that determines the\\central extension of $G$ by $\Z^f_2$} \\
  \hline
  $G^f$ & Total fermionic symmetry group \\
    \hline
  $H^n(G,M)$ & \makecell{$n$-th group cohomology of $G$ valued in $M$,\\where $M$ is an Abelian group} \\
  \hline
  $C^n(G,M)$ & Group of $n$-cochains of $G$ valued in $M$ \\
  \hline
  $B^n(G,M)$ & Group of $n$-coboundaries of $G$ valued in $M$ \\
  \hline
    $f$ & Physical fermion\\
  \hline
  $\cC$ & General fusion/tensor category \\
  \hline
  $\mathbb{1}$ & Tensor unit in $\cC$ \\
  \hline
  $\cZ_1(\cC)$ & Drinfeld center of $\cC$ \\
  \hline
  $\cZ_2(\cC)$ & M\"{u}ger center of $\cC$ \\
  \hline
  $\boxtimes$ & Deligne tensor product \\
  \hline
  $\textbf{Vec}_N$ & \makecell{Category of $N$-graded vector spaces,\\where $N$ is a group} \\
  \hline
  $\C^{p|q}$ & \makecell{Super vector space with grade-0 dimension\\$p$ and grade-1 dimension $q$} \\
  \hline
  $\svec$ & Category of super vector spaces \\
  \hline
  \makecell{${^\svec\cC}\equiv$\\$\cC/f$} & \makecell{Super fusion category,\\obtained by canonical construction of\\enriched monoidal category,\\or equivalently, by condensing the\\fermion $f$ in $\cZ_1(\cC)$} \\
  \hline
  $\text{Hom}(X,Y)$ & \makecell{Hom-space of $X,Y\in\cC$, which is the\\$\C$-(super) vector space consists of\\morphisms from $X$ to $Y$} \\
  \hline
  $\makecell{\text{End}(X):=\\\text{Hom}(X,X)}$ & \makecell{Endomorphism space of $X$, $X\in\cC$} \\
  \hline
  \makecell{$\cS_G$} & $G$-graded super fusion category \\
  \hline
  ${^\svec\cC_G}$ & \makecell{$G$-graded super fusion category\\constructed by a fermion condensation\\from a $G$-graded unitary fusion category $\cC_G$} \\
  \hline
  \makecell{$a_\Bg$} & \makecell{String type, or simple object in $\cS_G$} \\
  \hline
  \makecell{m-type\\string} & $\text{End}(a_\Bg)=\C$ \\
  \hline
  \makecell{q-type\\string} & $\text{End}(a_\Bg)=\C^{1|1}$ \\
  \hline
  \makecell{$\sigma$-type\\string} & $\Z_2$-conserved q-type string \\
  \hline
  $\alpha$ or $|\alpha\ra$ & \makecell{A fusion state in hom-space\\$\text{Hom}(a_\Bg\ot b_\Bh,c_{\Bg\Bh})$} \\
  \hline
  $\underline{\alpha}$ & \makecell{The vertex supporting fusion state $\alpha$} \\
  \hline
  $s(\alpha)$ & fermion-parity of fusion state $\alpha$ \\
   \hline
  $s'(\Bg_0,\alpha)$ & \makecell{Total surface fermion-parity of fusion state\\$\alpha$ with leftmost group element $\Bg_0$} \\
  \hline
  $S(\Bg)$ & Anti-unitarity of $\Bg\in G$ \\
  \hline
  $U(\Bg)$ & Global symmetry transformation of $\Bg\in G$ \\
  \hline
  $c_{\underline{\alpha}},
  c^\dagger_{\underline{\alpha}}$ & \makecell{Fermion annihilation and creation\\operators on vertex $\underline{\alpha}$\\in 2+1D system} \\
  \hline
  $C_{ijkl},
  C^\dagger_{ijkl}$ & \makecell{Bulk fermion annihilation and creation\\operators in tetrahedron $\la ijkl \ra$\\in 3+1D system} \\
  \hline
\end{tabular}

\begin{tabular}{ |c|c| } 
  \hline
  $\quad o_4\quad$ & \makecell{Obstruction on $G$-crossed extension,\\appearing when there is\\$H^4(G,U(1)_T)$ 't Hooft anomaly
  }  \\
  \hline
  $\quad o_3\quad$ & \makecell{Obstruction to extending\\symmetry fractionalization to\\its $\Z_2^f$-modular extended category,\\appearing when there is $H^3(G,\Z_2)$\\fermionic 't Hooft anomaly
  }  \\
 \hline
  $\quad o_2\quad$ & \makecell{Obstruction to extending\\$G$-action to\\its $\Z_2^f$-modular extended category,\\appearing when there is $H^2(G,\Z_2)$\\fermionic 't Hooft anomaly
  }  \\
  \hline
  $\Theta$ & \makecell{Fermionic obstruction on surface\\pentagon equation, defined in Eq.~(\ref{eq.omega}),\\appearing when there is $H^3(G,\Z_2)$\\fermionic 't Hooft anomaly}  \\
  \hline
\end{tabular}
\end{center}
\begin{center}\label{table.notation}
Table I. Table of notations
\end{center}

\subsection{Summary of major results from mathematical perspective}
In Table II, we summarize non-chiral non-anomalous 2+1D fermionic topological phases, i.e., fermionic topological orders and fSET phases. The exactly solvable model of any non-chiral 2+1D fermionic topological order is a fermionic string-net model\cite{2dtopo,fSN}, whose input category is a super fusion category. A super fusion category can always be obtained by a fermion condensation (or mathematically canonical construction of an enriched monoidal category, introduced in detail in Appendix \ref{appen.sfc}) from a unitary fusion category. Denote a unitary fusion category as $\cC$ whose Drinfeld center $\Z_1(\cC)$ contains a fermion, its fermion condensed category is denoted as $\cC/f\equiv{^\svec\cC}$, where ${^\svec\cC}$ denotes for category $\cC$ enriched over $\svec$, which is a super fusion category. Then we define the Drinfeld center of a super fusion category that describes the anyon excitations of a fermionic string-net model in Subsection \ref{subsec.dcsfc} as $\mathcal{Z}_1({^\svec\cC})\cong{^\svec \cZ_1(\cC)}_0$. Here $\cZ_1(\cC)\cong \cZ_1(\cC)_0\oplus \cZ_1(\cC)_1$ ($0,1$ are group elements of $\Z_2^f$) is a spin modular category, where its grade-0 sector $\cZ_1(\cC)_0$ is a super modular category, and its grade-1 sector $\cZ_1(\cC)_1$ is the fermion-parity vortex/flux sector. In other words, a spin modular category is a unitary modular category that contains a fermion $f$, such that $f$ is transparent (up to double-braiding) in its grade-0 sector, and ${^\svec \cZ_1(\cC)}_0$ denotes for $\cZ_1(\cC)_0$ enriched over $\svec$. Gauging the $\Z_2^f$ fermion-parity symmetry for ${^\svec \cZ_1(\cC)}_0$ recovers the bosonic string-net model inputted by unitary fusion category $\cC$, whose anyon excitations are described by $\cZ_1(\cC)$. On the other hand, the exactly solvable model of an fSET phase with total symmetry $G^f=\mathbb{Z}_2^{f}\times_{\omega_2}G$ is a fermionic symmetry-enriched string-net model, whose input category is a $G$-graded super fusion category $\cS_G$, introduced in Subsection \ref{subsec.gsuperfusion}, and partially defined (when $\omega_2$ is trivial) in Subsection \ref{subsec.gsuper}. Its Drinfeld center that describes the anyon excitations of this fSET phase is conjectured to be $\cZ_1(\cS_\one)$ ($\one$ is the trivial group element in $G$), and the symmetry acts on these anyons through the so-called $G^f$-action (anyon permutation) and symmetry fractionalization\cite{bulmash2022fermionic,FSET}. Then performing a $G^f$-crossed extension\cite{FSET}, i.e., coupling the anyon excitations with $G^f$-symmetry defects consistently, we obtain a $G^f$-crossed super modular category $\cZ_1({\cS_\one})_{G^f}^\times$. Then gauging the $\Z_2^f$ symmetry when $\omega_2$ is trivial (we do not know yet how to gauge $\Z_2^f$ when $\omega_2$ is nontrivial), we obtain a bosonic SET phase whose exactly solvable model is a symmetry-enriched string-net model\cite{Cheng2017,heinrich2016symmetry} inputted by a $G$-graded unitary fusion category $\cC_G$ such that its center $\cZ_1(\cC_G)$ is a spin modular category, which is related to $\cS_G$ by ${^\svec\cC_G}=\cS_G$, i.e., $\cS_G$ is obtained by a fermion condensation on $\cC_G$. Here ${^\svec\cC_\Bg}=\cS_\Bg$, $\forall \Bg\in G$ (see details in Subsection \ref{subsec.gsuper}). Briefly, for the trivial group element sector ${^\svec\cC_\one}=\cS_\one$, $\cS_\one$ is obtained by a fermion condensation on $\cC_\one$, as $\cZ_1(\cC_G)$ is a spin modular category implies that $\cZ_1(\cC_\one)$ is also a spin modular category (see full procedures of fermion condensation in Diagram (\ref{diag.centersfc})); while for other $\Bg\neq \one$ sectors, $\cS_\Bg$ is not obtained by fermion condensation, but a so-called canonical construction of enriched category\cite{kong2021enriched} by our constraint that each $\cC_\Bg$ sector is closed under $\svec$-action. 

\begin{widetext}

\begin{center}
\label{table.1}
\begin{tabular}{ |c|c|c| } 
  \hline
  & \makecell{Non-anomalous non-chiral\\2+1D fermionic\\topological order} & \makecell{Non-anomalous non-chiral\\2+1D fSET phase\\with symmetry $G^f$} \\
  \hline
  Exactly solvable model & \makecell{Fermionic string-net model\\with input super fusion\\category ${^\svec\cC}$}& 
  \makecell{Fermionic symmetry-enriched\\string-net model with input\\$G$-graded super fusion category ${\cS_G}$} 
  \\ 
  \hline
  \makecell{Quasiparticles/Anyon\\excitations}  & 
  \makecell{$\mathcal{Z}_1({^\svec\cC})\cong{^\svec \cZ_1(\cC)}_0$\\(see Eq.~(\ref{eq.dcsfc}))}
  & \makecell{$\mathcal{Z}_1({\cS_\one})$ with $G^f$-action\\(see Conjecture \ref{conj.anyon})\\ \\Coupling with $G^f$-symmetry defects:\\ {$G^f$-crossed super modular category} $\cZ_1({\cS_\one})_{G^f}^\times$}
  \\ 
 \hline
  \makecell{Exactly solvable model\\after gauging $\Z_2^f$\\when $\omega_2$ is trivial} & \makecell{String-net model with input\\unitary fusion category $\cC$} & \makecell{Symmetry-enriched string-net\\model with input $G$-graded\\unitary fusion category $\cC_G$,\\where ${^\svec\cC_\Bg}=\cS_\Bg$, $\forall \Bg\in G$} 
  \\ 
  \hline
  \makecell{Anyon excitations\\after gauging $\Z_2^f$\\when $\omega_2$ is trivial}  & \makecell{$\mathcal{Z}_1(\cC)$, which is\\a spin modular category} 
  & \makecell{$\mathcal{Z}_1(\cC_\one)$ with $G$-action\\ \\Coupling with $G$-symmetry defects:\\ {$G$-crossed spin modular category $\cZ_1(\cC_\one)_G^\times$}}
  \\ \hline
  \end{tabular}
\end{center}
\begin{center}
Table II. Summary of contents on non-chiral non-anomalous 2+1D fermionic topological phases. 
\end{center}

\begin{center}
\begin{tabular}{ |c|c|c| } 
  \hline
  & \makecell{Non-chiral 2+1D SET phase\\with symmetry $G$\\with $H^4(G,U(1)_T)$ 't Hooft anomaly} & \makecell{Non-chiral 2+1D fSET phase\\with symmetry $G^f$ ($\omega_2$ is trivial)\\with $H^3(G,\Z_2)$ fermionic 't Hooft anomaly}  \\
  \hline
  \makecell{Bulk topological\\phase}
  & 
  \makecell{3+1D SPT phase with given\\$[\nu_4]\in H^4(G,U(1)_T)$}
  &
  \makecell{3+1D fSPT phase with given\\$[n_3]\in H^3(G,\Z_2)$ and\\$[\nu_4]\in C^4(G,U(1)_T)/B^4(G,U(1)_T)$\\satisfying $\dd \nu_4=(-1)^{n_3\smile_1 n_3}$}
  \\ \hline
  \makecell{Surface exactly\\solvable model}
  & 
  \makecell{Surface symmetry-enriched\\string-net model\cite{Cheng2017}, where\\ 
  the pentagon equation is obstructed\\by $\nu_4$}
  &
   \makecell{Fermionic surface symmetry-enriched\\string-net model, where\\$\bullet$ $\cF$-moves can violate fermion-parity\\conservation by $n_3$;\\
   $\bullet$ The fermionic pentagon equation is\\obstructed by $\Theta\nu_4$\\(see Section \ref{sec.fthooft}, and $\Theta$ is a phase factor\\defined in Eq.~(\ref{eq.omega}))} 
  \\ \hline
  \makecell{Surface anyon\\excitations}
  & 
  \makecell{$\mathcal{Z}_1(\cC_\one)$ with $G$-action\\and with $[o_4]\in H^4(G,U(1)_T)$\\obstruction on $G$-crossed extension\\(also the obstruction to gauging $G$),\\where $[o_4]=[\nu_4]$}
  &
   \makecell{$\mathcal{Z}_1({\cS_\one})$ with $G^f$-action and\\with $[o_3]\in H^3(G,\Z_2)$ obstruction on\\extending symmetry fractionalization\\from $\mathcal{Z}_1({\cS_\one})$ to $\mathcal{Z}_1(\cC_\one)$\\(also the obstruction to gauging $\Z^f_2$),\\where ${^\svec\cC_\one}=\cS_\one$ and $[o_3]=[n_3]$} 
  \\ \hline
\end{tabular}
\end{center}
\begin{center}\label{table.2}
Table III. Summary of contents on non-chiral 2+1D topological phases with 't Hooft anomaly. $[f_n]$ denotes the cohomology class of $n$-cocycle $f_n$. The 't Hooft anomalies are in a structure of layers. That is to say, if the first layer fermionic $H^3(G,\Z_2)$ anomaly vanishes, we then consider whether there is second layer bosonic $H^4(G,U(1)_T)$ anomaly.
\end{center}

\end{widetext}

In Table III, we summarize non-chiral 2+1D topological phases with 't Hooft anomaly. Since the 't Hooft anomalies are in a structure of layers, given an anomalous fSET phase for example, we should at first check whether it has fermionic $H^2(G,\Z_2)$ anomaly (the fermionic $H^1(G,\Z_T)$ anomaly corresponds to chiral phases). If the fermionic $H^2(G,\Z_2)$ anomaly vanishes, we check whether there is fermionic $H^3(G,\Z_2)$ anomaly. Again if the fermionic $H^3(G,\Z_2)$ anomaly vanishes, we check whether there is bosonic $H^4(G,U(1)_T)$ anomaly. Let us consider the simplest case first, i.e., a bosonic SET phases with symmetry $G$ with $H^4(G,U(1)_T)$ anomaly, and we denote its obstruction function as $o_4\in H^4(G,U(1)_T)$. In the content of surface anyons together with $G$-action of this anomalous bosonic SET phase, $o_4$ corresponds to the obstruction to perform a $G$-crossed extension on surface anyons\cite{BSET,barkeshli2020relative}. This $H^4(G,U(1)_T)$ anomaly can be compensated by a bulk 3+1D SPT phase characterized by a 4-cocycle $\nu_4\in H^4(G,U(1)_T)$, where $\nu_4$ is in the same cohomology class as $o_4$, i.e., $[o_4]=[\nu_4]$. The surface exactly solvable model of this SET phase with $H^4(G,U(1)_T)$ anomaly is a so-called surface symmetry-enriched string-net model\cite{Cheng2017}, whose pentagon equation is exactly obstructed by $\nu_4$. On the other hand, for an fSET phase with symmetry $G^f$ (when $\omega_2$ is trivial, as we do not know how to deal with cases with nontrivial $\omega_2$ yet) with $H^3(G,\Z_2)$ fermionic 't Hooft anomaly, we denote its obstruction function as $o_3\in H^3(G,\Z_2)$. Since the anomaly is totally due to the anomalous symmetry action, while the surface topological order itself is anomaly-free, the fixed-point ground state wavefunctions of surface topological order can be described by a super fusion category $\cS_\one$, whose Drinfeld center that describes the surface anyons is $\cZ_1(\cS_\one)$. In the content of surface anyons $\cZ_1(\cS_\one)$ together with $G^f$-action of this anomalous fSET phase, $o_3$ corresponds on extending symmetry fractionalization from a super modular category $\cZ_1(\cS_\one)$ to its $\Z_2^f$-modular extension \cite{bulmash2022anomaly}, i.e., $\cZ_1(\cC_\one)$ as a spin modular category, where ${^\svec\cC_\one}=\cS_\one$. This fermionic $H^3(G,\Z_2)$ anomaly can be compensated by a bulk 3+1D fSPT phase characterized by a pair $(n_3,\nu_4)$, where $n_3$ is a 3-cocycle in $H^3(G,\Z_2)$, which is in the same cohomology class as $o_3$, i.e., $[o_3]=[n_3]$, and $\nu_4$ is a 4-cochain satisfying $\dd \nu_4=(-1)^{n_3\smile_1 n_3}$. The surface exactly solvable model of this fSET phase with fermionic $H^3(G,\Z_2)$ anomaly is a fermionic surface symmetry-enriched string-net model, whose pentagon equation is obstructed by $\Theta\nu_4$, where $\Theta$ is a phase factor induced by fermion creation and annihilation operators from the bulk 3+1D fSPT phase, defined in Eq.~(\ref{eq.omega}). $\Theta$ appears as long as there is fermionic 't Hooft anomaly, i.e., $H^3(G,\Z_2)$, $H^2(G,\Z_2)$ or $H^1(G,\Z_T)$ fermionic 't Hooft anomaly.

\subsection{Organization of this paper}
The rest of the paper is organized as follows: Firstly, we construct the so-called fermionic symmetry-enriched string-net models in Section \ref{sec.fsesn}, generalizing both symmetry-enriched string-net models and fermionic string-net models\cite{2dtopo,gu2014lattice,fSN}, summarized in Subsection \ref{summary}. They are exactly solvable models for 2+1D non-anomalous non-chiral fSET phases. Secondly, we construct the so-called surface fermionic symmetry-enriched string-net models in Section \ref{sec.fthooft}, summarized in Subsection \ref{summary.surf}. They are exactly solvable models for 2+1D non-chiral surface fSET phases with fermionic 't Hooft anomaly, i.e., fSET phases that live on the surface of bulk 3+1D fSPT phases. Since only non-chiral surface states can be realized by our surface model, our approach can only cover the $H^3(G,\Z_2)$ and $H^2(G,\Z_2)$ fermionic 't Hooft anomalies. We conclude that $H^3(G,\Z_2)$ fermionic 't Hooft anomaly corresponds to a violation of fermion-parity conservation in certain surface $\cF$-moves defined in Subsection \ref{subsec.fa}, and also a $\Theta$ obstruction in the fermionic pentagon equation. Finally, we discuss the mathematical framework that arises from our lattice model construction for fSET phases in Section \ref{sec.math}. Further, in Appendix \ref{appen.sfc}, we review basic definitions of super fusion category, super modular category and spin modular category, which are broadly used in this paper. In Appendix \ref{appen.anyonbset}, we review the categorical description of the anyons under symmetry action (or $G$-action) in bosonic SET phases\cite{BSET}. Given a unitary modular tensor category $\cM$ that describes anyons, and a symmetry $G$, there are three pieces of data: $G$-action on $\cM$, symmetry fractionalization, and gauging $G$ on $\cM$ (i.e., a $G$-crossed extension followed by a $G$-equivariantization on $\cM$). In Appendix \ref{appen.relating}, we review the relation between symmetry-enriched string-net model data and anyon symmetry fractionalization data for bosonic/fermionic SET phases whose intrinsic topological orders are Abelian gauge theories\cite{Cheng2017}, by which we can show that the $o_3$ obstruction on anyon level exactly matches with the $n_3$ obstruction in lattice model level for an example in Remark. \ref{rmk.ego3n3}. In Appendix \ref{appen.h2h3}, we review the definitions of obstruction functions $o_2\in H^2(G,\Z_2)$ and $o_3\in H^3(G,\Z_2)$\cite{bulmash2022fermionic,bulmash2022anomaly}, which are the obstructions to extending $G$-action and symmetry fractionalization from the super modular category (that describes the anyon excitations) to its $\Z_2^f$-modular extended category (as a spin modular category) respectively. In Appendix \ref{eg.Z4anyon}, we review the anyon data of surface $\cZ_1(\textbf{Vec}_{\mathbb{Z}_4})\boxtimes\{\mathbb{1},f\}$ topological order with $\mathbb{Z}_2^f\times \mathbb{Z}_2\times \mathbb{Z}_4$ symmetry\cite{Fid2018}, for which we construct its surface fermionic symmetry-enriched string-net model in Subsection \ref{subsec.eg}. In Appendix \ref{appen.part}, we use another approach called partition function approach to derive the explicit expressions of bulk $\widetilde{\cF}$-move and boundary $\cF$-move for surface fSET phases with $H^3(G,\Z_2)$ fermionic 't Hooft anomaly, and we obtain exactly the same expressions as Eq.~(\ref{eq.bulkfmove}) and Eq.~(\ref{eq.surfacefmove}). In Appendix \ref{Appendix.cochain}, we list canonical choices of the cochains in 3+1D fSPT phases, which will be used when we extend our exactly solvable model construction to $H^2(G,\Z_2)$ fermionic 't Hooft anomaly in the future.

\section{Fermionic symmetry-enriched string-net models}\label{sec.fsesn}

\subsection{Fermionic global symmetry}\label{subsec.gf}
Denote the fermion-parity symmetry group as $\mathbb{Z}_2^{f}=\{1,P_f=(-1)^{N_f}\}$, where $N_f$ is the total fermion number operator. The fermionic symmetry group $G^f$ in general can be any central extension of a bosonic symmetry group $G$ by $\mathbb{Z}_2^{f}$, written in a short exact sequence as
\begin{align}
    1\to \Z_2^f\to G^f \to G \to 1,
\end{align}
which is specified by a 2-cocycle $\omega_2 \in Z^2(G,\mathbb{Z}_2)$, where $G$ is a discrete finite group. We simply denote $G^f$ as
\begin{align}
G^f=
\mathbb{Z}_2^{f}
\times_{\omega_2}
G.
\end{align}
Denote a group element in $G^f$ by $(x,\Bg)$, where $x\in\Z_2^f$ and $\Bg\in G$. Group multiplication in $G^f$ is given by
\begin{align}
    (x,\Bg)(y,\Bh)=(x+y+\omega_2(\Bg,\Bh),\Bg\Bh).
\end{align}
And there might be time-reversal symmetry in $G$, which is anti-unitary. Therefore, we define a $\mathbb{Z}_2$-grading on $G$: for any $\textbf{g}\in G$,
\begin{equation}
S(\textbf{g})=
\left\{\begin{array}{l}
1
\text{, \ \ if }\textbf{g}\text{ is unitary}\\ 
*
\text{, \ \ if }\textbf{g}\text{ is anti-unitary}
\end{array}\right..
\end{equation}
Denote the global symmetry transformation of $\textbf{g}$ as $U(\textbf{g})$. $U$ forms a linear representation of $G^f$, and therefore it forms a projective representation of $G$:
\begin{align}
    U(\Bg)U(\Bh)=P_f^{\omega_2(\Bg,\Bh)}U(\Bg\Bh).
\end{align}
We note that changing $\omega_2\to\omega_2+\dd\phi_1$, where $\phi_1\in C^1(G,\Z_2)$ is a 1-cochain and $\dd\phi_1$ is a 2-coboundary, corresponds to changing the form of symmetry transformation by
\begin{align}
    U(\Bg)\to P_f^{\phi_1(\Bg)}U(\Bg).
\end{align}
Therefore, defining the symmetry transformation of $G^f$ needs to specify a particular cocycle $\omega_2$, but not the cohomology class $[\omega_2]$\cite{bulmash2022fermionic}.

\subsection{\texorpdfstring{$G$}{}-graded super fusion category}
\label{subsec.gsuperfusion}
To define a fermionic symmetry-enriched string-net model, we need to input a $G$-graded super fusion category, denoted as $\cS_G$. Mathematically, the general definition of $G$-graded super fusion category is still unknown. However, we can still say that $\cS_G$ must have the following two properties:
\begin{itemize}
    \item $\cS_G$ is a super fusion category (Definition \ref{def.superfusion}), i.e., it is a unitary semisimple rigid monoidal category that is enriched over $\svec$ (Definition \ref{def.supervector}), where enriching over $\svec$ means that hom-spaces in $\cS_G$ are super vector spaces.
    \item $\cS_G$ has a $G$-grading structure, i.e., $\cS_G=\underset{\Bg\in G}{\oplus}\cS_\Bg$ and $\forall a_\Bg \in\cS_\Bg, b_\Bh \in\cS_\Bh, a_\Bg\ot b_\Bh\in \cS_{\Bg\Bh}$. Here $\cS_\mathbf{1}$ is always a super fusion category, where $\one$ is the trivial element in $G$.
\end{itemize}
In particular, when $\omega_2$ is trivial ($\omega_2$ is introduced in Subsection \ref{subsec.gf}), $\cS_G$ can always be obtained by a fermion condensation (or a canonical construction of enriched monoidal category over $\svec$) from a $G$-graded unitary fusion category $\cC_G$\cite{etingof2010fusion,gelaki2009centers}, i.e., $\cS_G:={^{\svec}\cC_G}$, explicitly defined in Subsection \ref{subsec.gsuper}. But when $\omega_2$ is nontrivial, we do not know how to define the $G$-graded super fusion category yet. Nonetheless, the two properties listed above are enough for us to define a fermionic symmetry-enriched string-net model. We expect that our physical model brings out the constraints in defining the $G$-graded super fusion category when $\omega_2$ is nontrivial.

\subsection{Degrees of freedom}\label{subsection.fusion}


Given a \text{finite} fermionic global symmetry group $G^f=
\mathbb{Z}_2^{f}
\times_{\omega_2}
G$ and a $G$-graded super fusion category $\cS_G=\underset{\Bg\in G}{\oplus}\cS_\Bg$ (though the definition of $G$-graded super fusion category is unknown, we only make use of the two properties listed in Subsection \ref{subsec.gsuperfusion}), a fermionic symmetry-enriched string-net model on a honeycomb lattice consists of the following degrees of freedom:
\begin{enumerate}
\item $\left\vert G \right\vert$-level group element labels $\left\vert \textbf{g}_i \right\rangle$ in each plaquette, where $\textbf{g}_i\in G$, $i\in\mathbb{N}$. Graphically,
\begin{equation}\label{graph.plaquette}
    \includegraphics[scale=.4]{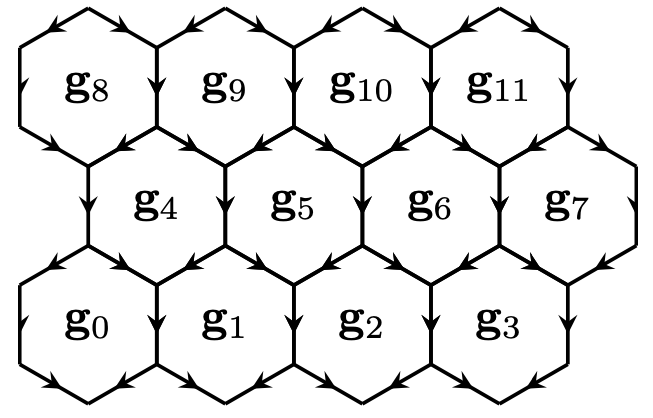},
\end{equation}
where the arrows on links represent the branching structure (arrows are always 2-in-1-out or 1-in-2-out on each vertex). Recall that $U(\textbf{g}_0)$ denotes for the global symmetry transformation of $\textbf{g}_0$. We choose the following convention:
\begin{align}
    U(\textbf{g}_0)\left\vert \textbf{g}_i \right\rangle
    =
    \left\vert \textbf{g}_0\textbf{g}_i \right\rangle.
\end{align}
The Hilbert space on plaquettes is 
\begin{align}\label{eq.hp}
    \cH_P=\bigotimes_{p\in P}\C[G],
\end{align}
where $P$ is the set of all plaquettes in the lattice, $p$ denotes for each plaquette, and $\C[G]$ is the Hilbert space of linear combinations of group elements in $G$ with coefficient in $\C$.

\item $G$-graded string types $a_{\textbf{g}},b_{\textbf{h}},...\in \cS_G$ on each link, where $a_\Bg \in\cS_\Bg$, $b_\Bh \in\cS_\Bh$, ... (A string type corresponds to a simple object in $\cS_G$. So that without causing any ambiguity, we may use the same notation $\cS_G$ to denote for the set of isomorphism classes of simple objects in $\cS_G$). Moreover, the sector $\cS_\one$, graded by the trivial group element $\one$, is always a super fusion category, and $\mathbb{1},a,b,...\in\cS_\mathbf{1}$, where $\mathbb{1}$ is the tensor unit in $\cS_\one$. Each $G$-graded string type $a_{\textbf{g}}$ represents a symmetry defect (or symmetry domain wall) between two plaquettes:
\begin{equation}
    \includegraphics[scale=.2]{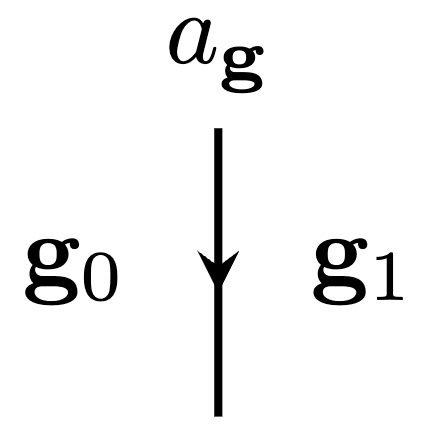},
\end{equation}
where $\Bg\equiv\Bg_0^{-1}\Bg_1$. 

Since hom-spaces in a $G$-graded super fusion category are in general super vector spaces (see Definition \ref{def.supervector}), each string type in $\cS_G$ can be either m-type or q-type \cite{fSN,fc}. A string type $a_\Bg$ is called \textit{m-type} if $\text{End}(a_\Bg):=\text{Hom}(a_\Bg,a_\Bg)=\C$; and is called \textit{q-type} if $\text{End}(a_\Bg)=\C^{1|1}$. We assign a number $n_{a_{\textbf{g}}}$ to each $a_{\textbf{g}}\in \cS_G$:
\begin{align}
    n_{a_{\textbf{g}}}
    :=\text{dim End}(a_{\textbf{g}})
    =
    \left\{\begin{array}{l}
    1
    \text{, if }a_{\textbf{g}}\text{ is m-type}\\ 
    2
    \text{, if }a_{\Bg}\text{ is q-type}
    \end{array}\right.,
\end{align}
which is the dimension of the endomorphism space of $a_\Bg$. 

The quantum dimension of $a_\Bg$ is denoted as $d_{a_\Bg}$, which can be defined through fusion multiplicity around Eq.~(\ref{eq.defqd}) or through a certain $F$-move as in Ref.~\onlinecite{Kitaev2006anyons}. Then the total quantum dimension for each $\cS_\Bg$-sector is defined as \begin{align}
    D^2_\Bg=\sum_{a_\Bg\in\cS_\Bg}\frac{d^2_{a_\Bg}}{n_{a_\Bg}}.
\end{align}
We require that for any $\Bg\in G$, the total quantum dimensions of all $\Bg$-sectors are equal, i.e.,
\begin{align}\label{eq.totalquantumdim}
    D^2_\one=D^2_\Bg, \ \ \forall \Bg\in G.
\end{align}
 
\item Fusion states $\alpha,\beta,...$ in fusion space $\text{Hom}(a_\Bg\ot b_\Bh,c_{\Bg\Bh})$ or $\text{Hom}(c_{\Bg\Bh},a_\Bg\ot b_\Bh)$ on each vertex, graphically represented as 
\begin{equation}\label{graph.fusion}
    \includegraphics[scale=.42]{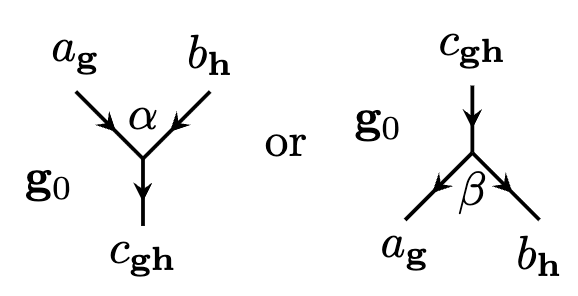}.
\end{equation}
Below let us consider only fusion space $\text{Hom}(a_\Bg\ot b_\Bh,c_{\Bg\Bh})$, and the property of $\text{Hom}(c_{\Bg\Bh},a_\Bg\ot b_\Bh)$ is similar. The hom-space $\text{Hom}(a_\Bg\ot b_\Bh,c_{\Bg\Bh})\in\svec$ is a super vector space, i.e.,
\begin{align}
    &\text{Hom}(a_\Bg\ot b_\Bh,c_{\Bg\Bh})
    \nonumber\\
    =&\text{Hom}(a_\Bg\ot b_\Bh,c_{\Bg\Bh})_0\oplus \text{Hom}(a_\Bg\ot b_\Bh,c_{\Bg\Bh})_1,
\end{align}
where $\text{Hom}(a_\Bg\ot b_\Bh,c_{\Bg\Bh})_0$ is the grade-0 (parity-even) part, $\text{Hom}(a_\Bg\ot b_\Bh,c_{\Bg\Bh})_1$ is the grade-1 (parity-odd) part. We define the so-called \textit{fusion multiplicity} by
\begin{align}
    N^{a_{\textbf{g}}b_{\textbf{h}}}
    _{c_{\textbf{g}\textbf{h}}}
    :=\text{dim Hom}(a_\Bg\ot b_\Bh,c_{\Bg\Bh}).
\end{align}
Different fusion states are labeled by $\alpha,\beta,...=1,...,N^{a_{\textbf{g}}b_{\textbf{h}}}_{c_{\textbf{g}\textbf{h}}}$. In other words, basis in $\text{Hom}(a_\Bg\ot b_\Bh,c_{\Bg\Bh})$ are $\{|\alpha\rangle|
\alpha=1,...,N^{a_{\textbf{g}}b_{\textbf{h}}}_{c_{\textbf{g}\textbf{h}}}\}$. Then we also define bosonic and fermionic fusion multiplicities respectively:
\begin{align}
    B^{a_{\textbf{g}}b_{\textbf{h}}}
    _{c_{\textbf{g}\textbf{h}}}:=
    \text{dim Hom}(a_\Bg\ot b_\Bh,c_{\Bg\Bh})_0, 
    \nonumber\\
    F^{a_{\textbf{g}}b_{\textbf{h}}}
    _{c_{\textbf{g}\textbf{h}}}:=
    \text{dim Hom}(a_\Bg\ot b_\Bh,c_{\Bg\Bh})_1,
\end{align}
and obviously,
\begin{align}\label{eq.n1}
    N^{a_{\textbf{g}}b_{\textbf{h}}}
    _{c_{\textbf{g}\textbf{h}}}
    =B^{a_{\textbf{g}}b_{\textbf{h}}}
    _{c_{\textbf{g}\textbf{h}}}+
    F^{a_{\textbf{g}}b_{\textbf{h}}}
    _{c_{\textbf{g}\textbf{h}}}.
\end{align}
Therefore, the string fusion rules are written as
\begin{align}\label{eq.stringfusion}
    a_{\textbf{g}}\otimes b_{\textbf{h}}
    =
    \underset{{c_{\textbf{g}\textbf{h}}\in \mathcal{S}_\mathbf{gh}}}{\bigoplus}
    \C^{B^{a_{\textbf{g}}b_{\textbf{h}}}
    _{c_{\textbf{g}\textbf{h}}}|
    F^{a_{\textbf{g}}b_{\textbf{h}}}
    _{c_{\textbf{g}\textbf{h}}}}
    c_{\textbf{g}\textbf{h}}.
\end{align} 
We define the \textit{quantum dimension} $d_{a_\Bg}$ of $a_\Bg$ as the largest eigenvalue of matrix $\hat{\eta}_{a_\Bg}$, where
\begin{align}\label{eq.defqd}
    \hat{\eta}_{a_\Bg}:=
    \begin{pmatrix}
    \frac{N^{a_\Bg b_\Bh}_{c_{\Bg\Bh}}}{n_{c_{\Bg\Bh}}}: b_\Bh,c_{\Bg\Bh}\in\cS_G
    \end{pmatrix}.
\end{align}
The Hilbert space of fusion spaces on vertices is
\begin{align}\label{eq.hv}
    \cH_V=\underset{v\in V}{\bigotimes}\cH^{\pm}_v,
\end{align}
where $V$ is the set of all vertices in the lattice, $v$ denotes for each vertex, and 
\begin{align}
    \cH^+_v:=\bigoplus_{a_\Bg,b_\Bh,c_{\Bg\Bh}\in\cS_G}
    \text{Hom}(a_\Bg\ot b_\Bh,c_{\Bg\Bh}),
    \nonumber\\
    \cH^-_v:=\bigoplus_{a_\Bg,b_\Bh,c_{\Bg\Bh}\in\cS_G}
    \text{Hom}(c_{\Bg\Bh},a_\Bg\ot b_\Bh).
\end{align}

\item $\left\vert G \right\vert$ species of physical fermions on each vertex. Labeling the corresponding vertices of fusion states $\alpha,\beta,...$ by $\underline{\alpha},\underline{\beta},...$, the annihilation and creation operators of species-$\Bg$ fermion are denoted as $c^\textbf{g}_{\underline{\alpha}}$ and $c^{\textbf{g}\dagger}_{\underline{\alpha}}$. They satisfy the anti-commutation relation:
\begin{align}
    c^{\Bg}_{\underline{\alpha}}
    c^{\Bh}_{\underline{\beta}}=
    -c^{\Bh}_{\underline{\beta}}
    c^{\Bg}_{\underline{\alpha}},\ \ 
    c^{\Bg}_{\underline{\alpha}}
    c^{\Bh\dagger}_{\underline{\beta}}=
    -c^{\Bh\dagger}_{\underline{\beta}}
    c^{\Bg}_{\underline{\alpha}},
\end{align}
for all $\underline{\alpha}\neq \underline{\beta}$, $\Bg,\Bh\in G$. A global symmetry transformation of $\Bg_0$ acts on $c^\textbf{g}_{\underline{\alpha}}$ as
\begin{align}\label{eq.symonfermion}
    U(\textbf{g}_0)c^\textbf{g}
    _{\underline{\alpha}}U^\dagger(\textbf{g}_0)
    =
    (-1)^{\omega_2( \textbf{g}_0,\textbf{g})}
    c^{\textbf{g}_0\textbf{g}}_{\underline{\alpha}}.
\end{align}

We define a \textit{fermion-parity function} $s(\alpha)=0,1$ to indicate the parity of a fusion state $\alpha$ is even or odd, i.e., 
\begin{equation}
    s(\alpha)
    =
    \left\{\begin{array}{l}
    0
    \text{, \ if }|\alpha\rangle \in\text{Hom}(a_\Bg\ot b_\Bh,c_{\Bg\Bh})_0\\ 
    1
    \text{, \ if }|\alpha\rangle\in\text{Hom}(a_\Bg\ot b_\Bh,c_{\Bg\Bh})_1
    \end{array}\right..
\end{equation}
Then we decorate fermions on all parity-odd fusion states by a sequence of fermion creation operators. When $s(\alpha)=1$, we choose the convention that the decorated fermion species depends on the group element in the left-most plaquette, graphically represented as
\begin{equation}\label{eq.fermionconven}
    \includegraphics[scale=.42]{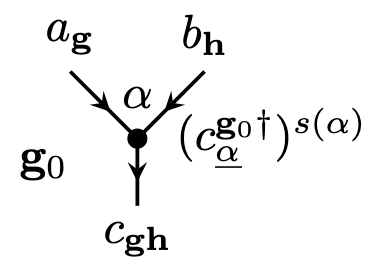},
\end{equation}
where the black dot denotes for the physical fermion when $s(\alpha)=1$.

The fermionic Hilbert space is the following Fock space:
\begin{align}\label{eq.hf}
    \cH_f={\bigoplus_{S\subset (V\times G)}}
    \underset{(v,\Bg)\in S}{\prod}  c_v^{\Bg\dagger}
    |0\rangle,
\end{align}
where $|0\rangle$ is the state of no fermion, and $S$ is any subset of the Cartesian product of set of all vertices $V$ and the group $G$.
\end{enumerate}

In conclusion, the total Hilbert space is
\begin{align}
    \cH=\cH_P\ot\cH_V\ot\cH_f,
\end{align}
where $\cH_P,\cH_V,\cH_f$ are defined in Eq.~(\ref{eq.hp}), (\ref{eq.hv}) and (\ref{eq.hf}) respectively. 

\begin{remark}
\label{rmk.q-type}
A q-type string behaves like a Majorana/Kitaev chain in  topological phases\cite{fc} (but more generalized in a way that q-type strings may not be $\Z_2$-conserved, while Kitaev chains are $\Z_2$-conserved as the fusion outcomes of two Kitaev chains is vacuum). Therefore, q-type strings have the following two properties: (1) Any q-type string configuration must be closed, or say, q-type string cannot terminate or emerge at any vertex in the lattice model. (2) Fermions can be created or annihilated by pairs on q-type strings, and a fermion can ``slide" along q-type strings freely without energy cost. Accordingly, when there is q-type string involved in the fusion vertex in Graph (\ref{graph.fusion}), i.e., at least one of $a_{\textbf{g}},b_{\textbf{h}}$ is q-type, we must have
${B^{a_{\textbf{g}}b_{\textbf{h}}}_{c_{\textbf{g}\textbf{h}}}}={F^{a_{\textbf{g}}b_{\textbf{h}}}_{c_{\textbf{g}\textbf{h}}}}$. In other words, it is guaranteed that for any bosonic fusion state, there always exists a corresponding fermionic fusion state. Therefore, for a particular fusion state $\alpha$ in $\text{Hom}(a_\Bg\ot b_\Bh,c_{\Bg\Bh})$, in such cases that there is q-type string involved, we introduce the notation $\alpha\times f$ (intuitively understood as sliding a fermion $f$ onto $\alpha$) without any ambiguity to denote for the corresponding fermionic state if $\alpha$ is bosonic, and the corresponding bosonic state if $\alpha$ is fermionic. If we choose the convention that all bosonic fusion states are ordered before fermionic fusion states, i.e., $\alpha=1,...,B^{a_{\textbf{g}}b_{\textbf{h}}}_{c_{\textbf{g}\textbf{h}}},B^{a_{\textbf{g}}b_{\textbf{h}}}_{c_{\textbf{g}\textbf{h}}}+1,...,2B^{a_{\textbf{g}}b_{\textbf{h}}}_{c_{\textbf{g}\textbf{h}}}$, then $\alpha\times f=\alpha+B^{a_{\textbf{g}}b_{\textbf{h}}}_{c_{\textbf{g}\textbf{h}}}$.
\end{remark}

\begin{remark}
    The assignment of $\lvert G\rvert$ species of physical fermions, together with Eq.~(\ref{eq.symonfermion}), is actually a generalization of the Kramers doublet. For example, let us consider a total symmetry $\Z_4^{Tf}=\Z_2^{f}\times_{\omega_2}\Z_2^{T}$, where $\Z_2^{T}=\{\one,T\}$ is the time-reversal symmetry and $\omega_2(T,T)=1$. Denote the Kramer‘s doublet as $c^\uparrow$ and $c^\downarrow$, the time-reversal symmetry acts as $U(T)c^\uparrow U(T)^\dagger= c^\downarrow$, $U(T)c^\downarrow U(T)^\dagger= -c^\uparrow$.
\end{remark}

\begin{definition}\label{def.sigmatype}
    A q-type string $a_\Bg$ is further called \textit{$\sigma$-type or Ising type} if it satisfies the $\Z_2$-conservation law as q-type string on fusion:
    \begin{align}\label{eq.sigmatype}
        a_\Bg\otimes a_\Bg=\underset{{c_{\textbf{g}\textbf{g}}\in \mathcal{S}_\mathbf{gg}}}{\oplus}
        \C^{n|n}
        c_{\textbf{g}\textbf{g}},
    \end{align}
    where all $c_{\textbf{g}\textbf{g}}$ are m-type strings and $n:={B^{a_{\textbf{g}}b_{\textbf{h}}}_{c_{\textbf{g}\textbf{h}}}}={F^{a_{\textbf{g}}b_{\textbf{h}}}_{c_{\textbf{g}\textbf{h}}}}$.
\end{definition}


\subsection{Generalized fermionic symmetric local unitary transformation and fixed-point wavefunction}

We define the following equivalence relation between two gapped fermionic quantum states:
\begin{align}
    |\Phi\ra \sim |\Phi'\ra \text{ iff. }
    |\Phi'\ra=\mathcal{T}[e^{i\int d\lambda H_f(\lambda)}]|\Phi\ra,
\end{align}
where the evolution does not close the energy gap, $\lambda$ is the order parameter and $\mathcal{T}$ is the path-ordering operator. $H_f(\lambda)=\sum_i \cO_i(\lambda)$ is the fermionic Hamiltonian, where $i$ denotes for each local region and $\cO_i(\lambda)$ is a product of even number of local fermionic operators and any number of local bosonic operators. Here $\mathcal{T}[e^{i\int d\lambda H_f(\lambda)}]$ is called a fermionic local unitary (fLU) transformation\cite{2dtopo}. We note that a bosonic local unitary transformation is a special case of fLU transformation when there is no local fermionic operator involved. 

However, requiring the transformation to be unitary is not general enough. We hereby define the so-called generalized fermionic local unitary (gfLU) transformation\cite{2dtopo,fSN}. It generalizes the concept of fLU transformation in a way that the transformation can be projective-unitary. Denote a gfLU transformation as $\tilde{U}$. $\tilde{U}$ is called \textit{projective-unitary} if
\begin{align}\label{eq.projuni}
    \tilde{U}^\dagger \tilde{U}=\tilde{P},\ \ 
    \tilde{U}\tilde{U}^\dagger=\tilde{P}',
\end{align}
where $\tilde{P}$ and $\tilde{P}'$ are two projectors, i.e., $\tilde{P}^2=\tilde{P}$ and $\tilde{P}'^2=\tilde{P}'$.

A generalized fermionic symmetric local unitary (gfSLU) transformation is a gfLU transformation that is invariant under any global symmetry action. In classifying 2+1D fSET phases, the wavefunction renormalizations are generated by gfSLU transformations. Then the equivalence classes of wavefunctions are defined up to gfSLU transformations, which leads to the concept of fixed-point wavefunction (as a characteristic of each equivalence class).

Recall that by a string-net model\cite{levin2005string,chen2010local}, we mean inputting a set of categorical data (for original string-net model, the input is a unitary fusion category; for the fermionic symmetry-enriched string-net model we define here, the mathematical definition of the input category is only partially known), a quantum lattice model with a full description on fixed-point ground state wavefunctions and Hamiltonian is outputted. Let $\Psi(-)$ denote for the fixed-point ground state wavefunction on a certain patch. The topological ground state (defined on a sphere) is a superposition of all possible string-net configurations:
\begin{align}
|\Psi\rangle=\sum_\text{all conf.}
\Psi\left(
\vcenter{\hbox{\includegraphics[scale=.3]{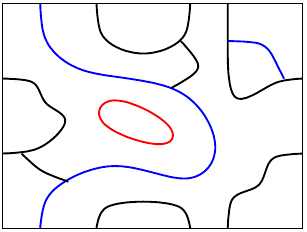}}}
\right)
\left
|\vcenter{\hbox{\includegraphics[scale=.3]{strnet}}}
\right \rangle.
\end{align}

\begin{widetext}
There exists a gfSLU transformation as a gauge transformation in our model:
\begin{equation}\label{eq.gauge}
    \includegraphics[scale=.42]{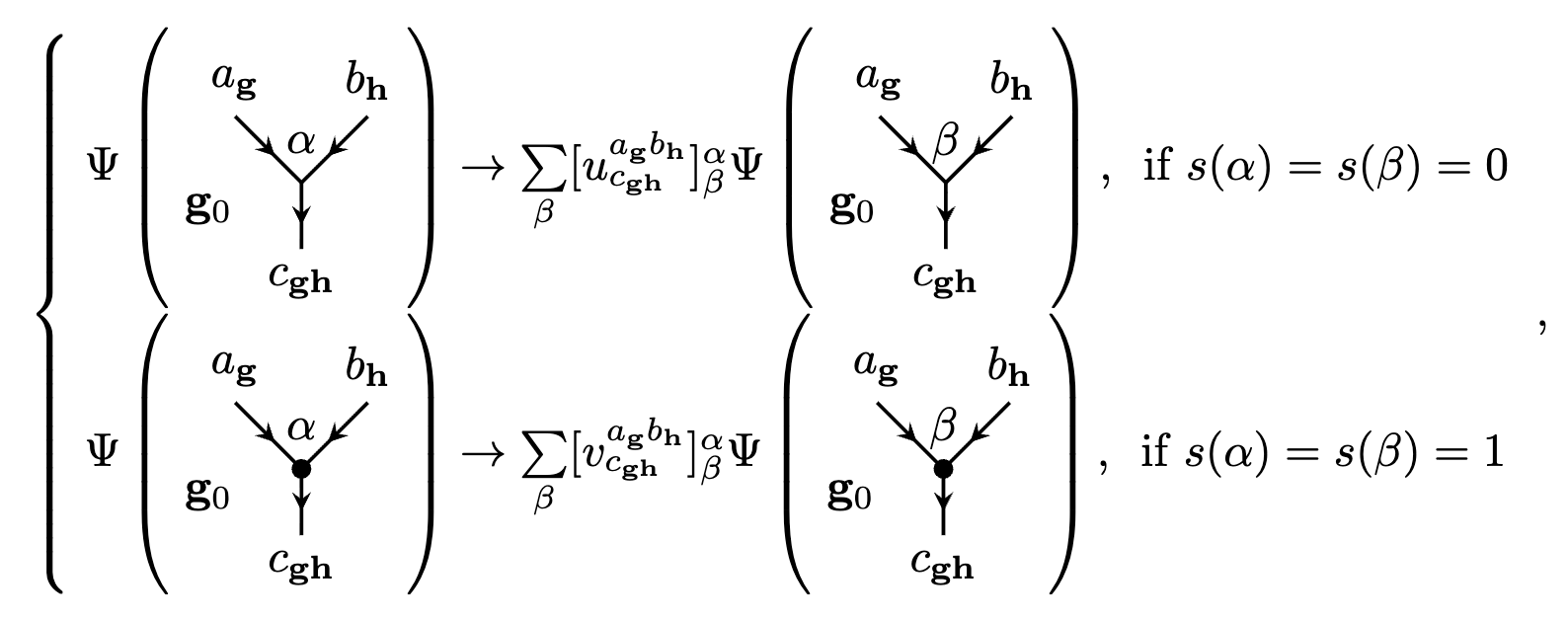}
\end{equation}
where $u^{a_\textbf{g}b_\textbf{h}}_{c_\textbf{gh}}$ and $v^{a_\textbf{g}b_\textbf{h}}_{c_\textbf{gh}}$ are two unitary matrices. This corresponds to a unitary transformation in the super vector space $\text{Hom}(a_\Bg\ot b_\Bh,c_{\Bg\Bh})$. When $s(\alpha)=s(\beta)=0$, $u^{a_\textbf{g}b_\textbf{h}}_{c_\textbf{gh}}$ is a unitary basis transformation in $\text{Hom}(a_\Bg\ot b_\Bh,c_{\Bg\Bh})_0$. When $s(\alpha)=s(\beta)=1$, $v^{a_\textbf{g}b_\textbf{h}}_{c_\textbf{gh}}$ is a unitary basis transformation in $\text{Hom}(a_\Bg\ot b_\Bh,c_{\Bg\Bh})_1$.

\subsection{Renormalization moves}\label{subsec.renorm}
The first type of renormalization move, called \textit{$\mathcal{F}$-move}, is a gfSLU transformation defined as
\begin{equation}\label{f1}
    \includegraphics[scale=.42]{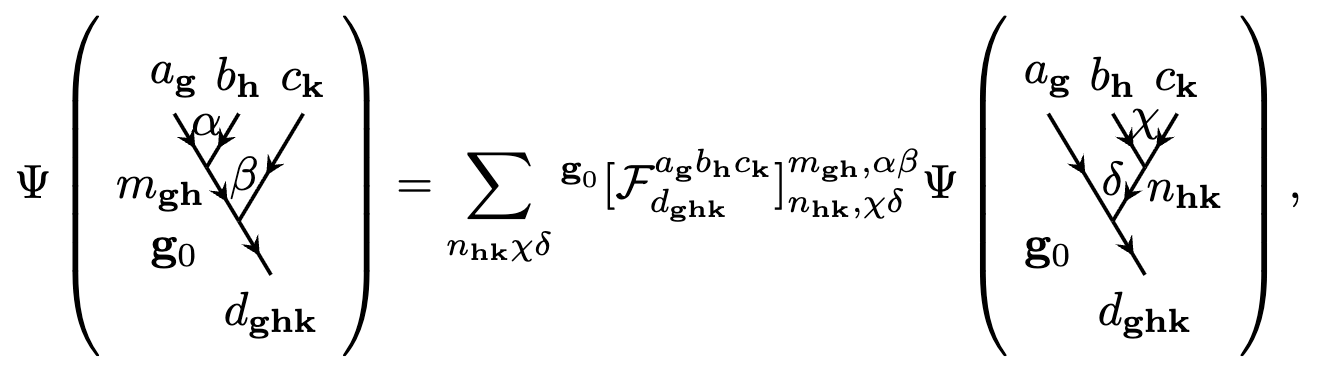}
\end{equation}
where the $\mathcal{F}$-move can be more simply denoted as $^{\textbf{g}_0}[{\mathcal{F}^{a_{\textbf{g}}b_{\textbf{h}}c_{\textbf{k}}}
_{d}]
^{m,\alpha\beta}
_{n,\chi\delta}}$. It is a combination of fermion creation and annihilation operators and a bosonic-part $F$-move (mathematically corresponds to the $F$-symbol or associator in the input $G$-graded super fusion category):
\begin{align}
^{\textbf{g}_0}[{\mathcal{F}^{a_{\textbf{g}}b_{\textbf{h}}c_{\textbf{k}}}
_{d}]^{m,\alpha\beta}_{n,\chi\delta}}
=
(c^{\textbf{g}_0 \dagger}_{\underline{\alpha}})^{s(\alpha)}
(c^{\textbf{g}_0 \dagger}_{\underline{\beta}})^{s(\beta)}
(c^{\textbf{g}_0}_{\underline{\delta}})^{s(\delta)}
{(c^{\textbf{g}_0\textbf{g}}_{\underline{\chi}})^{s(\chi)}}
\cdot
{^{\Bg_0}[{F^{a_\Bg b_\Bh c_\Bk}
_{d}]^{m,\alpha\beta}_{n,\chi\delta}}},
\label{ffmove}
\end{align}
where a gfSLU transformation should preserve fermion-parity, i.e., 
\begin{align}\label{eq.s}
    s(\alpha)+s(\beta)+s(\delta)+s(\chi)
    =0\text{ (mod 2).}
\end{align}
Associativity of fusion of strings implies
\begin{align}\label{eq.n2}
\sum_{e_{\mb g \mb h}} 
\frac{N^{a_{\mb g}b_{\mb h}}_{e_{\mb g\mb h}} N^{e_{\mb g \mb h}c_{\mb k}}_{d_{\mb g \mb h \mb k}}}{n_{e_{\mb g \mb h}}}
=
\sum_{f_{\mb h \mb k}} 
\frac{N^{b_{\mb h }c_{\mb k}}_{f_{\mb h \mb k}}N^{a_{\mb g}f_{\mb h \mb k}}_{d_{\mb g \mb h \mb k}}}{n_{f_{\mb h \mb k}}}
.
\end{align}
Since the fermion-parity-odd sector and the parity-even sector are independent (due to fermion-parity conservation), this relation can be further split as

\begin{align}\label{eq.n3}
    \sum_{e_{\mb g \mb h}} 
    \frac{B^{a_{\mb g}b_{\mb h}}_{e_{\mb g\mb h}} B^{e_{\mb g \mb h}c_{\mb k}}_{d_{\mb g \mb h \mb k}}+F^{a_{\mb g}b_{\mb h}}_{e_{\mb g\mb h}} F^{e_{\mb g \mb h}c_{\mb k}}_{d_{\mb g \mb h \mb k}}}
    {n_{e_{\mb g \mb h}}}
    =
    \sum_{f_{\mb h \mb k}} 
    \frac{B^{b_{\mb h }c_{\mb k}}_{f_{\mb h \mb k}}B^{a_{\mb g}f_{\mb h \mb k}}_{d_{\mb g \mb h \mb k}}+F^{b_{\mb h }c_{\mb k}}_{f_{\mb h \mb k}}F^{a_{\mb g}f_{\mb h \mb k}}_{d_{\mb g \mb h \mb k}}}
    {n_{f_{\mb h \mb k}}},
\end{align}
\begin{align}
    \sum_{e_{\mb g \mb h}} 
    \frac{B^{a_{\mb g}b_{\mb h}}_{e_{\mb g\mb h}} F^{e_{\mb g \mb h}c_{\mb k}}_{d_{\mb g \mb h \mb k}}+F^{a_{\mb g}b_{\mb h}}_{e_{\mb g\mb h}} B^{e_{\mb g \mb h}c_{\mb k}}_{d_{\mb g \mb h \mb k}}}
    {n_{e_{\mb g \mb h}}}
    =
    \sum_{f_{\mb h \mb k}} 
    \frac{B^{b_{\mb h }c_{\mb k}}_{f_{\mb h \mb k}}F^{a_{\mb g}f_{\mb h \mb k}}_{d_{\mb g \mb h \mb k}}+F^{b_{\mb h }c_{\mb k}}_{f_{\mb h \mb k}}B^{a_{\mb g}f_{\mb h \mb k}}_{d_{\mb g \mb h \mb k}}}
    {n_{f_{\mb h \mb k}}}.
\end{align}

\begin{remark}\label{rmk.othermoves}
   There are other types of renormalization moves as gfSLU transformations: the $\cO$-move and the $\cY$-move that change the number of vertices, and the dual $\cF$-move, $\cH$-move and dual $\cH$-move that are closely related to the $\cF$-move. The definitions of these renormalization moves are listed in Appendix \ref{appen.othermove}.
\end{remark}

Derived from Eq.~(\ref{eq.s}), the quantum dimensions further satisfy the following equation:
\begin{align}\label{eq.d}
    d_{a_\Bg}d_{b_\Bh}=
    \sum_{c_{\Bg\Bh}}\frac{N^{a_\Bg b_\Bh}_{c_{\Bg\Bh}}d_{c_{\Bg\Bh}}}{n_{c_{\Bg\Bh}}}
    \text{ or }
    \sum_{a_\Bg b_\Bh}
    \frac{N^{a_\Bg b_\Bh}_{c_{\Bg\Bh}} d_{a_\Bg}d_{b_\Bh}}
    {n_{a_\Bg} n_{b_\Bh}}
    =
    d_{c_{\Bg\Bh}}D^2,
\end{align}
where $D^2=\sum_{\Bg\in G}D^2_\Bg=\sum_{a_\Bg\in\cS_G}\frac{d^2_{a_\Bg}}{n_{a_\Bg}}$, and the above two equations are equivalent (i.e., we can derive either one from another). The proof of either of the above two equations is similar to that in Ref.~\onlinecite{fSN}. We note that the condition on $\cO$-move in Eq.~(\ref{eq.o3}) also recovers Eq.~(\ref{eq.d}), which can be shown by substituting the expression in Eq.~(\ref{eq.o}) into Eq.~(\ref{eq.o3}).

The symmetric condition on $\cF$-move requires the following commutative diagram:
\begin{equation}\label{eq.Fsymmetric}
    \includegraphics[scale=.42]{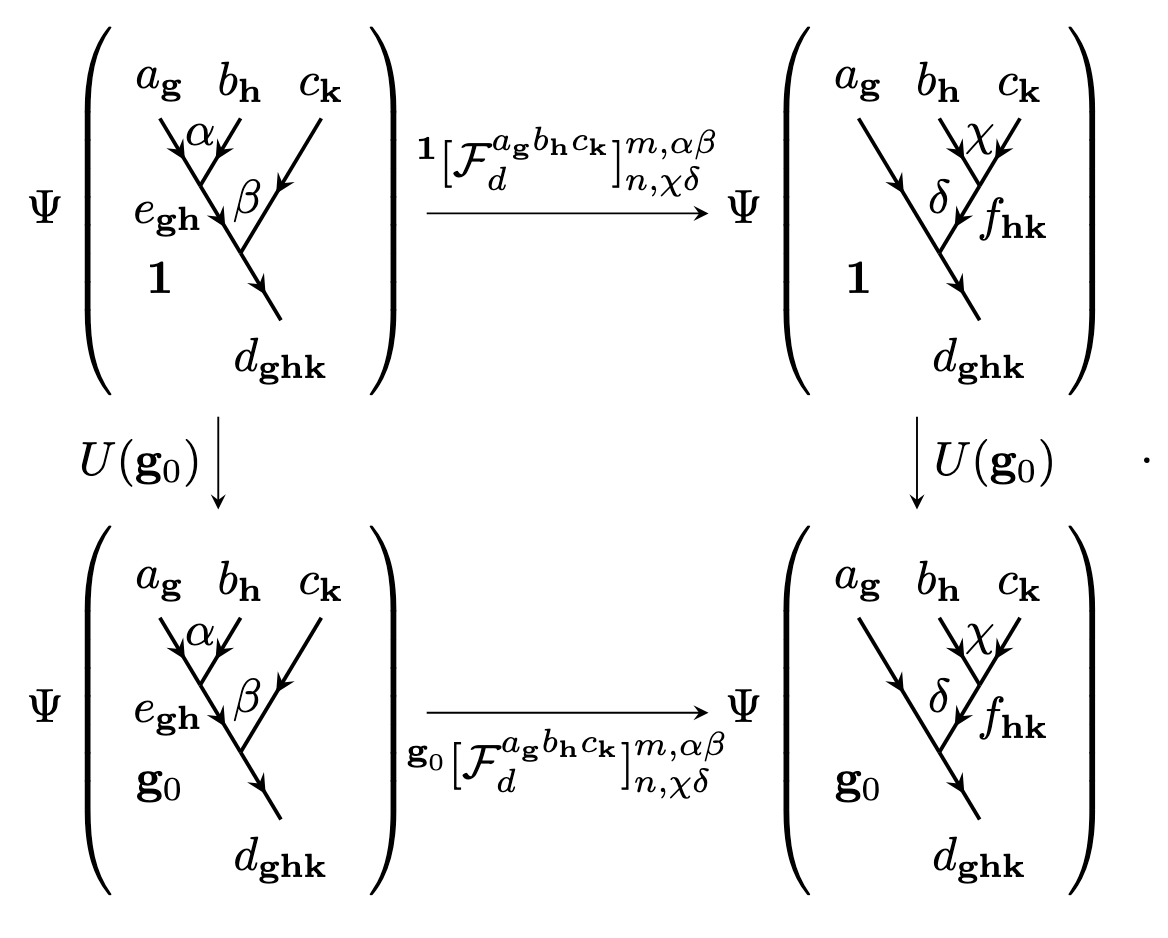}
\end{equation}
It implies
\begin{align}
^{\textbf{g}_0}[{\mathcal{F}^{a_{\textbf{g}}b_{\textbf{h}}c_{\textbf{k}}}_{d}]
^{m,\alpha\beta}
_{n,\chi\delta}}
&=
U(\textbf{g}_0)
{^{\textbf{1}}[{\mathcal{F}^{a_{\textbf{g}}b_{\textbf{h}}c_{\textbf{k}}}_{d}]
^{m,\alpha\beta}
_{n,\chi\delta}}}
U^\dagger(\textbf{g}_0)
\nonumber\\
&=
(-1)^{\omega_2( \textbf{g}_0,\textbf{g})s(\chi)}
(c^{\textbf{g}_0 \dagger}_{\underline{\alpha}})^{s(\alpha)}
(c^{\textbf{g}_0 \dagger}_{\underline{\beta}})^{s(\beta)}
(c^{\textbf{g}_0}_{\underline{\delta}})^{s(\delta)}
{(c^{\textbf{g}_0\textbf{g}}_{\underline{\chi}})^{s(\chi)}}
({^{\textbf{1}}[{F^{a_{\textbf{g}}b_{\textbf{h}}c_{\textbf{k}}}_{d}]^{m,\alpha\beta}
_{n,\chi\delta}}})^{S(\textbf{g}_0)},
\end{align}
where the bosonic-part $F$-move satisfies 
\begin{align}
{^{\mathbf{g}_0}[{F^{a_{\textbf{g}}b_{\textbf{h}}c_{\textbf{k}}}_{d}]^{m,\alpha\beta}
_{n,\chi\delta}}}=
(-1)^{\omega_2( \textbf{g}_0,\textbf{g})s(\chi)}
({^{\textbf{1}}[{F^{a_{\textbf{g}}b_{\textbf{h}}c_{\textbf{k}}}_{d}]^{m,\alpha\beta}
_{n,\chi\delta}}})^{S(\textbf{g}_0)}. 
\label{sc}
\end{align}

The fermionic pentagon equation is written as
\begin{equation}
\underset{\epsilon}{\sum} 
{^{\textbf{1}}[{\mathcal{F}^{j_{\textbf{gh}}c_{\textbf{k}}d_{\textbf{l}}}_{e}]^{m,\beta\chi}_{q,\delta\epsilon}}}
{^{\textbf{1}}[{\mathcal{F}^{a_{\textbf{g}}b_{\textbf{h}}q_{\textbf{kl}}}_{e}]^{j,\alpha\epsilon}_{p,\phi\gamma}}}
=
\underset{n_{\Bh\Bk}\eta\psi\kappa}{\sum} 
{^{\textbf{1}}[{\mathcal{F}^{a_{\textbf{g}}b_{\textbf{h}}c_{\textbf{k}}}_{m}]^{j,\alpha\beta}_{n,\eta\psi}}}
{^{\textbf{1}}[{\mathcal{F}^{a_{\textbf{g}}n_{\textbf{hk}}d_{\textbf{l}}}_{e}]^{m,\psi\chi}_{p,\kappa\gamma}}}
^{\textbf{g}}[{\mathcal{F}^{b_{\textbf{h}}c_{\textbf{k}}d_{\textbf{l}}}_{p}]^{n,\eta\kappa}_{q,\delta\phi}}
.
\label{fpenta1}
\end{equation}
Reordering and canceling the fermion creation and annihilation operators gives a sign factor $(-1)^{s(\alpha)s(\delta)}$ as shown in Ref. \onlinecite{2dtopo}, and by Eq.~(\ref{sc}), 
\begin{equation}
\underset{\epsilon}{\sum} 
{^{\textbf{1}}[{F^{j_{\textbf{gh}}c_{\textbf{k}}d_{\textbf{l}}}_{e}]^{m,\beta\chi}_{q,\delta\epsilon}}}
{^{\textbf{1}}[{F^{a_{\textbf{g}}b_{\textbf{h}}q_{\textbf{kl}}}_{e}]^{j,\alpha\epsilon}_{p,\phi\gamma}}}
=
(-1)^{(s(\alpha)+\omega_2( \textbf{g},\textbf{h}))s(\delta)}
\underset{n_{\Bh\Bk}\eta\psi\kappa}{\sum} 
{^{\textbf{1}}[{F^{a_{\textbf{g}}b_{\textbf{h}}c_{\textbf{k}}}_{m}]^{j,\alpha\beta}_{n,\eta\psi}}}
{^{\textbf{1}}[{F^{a_{\textbf{g}}n_{\textbf{hk}}d_{\textbf{l}}}_{e}]^{m,\psi\chi}_{p,\kappa\gamma}}}
({^{\textbf{1}}[{F^{b_{\textbf{h}}c_{\textbf{k}}d_{\textbf{l}}}_{p}]^{n,\eta\kappa}_{q,\delta\phi}}})^{S(\textbf{g})}
.
\label{fpenta2}
\end{equation}

\begin{remark}\label{rmk.some}
    If $s(\alpha)=\omega_2(\Bg,\Bh)$, Eq.~(\ref{fpenta2}) reduces to bosonic pentagon equation with vanishing sign factor.
\end{remark}

\begin{remark}
    {The most general fermionic pentagon equation is}
    \begin{equation}
    {\underset{\epsilon}{\sum} 
        {^{\Bg_0}[{\mathcal{F}^{j_{\textbf{gh}}c_{\textbf{k}}d_{\textbf{l}}}_{e}]^{m,\beta\chi}_{q,\delta\epsilon}}}
        {^{\Bg_0}[{\mathcal{F}^{a_{\textbf{g}}b_{\textbf{h}}q_{\textbf{kl}}}_{e}]^{j,\alpha\epsilon}_{p,\phi\gamma}}}
        =
        \underset{n_{\Bh\Bk}\eta\psi\kappa}{\sum} 
        {^{\Bg_0}[{\mathcal{F}^{a_{\textbf{g}}b_{\textbf{h}}c_{\textbf{k}}}_{m}]^{j,\alpha\beta}_{n,\eta\psi}}}
        {^{\Bg_0}[{\mathcal{F}^{a_{\textbf{g}}n_{\textbf{hk}}d_{\textbf{l}}}_{e}]^{m,\psi\chi}_{p,\kappa\gamma}}}
        ^{\Bg_0\Bg}[{\mathcal{F}^{b_{\textbf{h}}c_{\textbf{k}}d_{\textbf{l}}}_{p}]^{n,\eta\kappa}_{q,\delta\phi}}.}
    \end{equation}
    {We can check that it reduces to Eq.~(\ref{fpenta2}) after canceling the fermion operators by Eq.~(\ref{sc}), where sign factor is}
    \begin{align}
        {(-1)^{\omega_2(\Bg_0,\Bg\Bh)s(\delta)+
        \omega_2(\Bg_0,\Bg)s(\phi)+\omega_2(\Bg_0,\Bg)s(\eta)+
        \omega_2(\Bg_0,\Bg)s(\kappa)+
        \omega_2(\Bg_0\Bg,\Bh)s(\delta)+s(\alpha)s(\delta)}
        =
        (-1)^{(s(\alpha)+\omega_2( \textbf{g},\textbf{h}))s(\delta)},}
    \end{align}
    {by $s(\phi)+s(\eta)+s(\kappa)=s(\delta)$ (mod 2) and $\omega_2(\Bg_0,\Bg)+\omega_2(\Bg_0,\Bg\Bh)+\omega_2(\Bg_0\Bg,\Bh)=\omega_2(\Bg,\Bh)$ (mod 2).}
\end{remark}

    The $\cF$-move as a gfSLU transformation is further required to satisfy the following two projective-unitary conditions\cite{fSN} (recall Eq.~(\ref{eq.projuni})):
\begin{equation}
    \underset{n_{\Bh\Bk}\chi\delta}{\sum} 
    {^{\Bg_0}[{F^{a_\Bg b_\Bh c_\Bk}
    _{d}]^{m',\alpha'\beta'}_{n,\chi\delta}}}
    ({^{\Bg_0}[{F^{a_\Bg b_\Bh c_\Bk}
    _{d}]^{m,\alpha\beta}_{n,\chi\delta}}})^{*}
    =
    \left\{\begin{array}{l}
    \delta_{m_{\Bg\Bh} m_{\Bg\Bh}'}\delta_{\alpha\alpha'}
    \delta_{\beta\beta'}
    \text{, \ \ if }m_{\Bg\Bh}\text{ is m-type
    }\\ 
    \frac{1}{2} [\delta_{m_{\Bg\Bh} m_{\Bg\Bh}'}\delta_{\alpha\alpha'}\delta_{\beta\beta'}+
    {^{\Bg_0}\Xi^{a_\Bg b_\Bh m,\alpha\beta}_{c_\Bk d}}
    \delta_{m_{\Bg\Bh} m_{\Bg\Bh}'}\delta_{(\alpha\times f)\alpha'}\delta_{(\beta\times f)\beta'}]
    \text{, \ \ if }m_{\Bg\Bh}\text{ is q-type}
    \end{array}\right.;
    \label{eq.fprojuni1}
\end{equation}
\begin{equation}
    \underset{m_{\Bg\Bh}\alpha\beta}{\sum} 
    ({^{\Bg_0}[{F^{a_\Bg b_\Bh c_\Bk}
    _{d}]^{m,\alpha\beta}_{n',\chi'\delta'}}})^{*} 
    {^{\Bg_0}[{F^{a_\Bg b_\Bh c_\Bk}
    _{d}]^{m,\alpha\beta}_{n,\chi\delta}}} 
    =
    \left\{\begin{array}{l}
    \delta_{n_{\Bh\Bk} n_{\Bh\Bk}'}\delta_{\chi\chi'}\delta_{\delta\delta'}
    \text{, \ \ if }n_{\Bh\Bk}\text{ is m-type
    }\\ 
    \frac{1}{2} [\delta_{n_{\Bh\Bk} n_{\Bh\Bk}'}\delta_{\chi\chi'}\delta_{\delta\delta'}
    +
    {^{\Bg_0}\Xi^{a_\Bg b_\Bh}_{c_\Bk d n,\chi\delta}}
    \delta_{n_{\Bh\Bk}n_{\Bh\Bk}'}\delta_{(\chi\times f)\chi'}\delta_{(\delta\times f)\delta'}]
    \text{, \ \ if }n_{\Bh\Bk}\text{ is q-type}
    \end{array}\right.,
    \label{eq.fprojuni2}
\end{equation}
where the notations $\alpha\times f$ and $\beta\times f$ are introduced in Remark \ref{rmk.q-type}, which is well-defined here as $m_{\Bg\Bh}$ is q-type for the case in the second line, and similarly for $\chi\times f$ and $\delta\times f$. Here ${^{\Bg_0}\Xi^{a_\Bg b_\Bh m,\alpha\beta}_{c_\Bk d}}$ and ${^{\Bg_0}\Xi^{a_\Bg b_\Bh}_{c_\Bk d n,\chi\delta}}$ are two phase factors satisfying\cite{fSN}
\begin{align}
    ({^{\Bg_0}\Xi^{a_\Bg b_\Bh m,\alpha\beta}_{c_\Bk d}})^*={^{\Bg_0}\Xi^{a_\Bg b_\Bh m,(\alpha\times f)(\beta\times f)}_{c_\Bk d}}, \ \ 
    ({^{\Bg_0}\Xi^{a_\Bg b_\Bh}_{c_\Bk d n,\chi\delta}})^*={^{\Bg_0}\Xi^{a_\Bg b_\Bh}_{c_\Bk d n,(\chi\times f)(\delta\times f)}}.
\end{align}

The two projective-unitary conditions of dual $\cH$-move implies the following two projective-unitary conditions of $\cF$-move (see Appendix \ref{appen.othermove2}):
\begin{align}\label{eq.hproj1}
    &\sum_{f_{\mb g \mb h \mb k}\chi\delta}
    \frac{d_{f_{\mb g \mb h \mb k}}}{n_{f_{\mb g \mb h \mb k}}} 
    {^{\mb g_0}}[F^{c_{\mb g} e'_{\mb h} b_{\mb k}}_{f}]
    ^{a,\alpha'\delta}
    _{d,\beta'\chi}
    ({^{\mb g_0}}[F^{c_{\mb g} e_{\mb h} b_{\mb k}}_{f}]^{a,\alpha\delta}
    _{d,\beta\chi})^*
    \nonumber\\
    =&
    \frac{d_{a_{\mb g \mb h}} d_{d_{\mb h \mb k}} n_{e_{\mb h}}}{n_{a_{\mb g \mb h}} n_{d_{\mb h \mb k}} d_{e_{\mb h}}}
    \left\{\begin{array}{l}
    \delta_{e_\Bh e_\Bh'}\delta_{\alpha\alpha'}
    \delta_{\beta\beta'}
    \text{, \ \ if }e_\Bh\text{ is m-type
    }\\ 
    \frac{1}{2} (\delta_{e_\Bh e_\Bh'}\delta_{\alpha\alpha'}\delta_{\beta\beta'}+
    {^{\Bg_0}\Xi^{c_\Bg s}_{t a e_\Bh,\eta\alpha}}
    (^{\Bg_0}\Xi^{st e_\Bh,\eta\beta}_{b_\Bk d})^*
    \delta_{e_\Bh e_\Bh'}\delta_{(\alpha\times f)\alpha'}\delta_{(\beta\times f)\beta'})
    \text{, \ \ if }e_\Bh\text{ is q-type}
    \end{array}\right.,
\end{align}
where $s_{\Bg'}$, $t_{\Bg'^{-1}\Bh}$ and the fusion state $\eta:s_{\Bg'}\ot t_{\Bg'^{-1}\Bh}\to e_\Bh$ can be arbitrarily chosen; and
\begin{align}\label{eq.hproj2}
    &\sum_{e_\Bh \alpha\beta}
    \frac{d_{e_\Bh}}{n_{e_\Bh}} 
    ({^{\mb g_0}}[F^{c_{\mb g} e_{\mb h} b_{\mb k}}_{f'}]
    ^{a,\alpha\delta'}
    _{d,\beta\chi'})^*
    {^{\mb g_0}}[F^{c_{\mb g} e_{\mb h} b_{\mb k}}_{f}]^{a,\alpha\delta}
    _{d,\beta\chi}
    \nonumber\\
    =&
    \frac{d_{a_{\mb g \mb h}} d_{d_{\mb h \mb k}} n_{f_{\Bg\Bh\Bk}}}{n_{a_{\mb g \mb h}} n_{d_{\mb h \mb k}} d_{f_{\Bg\Bh\Bk}}}
    \left\{\begin{array}{l}
    \delta_{f_{\Bg\Bh\Bk}f_{\Bg\Bh\Bk}'}
    \delta_{\alpha\alpha'}
    \delta_{\beta\beta'}
    \text{, \ \ if }f_{\Bg\Bh\Bk}\text{ is m-type
    }\\ 
    \frac{1}{2} (\delta_{f_{\Bg\Bh\Bk}f_{\Bg\Bh\Bk}'}\delta_{\alpha\alpha'}\delta_{\beta\beta'}+
    {^{\Bg_0}\Xi^{s c_\Bg}_{d t f,\delta\eta}}
    (^{\Bg_0}\Xi^{s a}_{b_\Bk t f,\chi\eta})^*
    \delta_{f_{\Bg\Bh\Bk}f_{\Bg\Bh\Bk}'}
    \delta_{(\alpha\times f)\alpha'}\delta_{(\beta\times f)\beta'})
    \text{, \ \ if }f_{\Bg\Bh\Bk}\text{ is q-type}
    \end{array}\right.,
\end{align}
where $s_{\Bg'}$, $t_{\Bg'\Bg\Bh\Bk}$ and the fusion state $\eta:s_{\Bg'}\ot f_{\Bg\Bh\Bk}\to t_{\Bg'\Bg\Bh\Bk}$ can be arbitrarily chosen. We note that the two projective-unitary conditions of $\cH$-move are algebraically equivalent to Eqs.~(\ref{eq.hproj1}) and (\ref{eq.hproj2}).


\subsection{Parent Hamiltonian}\label{subsec.hamil}
The parent commuting-projector Hamiltonian of the fermionic symmetry-enriched string-net model defined above is
\begin{align}\label{eq.hamil}
    H=-\sum_\nu C_\nu-\sum_\nu A_\nu- \sum_l D_l- \sum_l Q_l- \sum_p B_p.
\end{align}
We introduce all the terms in the following:
\begin{itemize}
    \item $C_\nu$ ensures that the fermion number at each vertex $\nu$ equals to the grading of the fusion space (or say, the parity of the fusion state) at that vertex for the ground state\cite{chenqi}:
    \begin{align}\label{eq.cnu}
        C_\nu
        \vast|\
        \vcenter{\hbox{\includegraphics[scale=.16]{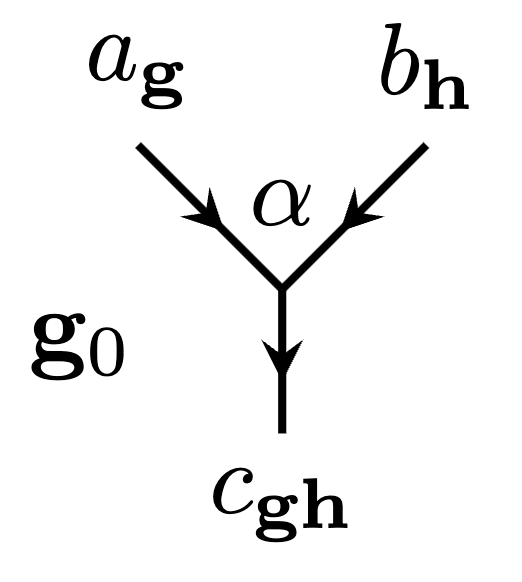}}}
        \vast\rangle
        =
        \delta_{\la N_f\ra,s(\alpha)}
        \vast|\
        \vcenter{\hbox{\includegraphics[scale=.16]{ingoing}}}
        \vast\rangle,
    \end{align}
    where $N_f$ is the fermion number operator, and 
    \begin{align}
        \la N_f\ra:= 
        \left \la
        \vcenter{\hbox{\includegraphics[scale=.16]{ingoing}}}
        \vast|\
        N_f
        \vast|\
        \vcenter{\hbox{\includegraphics[scale=.16]{ingoing}}}
        \right\rangle\in\{0,1\}.
    \end{align}
    
    \item $A_\nu$ ensures that the string fusion rules are obeyed at each vertex $\nu$ for ground state:
    \begin{align}\label{eq.anu}
        A_\nu
        \vast|\
        \vcenter{\hbox{\includegraphics[scale=.16]{ingoing}}}
        \vast\rangle
        =
        \left\{\begin{array}{l}
        \vast|\
        \vcenter{\hbox{\includegraphics[scale=.16]{ingoing}}}
        \vast \rangle
        \text{, \ \ if }N^{a_\Bg b_\Bh}_{c_{\Bg\Bh}}>0
        \\ 
        0
        \text{,  \ \ if }N^{a_\Bg b_\Bh}_{c_{\Bg\Bh}}=0
        \end{array}\right.
    \end{align}
    \item $Q_l$ ensures that the $G$-grading structure at each link $l$ for ground state:
    \begin{align}\label{eq.ql}
        Q_l
        \vast|\
        \vcenter{\hbox{\includegraphics[scale=.16]{defect}}}
        \vast\rangle
        =
        \delta_{\Bg,\Bg_0^{-1}\Bg_1}
        \vast|\
        \vcenter{\hbox{\includegraphics[scale=.16]{defect}}}
        \vast\rangle.
    \end{align}
    
    \item $D_l$ ensures that when link $l$ is decorated by a q-type string, there exist two equivalent fixed-point ground states by ``sliding'' fermions (recall Remark. \ref{rmk.q-type}), the form of which is similar to that in Ref.~\onlinecite{fSN}. {$D_l$ is determined by the dimensions of endomorphism space $n_{a_\Bg}$ and the phase factor between all possible equivalent fixed-point states (two fixed-point states are called equivalent if they differ only by a phase factor) on the honeycomb lattice, which can be derived from $\cY$-moves, $\cH$-moves, dual $\cH$-moves (these renormalization moves are introduced in Appendix \ref{appen.othermove}).}
    
    \item $B_p$ ensures that the fixed-point ground state is invariant under fusing loops onto the edges of each plaquette $p$, written as
    \begin{align}\label{eq.bp}
            B_p=\frac{1}{\cD^2}
            \underset{\Bg\in G}{\sum}|\Bg_p\Bg\ra\la\Bg_p|
            \underset{s_\Bg\in\cS_G}{\sum}
            \frac{d_{s_\Bg}}{n_{s_\Bg}}
            {B_p^{s_\Bg}}.
        \end{align}
    {$B_p^{s_\Bg}$ fuses a loop $s_\Bg$ on the edges ($\Bg=\Bg_0^{-1}\Bg_1$):}
        \begin{equation}
	{B_p^{s_\Bg}
	\vast|\,
	  \vcenter{\hbox{\includegraphics[scale=.04]{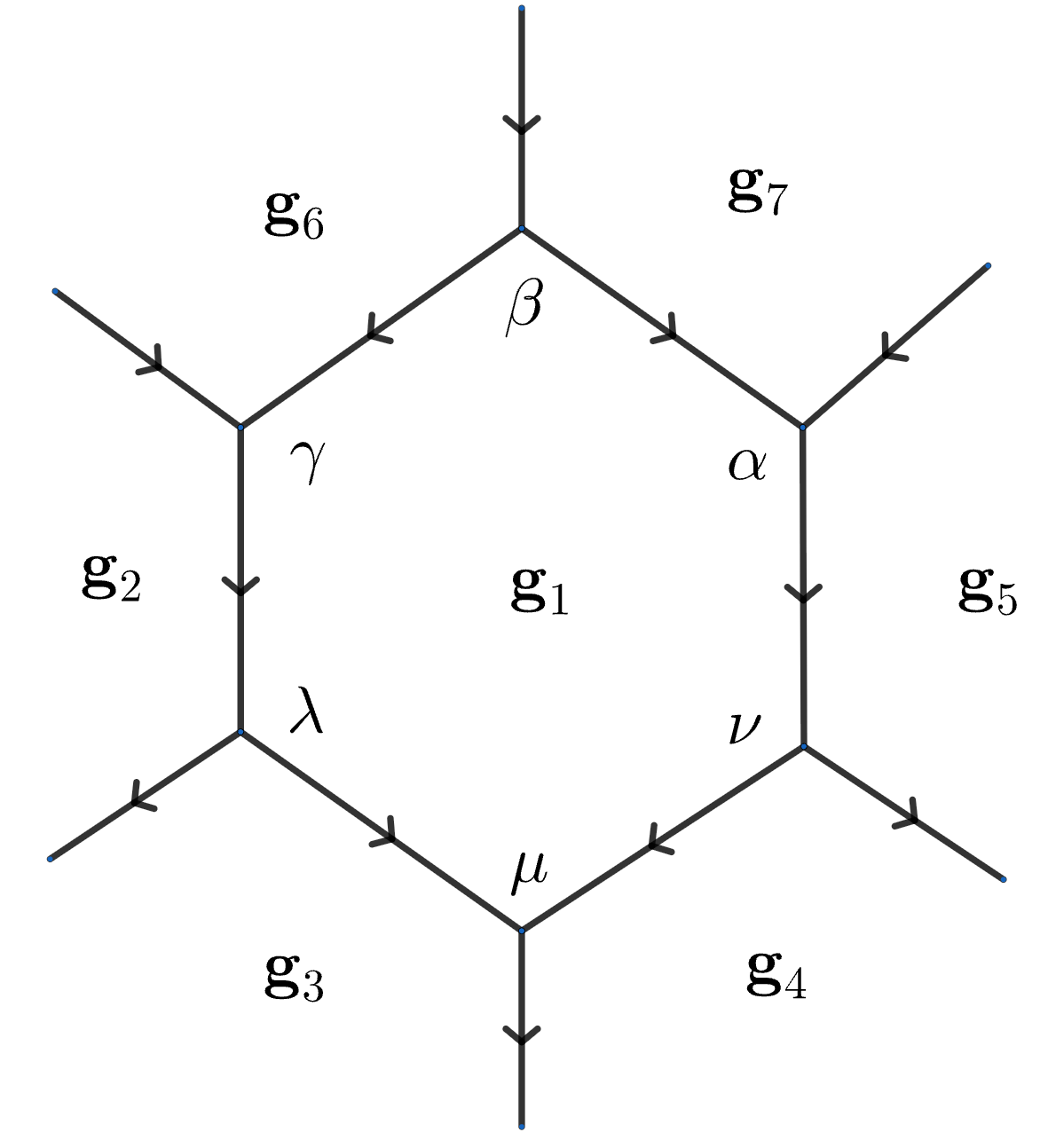}}}
	\vast\rangle
	=
	\vast|\,
	  \vcenter{\hbox{\includegraphics[scale=.04]{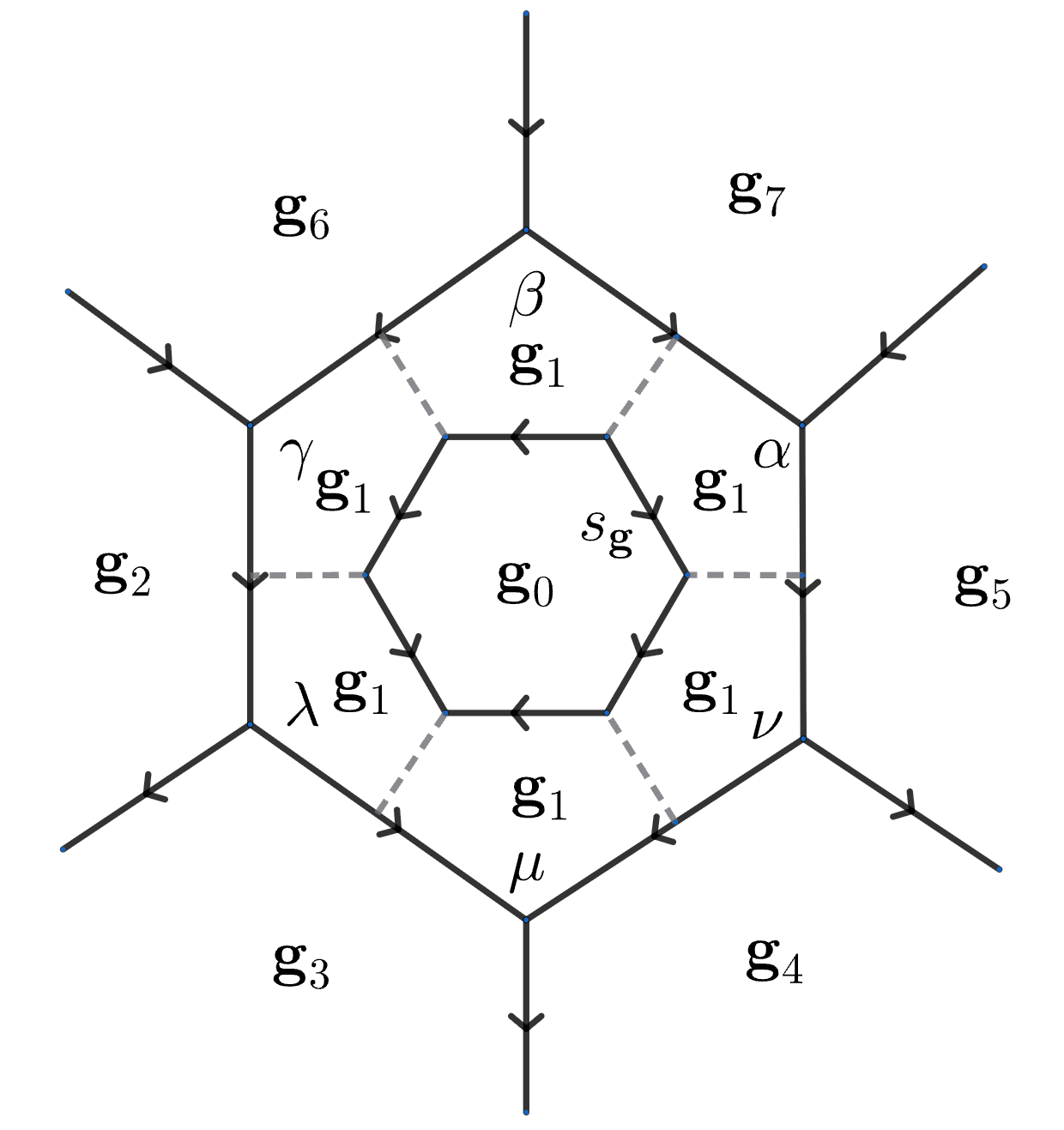}}}
	\vast\rangle
        =
        \cM
        \vast|\,
	  \vcenter{\hbox{\includegraphics[scale=.04]{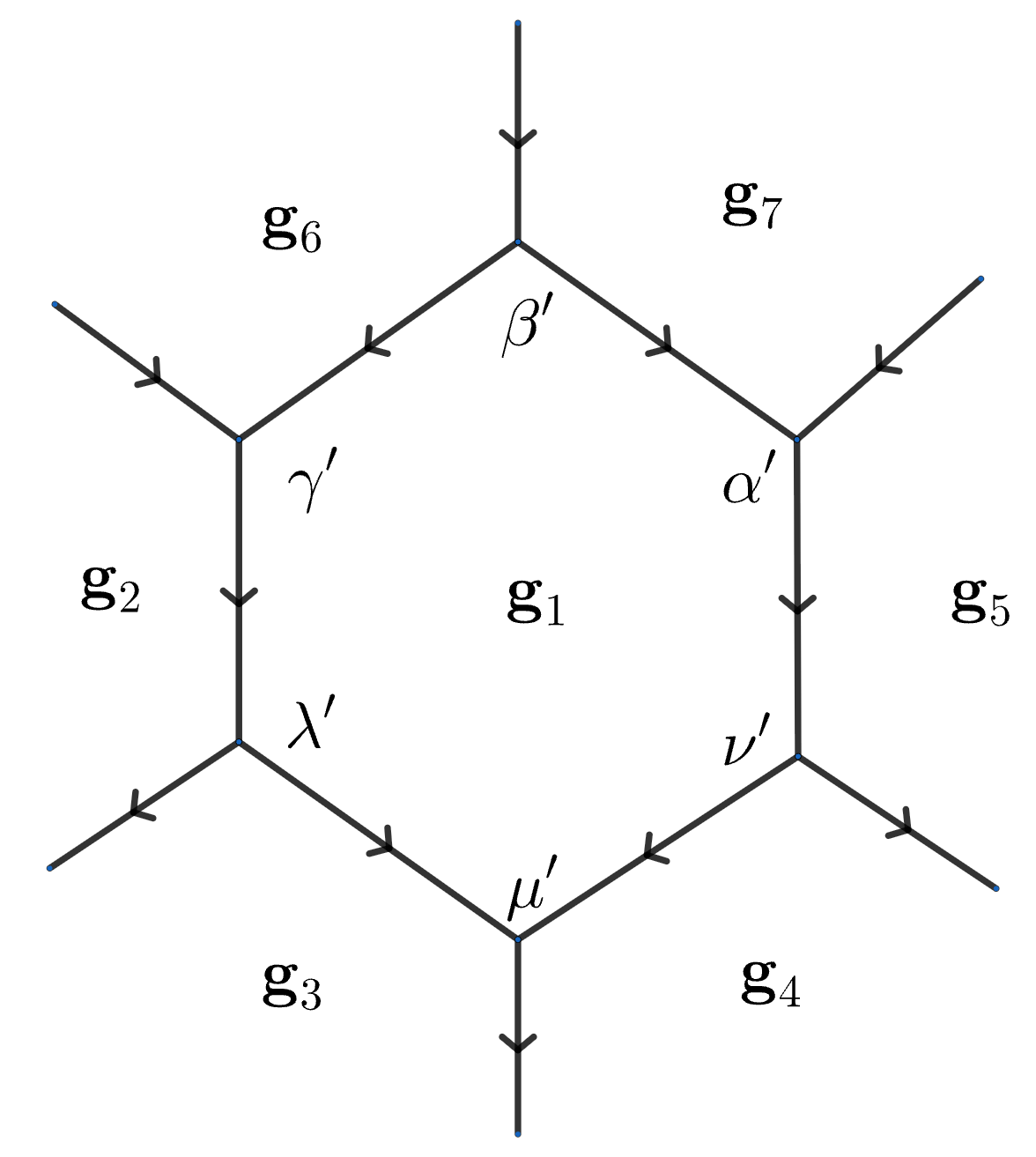}}}
	\vast\rangle,}
	\label{eq:bp}
    \end{equation}
    {where $\cM$ is a sequence of $\cF$-moves, $\cH$-moves, dual $\cH$-moves and $\cO$-moves, similarly to the construction in Ref.~\onlinecite{Cheng2017}.}
\end{itemize}
It is easy to check that all above terms are projectors, e.g., $A_\nu^2=A_\nu$, and commute with each other, e.g., $A_\nu B_p=B_p A_\nu$. 

\begin{remark}
    {Once the following data in a fermionic symmetry-enriched string-net model is given: the symmetry group $G^f$, string types $a_\Bg$, dimensions of endomorphism space $n_{a_\Bg}$, fusion rules $N^{a_\mb g b_{\mb h }}_{c_{\mb k}},F^{a_\mb g b_{\mb h }}_{c_{\mb k}}$ (which imply quantum dimensions $d_{a_\Bg}$), bosonic-part $F$-moves ${^{\mathbf{g}_0}[{F^{a_{\textbf{g}}b_{\textbf{h}}c_{\textbf{k}}}_{d}]^{m,\alpha\beta}
    _{n,\chi\delta}}}$, in total denoted as}
    \begin{equation}
        {\{G^f,a_\Bg,n_{a_\Bg},d_{a_\Bg},N^{a_\mb g b_{\mb h }}_{c_{\mb k}},F^{a_\mb g b_{\mb h }}_{c_{\mb k}},{^{\mathbf{g}_0}[{F^{a_{\textbf{g}}b_{\textbf{h}}c_{\textbf{k}}}_{d}]^{m,\alpha\beta}_{n,\chi\delta}}}\},}
    \end{equation}
    {the form of the parent Hamiltonian is directly and uniquely determined (up to gauge transformations in Eq.~(\ref{eq.gauge})):}
    \begin{itemize}
        \item {$Q_l$ is determined by $G$ through Eq.~(\ref{eq.ql});}
        
        \item {$C_\nu$ and $A_\nu$ are determined by fusion rules encoded in $N^{a_\mb g b_{\mb h }}_{c_{\mb k}},F^{a_\mb g b_{\mb h }}_{c_{\mb k}}$ through Eqs.~(\ref{eq.cnu}) and (\ref{eq.anu}) respectively;}
        
        \item {$D_l$ is determined by $n_{a_\Bg}$, $\cY$-moves, $\cH$-moves, dual $\cH$-moves, where by Eqs.~(\ref{eq.ymove}), (\ref{eq.o}), (\ref{eq.hmove}) and (\ref{eq.dualhmove}), the bosonic-part $Y$-moves, $H$-moves, dual $H$-moves are all determined by $\{n_{a_\Bg},d_{a_\Bg},N^{a_\mb g b_{\mb h }}_{c_{\mb k}},{^{\mathbf{g}_0}[{F^{a_{\textbf{g}}b_{\textbf{h}}c_{\textbf{k}}}_{d}]^{m,\alpha\beta}
        _{n,\chi\delta}}}\}$, while the Grassmann numbers are determined by $F^{a_\mb g b_{\mb h }}_{c_{\mb k}}$;}
        
        \item {$B_p$ is determined by $\cF$-moves, $\cH$-moves, dual $\cH$-moves, $\cO$-moves through Eq.~(\ref{eq.bp}), thus also determined by $\{n_{a_\Bg},d_{a_\Bg},N^{a_\mb g b_{\mb h }}_{c_{\mb k}},F^{a_\mb g b_{\mb h }}_{c_{\mb k}},{^{\mathbf{g}_0}[{F^{a_{\textbf{g}}b_{\textbf{h}}c_{\textbf{k}}}_{d}]^{m,\alpha\beta}_{n,\chi\delta}}}\}$.}
    \end{itemize}
    {Therefore, in most of the examples in Subsection \ref{subsec.ex}, we will only list the set of data $\{G,a_\Bg,n_{a_\Bg},d_{a_\Bg},N^{a_\mb g b_{\mb h }}_{c_{\mb k}},F^{a_\mb g b_{\mb h }}_{c_{\mb k}},$ ${^{\mathbf{g}_0}[{F^{a_{\textbf{g}}b_{\textbf{h}}c_{\textbf{k}}}_{d}]^{m,\alpha\beta}_{n,\chi\delta}}}\}$, which is complete, and the form of parent Hamiltonian written in renormalization moves can be directly obtained, thus omitted. However, presenting the Hamiltonian in terms of expressible operators is in general complicated, which will only be shown in a simple example in Remark \ref{rmk.hamiltoric}.}
\end{remark}

\subsection{Summary}\label{summary}
\begin{definition}[Fermionic symmetry-enriched string-net model]
    {Inputting a finite fermionic symmetry group $G^f=\Z_2^f\times_{\omega_2}G$ and a $G$-graded super fusion category $\cS_G$, a fermionic symmetry-enriched string-net model is an exactly solvable model for an fSET phase, constructed on a honeycomb lattice, consists of}
    \begin{itemize}
        \item Degrees of freedom introduced in Subsection \ref{subsection.fusion};
        \item Renormalization moves on fixed-point ground states satisfying the symmetric condition, introduced in Subsection \ref{subsec.renorm};
        \item A parent commuting-projector Hamiltonian, introduced in Subsection \ref{subsec.hamil},
    \end{itemize}
    {where the full data of fermionic symmetry-enriched string-net model is denoted as $\{G^f,a_\Bg,n_{a_\Bg},d_{a_\Bg}, N^{a_\mb g b_{\mb h }}_{c_{\mb k}}, F^{a_\mb g b_{\mb h }}_{c_{\mb k}},$ ${^{\mathbf{g}_0}[{F^{a_{\textbf{g}}b_{\textbf{h}}c_{\textbf{k}}}_{d}]^{m,\alpha\beta}_{n,\chi\delta}}}\}$, satisfying the following consistency equations:}
    \begin{align}
        N^{a_{\textbf{g}}b_{\textbf{h}}}
        _{c_{\textbf{g}\textbf{h}}}
        =B^{a_{\textbf{g}}b_{\textbf{h}}}
        _{c_{\textbf{g}\textbf{h}}}+
        F^{a_{\textbf{g}}b_{\textbf{h}}}
        _{c_{\textbf{g}\textbf{h}}},
    \end{align}

    \begin{align}
\sum_{e_{\mb g \mb h}} 
\frac{N^{a_{\mb g}b_{\mb h}}_{e_{\mb g\mb h}} N^{e_{\mb g \mb h}c_{\mb k}}_{d_{\mb g \mb h \mb k}}}{n_{e_{\mb g \mb h}}}
=
\sum_{f_{\mb h \mb k}} 
\frac{N^{b_{\mb h }c_{\mb k}}_{f_{\mb h \mb k}}N^{a_{\mb g}f_{\mb h \mb k}}_{d_{\mb g \mb h \mb k}}}{n_{f_{\mb h \mb k}}}
,
\end{align}
    
    \begin{align}
        \sum_{e_{\mb g \mb h}} 
        \frac{B^{a_{\mb g}b_{\mb h}}_{e_{\mb g\mb h}} B^{e_{\mb g \mb h}c_{\mb k}}_{d_{\mb g \mb h \mb k}}+F^{a_{\mb g}b_{\mb h}}_{e_{\mb g\mb h}} F^{e_{\mb g \mb h}c_{\mb k}}_{d_{\mb g \mb h \mb k}}}
        {n_{e_{\mb g \mb h}}}
        =
        \sum_{f_{\mb h \mb k}} 
        \frac{B^{b_{\mb h }c_{\mb k}}_{f_{\mb h \mb k}}B^{a_{\mb g}f_{\mb h \mb k}}_{d_{\mb g \mb h \mb k}}+F^{b_{\mb h }c_{\mb k}}_{f_{\mb h \mb k}}F^{a_{\mb g}f_{\mb h \mb k}}_{d_{\mb g \mb h \mb k}}}
        {n_{f_{\mb h \mb k}}},
    \end{align}

    \begin{align}
         d_{a_\Bg}d_{b_\Bh}=
    \sum_{c_{\Bg\Bh}}\frac{N^{a_\Bg b_\Bh}_{c_{\Bg\Bh}}d_{c_{\Bg\Bh}}}{n_{c_{\Bg\Bh}}},
    \end{align}

    \begin{align}
        {^{\mathbf{g}_0}[{F^{a_{\textbf{g}}b_{\textbf{h}}c_{\textbf{k}}}_{d}]^{m,\alpha\beta}
        _{n,\chi\delta}}}=
        (-1)^{\omega_2( \textbf{g}_0,\textbf{g})s(\chi)}
        ({^{\textbf{1}}[{F^{a_{\textbf{g}}b_{\textbf{h}}c_{\textbf{k}}}_{d}]^{m,\alpha\beta}
        _{n,\chi\delta}}})^{S(\textbf{g}_0)},
    \end{align}

    \begin{align}
        s(\alpha)+s(\beta)+s(\delta)+s(\chi)
        =0\text{ (mod 2)},
    \end{align}

    \begin{equation}
        \underset{\epsilon}{\sum} 
        {^{\textbf{1}}[{F^{j_{\textbf{gh}}c_{\textbf{k}}d_{\textbf{l}}}_{e}]^{m,\beta\chi}_{q,\delta\epsilon}}}
        {^{\textbf{1}}[{F^{a_{\textbf{g}}b_{\textbf{h}}q_{\textbf{kl}}}_{e}]^{j,\alpha\epsilon}_{p,\phi\gamma}}}
        =
        (-1)^{(s(\alpha)+\omega_2( \textbf{g},\textbf{h}))s(\delta)}
        \underset{n_{\Bh\Bk}\eta\psi\kappa}{\sum} 
        {^{\textbf{1}}[{F^{a_{\textbf{g}}b_{\textbf{h}}c_{\textbf{k}}}_{m}]^{j,\alpha\beta}_{n,\eta\psi}}}
        {^{\textbf{1}}[{F^{a_{\textbf{g}}n_{\textbf{hk}}d_{\textbf{l}}}_{e}]^{m,\psi\chi}_{p,\kappa\gamma}}}
        ({^{\textbf{1}}[{F^{b_{\textbf{h}}c_{\textbf{k}}d_{\textbf{l}}}_{p}]^{n,\eta\kappa}_{q,\delta\phi}}})^{S(\textbf{g})},
    \end{equation}
    and the four projective-unitary conditions in Eqs.~(\ref{eq.fprojuni1})-(\ref{eq.hproj2}).
\end{definition}

\begin{remark}
    When there is no nontrivial string type except from symmetry defects, i.e., $\cS_\one=\{\mathbb{1}\}$, the fermionic symmetry-enriched string-net model defined here reduces to the fixed-point wavefunction construction of 2+1D fSPT phases\cite{Wang2020}.
\end{remark}

\begin{remark}
    Given a bosonic string-net model with input unitary fusion category $\cC$, the anyons (or quasiparticle excitations) in the model are described by the Drinfeld center $\cZ_1(\cC)$ of $\cC$, which can be shown by calculating the so-called $Q$-algebra or tube algebra\cite{PhysRevB.90.115119} defined on the string-net model.
\end{remark}

\begin{remark}
    Given a symmetry-enriched string-net model with input $G$-graded unitary fusion category $\cC_G$, the anyons in the model are described by $\cZ_1(\cC_\one)$\cite{Cheng2017,heinrich2016symmetry}, which is a unitary modular tensor/fusion category (UMTC).
\end{remark}

\begin{conjecture}\label{conj.anyon}
    Given a fermionic symmetry-enriched string-net model with input $G$-graded super fusion category $\cS_G$, the quasiparticle/anyon excitations in the model are described by $\cZ_1(\cS_\one)$, where $\one\in G$ is the trivial element, and $\cZ_1(\cS_\one)$ is a super modular category.
\end{conjecture}

\begin{remark}
    {Physically, a fermionic symmetry-enriched string-net model can be viewed as constructed by the following two steps:}
    \begin{enumerate}
        \item {Constructing a fermionic string-net model\cite{2dtopo,fSN} with input $\cS_G$, where the strings are labeled by simple objects $a_\Bg,b_\Bh,...\in\cS_G$ ($a,b,...\in\cS_\one\text{ and }\Bg,\Bh,...\in G$). It corresponds to the exactly solvable model for the $G$-gauged topological phase, i.e., a fermionic topological order without any symmetry;}

        \item {Ungauging $G$ for the above string-net model by adding degrees of freedom labeled by $G$ to each plaquette (strings $a_\Bg,b_\Bh,...$ become also symmetry defects), as shown in Graph (\ref{graph.plaquette}), we obtain the fermionic symmetry-enriched string-net model.}
    \end{enumerate}
\end{remark}

\begin{remark}
    {For any $G$-graded super fusion category, there is no anti-unitary symmetry structure (by anti-unitary symmetry structure, we mean that if $\Bg_0\in G$ is anti-unitary, the $F$-move is complex-conjugated under the symmetry action of $\Bg_0$, as shown in Eq.~(\ref{sc})), otherwise it will not be a well-defined fusion category. Nevertheless, though any $G$-graded super fusion category does not contain anti-unitary symmetry structure, we can add it in the constructed symmetry-enriched string-net model, which is a physical model that cannot be described by any fusion category.}
\end{remark}
\end{widetext}

\subsection{Examples}\label{subsec.ex}
{In this Subsection, we present three examples of fermionic symmetry-enriched string-net models. Recall Eq.~(\ref{sc}), if $\omega_2$ is nontrivial or $G$ contains anti-unitary symmetry, the $F$-moves ${^{\mathbf{g}_0}[{F^{a_{\textbf{g}}b_{\textbf{h}}c_{\textbf{k}}}_{d}]^{m,\alpha\beta}
_{n,\chi\delta}}}$ depend on the leftmost group element $\Bg_0$, otherwise they do not.}
\begin{convention}\label{conv.indep}
        {For the $F$-move ${^{\mathbf{g}_0}[{F^{a_{\textbf{g}}b_{\textbf{h}}c_{\textbf{k}}}_{d}]^{m,\alpha\beta}
_{n,\chi\delta}}}$, if the fusion outcome of any $G$-graded string types is unique, it is simplified as ${^{\Bg_0}F^{a_{\textbf{g}}b_{\textbf{h}}c_{\textbf{k}}}}$. If the $F$-move is independent of the group element $\Bg_0$ in the leftmost plaquette, it is simplified as ${[{F^{a_{\textbf{g}}b_{\textbf{h}}c_{\textbf{k}}}_{d}]^{m,\alpha\beta}
_{n,\chi\delta}}}$.}
\end{convention}

\subsubsection{Toric code topological order stacking with a physical fermion with \texorpdfstring{$\Z_4^f$}{} symmetry}
    Let us consider the $\cZ_1(\textbf{Vec}_{\mathbb{Z}_2})\boxtimes\{\mathbb{1},f\}$ topological order with $\Z_4^f$ symmetry. Here $\cZ_1(\textbf{Vec}_{\mathbb{Z}_2})\boxtimes\{\mathbb{1},f\}=\{\unit,e,m,\psi\}\boxtimes\{\mathbb{1},f\}$ describes anyons of the toric code topological order stacking with a physical fermion $f$, where $\boxtimes$ is the Deligne tensor product, physically understood as stacking of topological phases. Denote group elements in the bosonic symmetry group as $G=\Z_2=\{0,1\}$. Then $\omega_2(\Bg,\Bh)=\Bg\Bh$ (mod 2). Given such a topological order and a fermionic global symmetry, there are several possibilities that how the symmetry interacts with the topological order, i.e., there are several possible fSET phases. 

    Let us further consider special cases that $s(\alpha)=\omega_2(\Bg,\Bh)=\Bg\Bh$ (mod 2) in the fermionic symmetry-enriched string-net models as stated in Remark \ref{rmk.some}, i.e., Eq.~(\ref{fpenta2}) reduces to bosonic pentagon equation. In classifying such fSET phases up to equivalence relations (i.e., gfSLU transformations), we divide these possible fSET phases into three classes\cite{Cheng2017}:
    \begin{enumerate}
        \item $\cS_{\Z_2}=\text{Vec}_{\Z_2\times\Z_2}=\cS_0\oplus\cS_1=\{\unit_0,e_0\}\oplus\{\unit_1,e_1\}$, where $\unit_1\ot\unit_1=e_1\ot e_1=\unit_0$. Since ${^0F^{a_{\textbf{g}}b_{\textbf{h}}c_{\textbf{k}}}}$ are the same as the bosonic solution in Ref.~\onlinecite{Cheng2017}, we direct conclude that the classification of this sub-case is $\Z_2\times\Z_2$. The first $\Z_2$ classification corresponds to
        \begin{equation}
            {{^0F^{e_0 e_0 e_0}={^1F^{e_0 e_0 e_0}=1}}},
        \end{equation}
        and
        \begin{equation}
            {{^0F^{e_0 e_0 e_0}={^1F^{e_0 e_0 e_0}=-1}}}.
        \end{equation}
        The second $\Z_2$ classification corresponds to
        \begin{align}
            {({{^0F^{\unit_1\unit_1\unit_1}}},{{^0F^{e_1 e_1 e_1}}})=(1,-1),}
            \nonumber\\
            {({{^1F^{\unit_1\unit_1\unit_1}}},{{^1F^{e_1 e_1 e_1}}})=(-1,1),}
        \end{align}
        and  
        \begin{align}
            {
            \begin{gathered}
                ({{^0F^{\unit_1\unit_1\unit_1}}},{{^0F^{e_1 e_1 e_1}}})=(-1,-1),
                \\
                ({{^1F^{\unit_1\unit_1\unit_1}}},{{^1F^{e_1 e_1 e_1}}})=(1,1),
            \end{gathered}
            }
        \end{align}
        For the second $\Z_2$ classification, the first solution corresponds to a trivial fSET phase, i.e., the $\cZ_1(\textbf{Vec}_{\mathbb{Z}_2})\boxtimes\{\mathbb{1},f\}$ topological order stacking with a $\Z_2$ SPT phase, where the fermion $f$ carries half $\Z_2$ charge, i.e.,
        \begin{align}
            U_f(1)U_f(1)=-U_f(0),
        \end{align}
        where $U_a(\Bg)$ is the localized symmetry operator of $\Bg$ near anyon $a$ defined in Appendix \ref{subsec.sf} (please do not mix it up with the global symmetry operator $U(\Bg)$). The second solution corresponds to a fSET phase where anyon $e$ (or $m$) and $f$ carry half $\Z_2$ charge.
        
        \item $\cS_{\Z_2}=\text{Vec}_{\Z_4}=\cS_0\oplus\cS_1=\{\unit_0,e_0\}\oplus\{\unit_1,e_1\}$, where $\unit_1\ot\unit_1=e_1\ot e_1=e_0$. There are two inequivalent solutions of $F$-moves: 
        \begin{align}
            {{^0F^{abc}}=1,\quad {^1F^{abc}}=(-1)^{abc},}
        \end{align}
        and
        \begin{align}
            {\begin{gathered}
                {^0F^{abc}}=e^{\frac{i\pi}{4}a(b+c-[b+c]_4)},
                \\
                {^1F^{abc}}=(-1)^{abc}e^{\frac{i\pi}{4}a(b+c-[b+c]_4)},
            \end{gathered}}
        \end{align}
        where we relabel $0:=\unit_0$, $2:=e_0$, $1:=\unit_1$, $3:=e_1$, and $[b+c]_4:=b+c$ (mod 4). The first solution corresponds to a fSET phase where anyon $e$ (or $m$) and $f$ carry half $\Z_2$ charge. The second solution corresponds to a fSET phase where all $e$, $m$ and $f$ carry half $\Z_2$ charge.
        
        \item $\cS_{\Z_2}=\textbf{Ising}=\cS_0\oplus\cS_1=\{\unit,e\}\oplus\{\sigma\}$, where $\sigma\ot\sigma=\unit\oplus e$. There are two inequivalent solutions of $F$-moves (distinguished by $\pm 1$ in ${^0[F^{\sigma\sigma\sigma}_\sigma]^a_b}$): 
        \begin{align}
		{\begin{gathered}
			{^0[F^{\sigma e\sigma}_{ e}]^\sigma_{\sigma}}={^0[F^{ e\sigma e}_\sigma]^\sigma_{\sigma}}={^1[F^{\sigma e\sigma}_{ e}]^\sigma_{\sigma}}={^1[F^{ e\sigma e}_\sigma]^\sigma_{\sigma}}=-1,\\
	    {^0[F^{\sigma\sigma\sigma}_\sigma]^a_b}
            =\frac{\pm 1}{\sqrt{2}}\begin{pmatrix}
			1 & 1 \\
			1 & -1
		\end{pmatrix},\\
	    {^1[F^{\sigma\sigma\sigma}_\sigma]^a_b}
            =-{^0[F^{\sigma\sigma\sigma}_\sigma]^a_b},
		\end{gathered}}
	\end{align}
        where $a,b=\unit,e$. Both solutions correspond to fSET phases where the $\Z_2$ symmetry permutes $e$ and $m$, and $f$ carries half $\Z_2$ charge.
    \end{enumerate}

\begin{remark}\label{rmk.hamiltoric}
    {We present the parent Hamiltonian explicitly in expressible operators for the case $\cS_{\Z_2}=\text{Vec}_{\Z_4}=\{\unit_0,e_0\}\oplus\{\unit_1,e_1\}\equiv\{0,2\}\oplus\{1,3\}$ on a honeycomb lattice in terms of Pauli operators and the following defined operators\cite{Cheng2017}:}
    \begin{align}
       { U_l|n\ra_l=i^n|n\ra_l,\ \ 
        V_l|n\ra_l=|n+1\ra_l,\ \ 
        \text{where }n\in\Z_4,}
    \end{align}
    {and $|n\ra_l$ denotes for the state of string type $n$ on link $l$. Since in this example all strings are $m$-type, the $D_l$ term vanishes. The Hamiltonian is given by}
    \begin{align}
     {H=-\sum_\nu C_\nu-\sum_\nu A_\nu- \sum_l Q_l- \sum_p B_p,}
    \end{align}
    {where we label the three links around each vertex $\nu$ as $l_1,l_2,l_3$, and the six links of each plaquette $p$ as $l_a,l_b,l_c,l_d,l_e,l_f$, as shown in the following graph (the lattice vertex, link, plaquette labels, differ from the state labels, are all colored in green):}
    \begin{equation}\label{graph.label}
    \includegraphics[scale=.03]{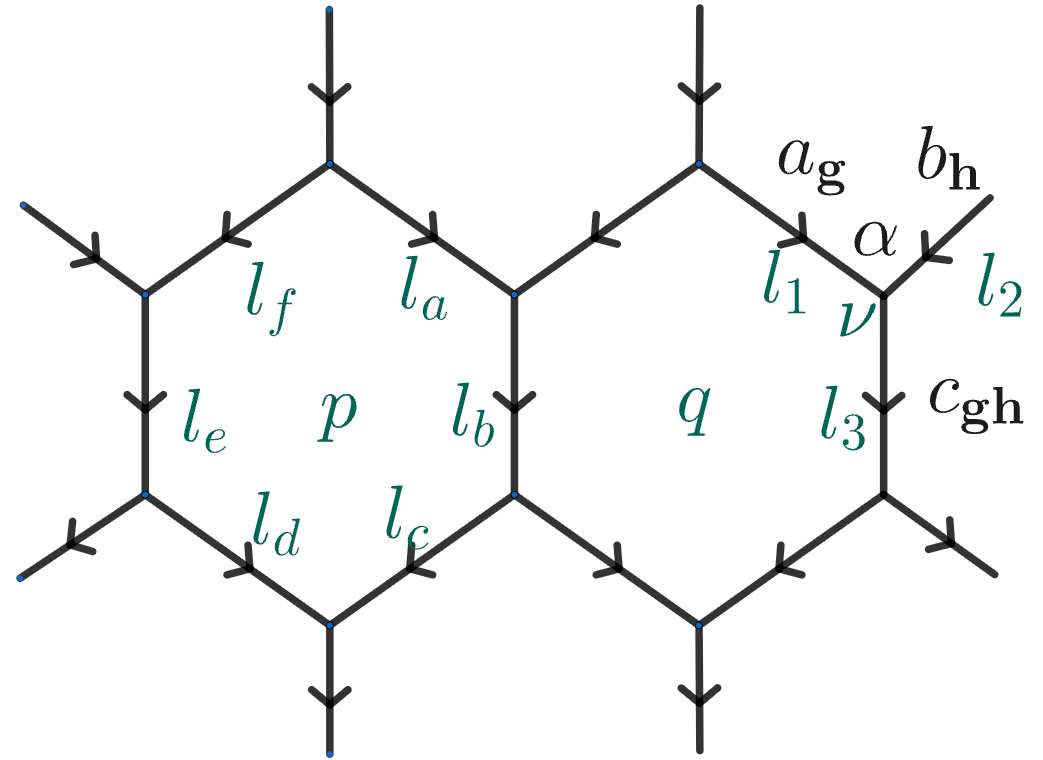}.
    \end{equation}
    {Then all the terms are written explicitly as}
    \begin{gather}
        {C_\nu=\frac{1+P_f(-1)^
        {\frac{1-U_{l_1}^2}{2}
        \frac{1-U_{l_2}^2}{2}
        }
        }{2},}\\
        {A_\nu=U^\dagger_{l_1}U^\dagger_{l_2}
        U_{l_3}+(U^\dagger_{l_1}U^\dagger_{l_2}U_{l_3})^\dagger,}\\
        {Q_l=\frac{1+U_l^2}{2}\frac{1+\sigma_p^z\sigma_q^z}{2}+\frac{1-U_l^2}{2}\frac{1-\sigma_p^z\sigma_q^z}{2},}\\
        {B_p=\frac{1}{4}\sum_{n=0}^3
        \begin{bmatrix}
        \sigma_p^x\begin{pmatrix}
        V_{l_a}V_{l_b}V_{l_c}
        V^\dagger_{l_d}V^\dagger_{l_e}
        V^\dagger_{l_f}+(V_{l_a}V_{l_b}V_{l_c}
        V^\dagger_{l_d}V^\dagger_{l_e}
        V^\dagger_{l_f})^\dagger\end{pmatrix}
        \end{bmatrix}^n
        ,}
    \end{gather}
    {where $P_f=(-1)^{N_f}$ is the fermion-parity operator acting on each vertex $\nu$, the fermion-parity function on each vertex $\nu$ is $s(\alpha)=\omega_2(\Bg,\Bh)=\Bg\Bh\text{ (mod 2)}=
        {\frac{1-U_{l_1}^2}{2}
        \frac{1-U_{l_2}^2}{2}
        }$, and $\sigma^z_p$ and $\sigma^x_p$ are Pauli matrices acting on plaquette state $|\Bg\ra$ ($\Bg\in\Z_2$) for each plaquette $p$. Note that the $B_p$ term can also be written as a sequence of renormalization moves. We can check that this Hamiltonian is invariant under the global $\Z_2$ symmetry operator $\prod_p\sigma_p^x$ and the fermion-parity operator $P_f=(-1)^{N_f}$ acting on all the vertices.}   
\end{remark}

\subsubsection{Super Tambara-Yamagami category as input with \texorpdfstring{$\Z_2^f\times\Z_2$}{} symmetry}\label{eg.TY}
    Let us consider the fermionic symmetry-enriched string-net with input $G$-graded super fusion category being the super Tambara-Yamagami category (as a $\Z_2$-graded super fusion category) defined in Example \ref{eg.nonano4}:
    \begin{align}
        {\cS_{\Z_2}}={\cS_0}\oplus{\cS_1}=\{0,...,N-1\}\oplus\{\sigma\},
    \end{align}
    where $N$ is an odd integer, $\sigma$ is a q-type string (and further a $\sigma$-type string). Let $a,b\in\{0,...,N-1\}$. Quantum dimensions are $d_a=1,d_\sigma=\sqrt{2N}$. Fusion rules are
    \begin{align}
        a\ot b
        &=
        \left\{\begin{array}{l}
        \C^{1|0} [a+b]_N
        \text{, \ \ if }a+b<N\\ 
        \C^{0|1} [a+b]_N
        \text{, \ \ if }a+b\geq N
        \end{array}\right.,
        \nonumber\\
        \sigma\ot a &=a\ot \sigma=\C^{1|1}\sigma,\ \ 
        \sigma\ot\sigma=\oplus_{a}\C^{1|1} a,
    \end{align}
    where $[a+b]_N:=a+b$ (mod $N$). {Recall Convention \ref{conv.indep}, the $F$-moves have no difference from those in super Tambara-Yamagami category, which are listed in Ref.~\onlinecite{fSN}.} This constructs the exactly solvable model of the following fSET phase: the $\cZ_1(\cS_0)$ topological order with $\Z_2^f\times\Z_2$ symmetry, where $\cZ_1(\cS_0)$ is the super modular category that describes the anyon excitations in this model.

\subsubsection{\texorpdfstring{${^\svec \text{SU}(2)_6}$}{} as input with \texorpdfstring{$\Z_2^f\times\Z_2$}{} symmetry}\label{eg.SU2}
    Let us consider the fermionic symmetry-enriched string-net with input $G$-graded super fusion category being ${^\svec \text{SU}(2)_6}$ (as a $\Z_2$-graded super fusion category) defined in Example \ref{eg.nonano5}:
    \begin{align}
        {\cS_{\Z_2}}={\cS_0}\oplus{\cS_1}=\{0,1\}\oplus\{\frac{1}{2},\frac{3}{2}\},
    \end{align}
    where $\frac{3}{2}$ is a q-type string (or further a $\sigma$-type string) and all other string types are m-type. Quantum dimensions are $d_0=1,d_1=1+\sqrt{2},d_{\frac{1}{2}}=\sqrt{2+\sqrt{2}},d_{\frac{3}{2}}=\sqrt{2}\sqrt{2+\sqrt{2}}$, which satisfy Eq.~(\ref{eq.totalquantumdim}). Fusion rules are
    \begin{align}
        1\ot 1=0\oplus\C^{1|1}1,&\ \ 
        \frac{1}{2}\ot \frac{1}{2}=0\oplus 1,
        \nonumber\\
        \frac{3}{2}\ot \frac{3}{2}=\C^{1|1}0\oplus \C^{1|1}1,&\ \ 
        \frac{1}{2}\ot\frac{3}{2}=\frac{3}{2}\ot \frac{1}{2}=\C^{1|1}1,
    \end{align}
    where for simplicity $\C^{1|0}$ is always omitted. {Recall Convention \ref{conv.indep}, the $F$-moves have no difference from those in ${^\svec \text{SU}(2)_6}\equiv \text{SU}(2)_6/f$, which can be calculated by the fermion condensation algorithm introduced in Ref.~\onlinecite{fc} or \onlinecite{fSN}, and $F$-moves in $\text{SU}(2)_6$ are listed in Ref.~\onlinecite{gils2013anyonic}. The calculation is omitted.} (The $F$-moves in $\cS_0={^\svec \text{SO}(3)_6}\equiv \text{SO}(3)_6/f$ are listed in Ref.~\onlinecite{fSN}.) This constructs the exactly solvable model of the following fSET phase: the $\cZ_1(\cS_0)\cong\cZ_1(^\svec \text{SO}(3)_6)$ topological order with $\Z_2^f\times\Z_2$ symmetry, where $\cZ_1(^\svec \text{SO}(3)_6)$ describes the anyon excitations in this model.

    
\begin{widetext}
    \begin{remark}
        {As calculated in Ref.~\onlinecite{fc} by tube algebra defined on fermionic string-net models, the anyon excitations $\cZ_1(^\svec \text{SO}(3)_6)=\{m_{00},m_{02},m_{20},m_{22}\}$ (the notations are from Ref.~\onlinecite{fc}). Performing a $Z_2^f$-crossed extension $\cZ_1(^\svec \text{SO}(3)_6)\cong {^\svec \cZ_1(\text{SO}(3)_6)}_0$ (details please refer to Subsection \ref{subsec.dcsfc} and $\cZ_1(\text{SO}(3)_6)$ is a spin modular category), we obtain}
        \begin{align}
            {[{^\svec \cZ_1(\text{SO}(3)_6)}_0]_{\Z_2^f}^\times
    \cong
    {^\svec \cZ_1(\text{SO}(3)_6)}_0\oplus
    {^\svec \cZ_1(\text{SO}(3)_6)}_1,}
        \end{align}
        {where ${^\svec \cZ_1(\text{SO}(3)_6)}_1=\{m_{11},q_{13},q_{31},m_{33}\}$ is understood as the sector of fermion-parity vortices.}

        {By Remark \ref{rmk.GSDtorus}, the ground state degeneracy (GSD) on torus $\mathbb{T}^2$ for four different spin structures are}
        \begin{align}
        {\text{GSD}(\mathbb{T}^2_{BB})=\text{GSD}(\mathbb{T}^2_{BN})=
        \text{GSD}(\mathbb{T}^2_{NB})=\text{GSD}(\mathbb{T}^2_{NN})=
        4,}
    \end{align}
    {where $B$ denotes for bounding (anti-periodic) and $N$ for non-bounding (periodic). Denote the ground state Hilbert spaces on torus with the four different spin structures as $A(\mathbb{T}^2_{BB})$, $A(\mathbb{T}^2_{BN})$, $A(\mathbb{T}^2_{NB})$, $A(\mathbb{T}^2_{NN})$ respectively, which are all 4-dimensional. The modular $S,T$ transformations on them have the form:}
    \begin{align}
        {\begin{pmatrix}
        A(\mathbb{T}^2_{BB}) \\
        A(\mathbb{T}^2_{BN}) \\
        A(\mathbb{T}^2_{NB})\\
        A(\mathbb{T}^2_{NN})
    \end{pmatrix}
    \overset{S}{\longrightarrow}
    \begin{pmatrix}
        S^{BB\to BB} & 0 & 0 & 0 \\
        0 & 0 & S^{NB\to BN} & 0\\
        0 & S^{BN\to NB} & 0  & 0 \\
        0 & 0 & 0  & S^{NN\to NN} \\
    \end{pmatrix}
    \begin{pmatrix}
        A(\mathbb{T}^2_{BB}) \\
        A(\mathbb{T}^2_{BN}) \\
        A(\mathbb{T}^2_{NB})\\
        A(\mathbb{T}^2_{NN})
    \end{pmatrix},}
    \end{align}
    
    \begin{align}
        {\begin{pmatrix}
        A(\mathbb{T}^2_{BB}) \\
        A(\mathbb{T}^2_{BN}) \\
        A(\mathbb{T}^2_{NB})\\
        A(\mathbb{T}^2_{NN})
    \end{pmatrix}
    \overset{T}{\longrightarrow}
    \begin{pmatrix}
        0 & T^{BN\to BB} & 0 & 0 \\
        T^{BB\to BN} & 0 & 0 & 0\\
        0 & 0 & T^{NB\to NB}  & 0 \\
        0 & 0 & 0  & T^{NN\to NN} \\
    \end{pmatrix}
    \begin{pmatrix}
        A(\mathbb{T}^2_{BB}) \\
        A(\mathbb{T}^2_{BN}) \\
        A(\mathbb{T}^2_{NB})\\
        A(\mathbb{T}^2_{NN})
    \end{pmatrix},}
    \end{align}
    {where all the $4\times 4$ matrices $S^{BB\to BB},S^{NB\to BN},...,T^{NN\to NN}$ are calculated and listed in Ref.~\onlinecite{fc}.}
    \end{remark}
\end{widetext}

\subsection{fSET phase obtained by partially-gauging an fSPT phase}
\label{subsec.partialgauge}
Let $\cG$ be a finite group and $N$ be an Abelian normal subgroup of $\mathcal{G}$. Given a 2+1D fSPT phase with $\Z_2^f\times\cG$ symmetry (i.e., $\omega_2$ is trivial), we can partially-gauge symmetry $N$ and obtain a fermionic $N$ gauge theory with symmetry $\Z_2^f\times G$. This corresponds to the following group extension written in a short exact sequence:
\begin{align}\label{eq.partialgroupext}
    1\to N \to \mathcal{G} \to G \to 1.
\end{align}
Let $a,b,...\in N$, $\Bg,\Bh,...\in G$ and $a_\Bg,b_\Bh,...\in\cG$. The group extension (in general may not be central extension) is specified by two pieces of data:
\begin{itemize}
    \item A group homomorphism $\varphi:G\to\text{Aut}(N)$, where $\text{Aut}(N)$ is the automorphism group of $N$.
    \item A cohomology class $[\mu]\in H^2_\varphi(G,N)$, where the 2-cocycle $\mu$ satisfies the following twisted cocycle condition: 
    \begin{equation}\label{eq.mucond}
        \varphi_\Bg(\mu(\Bh,\Bk))\mu(\Bg,\Bh\Bk)=
        \mu(\Bg,\Bh)\mu(\Bg\Bh,\Bk).
    \end{equation}
\end{itemize}
Then the group multiplication rule in $\mathcal{G}$ is
\begin{align}
    a_\Bg\times b_\Bh=[a\varphi_\Bg(b)\mu(\Bg,\Bh)]_{\Bg\Bh},
\end{align}

\begin{remark}
    {
    The partially-gauging procedure introduced above cannot be directly generalized to cases where $N$ is non-Abelian, as the 2-cocycle $\mu\in Z^2_\varphi(G,N)$ is only well-defined when $N$ is Abelian. Therefore, fSET phases obtained by such partially-gauged fSPT phases only cover the simple cases that the intrinsic topological order of an fSET phase is described by an Abelian gauge theory. Nevertheless, our fermionic symmetry-enriched string-net model framework describes general non-chiral fSET phases.
}
\end{remark}

\begin{widetext}
\begin{remark}
    {  
    In 2+1D, the relations among $\cG\times\Z_2^f$-fSPT phase, the fSET phase obtained by partially gauging $N$ for the $\cG\times\Z_2^f$-fSPT phase (which is a $N$ gauge theory with $G\times\Z_2^f$ symmetry), the fermoinic topological order (fTO) obtained by fully gauging $G$ for the $\cG\times\Z_2^f$-fSPT phase (which is a fermionic $\cG$ gauge theory\cite{gu2014lattice}), and their bosonic counterparts related by fermion condensation (or inversely, gauging $\Z_2^f$) are summarized in the following commutative diagram:
    }
    \begin{equation}\label{diag.partialgauge}
        \begin{tikzcd}
	\begin{array}{c} \text{SET phase: }\Z_2\text{ gauge theory}\\ \text{with }\cG\text{ symmetry} \end{array} && {\cG\times\Z_2^f-\text{fSPT phase}} \\
	\\
	\begin{array}{c} \text{SET phase: }N\times\Z_2\text{ gauge theory}\\ \text{with }G\text{ symmetry} \end{array} && \begin{array}{c} \text{fSET phase: }N \text{ gauge theory}\\ \text{with }G\times\Z_2^f \text{ symmetry} \end{array} \\
	\\
	{\text{TO: }\cG\times\Z_2\text{ gauge theory}} && {\text{fTO: fermionic }\cG\text{ gauge theory}}
	\arrow["{\text{F.c.}}", shift left, from=1-1, to=1-3]
	\arrow["{\text{Partially gauging } N}"', from=1-1, to=3-1]
	\arrow["{\text{Gauging }\Z_2^f}", shift left, from=1-3, to=1-1]
	\arrow["{\text{Partially gauging }N}", from=1-3, to=3-3]
	\arrow["{\text{F.c.}}", shift left, from=3-1, to=3-3]
	\arrow["{\text{Gauging }G}"', from=3-1, to=5-1]
	\arrow["{\text{Gauging }\Z_2^f}", shift left, from=3-3, to=3-1]
	\arrow["{\text{Gauging }G}", from=3-3, to=5-3]
	\arrow["{\text{F.c.}}", shift left, from=5-1, to=5-3]
	\arrow["{\text{Gauging }\Z_2^f}", shift left, from=5-3, to=5-1]
\end{tikzcd},
    \end{equation}
    {where f.c. is short for fermion condensation.}
\end{remark}

Let us briefly review the classification of 2+1D fSPT phases (without $p+ip$ superconductor layer) with $\Z_2^f\times\cG$ symmetry, and $\cG$ does not contain time-reversal symmetry (as we do not know how to gauging time-reversal symmetry yet). The classification is specified by a triplet $(n_1,n_2,\nu_3)$\cite{Wang2018,Wang2020}, where $n_1\in H^1(\cG,\Z_2)$ is the 1-cocycle specifying Kitaev chain decoration, $n_2\in H^2(\cG,\Z_2)$ the 2-cocycle specifying complex fermion decoration, and $\nu_3\in C^3(\cG,U(1))$ is the 3-cochain describing a 2+1D bosonic 2-2 move\cite{Wang2018}, satisfying the following consistency equations:
\begin{align}
&\dd n_1 = 0 \text{ (mod 2)},\\
&\dd n_2 = 0 \text{ (mod 2)},\\
&\dd \nu_3 = (-1)^{n_2\smile n_2}.\label{eq.2dnu3}
\end{align}

Then the full data in the fermionic symmetry-enriched string-net model obtained by partially-gauging can be directly read from the data of the original fSPT phase:
\begin{itemize}
    \item String types: $a_\Bg,b_\Bh,...\in\cS_G:=\cG$, which means that the set of isomorphism classes of simple objects in $\cS_G$ equals to the set of group elements in $\cG$ (recall Convention \ref{conv.iso}). A Kitaev chain becomes a $\sigma$-type string (Definition \ref{def.sigmatype}) after the partial-gauging, i.e., 
    whether a string type $a_\Bg$ is a $\sigma$-type or not is specified by: 
    \begin{align}
        n_{a_\Bg}=n_1(a_\Bg)+1.
    \end{align}
    \item Fusion rules (see Graph (\ref{graph.fusion})): 
    \begin{equation}
        a_\Bg\otimes b_\Bh=
        \left\{\begin{array}{l}
        \C^{1|0}[a\varphi_\Bg(b)\mu(\Bg,\Bh)]_{\Bg\Bh}
        \text{, \ \ if }a_\Bg, b_\Bh\text{ are all m-type and }n_2(a_\Bg,b_\Bh)=0\\ 
        \C^{0|1}[a\varphi_\Bg(b)\mu(\Bg,\Bh)]_{\Bg\Bh}.
        \text{, \ \ if }a_\Bg, b_\Bh\text{ are all m-type and }n_2(a_\Bg,b_\Bh)=1\\
        \C^{1|1}[a\varphi_\Bg(b)\mu(\Bg,\Bh)]_{\Bg\Bh}
        \text{, \ \ if at least one of }a_\Bg\text{ and }b_\Bh\text{ is }\sigma\text{-type}
        \end{array}\right..
    \end{equation}
    Due to this constrained fusion rule, quantum dimension of a $\sigma$-type string is $\sqrt{2}$, and all other string types are m-type with quantum dimension $1$.
    \item The bosonic-part $F$-move: (1) If all string types in $\cG$ are m-type, or say, $n_1$ is always trivial in the original $\cG\times\Z_2^f$-fSPT phase, we have
    \begin{align}
        {^\one F^{a_\Bg b_\Bh c_\Bk}}=\nu_3(a_\Bg,b_\Bh,c_\Bk),
    \end{align}
    where the fusion outcome of any strings is always unique and therefore we omit all indexes related to multiple fusion outcomes and simplify it as ${^\one F^{a_\Bg b_\Bh c_\Bk}}$. (2) {If there exist $\sigma$-type strings in $\cG$, the $F$-move is obtained from the fSPT phase data $\{n_1,n_2,\nu_3\}$ by firstly partially gauging $\Z_2^f\times N$ and then performing a fermion condensation ($F$-moves after fermion condensation can be explicitly calculated by the algorithm illustrated in Refs.~\cite{fc,fSN}), as shown in the following commutative diagram:}
    \begin{equation}
    \begin{tikzcd}
	&& \begin{array}{c} \Z_2^f\times \mathcal{G}-\text{fSPT phase}\\ \text{(with Kitaev-chain layer)}\\\text{described by }\{n_1,n_2,\nu_3\} \end{array} \\
	\\
	\begin{array}{c} \text{Symmetry-enriched string model}\\ \text{ with input }\mathcal{C}_G \end{array} && \begin{array}{c} \text{Fermionic symmetry-enriched string model}\\ \text{ with input }\mathcal{S}_G:={^\textbf{SVec}\mathcal{C}_G} \end{array}
	\arrow["{\text{Partially gauging }\Z_2^f\times N}"', from=1-3, to=3-1]
	\arrow["{\text{Partially gauging }N}", from=1-3, to=3-3]
	\arrow["{\text{F.c.}}", shift left, from=3-1, to=3-3]
	\arrow["{\text{Gauging }\Z_2^f}", shift left, from=3-3, to=3-1]
    \end{tikzcd},
    \end{equation}
    {which is part of Diagram (\ref{diag.partialgauge}).} In both cases, the fermionic pentagon equation in Eq.~(\ref{fpenta2}) reduces exactly to Eq.~(\ref{eq.2dnu3}):
    \begin{align}
            {^{\one}F^{(a_\Bg\times b_\Bh)c_{\textbf{k}}d_{\textbf{l}}}}\cdot
            {^{\one}F^{a_{\textbf{g}}b_{\textbf{h}}(c_\Bk\times d_\Bl)}}
            =
           (-1)^{n_2(a_\Bg,b_\Bh)n_2(c_\Bk,d_\Bl)}\cdot
            {^{\one}F^{a_{\textbf{g}}
            b_{\textbf{h}}c_{\textbf{k}}}}\cdot
            {^{\one}F^{a_{\textbf{g}}(b_\Bh\times c_\Bk)d_{\textbf{l}}}}\cdot
            ({^{\one}F^{b_{\textbf{h}}
            c_{\textbf{k}}d_{\textbf{l}}}})^{S(\Bg)}.
    \end{align}
\end{itemize}

\begin{remark}
    Only a subset of fSET phases can be obtained by partially-gauging 2+1D fSPT phases. There exist a large class of fSET phases that cannot be realized by partially-gauging.
\end{remark}

\begin{remark}
    To help understanding the gauging of 2+1D fSPT phases, we give a simple fully-gauged example. Let us consider 2+1D fSPT phases with $\Z_2^f\times\Z_2$ symmetry, the classification of which is $\Z_8$\cite{wang2017interacting}, denoted by $\nu=0,...,7$. Gauging the $\Z_2$ symmetry for the $\nu=2$ phase, we obtain the fermionic toric code topological order, which can be realized by a fermionic string-net model (Example \ref{eg.ftc}). Gauging the $\Z_2$ symmetry for the $\nu=1$ phase, we obtain the Majorana toric code topological order. Bosonic-part $F$-moves in the fermionic string-net model of Majorana toric code can be obtained by firstly gauging $\Z_2^f\times\Z_2$ and then performing a fermion condensation, shown in the following commuting diagram:
    \begin{equation}
    \begin{tikzcd}
	&& {\Z_2^f\times\Z_2-\text{fSPT phase }(\nu=1)} \\
	\\
	{\text{Ising topological order}} && {\text{Majorana toric code }\text{topological order}}
	\arrow["{\text{Gauging }\Z_2^f\times\Z_2}"', from=1-3, to=3-1]
	\arrow["{\text{Gauging }\Z_2}", from=1-3, to=3-3]
	\arrow["{\text{F.c.}}", shift left, from=3-1, to=3-3]
	\arrow["{\text{Gauging }\Z_2^f}", shift left, from=3-3, to=3-1]
    \end{tikzcd}.
    \end{equation}
    We recall that the Ising topological order can be realized by a string-net model inputted by the Ising fusion category $\textbf{Ising}=\{\mathbb{1},f,\sigma\}$, where $\sigma\ot\sigma=\mathbb{1}\oplus f$. The Majorana toric code can be realized by a fermionic string-net model inputted by the super fusion category ${^\svec\textbf{Ising}}=\{\mathbb{1},\sigma\}$ (Example \ref{eg.mtc}), $F$-moves of which are listed in Ref~\onlinecite{fSN}.
\end{remark}

\begin{remark}
    {
    Consider a 2+1D $\cG\times\Z_2^f$-fSPT phase without Kitaev chain layer, i.e., it is described by $\{\nu_3,n_2\}$. It is known that gauging the $\Z_2^f$ symmetry for any of such a $\cG\times\Z_2^f$-fSPT phase gives an SET phase: a toric code with $\cG$ symmetry, called a bosonic ``shadow'' model. This shadow model can also be obtained by gauging the $\Z_2$ symmetry for an auxiliary bosonic SPT phase with symmetry $\tilde{\cG}$, where $\tilde{\cG}$ is a central extension of $\cG$ over $\Z_2$ determined by the 2-cocycle $n_2$\cite{ellison2019disentangling,chen2021disentangling}.}
\end{remark}

\end{widetext}
\section{Surface fermionic symmetry-enriched string-net models}\label{sec.fthooft}

It is well known that the boundaries of 2+1D SPT phases are either gapless or symmetry broken\cite{jiang2021generalized}. However, in 3+1D or higher dimensions, aside from the mentioned two possibilities, the boundaries can be gapped and exhibit intrinsic topological order with all symmetry preserved. That is to say, if the bulk $(n+1)$D SPT phase has symmetry $G$, where $n\geq 3$, its boundary can be an anomalous SET phase also with symmetry $G$\cite{wang2021domain}. Here we focus on exactly this case for bulk 3+1D fermionic SPT (fSPT) phases with on-site symmetry $G^f=
\mathbb{Z}_2^{f}
\times
G$ ($\omega_2$ is trivial), i.e., we study the $G^f$-preserving and gapped boundaries of 3+1D fSPT phases, which belong to 2+1D anomalous fSET phases\cite{Fid2018,cheng2019fermionic,loo2025systematic}. For such anomalous fSET phases, their symmetry actions cannot be defined on-site purely on the 2+1D surface, therefore cannot be gauged within 2+1D\cite{wang2018symmetric}. Such obstruction to gauging the symmetry is called 't Hooft anomaly\cite{cheng2023lieb}. We further distinguish 't Hooft anomalies into bosonic ones and fermionic ones, corresponding to the obstruction to gauging $G$ and gauging $\Z_2^f$ respectively. While since our exactly solvable model construction only applies to non-chiral phases, our construction limits to non-chiral 2+1D fSET phases with 't Hooft anomaly.

We recall that given a bosonic topological order whose anyon excitations are described by a unitary modular tensor category (UMTC), and given a symmetry group $G$, 2+1D bosonic $G$-SET phases are characterized by the following data\cite{BSET}: (1) How the symmetry permutes anyons, called symmetry action. (2) The fractional quantum numbers carried by the anyons under the symmetry, called symmetry fractionalization. (3) The structure of symmetry defects, where introducing symmetry defects to the system consistently mathematically corresponds to a $G$-crossed extension of the original UMTC. However, some of the symmetry fractionalization classes are not consistent with symmetry defects, which leads to anomalous symmetry fractionalization that can only exist at the surface of some 3+1D bulk. The anomaly for a given symmetry fractionalization class is specified by an obstruction function $o_4\in H^4(G,U(1)_T)$. While 3+1D bosonic SPT phases are also classified by the cohomology class $H^4(G,U(1)_T)$. So that intuitively a 2+1D SET phase with nontrivial anomaly $o_4$ should live on the surface of certain bulk 3+1D bosonic SPT phase, such that the obstruction is compensated by the bulk SPT phase.

In fermionic cases, things are getting more involved. For non-anomalous fSET phases, it is required that all symmetries can be gauged, including the fermion-parity symmetry $\mathbb{Z}_2^f$. While for anomalous phases, there can be obstruction to extending symmetry fractionalization to the fermion-parity vortices/fluxes when trying gauging the $\mathbb{Z}_2^f$ symmetry, where the anomaly is specified by an obstruction $o_3\in H^3(G,\mathbb{Z}_2)$\cite{FSET}. We recall that 3-cocycles in $H^3(G,\mathbb{Z}_2)$ describe complex fermion decoration in 3+1D group super-cohomology model\cite{Gu2014} (without Kitaev chain decoration). So that it is natural to put a 2+1D fSET phase with $ H^3(G,\mathbb{Z}_2)$ anomaly as the surface of some 3+1D fSPT phase with complex fermion decoration. Furthermore, there may even be obstruction to extending symmetry action to fermion-parity vortices when trying gauging $\mathbb{Z}_2^f$, specified by any nontrivial $o_2\in H^2(G,\mathbb{Z}_2)$. We conjecture that such anomalous state corresponds to the surface of some bulk 3+1D fSPT phase with Kitaev chain layer, as the Kitaev chain decoration is exactly described by 2-cocycles in $H^2(G,\mathbb{Z}_2)$ in the general classification of 3+1D fSPT phases\cite{Wang2018,Wang2020}. While the $o_1\in H^1(G,\mathbb{Z}_T)$ obstruction corresponds to chiral anomalous fSET phases, which are beyond the scope of our surface exactly solvable models for non-chiral phases at present.

\begin{definition}
    A 2+1D fermionic symmetry-enriched topological phase with symmetry $G^f=
    \mathbb{Z}_2^{f}
    \times_{\omega_2}
    G$ is said to have \textit{fermionic 't Hooft anomaly} if there is an obstruction to gauging $\mathbb{Z}_2^{f}$.
\end{definition}

{There is a well-known example of non-chiral surface fSET phase: $\cZ_1(\textbf{Vec}_{\mathbb{Z}_4})\boxtimes\{\mathbb{1},f\}$ ($\Z_4$ gauge theory stacking with a physical fermion $f$) with $\mathbb{Z}_2^f\times \mathbb{Z}_2\times \mathbb{Z}_4$ symmetry. It is a surface fSET phase with $H^3(G,\Z_2)$ fermionic 't Hooft anomaly, living on the surface of the root phase of bulk 3+1D $\mathbb{Z}_2^f\times \mathbb{Z}_2\times \mathbb{Z}_4$-fSPT phase. The anyon excitations (together with $G$-action and symmetry fractionalization) in this example were studied in Refs.~\onlinecite{Fid2018,cheng2019fermionic}. The classification of gapped boundary of the root phase of 3+1D $\mathbb{Z}_2^f\times \mathbb{Z}_2\times \mathbb{Z}_4$-fSPT phase by group extension (such that the group super-cohomology data $n_3\in H^3(G,\Z_2)$, $n_2\in H^2(G,\Z_2)$ are trivialized) is studied in Refs.~\onlinecite{chen2021disentangling,loo2025systematic}. While it lacks an exactly solvable model description, which we will construct as a surface fermionic symmetry-enriched string-net model in the following, summarized in Subsection \ref{summary.surf}. We will also present the fully-solved data (string types, fusion rules and surface $F$-moves) for the exactly solvable model of this example in Subsection \ref{subsec.eg}.}


\subsection{Group super-cohomology classification of bulk 3+1D fSPT phase}\label{subsec.fa}

First we briefly review the classification of bulk 3+1D fSPT phases without $p+ip$ superconductor decoration or Kitaev chain decoration and when $\omega_2$ is trivial (we do not know how to include nontrivial $\omega_2$ in the surface exactly solvable model yet), via the fixed-point wavefunction approach, or say group super-cohomology approach\cite{Gu2014,Wang2018,Wang2020}. We construct fixed-point ground state wavefunctions for 3+1D fSPT phases on a triangulated and oriented lattice with the following two layers of degrees of freedom: 
\begin{enumerate}
\item
Complex fermion layer: 
Each complex fermion $c_{ijkl}$ is decorated at the center of each tetrahedron $\langle ijkl \rangle$. Existence of a complex fermion in tetrahedron $\langle ijkl \rangle$ is specified by a 3-cocycle $n_3(\Bg_i,\Bg_j,\Bg_k,\Bg_l)\in H^3(G,\mathbb{Z}_2)$, satisfying the symmetric condition (or say, the transformation rule between homogeneous and inhomogeneous cochains):
\begin{align}\label{eq.n3sym}
    &n_3(\textbf{g}_0,\textbf{g}_1,\textbf{g}_2,
    \textbf{g}_3)
    \nonumber\\
    =&
    U(\Bg_0)n_3(\one,\Bg^{-1}_0\Bg_1,
    \Bg^{-1}_0\Bg_2,\Bg^{-1}_0\Bg_3)
    U^\dagger(\Bg_0)
    \nonumber\\
    =&
    n_3(\Bg^{-1}_0\Bg_1,\Bg^{-1}_0\Bg_2,
    \Bg^{-1}_0\Bg_3),
\end{align}
and the 3-cocycle condition (recall Convention \ref{conv.cochain}):
\begin{align}
    \dd n_3(01234)
    =&n_3(1234)+n_3(0234)+n_3(0134)
    \nonumber\\
    &+n_3(0124)+n_3(0123)
    \nonumber\\
    =&0 \text{ (mod 2)}.
\end{align}
\item Bosonic SPT layer: $\vert G \vert$ level bosonic state $\vert g_i \rangle$ ($g_i\in G$) on each vertex $i$. A 3+1D 2-3 move (as a retriangulation or renormalization) is specified by a 4-cochain $\nu_4(\Bg_i,\Bg_j,\Bg_k,\Bg_l,\Bg_m)\in C^4(G,U_{T}(1))/B^4(G,U_{T}(1))$, satisfying the symmetric condition:
\begin{align}\label{eq.nu4sym}
    &\nu_4(\textbf{g}_0,\textbf{g}_1,\textbf{g}_2,
    \textbf{g}_3,\textbf{g}_4)
    \nonumber\\
    =&
    U(\Bg_0)\nu_4(\one,\Bg^{-1}_0\Bg_1,
    \Bg^{-1}_1\Bg_2,\Bg^{-1}_2\Bg_3,\Bg^{-1}_3\Bg_4)
    U^\dagger(\Bg_0)
    \nonumber\\
    =&
    \nu_4(\Bg^{-1}_0\Bg_1,\Bg^{-1}_0\Bg_2,
    \Bg^{-1}_0\Bg_3,\Bg^{-1}_0\Bg_4)^{S(\Bg_0)},
\end{align}
and the twisted 4-cocycle condition:
\begin{align}
    \dd \nu_4(012345)
    =& \frac{\nu_4(12345)\nu_4(01345)\nu_4(01235)}{\nu_4(02345)\nu_4(01245)\nu_4(01234)}
    \nonumber\\
    =&(-1)^{n_3\smile_1 n_3(012345)}.
\end{align}
\end{enumerate}
In this simple case, 3+1D fSPT phases are classified by the doublet $(n_3,\nu_4)$.

\begin{widetext}
\subsection{Surface renormalization moves}
\label{subsec.bbrenorm}
Given a global symmetry group $G^f=Z_2^f\times G$ ($\omega_2$ is trivial), a bulk 3+1D fSPT phase specified by $(n_3,\nu_4)$, and a surface/anomalous 2+1D fSET phase specified by a set of $G$-graded string types $a_\Bg,b_\Bh,...$ with given fusion rules (the $G$-graded string types no longer form a $G$-graded super fusion category as the fermionic pentagon equation is obstructed, as we will see below), we define the surface fermionic symmetry-enriched string-net model in this Section. gfSLU transformations in this bulk-boundary system are not well-defined purely on the surface, 
but only well-defined if considering the corresponding bulk transformation and the surface transformation as a whole. A renormalization move is a special kind of gfSLU transformation that changes the state configuration or the number of degrees of freedom. Therefore, renormalization moves on surface are also well-defined only when their corresponding bulk renormalization moves are put together. Below let $\Psi$ denote for the fixed-point ground state wavefunction of the whole bulk-boundary system. We define the \textit{surface $\mathcal{F}$-move} and the \textit{bulk $\widetilde{\mathcal{F}}$-move} together as a renormalization move, 
called \textit{bulk-boundary $\cF$-move}, which is also a gfSLU transformation:
\begin{align}\label{F1}
    \Psi\begin{pmatrix} \includegraphics[scale=.25]{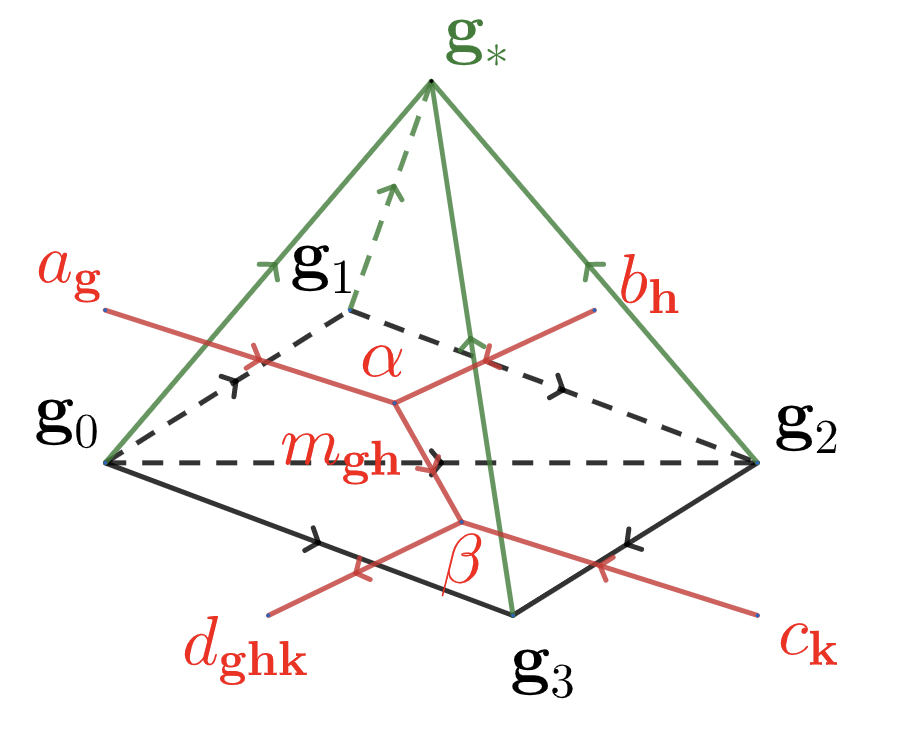} \end{pmatrix}
    =
    \sum_{n_{\Bh\Bk}\chi\delta}
    {^{\textbf{g}_0}[{\mathcal{F}^{a_{\textbf{g}}b_{\textbf{h}}c_{\textbf{k}}}
    _{d}]^{m,\alpha\beta}_{n,\chi\delta}}}\cdot
    {^{\textbf{g}_0}[\widetilde{\mathcal{F}}^{\textbf{g}\textbf{h}\textbf{k}}]
    (\textbf{g}_{*})}
    \Psi
    \begin{pmatrix} \includegraphics[scale=.25]{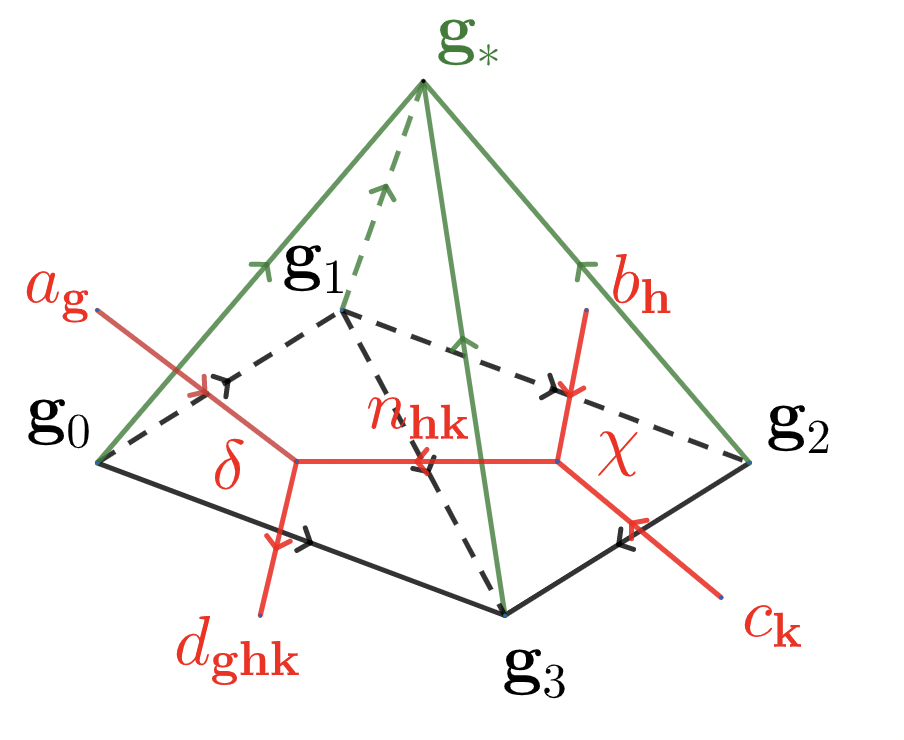} \end{pmatrix},
\end{align}
where ${^{\textbf{g}_0}[{\mathcal{F}^{a_{\textbf{g}}b_{\textbf{h}}c_{\textbf{k}}}
_{d}]^{m,\alpha\beta}_{n,\chi\delta}}}$ is the surface $\cF$-move, it transforms on the surface $G$-graded string types and fusion states (colored in red), and the bulk $\widetilde{\mathcal{F}}$-move ${^{\textbf{g}_0}[\widetilde{\mathcal{F}}^{\textbf{g}\textbf{h}\textbf{k}}]
(\textbf{g}_{*})}$ transforms on the fixed-point ground states of the bulk 3+1D fSPT phase (colored in green and black). To obtain an explicit expression for the bulk move ${^{\textbf{g}_0}[\widetilde{\mathcal{F}}^{\textbf{g}\textbf{h}\textbf{k}}]
(\textbf{g}_{*})}$, we put an auxiliary vertex 
$e$ with trivial group element $\Bg_e=\one$ on the other side of the bulk. We note that this side of bulk (colored in blue below) is virtual, and this trivial vertex $\Bg_e=\one$ is invariant under any symmetry action, i.e., it is always trivial. We find that the above bulk-boundary renormalization move can be divided into the following three steps:
\begin{align}
\Psi\begin{pmatrix} \includegraphics[scale=.35]{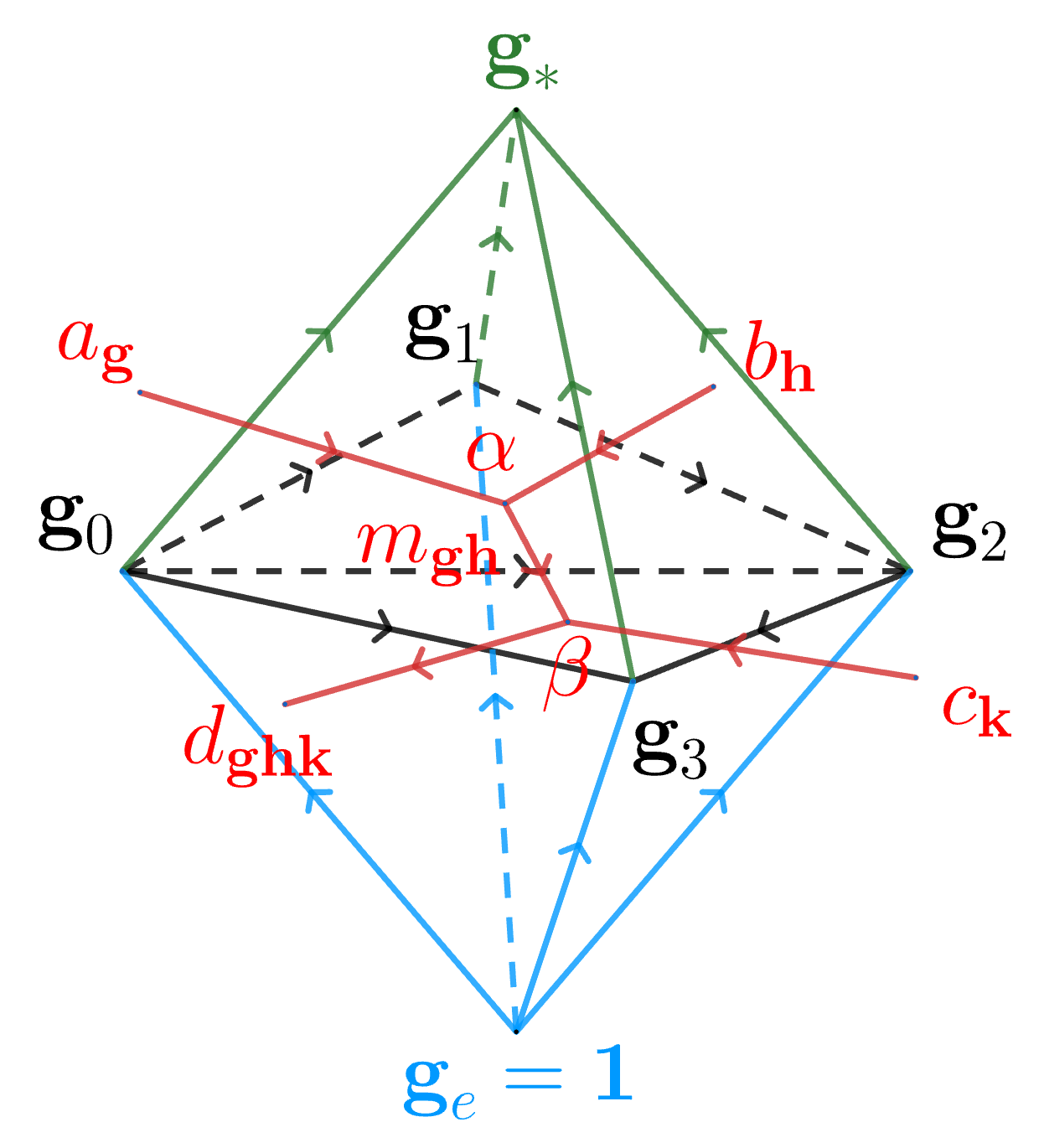} \end{pmatrix}
=&
\textcolor{Green}{\widetilde{\mathcal{F}}}(0123*)
\Psi
\begin{pmatrix} \includegraphics[scale=.35]{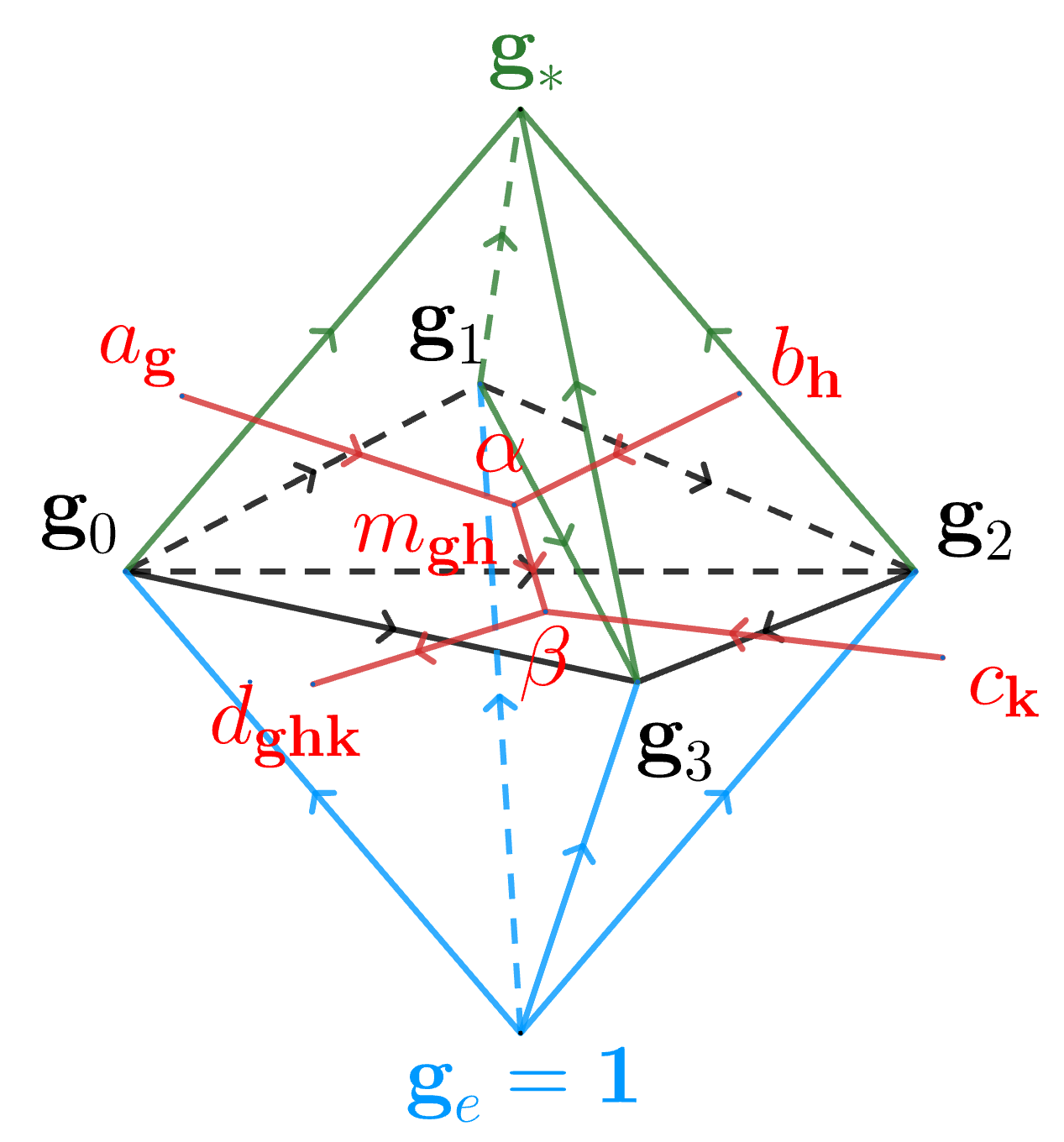} \end{pmatrix}
\nonumber\\
=&
\sum_{n_{\Bh\Bk}\chi\delta}
\textcolor{Green}{\widetilde{\mathcal{F}}}
(0123*)
c^{\dagger s(\alpha)}_{\underline{\alpha}}
c^{\dagger s(\beta)}_{\underline{\beta}}
c^{s(\delta)}_{\underline{\delta}}
c^{s(\chi)}_{\underline{\chi}}
{^{\textbf{g}_0}[{F^{a_{\textbf{g}}b_{\textbf{h}}c_{\textbf{k}}}
_{d}]^{m,\alpha\beta}_{n,\chi\delta}}}
\Psi
\begin{pmatrix} \includegraphics[scale=.35]{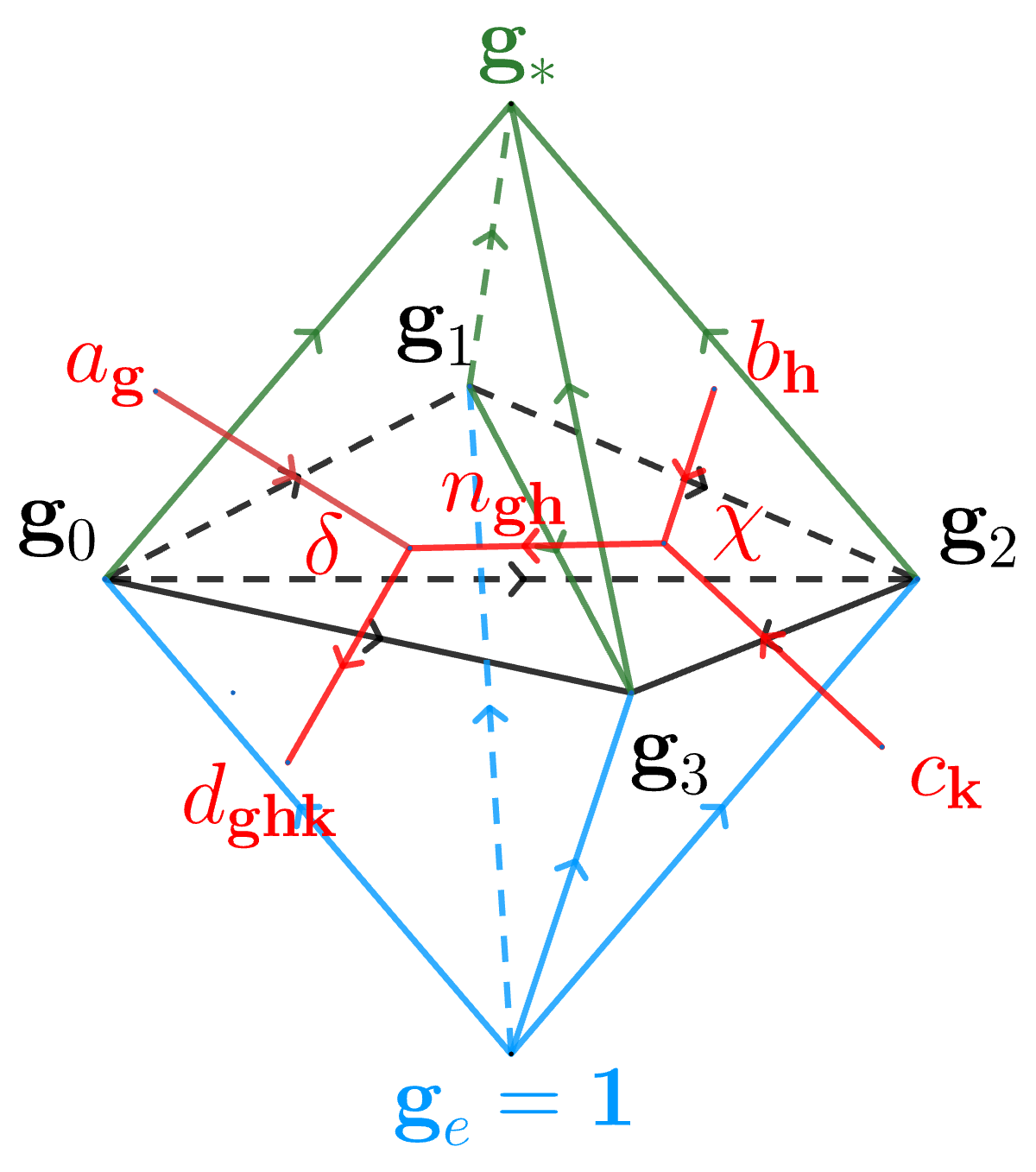} \end{pmatrix}
\nonumber\\
=&
\sum_{n_{\Bh\Bk}\chi\delta}
\textcolor{Green}{\widetilde{\mathcal{F}}}(0123*)
c^{\dagger s(\alpha)}_{\underline{\alpha}}
c^{\dagger s(\beta)}_{\underline{\beta}}
c^{s(\delta)}_{\underline{\delta}}
c^{s(\chi)}_{\underline{\chi}}
{^{\textbf{g}_0}[{F^{a_{\textbf{g}}b_{\textbf{h}}c_{\textbf{k}}}
_{d}]^{m,\alpha\beta}_{n,\chi\delta}}}
\textcolor{blue}{\widetilde{\mathcal{F}}}(e0123)^{-1}
\Psi
\begin{pmatrix} \includegraphics[scale=.35]{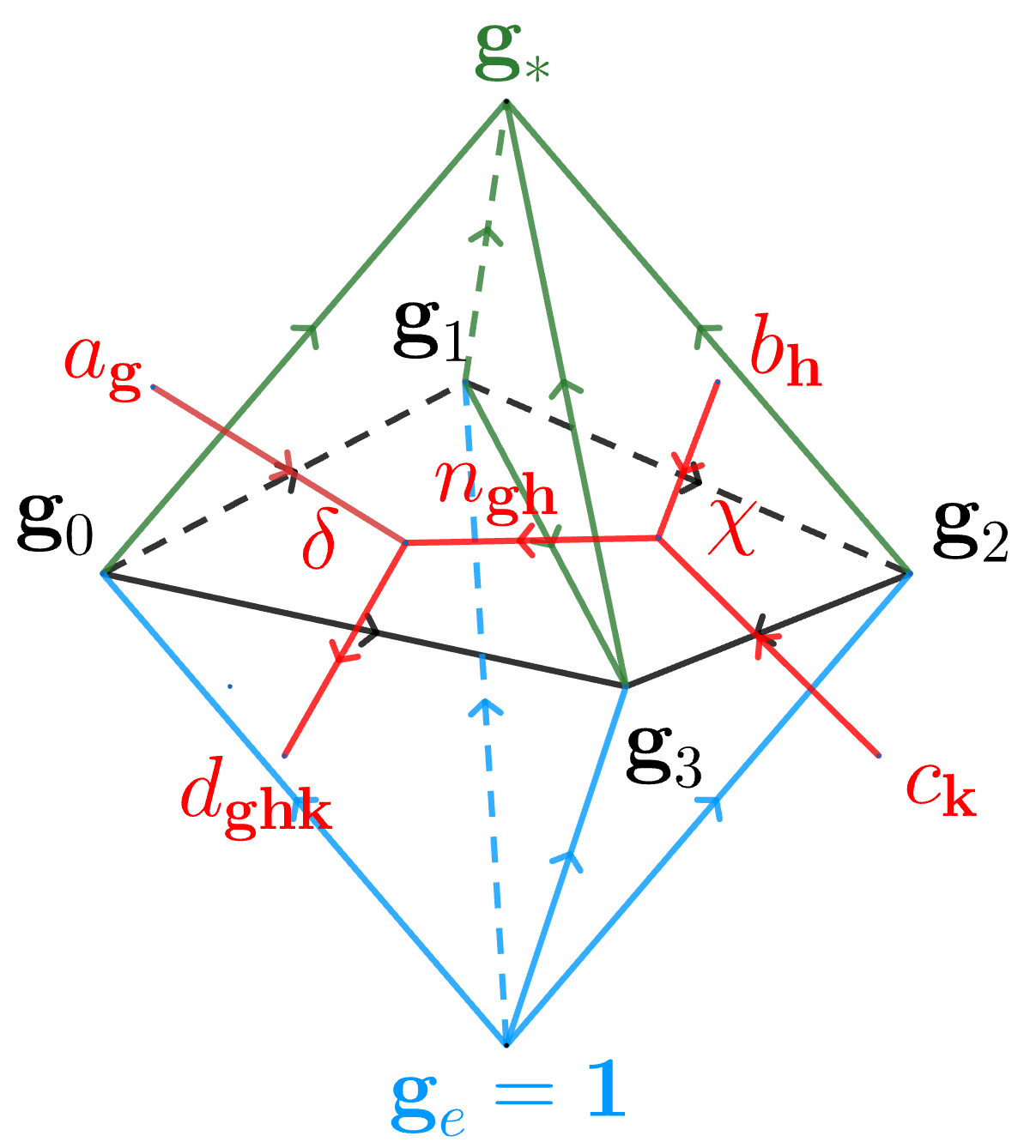} \end{pmatrix},
\label{aF1}
\end{align}
where $s(\alpha)+s(\beta)+s(\delta)+s(\chi)=0$ (mod 2), $c_{\underline{\alpha}}$ and $c_{\underline{\alpha}}^\dagger$ are surface fermion annihilation and creation operators on surface vertex $\underline{\alpha}$, and $\textcolor{Green}{\widetilde{\mathcal{F}}}(0123*)$ is the 2-3 move on the real bulk and $\textcolor{blue}{\widetilde{\mathcal{F}}}(e0123)^{-1}$ is the inverse 2-3 move on the virtual bulk. The 2-3 moves are explicitly expressed as\cite{Wang2018}
\begin{align}
\textcolor{Green}{\widetilde{\mathcal{F}}}(0123*)
=&
\nu_4(0123*)
C_{012*}^{\dagger n_3(012*)}
C_{023*}^{\dagger n_3(023*)}
C_{0123}^{n_3(0123)}
C_{013*}^{n_3(013*)}
C_{123*}^{n_3(123*)}
,
\label{bulk1}
\end{align}
\begin{align}
\textcolor{blue}{\widetilde{\mathcal{F}}}(e0123)^{-1}
=&
\nu_4(e0123)^{-1}
C_{0123}^{\dagger n_3(0123)}
C_{e023}^{\dagger n_3(e023)}
C_{e012}^{\dagger n_3(e012)}
C_{e123}^{n_3(e123)}
C_{e013}^{n_3(e013)},
\label{bulk2}
\end{align}
where $\nu_4(ijklm):=\nu_4(\Bg_i,\Bg_j,\Bg_k,\Bg_l,\Bg_m)$ (recall Convention \ref{conv.cochain}) is the bosonic-part of the 2-3 move, $C_{ijkl}$ and $C_{ijkl}^\dagger$ are bulk fermion annihilation and creation operators in tetrahedron $\la ijkl \ra$, and $n_3(ijkl):=n_3(\Bg_i,\Bg_j,\Bg_k,\Bg_l)=0,1$ specifies the fermion-parity in $\la ijkl \ra$. Let $\textbf{g}=\Bg^{-1}_0\textbf{g}_1$, $\textbf{h}=\textbf{g}_{1}^{-1}\textbf{g}_2$, $\textbf{k}=\textbf{g}_{2}^{-1}\textbf{g}_3$ and $\textbf{m}=\textbf{g}_{3}^{-1}\textbf{g}_*$. Then eliminating $C_{0123}^{\dagger n_3(0123)}$ with $C_{0123}^{n_3(0123)}$, the bulk-boundary $\cF$-move has the following expression:
\begin{align}\label{eq.bbfexp}
    {^{\textbf{g}_0}[{\mathcal{F}^{a_{\textbf{g}}b_{\textbf{h}}c_{\textbf{k}}}
    _{d}]^{m,\alpha\beta}_{n,\chi\delta}}}
    {^{\textbf{g}_0}[\widetilde{\mathcal{F}}^{\textbf{g}\textbf{h}\textbf{k}}]
    (\textbf{g}_{*})}
    =&
    (-1)^{n_3(0123)(n_3(013*)+n_3(123*))}
    C_{012*}^{\dagger n_3(012*)}
    C_{023*}^{\dagger n_3(023*)}
    C_{013*}^{n_3(013*)}
    C_{123*}^{n_3(123*)}
    \nonumber\\&
    C_{e023}^{\dagger n_3(e023)}
    C_{e012}^{\dagger n_3(e012)}
    C_{e123}^{n_3(e123)}
    C_{e013}^{n_3(e013)}
    \frac{\nu_4(\Bg,\Bh,\Bk,\Bm)^{S(\Bg_0)}}{\nu_4(\Bg_0,\Bg,\Bh,\Bk)}
    \nonumber\\&
    c^{\dagger s(\alpha)}_{\underline{\alpha}}
    c^{\dagger s(\beta)}_{\underline{\beta}}
    c^{s(\delta)}_{\underline{\delta}}
    c^{s(\chi)}_{\underline{\chi}}
    {^{\textbf{g}_0}
    [{F^{a_{\textbf{g}}b_{\textbf{h}}c_{\textbf{k}}}
    _{d}]^{m,\alpha\beta}_{n,\chi\delta}}},
\end{align}
where all surface fermion operators commute with bulk fermion operators, e.g., $c^\dagger_{\underline{\alpha}}C_{123*}=C_{123*}c^\dagger_{\underline{\alpha}}$, and the homogeneous $\nu_4$ are changed to inhomogeneous ones by Eq.~(\ref{eq.nu4sym}).

\begin{remark}
    We can obtain the same expression of bulk-boundary $\cF$-move in Eq.~(\ref{eq.bbfexp}) by another approach called partition function approach, shown in Appendix \ref{appen.part}.
\end{remark}

\subsubsection{\texorpdfstring{$H^3(G,\Z_2)$}{} fermionic 't Hooft anomaly}
Let us carefully examine Eq.~(\ref{eq.bbfexp}) that which terms belong to the boundary and which belong to the bulk. Recall Eq.~(\ref{F1}), the bulk $\widetilde{\mathcal{F}}$-move has expression:
\begin{align}\label{eq.bulkfmove}
    {^{\textbf{g}_0}[\widetilde{\mathcal{F}}^{\textbf{g}\textbf{h}\textbf{k}}]
    (\textbf{g}_{*})}
    =
    (-1)^{n_3(0123)(n_3(013*)+n_3(123*))}
    C_{012*}^{\dagger n_3(012*)}
    C_{023*}^{\dagger n_3(023*)}
    C_{013*}^{n_3(013*)}
    C_{123*}^{n_3(123*)}
    \frac{\nu_4(\Bg,\Bh,\Bk,\Bm)^{S(\Bg_0)}}{\nu_4(\Bg_0,\Bg,\Bh,\Bk)}.
\end{align}
While the surface $\cF$-move ${^{\textbf{g}_0}[{\mathcal{F}^{a_{\textbf{g}}b_{\textbf{h}}c_{\textbf{k}}}
_{d}]^{m,\alpha\beta}_{n,\chi\delta}}}$ should involve the fermion operators from the virtual bulk, i.e., fermion operators on tetrahedrons $\la e023 \ra$, $\la e012 \ra$, $\la e123 \ra$ and $\la e013 \ra$, and therefore it generally may not preserve fermion-parity, with the following expression:
\begin{align}\label{eq.surfacefmove}
    ^{\textbf{g}_0}[{\mathcal{F}^{a_{\textbf{g}}b_{\textbf{h}}c_{\textbf{k}}}
    _{d}]
    ^{m,\alpha\beta}
    _{n,\chi\delta}}
    =
    c^{\dagger s(\alpha)}_{\underline{\alpha}}
    c^{\dagger s(\beta)}_{\underline{\beta}}
    c^{s(\delta)}_{\underline{\delta}}
    c^{s(\chi)}_{\underline{\chi}}
    C_{\underline{\beta}}^{\dagger n_3(e023)}
    C_{\underline{\alpha}}^{\dagger n_3(e012)}
    C_{\underline{\chi}}^{n_3(e123)}
    C_{\underline{\delta}}^{n_3(e013)}
    {^{\textbf{g}_0}[{F^{a_{\textbf{g}}b_{\textbf{h}}c_{\textbf{k}}}
    _{d}]^{m,\alpha\beta}_{n,\chi\delta}}},
\end{align}
where $C^\dagger_{\underline{\alpha}}:=C^\dagger_{e012}$, $C^\dagger_{\underline{\beta}}:=C^\dagger_{e023}$, $C_{\underline{\delta}}:=C_{e013}$, $C_{\underline{\chi}}:=C_{e123}$, which are understood as surface fermions that are pumped from the bulk (therefore they still satisfy the anti-commutation relation of bulk fermions, and commute with all surface fermions). 

\begin{definition}[Total surface fermion-parity function]
    {The total surface fermion-parity function for each fusion state $\alpha$ in Graph (\ref{graph.surfacefusion}) is defined as}
    \begin{equation}
        {s'(\Bg_0,\alpha)=
        s(\alpha)+n_3(\Bg_0,\Bg,\Bh).}
    \end{equation}
\end{definition}

Then for the surface $\cF$-move above, we have $s'(\Bg_0,\alpha)=s(\alpha)+n_3(e012)$, $s'(\Bg_0,\beta)=s(\beta)+n_3(e023)$, $s'(\Bg_0,\delta)=s(\delta)+n_3(e013)$, $s'(\Bg_0\Bg,\chi)=s(\chi)+n_3(e123)$, and they satisfy
\begin{align}\label{eq.n3anomaly}
    s'(\Bg_0,\alpha)+s'(\Bg_0,\beta)+s'(\Bg_0,\delta)
    +s'(\Bg_0\Bg,\chi)\text{ (mod 2)}=n_3(0123)\equiv n_3(\Bg,\Bh,\Bk),
\end{align}
i.e., \textit{the surface $\cF$-move in general violates the fermion-parity conservation by $n_3(\Bg,\Bh,\Bk)\in H^3(G,\Z_2)$}. This violation of fermion-parity conservation on surface $\cF$-move exactly counts for the $H^3(G,\Z_2)$ fermionic 't Hooft anomaly.

\begin{remark}
    Since 't Hooft anomalies are anomalies on symmetry actions, in our surface fermionic symmetry-enriched string-net model, string types $a_\one,b_\one,...$ with trivial group element grading still form a super fusion category $\cS_\one$, i.e., $\cS_\one=\{a_\one,b_\one,...\}$.
\end{remark}

\begin{remark}\label{rmk.n3o3}
    In the context of excitations/anyons on surface. 
    Let $o_3\in H^3(G,\Z_2)$ be the obstruction on extending symmetry fractionalization from $\mathcal{Z}_1({\cS_\one})$ to $\mathcal{Z}_1(\cC_\one)$. The $n_3$ obstruction on fermion-parity conservation for a surface $\cF$-move matches with the $o_3$ anomaly, i.e.,
    \begin{equation}
        [o_3]=[n_3],
    \end{equation}
    where $[\ ]$ denotes for the cohomology class.
\end{remark}

A surface string fusion is graphically represented as
\begin{equation}\label{graph.surfacefusion}
    \includegraphics[scale=.35]{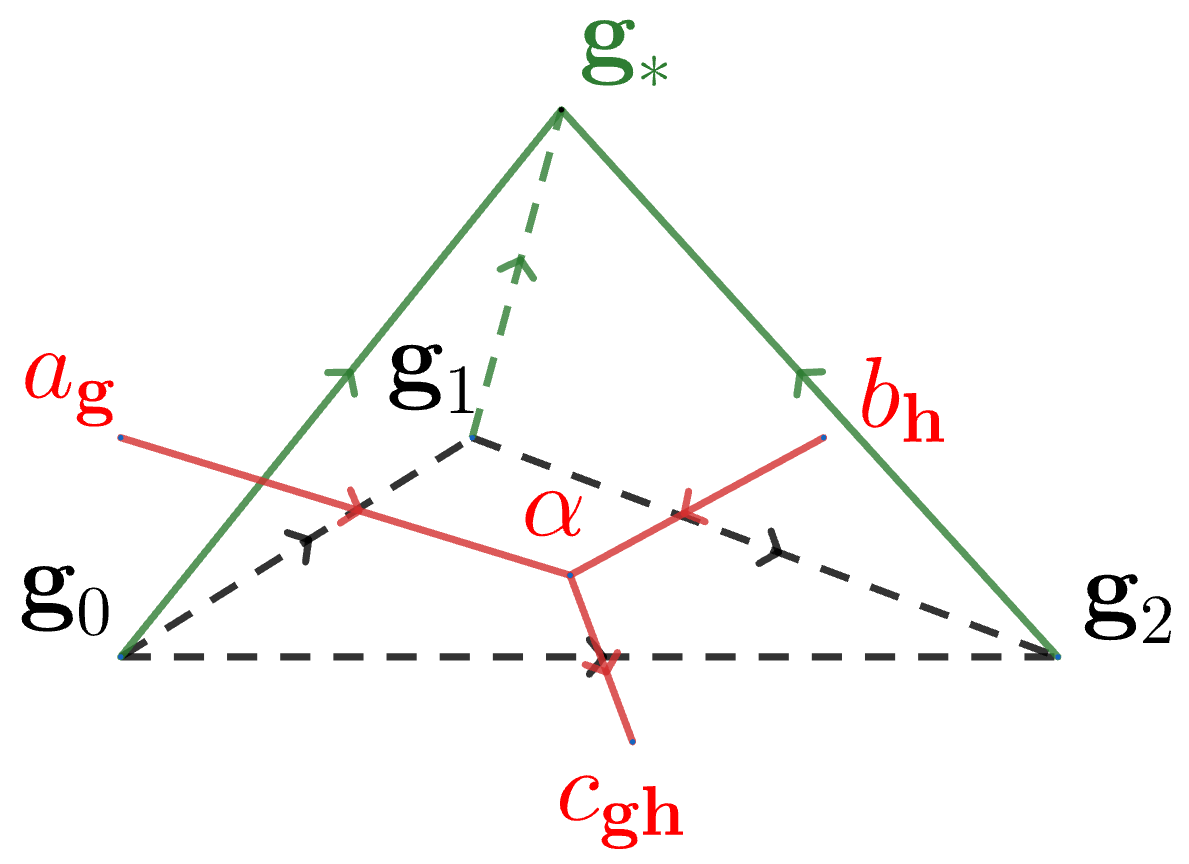}.
\end{equation}

The surface string fusion rules now depend on the leftmost group element $\Bg_0$:
\begin{equation}
    {^{\Bg_0}[a_\Bg\otimes b_\Bh]}=
    \left\{\begin{array}{l}
    \underset{{c_{\textbf{g}\textbf{h}} }}{\bigoplus}
    \C^{B^{a_{\textbf{g}}b_{\textbf{h}}}
    _{c_{\textbf{g}\textbf{h}}}|
    F^{a_{\textbf{g}}b_{\textbf{h}}}
    _{c_{\textbf{g}\textbf{h}}}}
    c_{\textbf{g}\textbf{h}}
    \text{, \ \ if }n_3(\Bg_0,\Bg,\Bh)=0\\ 
    \underset{{c_{\textbf{g}\textbf{h}} }}{\bigoplus}
    \C^{F^{a_{\textbf{g}}b_{\textbf{h}}}
    _{c_{\textbf{g}\textbf{h}}}|
    B^{a_{\textbf{g}}b_{\textbf{h}}}
    _{c_{\textbf{g}\textbf{h}}}}
    c_{\textbf{g}\textbf{h}}
    \text{, \ \ if }n_3(\Bg_0,\Bg,\Bh)=1
    \end{array}\right.,
\end{equation}
where when $n_3(e012):=n_3(\Bg_0,\Bg,\Bh)=1$, a bulk fermion is attached to the surface and the surface fermion-parity on vertex $\underline{\alpha}$ is changed, and thereby the original bosonic fusion states become fermionic, while the original fermionic fusion states become bosonic.



\subsubsection{'t Hooft anomalous fermionic pentagon equation}
The consistency equation for the bulk-boundary $\cF$-move in Eq.~(\ref{F1}) is a surface pentagon equation together with the corresponding bulk moves:
\begin{align}\label{eq.ferano}
&\underset{\epsilon}{\sum} 
{^{\Bg_0}[{\mathcal{F}^{j_{\textbf{gh}}c_{\textbf{k}}d_{\textbf{l}}}_{e}]^{m,\beta\chi}_{q,\delta\epsilon}}}
{^{\Bg_0}[\widetilde{\mathcal{F}}^{(\textbf{g}\textbf{h})\textbf{k}\textbf{l}}]
(\textbf{g}_{*})}
{^{\Bg_0}[{\mathcal{F}^{a_{\textbf{g}}b_{\textbf{h}}q_{\textbf{kl}}}_{e}]^{j,\alpha\epsilon}_{p,\phi\gamma}}}
{^{\Bg_0}[\widetilde{\mathcal{F}}^{\textbf{g}\textbf{h}(\textbf{k}\textbf{l})}]
(\textbf{g}_{*})}
\nonumber\\
=&
\underset{n_{\Bh\Bk}\eta\psi\kappa}{\sum} 
{^{\Bg_0}[{\mathcal{F}^{a_{\textbf{g}}b_{\textbf{h}}c_{\textbf{k}}}_{m}]^{j,\alpha\beta}_{n,\eta\psi}}}
{^{\Bg_0}[\widetilde{\mathcal{F}}^{\textbf{g}\textbf{h}\textbf{k}}]
(\textbf{g}_{*})}
{^{\Bg_0}[{\mathcal{F}^{a_{\textbf{g}}n_{\textbf{hk}}d_{\textbf{l}}}_{e}]^{m,\psi\chi}_{p,\kappa\gamma}}}
{^{\Bg_0}[\widetilde{\mathcal{F}}^{\textbf{g}(\textbf{h}\textbf{k})\textbf{l}}]
(\textbf{g}_{*})}
{^{\Bg_0\textbf{g}}[{\mathcal{F}^{b_{\textbf{h}}c_{\textbf{k}}d_{\textbf{l}}}_{p}]^{n,\eta\kappa}_{q,\delta\phi}}}
{^{\Bg_0\textbf{g}}[\widetilde{\mathcal{F}}^{\textbf{h}\textbf{k}\textbf{l}}]
(\textbf{g}_{*})}
.
\end{align}
Using the expression in Eq.~(\ref{aF1}), eliminating the fermion creation and annihilation operators, we find that the above equation splits into the following two equations: 
\begin{align}\label{eq.doublehex}
    \frac{\nu_4(0234*)}{\nu_4(e0234)}
    \frac{\nu_4(0124*)}{\nu_4(e0124)}
    =
    \frac{(-1)^{n_3\smile_1 n_3(01234*)}}
    {(-1)^{n_3\smile_1 n_3(e01234)}}
    \frac{\nu_4(0123*)}{\nu_4(e0123)}
    \frac{\nu_4(0134*)}{\nu_4(e0134)}
    \frac{\nu_4(1234*)}{\nu_4(e1234)},
\end{align}
\begin{align}\label{eq.ano}
    \underset{\epsilon}{\sum} 
    {^{\Bg_0}[{F^{j_{\textbf{gh}}c_{\textbf{k}}d_{\textbf{l}}}_{e}]^{m,\beta\chi}_{q,\delta\epsilon}}}
    {^{\Bg_0}[{F^{a_{\textbf{g}}b_{\textbf{h}}q_{\textbf{kl}}}_{e}]^{j,\alpha\epsilon}_{p,\phi\gamma}}}
    =
    (-1)^{s(\alpha)s(\delta)}
    \underset{n_{\Bh\Bk}\eta\psi\kappa}{\sum} 
    {^{\Bg_0}[{F^{a_{\textbf{g}}b_{\textbf{h}}c_{\textbf{k}}}_{m}]^{j,\alpha\beta}_{n,\eta\psi}}}
    {^{\Bg_0}[{F^{a_{\textbf{g}}n_{\textbf{hk}}d_{\textbf{l}}}_{e}]^{m,\psi\chi}_{p,\kappa\gamma}}}
    ^{\Bg_0\textbf{g}}[{F^{b_{\textbf{h}}c_{\textbf{k}}d_{\textbf{l}}}_{p}]^{n,\eta\kappa}_{q,\delta\phi}},
\end{align}
where we see that Eq.~(\ref{eq.doublehex}) is a combination of two hexagon equations\cite{Wang2020} (one for the real bulk, one for the virtual bulk) of 3+1D fSPT phases by adding a trivial term $\frac{\nu_4(01234)}{\nu_4(01234)}=1$, and Eq.~(\ref{eq.ano}) is the surface fermionic pentagon equation.

    


The symmetric condition on the bulk-boundary $\cF$-move requires the following commutative diagram:
\begin{equation}
    \includegraphics[scale=.42]{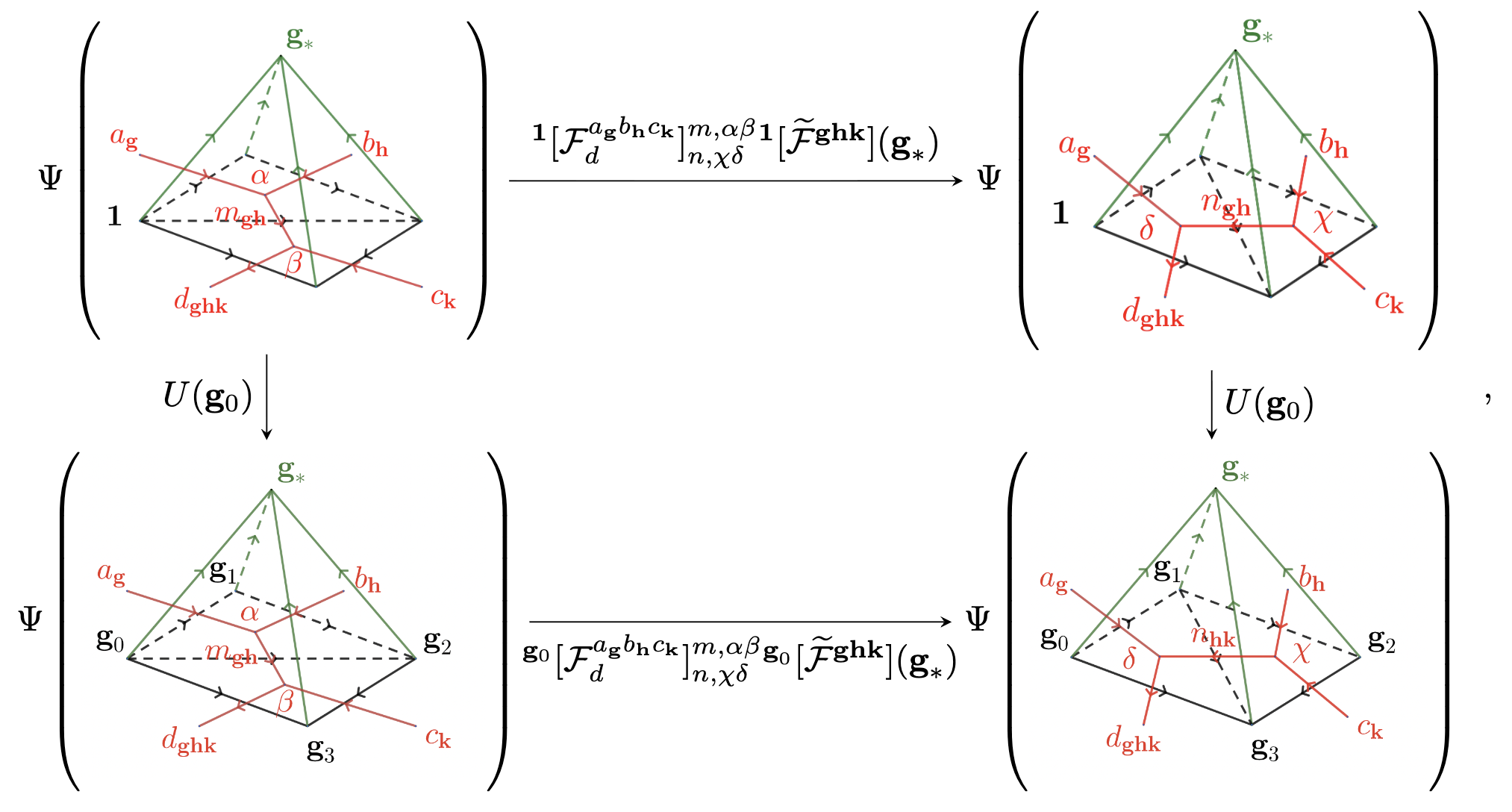}
\end{equation}
where by the expression of bulk $\widetilde{\cF}$-move in Eq.~(\ref{eq.bulkfmove}) and recall that $n_3$ is invariant under symmetry action as written in Eq.~(\ref{eq.n3sym}), we have $
    {^{\textbf{g}_0}
    [\widetilde{\mathcal{F}}^
    {\textbf{g}\textbf{h}\textbf{k}}]
    (\textbf{g}_{*})}
    =\nu_4(\Bg_0,\Bg,\Bh,\Bk)^{-1}\cdot
    {^{\one}[\widetilde{\mathcal{F}}^{\textbf{g}\textbf{h}\textbf{k}}]
    (\textbf{g}_{*})}$. 
It implies
\begin{align}\label{eq.fsymcond}
    {^{\textbf{g}_0}[{\mathcal{F}^{a_{\textbf{g}}b_{\textbf{h}}c_{\textbf{k}}}
    _{d}]^{m,\alpha\beta}_{n,\chi\delta}}}
    =
    \nu_4(\Bg_0,\Bg,\Bh,\Bk)
    C_{e023}^{\dagger n_3(e023)}
    C_{e012}^{\dagger n_3(e012)}
    C_{e123}^{n_3(e123)}
    C_{e013}^{n_3(e013)}
    {^{\one}[{\mathcal{F}^{a_{\textbf{g}}b_{\textbf{h}}c_{\textbf{k}}}
    _{d}]^{m,\alpha\beta}_{n,\chi\delta}}},
\end{align}
where  
\begin{align}\label{eq.symaction}
    {^{\Bg_0}[{F^{a_{\textbf{g}}b_{\textbf{h}}c_{\textbf{k}}}
    _{d}]^{m,\alpha\beta}_{n,\chi\delta}}}
    =
    {^{\Bg_0}
    \Theta(a_{\textbf{g}},b_{\textbf{h}},c_{\textbf{k}})}
    {\nu_4 (\textbf{g}_0,\textbf{g},\textbf{h},\textbf{k})}
    ({^{\one}[{F^{a_{\textbf{g}}b_{\textbf{h}}c_{\textbf{k}}}
    _{d}]^{m,\alpha\beta}_{n,\chi\delta}}})^{S(\Bg_0)}.
\end{align}
Here the phase factor ${^{\Bg_0}\Theta(a_{\textbf{g}},b_{\textbf{h}},c_{\textbf{k}})}$, called \textit{fermionic obstruction}, is defined as a solution of the following equation:
\begin{align}\label{eq.omega}
    \frac{{^{\Bg_0}\Theta(j_{\Bg\Bh},c_\Bk,d_\Bl)}{^{\Bg_0}\Theta(a_\Bg,b_\Bh,q_{\Bk\Bl})}}
    {{^{\Bg_0}\Theta(a_{\textbf{g}},b_{\textbf{h}},c_{\textbf{k}}}){^{\Bg_0}\Theta(a_{\textbf{g}},n_{\textbf{hk}},d_{\textbf{l}}}){^{\Bg_0}\Theta(b_{\textbf{h}},c_{\textbf{k}},d_{\textbf{l}}})}
    =& \dd \nu_4(\Bg_0,\Bg,\Bh,\Bk,\Bl),
\end{align}
together with a required property ${^{\Bg_0}\Theta(a_\Bg,b_\Bh,c_\Bk)}{^{\Bg}\Theta(a_\Bg,b_\Bh,c_\Bk)}^{S(\Bg_0)}={^{\Bg_0\Bg}\Theta(a_\Bg,b_\Bh,c_\Bk)}$. We check in Appendix \ref{appen.check} that the anomalous symmetry action in Eq.~(\ref{eq.symaction}) and the surface fermionic pentagon equation in Eq.~(\ref{eq.ano}).

\begin{remark}
    The phase factor ${^{\Bg_0}\Theta(a_{\textbf{g}},b_{\textbf{h}},c_{\textbf{k}})}$ counts for the contribution of the fermion creation and annihilation operators $C_{e023}^{\dagger n_3(e023)}
    C_{e012}^{\dagger n_3(e012)}
    C_{e123}^{n_3(e123)}
    C_{e013}^{n_3(e013)}$ from the bulk 3+1D fSPT phase appearing in Eq.~(\ref{eq.fsymcond}), by first eliminating all fermion operators in the consistency equation in Eq.~(\ref{eq.ferano}) and obtaining total phase factor $\dd \nu_4(e01234):=\dd \nu_4(\Bg_0,\Bg,\Bh,\Bk,\Bl)$, and then distributing the total phase factor for each surface $F$-move, which is the meaning of Eq.~(\ref{eq.omega}). We show in Appendix~\ref{appen.d2} that only with this phase factor ${^{\Bg_0}\Theta(a_{\textbf{g}},b_{\textbf{h}},c_{\textbf{k}})}$, the anomalous symmetry condition in Eq.~(\ref{eq.symaction}) is consistent with the surface fermionic pentagon equation in Eq.~(\ref{eq.ano}).
\end{remark}

Replacing Eq.~(\ref{eq.symaction}) into Eq.~(\ref{eq.ano}) and let $\Bg_0=\one$, we obtained the so-called \textit{'t Hooft anomalous fermionic pentagon equation}:
\begin{align}\label{eq.anopenta}
    \underset{\epsilon}{\sum} 
    {^{\textbf{1}}[{F^{j_{\textbf{gh}}c_{\textbf{k}}d_{\textbf{l}}}_{e}]^{m,\beta\chi}_{q,\delta\epsilon}}}
    {^{\textbf{1}}[{F^{a_{\textbf{g}}b_{\textbf{h}}q_{\textbf{kl}}}_{e}]^{j,\alpha\epsilon}_{p,\phi\gamma}}}
    =&
    {^{\Bg}\Theta(b_\Bh,c_\Bk,d_\Bl)}
    {\nu_4 (\textbf{g},\textbf{h},\textbf{k},\textbf{l})}
    (-1)^{s(\alpha)s(\delta)}
    \nonumber\\&
    \underset{n_{\Bh\Bk}\eta\psi\kappa}{\sum} 
    {^{\textbf{1}}[{F^{a_{\textbf{g}}b_{\textbf{h}}c_{\textbf{k}}}_{m}]^{j,\alpha\beta}_{n,\eta\psi}}}
    {^{\textbf{1}}[{F^{a_{\textbf{g}}n_{\textbf{hk}}d_{\textbf{l}}}_{e}]^{m,\psi\chi}_{p,\kappa\gamma}}}
    ({^{\textbf{1}}[{F^{b_{\textbf{h}}c_{\textbf{k}}d_{\textbf{l}}}_{p}]^{n,\eta\kappa}_{q,\delta\phi}}})^{S(\textbf{g})},
\end{align}
where ${^{\Bg}\Theta(b_\Bh,c_\Bk,d_\Bl)}$ is the fermionic obstruction, ${\nu_4 (\textbf{g},\textbf{h},\textbf{k},\textbf{l})}$ is hereby called the bosonic obstruction.

\begin{remark}
    Other types of bulk-boundary renormalization moves, such as bulk-boundary $\cO$-move, $\cY$-move, $\cH$-move and so on can be defined following a similar spirit combining this Subsection and Appendix \ref{appen.othermove} and omitted here.
\end{remark}

\begin{remark}
    Applying Eq.~(\ref{eq.symaction}) in the last $F$-move in Eq.~(\ref{eq.ano}), full 't Hooft anomalous fermionic pentagon equations for surface $F$-moves with general leftmost group element $\Bg_0$ are
\begin{align}\label{eq.redun}
    \underset{\epsilon}{\sum} 
    {^{\Bg_0}[{F^{j_{\textbf{gh}}c_{\textbf{k}}d_{\textbf{l}}}_{e}]^{m,\beta\chi}_{q,\delta\epsilon}}}
    {^{\Bg_0}[{F^{a_{\textbf{g}}b_{\textbf{h}}q_{\textbf{kl}}}_{e}]^{j,\alpha\epsilon}_{p,\phi\gamma}}}
    =&
    {^{\Bg}\Theta(b_\Bh,c_\Bk,d_\Bl)}^{S(\Bg)}
    \frac
    {\nu_4 (\textbf{g}_0\Bg,\textbf{h},\textbf{k},\textbf{l})}
    {\nu_4 (\textbf{g}_0,\textbf{h},\textbf{k},\textbf{l})^{S(\Bg)}}
    (-1)^{s(\alpha)s(\delta)}
    \nonumber\\&
    \underset{n_{\Bh\Bk}\eta\psi\kappa}{\sum} 
    {^{\Bg_0}[{F^{a_{\textbf{g}}b_{\textbf{h}}c_{\textbf{k}}}_{m}]^{j,\alpha\beta}_{n,\eta\psi}}}
    {^{\Bg_0}[{F^{a_{\textbf{g}}n_{\textbf{hk}}d_{\textbf{l}}}_{e}]^{m,\psi\chi}_{p,\kappa\gamma}}}
    (^{\Bg_0}[{F^{b_{\textbf{h}}c_{\textbf{k}}d_{\textbf{l}}}_{p}]^{n,\eta\kappa}_{q,\delta\phi}})^{S(\Bg)}
    ,
\end{align}
    where all these equations with nontrivial $\Bg_0$ are actually redundant, as Eq.~(\ref{eq.symaction}) and Eq.~(\ref{eq.anopenta}) completely determine all surface $F$-moves.
\end{remark}

\subsection{Surface parent Hamiltonian}\label{subsec.surfhamil}
{The parent Hamiltonian of a surface fermionic symmetry-enriched string-net model with $H^3(G,\Z_2)$ fermionic 't Hooft anomaly is}
\begin{align}
    {H'=-\sum_\nu C^{\text{surface}}_\nu-\sum_\nu A_\nu-\sum_l D_l- \sum_l Q_l-\sum_p B_p^{\text{surface}},}
\end{align}
{where the terms $A_\nu,D_l,Q_l$ are the same as the non-anomalous case in Eq.~(\ref{eq.hamil}), while $C^{\text{surface}}_\nu$ and $B^{\text{surface}}_p$ are modified as}
\begin{itemize}
    \item {$C^{\text{surface}}_\nu$ equalizes the fermion number $\la N_f\ra$ with the total fermion-parity function on surface $s'(\Bg_0,\alpha)$, written as}
    \begin{align}
        C^{\text{surface}}_\nu
        \vast|\
        \vcenter{\hbox{\includegraphics[scale=.16]{ingoing}}}
        \vast\rangle
        =
        \delta_{\la N_f\ra,s'(\Bg_0,\alpha)}
        \vast|\
        \vcenter{\hbox{\includegraphics[scale=.16]{ingoing}}}
        \vast\rangle,
    \end{align}

    \item {$B_p^{\text{surface}}$ may change the fermion-parity in general, written as}
    \begin{align}
            B_p^{\text{surface}}=\frac{1}{\cD^2}
            \underset{\Bg\in G}{\sum}|\Bg_p\Bg\ra\la\Bg_p|
            \underset{s_\Bg}{\sum}
            \frac{d_{s_\Bg}}{n_{s_\Bg}}
            [{B_p^{\text{surface}}]^{s_\Bg}}.
        \end{align}
    {$[{B_p^{\text{surface}}]^{s_\Bg}}$ fuses a loop $s_\Bg$ on the edges:}
        \begin{equation}
	{[{B_p^{\text{surface}}]^{s_\Bg}}
	\vast|\,
	  \vcenter{\hbox{\includegraphics[scale=.04]{Bp1}}}
	\vast\rangle
	=
	\vast|\,
	  \vcenter{\hbox{\includegraphics[scale=.04]{Bp2}}}
	\vast\rangle
        =
        \cM_{\text{surface}}
        \vast|\,
	  \vcenter{\hbox{\includegraphics[scale=.04]{Bp3}}}
	\vast\rangle,}
    \end{equation}
    {where $\cM_{\text{surface}}$ is a sequence of surface $\cF$-moves, $\cH$-moves, dual $\cH$-moves and $\cO$-moves. It in general changes the fermion-parity by}
    \begin{align}
        \sum_{\substack{
        \kappa = \alpha', \beta', \gamma' \\
        \lambda', \mu', \nu'
        }} s'(\Bg_0,\kappa)-\sum_{\substack{
        \eta = \alpha, \beta, \gamma \\
        \lambda, \mu, \nu
        }} s'(\Bg_1,\eta)
        =&
        n_3(\Bg_0,\Bg_1,\Bg_2,\Bg_6)+
        n_3(\Bg_0,\Bg_1,\Bg_6,\Bg_7)
        +n_3(\Bg_0,\Bg_1,\Bg_7,\Bg_5)+
        \nonumber\\&
        n_3(\Bg_0,\Bg_1,\Bg_4,\Bg_5)+
        n_3(\Bg_0,\Bg_1,\Bg_3,\Bg_4)
        +n_3(\Bg_0,\Bg_1,\Bg_2,\Bg_3),
    \end{align}
    {obtained by counting the fermion-parity change in Eq.~(\ref{eq.n3anomaly}) for all involved surface $\cF$-moves, $\cH$-moves, dual $\cH$-moves.}
\end{itemize}
{This surface parent Hamiltonian $H'$ is still a commuting-projector Hamiltonian.}

\subsection{Summary}\label{summary.surf}
\begin{definition}[Surface fermionic symmetry-enriched string-net model with $H^3(G,\Z_2)$ fermionic 't Hooft anomaly]
    {Inputting a finite fermionic symmetry group $G^f=\Z_2^f\times G$, a bulk 3+1D $G^f$-fSPT phase (with up to complex fermion layer) described by $\{n_3,\nu_4\}$, and a set of $G$-graded string types $\{a_\Bg,b_\Bh,...\}$ with given fusion rules such that $\cS_\one=\{a_\one,b_\one,...\}$ is a super fusion category, a surface fermionic symmetry-enriched string-net model with $H^3(G,\Z_2)$ fermionic 't Hooft anomaly on a surface 2+1D honeycomb lattice consists of}
    \begin{itemize}
        \item Degrees of freedom:
        \begin{enumerate}
            \item Group element labels $|\Bg_i\ra$ in each plaquette, where $\Bg_i\in G$.
            \item $G$-graded string types $a_\Bg,b_\Bh,...$ on each link, where $a_\one,b_\one,...\in\cS_\one$ and $\Bg,\Bh,...\in G$. The dimension of endomorphism space $n_{a_\Bg}$ and quantum dimension $d_{a_\Bg}$ are defined similarly as in Subsection \ref{subsection.fusion}.
            \item Fusion states $\alpha,\beta,...$ in fusion space ${\text{Hom}(a_\Bg\otimes b_\Bh,c_{\Bg\Bh})}$ on each vertex. Fusion multiplicities $N^{a_\mb g b_{\mb h }}_{c_{\mb k}},B^{a_\mb g b_{\mb h }}_{c_{\mb k}}$, $F^{a_\mb g b_{\mb h }}_{c_{\mb k}}$ are defined similarly as in Subsection \ref{subsection.fusion}.
            \item $|G|$ species of physical fermions on each vertex. Still we define
            \begin{equation}
                s(\alpha)
                =
                \left\{\begin{array}{l}
                0
                \text{, \ if }|\alpha\rangle \in\text{Hom}(a_\Bg\ot b_\Bh,c_{\Bg\Bh})_0\\ 
                1
                \text{, \ if }|\alpha\rangle\in\text{Hom}(a_\Bg\ot b_\Bh,c_{\Bg\Bh})_1
                \end{array}\right..
            \end{equation}
            While the total surface parity function of fusion state $\alpha$ with leftmost group element $\Bg_0$ (as shown in Graph (\ref{graph.surfacefusion})) is
            \begin{equation}
                s'(\Bg_0,\alpha)=s(\alpha)+n_3(\Bg_0,\Bg,\Bh).
            \end{equation}
            The string fusion rules are therefore $\Bg_0$-dependent:
            \begin{equation}
                {^{\Bg_0}[a_\Bg\otimes b_\Bh]}=
                \left\{\begin{array}{l}
                \underset{{c_{\textbf{g}\textbf{h}} }}{\bigoplus}
                \C^{B^{a_{\textbf{g}}b_{\textbf{h}}}
                _{c_{\textbf{g}\textbf{h}}}|
                F^{a_{\textbf{g}}b_{\textbf{h}}}
                _{c_{\textbf{g}\textbf{h}}}}
                c_{\textbf{g}\textbf{h}}
                \text{, \ \ if }n_3(\Bg_0,\Bg,\Bh)=0\\ 
                \underset{{c_{\textbf{g}\textbf{h}} }}{\bigoplus}
                \C^{F^{a_{\textbf{g}}b_{\textbf{h}}}
                _{c_{\textbf{g}\textbf{h}}}|
                B^{a_{\textbf{g}}b_{\textbf{h}}}
                _{c_{\textbf{g}\textbf{h}}}}
                c_{\textbf{g}\textbf{h}}
                \text{, \ \ if }n_3(\Bg_0,\Bg,\Bh)=1
                \end{array}\right..
            \end{equation}
            
        \end{enumerate}
        \item Surface renormalization moves on surface fixed-point ground states satisfying 't Hooft anomalous symmetric condition, introduced in Subsection \ref{subsec.bbrenorm};
        \item A surface parent commuting-projector Hamiltonian, introduced in Subsection \ref{subsec.surfhamil},
    \end{itemize}
    {where the full data this surface exactly solvable model is denoted as $\{G^f,n_3,\nu_4,a_\Bg,n_{a_\Bg},d_{a_\Bg}, N^{a_\mb g b_{\mb h }}_{c_{\mb k}}, F^{a_\mb g b_{\mb h }}_{c_{\mb k}},$ ${^{\mathbf{g}_0}[{F^{a_{\textbf{g}}b_{\textbf{h}}c_{\textbf{k}}}_{d}]^{m,\alpha\beta}_{n,\chi\delta}}}\}$, satisfying the following consistency equations:}
    \begin{align}
        N^{a_{\textbf{g}}b_{\textbf{h}}}
        _{c_{\textbf{g}\textbf{h}}}
        =B^{a_{\textbf{g}}b_{\textbf{h}}}
        _{c_{\textbf{g}\textbf{h}}}+
        F^{a_{\textbf{g}}b_{\textbf{h}}}
        _{c_{\textbf{g}\textbf{h}}},
    \end{align}

     \begin{align}
\sum_{e_{\mb g \mb h}} 
\frac{N^{a_{\mb g}b_{\mb h}}_{e_{\mb g\mb h}} N^{e_{\mb g \mb h}c_{\mb k}}_{d_{\mb g \mb h \mb k}}}{n_{e_{\mb g \mb h}}}
=
\sum_{f_{\mb h \mb k}} 
\frac{N^{b_{\mb h }c_{\mb k}}_{f_{\mb h \mb k}}N^{a_{\mb g}f_{\mb h \mb k}}_{d_{\mb g \mb h \mb k}}}{n_{f_{\mb h \mb k}}}
,
\end{align}
    
    \begin{align}
        \sum_{e_{\mb g \mb h}} 
        \frac{B^{a_{\mb g}b_{\mb h}}_{e_{\mb g\mb h}} B^{e_{\mb g \mb h}c_{\mb k}}_{d_{\mb g \mb h \mb k}}+F^{a_{\mb g}b_{\mb h}}_{e_{\mb g\mb h}} F^{e_{\mb g \mb h}c_{\mb k}}_{d_{\mb g \mb h \mb k}}}
        {n_{e_{\mb g \mb h}}}
        =
        \sum_{f_{\mb h \mb k}} 
        \frac{B^{b_{\mb h }c_{\mb k}}_{f_{\mb h \mb k}}B^{a_{\mb g}f_{\mb h \mb k}}_{d_{\mb g \mb h \mb k}}+F^{b_{\mb h }c_{\mb k}}_{f_{\mb h \mb k}}F^{a_{\mb g}f_{\mb h \mb k}}_{d_{\mb g \mb h \mb k}}}
        {n_{f_{\mb h \mb k}}},
    \end{align}

    \begin{align}
         d_{a_\Bg}d_{b_\Bh}=
    \sum_{c_{\Bg\Bh}}\frac{N^{a_\Bg b_\Bh}_{c_{\Bg\Bh}}d_{c_{\Bg\Bh}}}{n_{c_{\Bg\Bh}}},
    \end{align}

    \begin{align}
        {^{\Bg_0}[{F^{a_{\textbf{g}}b_{\textbf{h}}c_{\textbf{k}}}
        _{d}]^{m,\alpha\beta}_{n,\chi\delta}}}
        =
        {^{\Bg_0}
        \Theta(a_{\textbf{g}},b_{\textbf{h}},c_{\textbf{k}})}
        {\nu_4 (\textbf{g}_0,\textbf{g},\textbf{h},\textbf{k})}
        ({^{\one}[{F^{a_{\textbf{g}}b_{\textbf{h}}c_{\textbf{k}}}
        _{d}]^{m,\alpha\beta}_{n,\chi\delta}}})^{S(\Bg_0)},
    \end{align}

    \begin{align}
        s'(\Bg_0,\alpha)+s'(\Bg_0,\beta)+s'(\Bg_0,\delta)
        +s'(\Bg_0\Bg,\chi)\text{ (mod 2)}= n_3(\Bg,\Bh,\Bk),
    \end{align}

    \begin{align}
        \underset{\epsilon}{\sum} 
        {^{\textbf{1}}[{F^{j_{\textbf{gh}}c_{\textbf{k}}d_{\textbf{l}}}_{e}]^{m,\beta\chi}_{q,\delta\epsilon}}}
        {^{\textbf{1}}[{F^{a_{\textbf{g}}b_{\textbf{h}}q_{\textbf{kl}}}_{e}]^{j,\alpha\epsilon}_{p,\phi\gamma}}}
        =&
        {^{\Bg}\Theta(b_\Bh,c_\Bk,d_\Bl)}
        {\nu_4 (\textbf{g},\textbf{h},\textbf{k},\textbf{l})}
        (-1)^{s(\alpha)s(\delta)}
        \nonumber\\&
        \underset{n_{\Bh\Bk}\eta\psi\kappa}{\sum} 
        {^{\textbf{1}}[{F^{a_{\textbf{g}}b_{\textbf{h}}c_{\textbf{k}}}_{m}]^{j,\alpha\beta}_{n,\eta\psi}}}
        {^{\textbf{1}}[{F^{a_{\textbf{g}}n_{\textbf{hk}}d_{\textbf{l}}}_{e}]^{m,\psi\chi}_{p,\kappa\gamma}}}
        ({^{\textbf{1}}[{F^{b_{\textbf{h}}c_{\textbf{k}}d_{\textbf{l}}}_{p}]^{n,\eta\kappa}_{q,\delta\phi}}})^{S(\textbf{g})},
    \end{align}

    \begin{align}
        \frac{{^{\Bg_0}\Theta(j_{\Bg\Bh},c_\Bk,d_\Bl)}{^{\Bg_0}\Theta(a_\Bg,b_\Bh,q_{\Bk\Bl})}}
        {{^{\Bg_0}\Theta(a_{\textbf{g}},b_{\textbf{h}},c_{\textbf{k}}}){^{\Bg_0}\Theta(a_{\textbf{g}},n_{\textbf{hk}},d_{\textbf{l}}}){^{\Bg_0}\Theta(b_{\textbf{h}},c_{\textbf{k}},d_{\textbf{l}}})}
        =& \dd \nu_4(\Bg_0,\Bg,\Bh,\Bk,\Bl),
    \end{align}
    where ${^{\Bg_0}\Theta(a_\Bg,b_\Bh,c_\Bk)}{^{\Bg}\Theta(a_\Bg,b_\Bh,c_\Bk)}^{S(\Bg_0)}={^{\Bg_0\Bg}\Theta(a_\Bg,b_\Bh,c_\Bk)}$, and the four projective-unitary conditions in Eqs.~(\ref{eq.fprojuni1})-(\ref{eq.hproj2}).
\end{definition}

\subsection{Abelian gauge theory surface topological order and surface anyons}
\label{rmk.surfaceabeliangauge}
    Let us consider the simple case that the surface topological order is an Abelian gauge theory (let $N$ be the Abelian gauge group) stacking with a physical fermion $f$, i.e., the surface anyon types are $\cZ_1(\cS_\one)\cong \cZ_1(\textbf{Vec}_N)\boxtimes\{\mathbb{1},f\}$, where $\cZ_1(\textbf{Vec}_N)$ describes anyons in the $N$ gauge theory, and given that the corresponding bulk 3+1D fSPT phase is classified by $(n_3,\nu_4)$, then the surface fermionic symmetry-enriched string-net model is constructed by the following data
    \begin{itemize}
        \item $\cS_\one=\textbf{Vec}_{N}=\{a,b,...|a,b,...=0,...,N-1\}$ is a unitary fusion category.
        
        \item Other $G$-graded string types $\{a_\Bg,b_\Bh,...,\text{for }\Bg,\Bh,...\neq\one\}$ are constructed by a group extension of $G$ by $N$, and we denote the extended group as $\mathcal{G}$, written in a short exact sequence as $1\to N \to \mathcal{G} \to G \to 1$, i.e., $a_\Bg,b_\Bh,...$ are elements in $\cG$. Similarly as in Subsection \ref{subsec.partialgauge}, the group extension is specified by a group homomorphism $\varphi:G\to\text{Aut}(N)$ and a cohomology class $[\mu]\in H^2_\varphi(G,N)$ satisfying Eq.~(\ref{eq.mucond}). Group multiplication rule in $\cG$ is 
        \begin{align}
            a_\Bg\times b_\Bh=[a+\varphi_\Bg(b)+\mu(\Bg,\Bh)]_{\Bg\Bh},
        \end{align}
        where since $N$ is Abelian, we denote the group multiplication in $N$ as $+$ here. All the $G$-graded string types are m-type.
        
        \item The surface string fusion rules when the  group element in the leftmost plaquette being $\Bg_0$ are
        \begin{equation}
            {^{\Bg_0}[a_\Bg\otimes b_\Bh]}=
            \left\{\begin{array}{l}
            \C^{1|0}[a+\varphi_\Bg(b)+\mu(\Bg,\Bh)]_{\Bg\Bh}
            \text{, \ \ if }n_3(\Bg_0,\Bg,\Bh)=0\\ 
            \C^{0|1}[a+\varphi_\Bg(b)+\mu(\Bg,\Bh)]_{\Bg\Bh}
            \text{, \ \ if }n_3(\Bg_0,\Bg,\Bh)=1
            \end{array}\right.. 
        \end{equation}
        The quantum dimension of any $G$-graded string type is $1$. We note that in this simple case $s(\alpha)=0$ for any fusion state $\alpha$, while $s'(\textbf{g}_0,\alpha)=n_3(\Bg_0,\Bg,\Bh)$.
        
        \item The phase factor ${^{\Bg_0}\Theta(a_\Bg,b_\Bh,c_\Bk)}$ is obtained by solving Eq.~(\ref{eq.omega}). In this simple case, when $\Bg_0$ is fixed, ${^{\Bg_0}\Theta(a_\Bg,b_\Bh,c_\Bk)}\in C^3(\mathcal{G},U(1))$ is an $U(1)$-valued 3-cochain, and Eq.~(\ref{eq.omega}) is a twisted 3-cocycle condition for each fixed $\Bg_0$.
        
        \item Surface $\cF$-moves ${^{\one}[{F^{a_{\textbf{g}}b_{\textbf{h}}c_{\textbf{k}}}
        _{d}]^{m,\alpha\beta}_{n,\chi\delta}}}$ with leftmost group element being $\one$, abbreviated as ${^{\one}F^{a_{\textbf{g}}b_{\textbf{h}}c_{\textbf{k}}}}$, are obtained by solving the 't Hooft anomalous fermionic pentagon equation in Eq.~(\ref{eq.anopenta}), i.e.,
        \begin{align}\label{eq.pentaabelian}
            {^{\one}F^{(a_\Bg\times b_\Bh)c_{\textbf{k}}d_{\textbf{l}}}}\cdot
            {^{\one}F^{a_{\textbf{g}}b_{\textbf{h}}(c_\Bk\times d_\Bl)}}
            =
           {^{\Bg}\Theta(b_\Bh,c_\Bk,d_\Bl)}
            {\nu_4 (\textbf{g},\textbf{h},
            \textbf{k},\textbf{l})} 
            {^{\one}F^{a_{\textbf{g}}
            b_{\textbf{h}}c_{\textbf{k}}}}\cdot
            {^{\one}F^{a_{\textbf{g}}(b_\Bh\times c_\Bk)d_{\textbf{l}}}}\cdot
            {^{\one}F^{b_{\textbf{h}}
            c_{\textbf{k}}d_{\textbf{l}}}}.
        \end{align}
        And surface $\cF$-moves ${^{\Bg_0}F^{a_{\textbf{g}}b_{\textbf{h}}c_{\textbf{k}}}}$ with general leftmost group element $\Bg_0$ directly obtained by Eq.~(\ref{eq.symaction}).
    \end{itemize}
\end{widetext}

    In this simple case, we can relate surface fermionic symmetry-enriched string-net model data with the surface anyon data, as introduced by Ref.~\onlinecite{Cheng2017}. We first write the surface $F$-moves in the following special form:
    \begin{align}\label{eq.specialform}
        {^{\one}F^{a_\Bg b_\Bh c_\Bk}}=
        \chi_c(\Bg,\Bh)\alpha_3(\Bg,\Bh,\Bk),
    \end{align}
    where Eq.~(\ref{eq.pentaabelian}) implies that $\chi$ and $\alpha_3$ should satisfy the following properties:
\begin{itemize}
    \item $\chi$ is a character on $N$:
    \begin{equation}\label{eq.charact}
        \chi_a(\Bg,\Bh)\chi_b(\Bg,\Bh)=\chi_{a+b}(\Bg,\Bh).
    \end{equation}
    \item Fixing $a$, $\chi_a\in C^2(G,U(1))$ is a 2-cochain satisfying 
    \begin{equation}\label{eq.chi}
    \frac{{\chi_d(\mb{gh,k})\chi_{d}(\mb{g,h})}}{{\chi_{d}(\mb{g,hk})\chi_d^{S(\Bg)}(\mb{h,k})}}
    ={^{\Bg}\Theta(b_\Bh,c_\Bk,d_\one)}.
    \end{equation}
    \item $\alpha_3\in C^3(G,U(1))$ is a 3-cochain satisfying 
    \begin{equation}\label{eq.alpha3}
        \chi_{\mu(\mb{k,l})}(\mb{g,h})=
        \nu_4(\Bg,\Bh,\Bk,\Bl)
        \dd \alpha_3(\Bg,\Bh,\Bk,\Bl).
\end{equation}
\end{itemize}
    We show derivations of the above properties in Appendix \ref{subsec.relating}. The surface anyons of Abelian $N$ gauge theory are described by $\cZ_1(\text{Vec}_N)\boxtimes\{\mathbb{1},f\}$. We denote the anyons by
    \begin{align}
    \{a=(a_m,a_e)| a_m, a_e\in N\},
\end{align}
    where $a_m\in N$ are gauge fluxes, $a_e:N\to U(1)$ are gauge charges that are one-dimensional irreducible representations of $N$, which also form a group $\hat{N}\cong N$ called character group, i.e., $a_e\in\hat{N}$. Recall Definition \ref{def.symaction}, assuming that the $G$-action (or anyon permutation) $\rho:G\to\text{Aut}(\cZ_1(\text{Vec}_N))$ is in the following relatively trivial form:
    \begin{align}\label{eq.trivialsymaction}
        {\rho_\Bg(a)=(a_m,\rho_\Bg(a_e)),}
    \end{align}
    {i.e., the $G$-action does not permute any gauge flux, where for non-anomalous Abelian $N$ gauge theory it reduces to $\rho_\Bg(a)=(a_m,a_e^{S(\Bg)})$.} Then the anyon valued 2-cocycle characterizing the symmetry fractionalization class $\textswab{w}\in H^2_\rho(G,\cZ_1(\text{Vec}_N))$ (see Eq.~(\ref{eq.omegaanyon}), and we note that in this trivial symmetry action case, $\textswab{w}$ is a 2-cocycle) can be separated into
    \begin{equation}
        \textswab{w}=\textswab{w}_e\ot \textswab{w}_m\equiv (\textswab{w}_e,\textswab{w}_m),
    \end{equation}
    where $\textswab{w}_e\in H^2_\rho(G,N)$ and $\textswab{w}_m\in H^2_\rho(G,\hat{N})$. As shown in Ref.~\onlinecite{Cheng2017}, we have the following relation between surface exactly solvable model data and anyon symmetry fractionalization data:
    \begin{align}\label{eq.relating}
        \mu(\Bg,\Bh)\equiv \textswab{w}_e(\Bg,\Bh),\ 
        \chi_{a_m}(\Bg,\Bh)\equiv R^{(-,\textswab{w}_m(\Bg,\Bh))(a_m,-)},
    \end{align}
    where $R^{(a_m,a_e)(b_m,b_e)}$ is the $R$-symbol (half-braiding) between two anyons $a=(a_m,a_e)$ and $b=(b_m,b_e)$.

\subsection{Example: Surface \texorpdfstring{$\cZ_1(\textbf{Vec}_{\mathbb{Z}_4})\boxtimes\{\mathbb{1},f\}$}{} topological order with \texorpdfstring{$\mathbb{Z}_2^f\times \mathbb{Z}_2\times \mathbb{Z}_4$}{} symmetry}\label{subsec.eg}

We briefly review anyon data studies in Refs.~\onlinecite{Fid2018,cheng2019fermionic} in Appendix \ref{eg.Z4anyon}. The surface anyon types are $\cZ_1(\textbf{Vec}_{\mathbb{Z}_4})\boxtimes\{\mathbb{1},f\}$, where $\cZ_1(\textbf{Vec}_{\mathbb{Z}_4})$ is the unitary modular tensor category of anyon types in the $\mathbb{Z}_4$ gauge theory, $f$ is the physical fermion, and $\cZ_1(\textbf{Vec}_{\mathbb{Z}_4})\boxtimes\{\mathbb{1},f\}$ is a super modular category (see Definition \ref{def.supermodular}). The corresponding bulk phase is the root phase of $\mathbb{Z}_2^f\times \mathbb{Z}_2\times \mathbb{Z}_4$-fSPT phase, as reviewed below. We construct the surface fermionic symmetry-enriched string-net model for this example according to Subsection \ref{rmk.surfaceabeliangauge}.

\subsubsection{Bulk 3+1D \texorpdfstring{$\mathbb{Z}_2^f\times \mathbb{Z}_2\times \mathbb{Z}_4$}{}-fSPT phases}\label{subsec.bulkfspt}

The classification of fSPT phases with $G^f=\mathbb{Z}_2^f\times \mathbb{Z}_2\times \mathbb{Z}_4$ symmetry is\cite{threeloop}
\begin{align}
\mathbb{Z}_2 \times \mathbb{Z}_4,
\end{align}
where the classification group structure is a $\mathbb{Z}_2$ complex fermion layer embedded in a $\mathbb{Z}_2 \times \mathbb{Z}_2$ BSPT phase layer. Denote the four phases in the $\Z_4$ classification as $\nu=0,1,2,3$. Denote group elements in $G=\mathbb{Z}_2 \times \mathbb{Z}_4$ as $\{\textbf{g}=(\textbf{g}_1,\textbf{g}_2)|\Bg_1=0,1,\Bg_2=0,1,2,3\}$, and the group multiplication rule is written as $\Bg\Bh=(\Bg_1+\Bh_1,\Bg_2+\Bh_2)$. 

For the $\nu=1$ root phase in the $\mathbb{Z}_4$ classification, the $(n_3,\nu_4)$ data is
\begin{align}
    n_3(\Bg,\Bh,\Bk)=[\Bg_2\Bh_1\Bk_1]_2,\ \ 
    \nu_4(\Bg,\Bh,\Bk,\Bl)=e^{\frac{\pi i}{2}\Bg_2\Bh_1\Bk_1\Bl_1},
\end{align}
where $[a]_2:=a$ (mod $2$), and they satisfy 
\begin{align}
    \dd \nu_4(\Bg_0,\Bg,\Bh,\Bk,\Bl)=&(-1)^{n_3\smile_1  n_3(\Bg_0,\Bg,\Bh,\Bk,\Bl)}
    \nonumber\\
    =&(-1)^{(\Bg_0)_2\Bg_1\Bh_1\Bk_1\Bl_1}.
\end{align}
(Please do not confuse the notation $\textbf{g}_0=((\textbf{g}_0)_1,(\textbf{g}_0)_2)\in \mathbb{Z}_2 \times \mathbb{Z}_4$.) We note that there can be other solutions up to coboundary transformations, e.g., $n_3(\Bg,\Bh,\Bk)=[\Bg_1\Bh_1\Bk_2]_2, 
    \nu_4(\Bg,\Bh,\Bk,\Bl)=e^{\frac{\pi i}{2}\Bg_1\Bh_1\Bk_1\Bl_2}$ is another solution.

For the $\nu=2$ phase, which is bosonic, the $\nu_4$ data is
\begin{align}
    \nu_4(\Bg,\Bh,\Bk,\Bl)=e^{\pi i \Bg_2\Bh_1\Bk_1\Bl_1},
\end{align}
which is a 4-cocycle.

\subsubsection{Surface fermionic symmetry-enriched string-net model for the \texorpdfstring{$\nu=1$}{} bulk fSPT phase}

According to Subsection \ref{rmk.surfaceabeliangauge}, we label the surface string types by
\begin{align}
    \cG:=\{a_\textbf{g}| a=0,1,2,3, \ \textbf{g}=(\textbf{g}_1,\textbf{g}_2)\in \mathbb{Z}_2 \times \mathbb{Z}_4 \}.
\end{align}
Group multiplication rule in $\cG$ is
\begin{align}
    a_\Bg\times b_\Bh=[a+b+\Bg_1\Bh_1]_4,
\end{align}
where $[a]_4=a$ (mod 4), and $\mu(\Bg,\Bh)=\Bg_1\Bh_1$, i.e.,
\begin{equation}
    \cG\cong\Z_8\times\Z_4.
\end{equation}
The surface string fusion rules are: 
\begin{equation}
    {^{\Bg_0}[a_\Bg\otimes b_\Bh]}=
    \left\{\begin{array}{l}
    \C^{1|0}([a+b+\Bg_1\Bh_1]_4)_\textbf{gh}
    \text{, \ \ if }n_3(\Bg_0,\Bg,\Bh)=0\\ 
    \C^{0|1}([a+b+\Bg_1\Bh_1]_4)_\textbf{gh}
    \text{, \ \ if }n_3(\Bg_0,\Bg,\Bh)=1
    \end{array}\right.,
\end{equation}
i.e., $s'(\Bg_0,\alpha)=n_3(\Bg_0,\Bg,\Bh)=[(\Bg_0)_2\Bg_1\Bh_1]_2$ while $s(\alpha)=0$. Solving Eq.~(\ref{eq.omega}), where $\dd \nu_4(\Bg_0,\Bg,\Bh,\Bk,\Bl)=(-1)^{(\Bg_0)_2\Bg_1\Bh_1\Bk_1\Bl_1}$, we obtain the fermionic obstruction
\begin{align}
    {^{\Bg_0}\Theta(a_\Bg,b_\Bh,c_\Bk)}=(-1)^{(\Bg_0)_2 \Bg_1 \Bh_1 c}.
\end{align}
Surface $F$-moves with leftmost group element $\one$ are
\begin{align}\label{eq.ex1F}
    {^{\one}F^{a_\Bg b_\Bh c_\Bk}}=e^{\frac{\pi i}{2}\Bg_2 \Bh_1 c}.
\end{align}
They satisfy Eq.~(\ref{eq.anopenta}), or Eq.~(\ref{eq.pentaabelian}) in surface Abelian gauge theory topological order case, i.e.,
\begin{align}\label{eq.expenta}
    &{^{\one}F^{(a_\Bg\times b_\Bh)c_{\textbf{k}}d_{\textbf{l}}}}\cdot
    {^{\one}F^{a_{\textbf{g}}b_{\textbf{h}}(c_\Bk\times d_\Bl)}}
    \nonumber\\
    =&
    (-1)^{\Bg_2 \Bh_1 \Bk_1 d}\cdot
    e^{\frac{\pi i}{2}\Bg_2\Bh_1\Bk_1\Bl_1}
    \nonumber\\&
    {^{\one}F^{a_{\textbf{g}}
    b_{\textbf{h}}c_{\textbf{k}}}}\cdot
    {^{\one}F^{a_{\textbf{g}}(b_\Bh\times c_\Bk)d_{\textbf{l}}}}\cdot
    {^{\one}F^{b_{\textbf{h}}
    c_{\textbf{k}}d_{\textbf{l}}}}.
\end{align}
Then by Eq.~(\ref{eq.symaction}), surface $F$-moves with general leftmost group element $\Bg_0$ are
\begin{align}\label{eq.exgf}
    {^{\Bg_0}F^{a_\Bg b_\Bh c_\Bk}}=
    (-1)^{(\Bg_0)_2 \Bg_1 \Bh_1 c}
    e^{\frac{\pi i}{2}(\Bg_0)_2\Bg_1\Bh_1\Bk_1}
    e^{\frac{\pi i}{2}\Bg_2 \Bh_1 c}.
\end{align}
They satisfy Eq.~(\ref{eq.redun}), i.e.,
\begin{align}\label{eq.exgpenta}
    &{^{\Bg_0}F^{(a_\Bg\times b_\Bh)c_{\textbf{k}}d_{\textbf{l}}}}\cdot
    {^{\Bg_0}F^{a_{\textbf{g}}b_{\textbf{h}}(c_\Bk\times d_\Bl)}}
    \nonumber\\
    =&
    (-1)^{\Bg_2 \Bh_1 \Bk_1 d}\cdot
    e^{\frac{\pi i}{2}\Bg_2\Bh_1\Bk_1\Bl_1}
    \nonumber\\&
    {^{\Bg_0}F^{a_{\textbf{g}}
    b_{\textbf{h}}c_{\textbf{k}}}}\cdot
    {^{\Bg_0}F^{a_{\textbf{g}}(b_\Bh\times c_\Bk)d_{\textbf{l}}}}\cdot
    {^{\Bg_0}F^{b_{\textbf{h}}
    c_{\textbf{k}}d_{\textbf{l}}}}.
\end{align}
We have checked by Mathematica that the surface $F$-moves with general leftmost group element $\Bg_0$ satisfy the above equation, where Eq.~(\ref{eq.expenta}) corresponds to the case that $(\Bg_0)_2=0$.

\begin{remark}
\label{rmk.ego3n3}
    {The anyon data (i.e., anyon types and symmetry fractionalization) obtained from the surface fermionic symmetry-enriched string-net model exactly matches with the anyon data in Refs.~\onlinecite{Fid2018,cheng2019fermionic}.} Since in this example, the surface topological order is simply an Abelian $\Z_4$ gauge theory, the relation in Eq.~(\ref{eq.relating}) that relates the above exactly solvable model data to anyon symmetry fractionalization data is satisfied. Let us rewrite the surface $F$-move by
    \begin{align}
        {^{\one}F^{a_\Bg b_\Bh c_\Bk}}=
        \chi_c(\Bg,\Bh)
        =e^{\frac{\pi i}{2}\Bg_2 \Bh_1 c},
    \end{align}
    where $\alpha_3=1$ is trivial here. We can check that the properties in Eq.~(\ref{eq.charact})-(\ref{eq.alpha3}) are satisfied. Then recall the anyon $R$-symbol in Eq.~(\ref{eq.exrsymbol}) and the symmetry fractionalization data in Eq.~(\ref{eq.exsf}) for this example, it is direct to see that Eq.~(\ref{eq.relating}) is satisfied, i.e.,
    \begin{align}
        {\textswab{w}_e(\textbf{g},\textbf{h})
        =\textbf{g}_1 \textbf{h}_1, \ \ 
        \textswab{w}_m(\textbf{g},\textbf{h})=\textbf{g}_2 \textbf{h}_1.}
    \end{align}

    Further, we see that in this example, the anyon-level obstruction function $o_3$ in Eq.~(\ref{eq.exo3}), which is the obstruction to extend the symmetry fractionalization in the super modular category to its minimal modular extended theory (as a spin modular category, see Definition \ref{def.spinmodular}), is exactly the same as the $n_3$ data in Eq.~(\ref{eq.n3anomaly}), i.e., we have
    \begin{align}
        o_3(\Bg,\Bh,\Bk)=n_3(\Bg,\Bh,\Bk)=[\Bg_2\Bh_1\Bk_1]_2.
    \end{align}
\end{remark}

\subsubsection{Surface symmetry-enriched string-net model for the \texorpdfstring{$\nu=2$}{} bulk SPT phase}

In this case, the bulk 3+1D fSPT phase and surface anomalous SET phase are both purely bosonic. Surface string types are the same as $\nu=1$ case. While the surface string fusion rules do not depend on the leftmost group element $\Bg_0$ anymore: 
\begin{equation}
    a_\Bg\otimes b_\Bh=
    ([a+b+\Bg_1\Bh_1]_4)_\textbf{gh}.
\end{equation}
Surface $F$-moves are:
\begin{align}
    {^{\one}F^{a_\Bg b_\Bh c_\Bk}}=e^{\pi i\Bg_2 \Bh_1 c}.
\end{align}
They satisfy Eq.~(\ref{eq.anopenta}) in the simple case that the phase factor ${\Theta}$ is trivial:
\begin{align}
    {^{\one}F^{j_{\Bg\Bh} c_\Bk d_\Bl}}\cdot
    {^{\one}F^{a_\Bg b_\Bh q_{\Bk\Bl}}}
    =&
    e^{\pi i\Bg_2\Bh_1\Bk_1\Bl_1}\cdot
    \nonumber\\&
    {^{\one}F^{a_{\textbf{g}} b_{\textbf{h}} c_{\textbf{k}}}}\cdot
    {^{\one}F^{a_{\textbf{g}} n_{\textbf{hk}} d_{\textbf{l}}}}\cdot
    {^{\one}F^{b_{\textbf{h}} c_{\textbf{k}} d_{\textbf{l}}}}.
\end{align}

\subsection{\texorpdfstring{$H^2(G,\Z_2)$}{} fermionic 't Hooft anomaly }
For the surface fermionic symmetry-enriched string-net models defined in Section \ref{sec.fthooft}, we only construct up to the $H^3(G,\Z_2)$ fermionic 't Hooft anomaly. While exactly solvable models for the $H^2(G,\Z_2)$ fermionic 't Hooft anomaly introduced in Appendix \ref{appen.h2h3} are left for future study. We conjecture that the $H^2(G,\Z_2)$ fermionic 't Hooft anomaly is characterized by a violation of $\Z_2$-conservation of $\sigma$-type strings (Definition \ref{def.sigmatype}) on each vertex. Formally, let $o_2\in H^2(G,\Z_2)$ be the obstruction to extend the $G$-action from a super modular category to its $\Z_2^f$ modular extension as a spin modular category (see details in Appendix \ref{appen.h2h3}), where the super modular category describes the surface anyons. Then in the following fusion rule that involves $\sigma$-type string on each vertex in the surface fermionic symmetry-enriched string-net model:
\begin{align}
        a_\Bg\otimes b_\Bh=\underset{{c_{\textbf{g}\textbf{h}}}}{\oplus}
        \C^{n|n}
        c_{\textbf{g}\textbf{h}}.
\end{align}
Assume that whether a string type $a_\Bg$ is m-type or q/$\sigma$-type is fully determined by the group element label $\Bg$, we denote the total number of $\sigma$-type strings that appear in the above fusion rule as $N_\sigma(\Bg,\Bh)$. We then conjecture that $N_\sigma(\Bg,\Bh)$ (mod 2) is in the same cohomology class as $o_2(\Bg,\Bh)$, i.e.,
\begin{equation}
    [N_\sigma \text{ (mod 2)}]=[o_2]\in H^2(G,\mathbb{Z}_2).
\end{equation}
Furthermore, there is also a limitation in our surface models that they can only deal with simple cases when $\omega_2$ is trivial. How to construct exactly solvable models for surface fSET phases with 't Hooft anomalies when $\omega_2$ is nontrivial is also left for future study.

A limitation of our constructed (surface) fermionic symmetry-enriched string-net models in this paper is that they only realize non-chiral 2+1D fermionic topological phases. However for example, the $H^1(G,\Z_T)$ fermionic 't Hooft anomaly, which corresponds to the $p+ip$ superconductor layer of bulk 3+1D fSPT phase, can only be realized by chiral surface fSET phases. There is an attempt to construct exactly solvable models for chiral topological phases in Ref.~\onlinecite{green2023enriched}, combining the ideas of which may generalize our construction to chiral cases. 

\begin{remark}
    {By Ref.~\onlinecite{barkeshli2024higher}, there is a 3-group structure in a 3+1D fSPT phase with $G^f=G\times\Z_2^f$ symmetry characterized by $(\nu_4,n_3,n_2)$ (we set $\omega_2$ to be trivial as our constructed exactly solvable models for fermionic 't Hooft anomalies are only for $\omega_2$ trivial cases). After gauging the fermion-parity symmetry $\Z_2^f$, the bosonic shadow has a 3-group structure that consists of:}
\begin{itemize}
    \item {A 0-form global symmetry $G$.}


    \item {A 1-form $\Z_2$ symmetry generated by the Kitaev chain surface (non-invertible by itself, but invertible in the 3-group), whose decoration is specified by $n_2\in H^2(G,\Z_2)$.}

    \item {A 2-form $\Z_2$ symmetry generated by the Wilson line of complex fermion, whose decoration is specified by $n_3\in H^3(G,\Z_2)$, which is constrained by $dn_3 = n_2 \cup n_2$.}
\end{itemize}
{Then let us consider a surface 2+1D fSET phase with $H^2$ fermionic 't Hooft anomaly on its boundary (before gauging $\Z_2^f$). The boundary $H^2$ anomaly is compensated by the bulk 1-form symmetry generated by the Kitaev chain. Additionally, when $n_2$ is a coboundary, the boundary $H^3$ anomaly is compensated by a bulk 2-form symmetry generated by the complex fermion, where $n_3$ now satisfies $dn_3=0$.}
\end{remark}

\section{Mathematical framework}
\label{sec.math}
\subsection{\texorpdfstring{$G$}{}-graded super fusion category when \texorpdfstring{$\omega_2$}{} is trivial}
\label{subsec.gsuper}
Here we give a partial definition of $G$-graded super fusion category.
\begin{definition}[$G$-graded super fusion category when $\omega_2$ is trivial]\label{def.gsfc}
    Let $\cC_G=\underset{\Bg\in G}{\oplus}\cC_\Bg$ be a $G$-graded (unitary) fusion category such that $\cZ_1(\cC_G)$ is a spin modular category (Definition \ref{def.spinmodular}) that contains a fermion $f$ (Definition \ref{def.fermion}), where $\cZ_1(\cC_G)$ denotes for the Drinfeld center of $\cC_G$. (Or equivalently, there exists a braided monoidal functor $\phi:\svec\to\cZ_1(\cC_G)$). Then there is a monoidal $\{\mathbb{1},f\}\equiv\svec$-action on $\cC_G$. If each sector $\cC_\Bg$ is closed under such $\svec$-action (each $\cC_\Bg$ forms an $\svec$-module category), we call 
    \begin{align}
        {^\svec\cC_G}:=\underset{\Bg\in G}{\oplus}{^\svec\cC_\Bg}
    \end{align}
    a $G$-graded (unitary) super fusion category when $\omega_2$ is trivial. 

    We can translate this definition as the following:
    \begin{itemize}
        \item Given that there exists a braided monoidal functor $\phi:\svec\to\cZ_1(\cC_G)$, ${^\svec\cC_G}$ is a super fusion category obtained by a canonical construction of enriched monoidal category (Proposition \ref{prop.canon} and Definition \ref{def.superfusion}).
        \item For the trivial group element $\one\in G$, $\cZ_1(\cC_G)$ is a spin modular category implies that $\cZ_1(\cC_\one)$ is also a spin modular category that contains the fermion $f$. Then ${^\svec\cC_\one}$ is a super fusion category obtained by a canonical construction of enriched monoidal category.
        \item For all nontrivial $\Bg\in G$, given each ${\cC_\Bg}$ is a $\svec$-module category, each ${^\svec\cC_\Bg}$ is a super category obtained by a canonical construction of enriched category (Definition \ref{def.supercat} and Proposition \ref{prop.canonenrich}). 
    \end{itemize}
    Or say, the $G$-graded super fusion category ${^\svec\cC_G}$ is obtained by a fermion condensation from a $G$-graded unitary fusion category ${\cC_G}$ such that the $G$-grading structure in ${\cC_G}$ is preserved. {We illustrate the difficulty on defining general $G$-graded super fusion category where $\omega_2$ can be nontrivial in Remark \ref{rmk.diff}.}
\end{definition}

\begin{remark}
    When $\omega_2$ is nontrivial, there may not exist a 
    $\Z_2$-conserved fermion ($f$ is called $\Z_2$-conserved if $f\ot f=\mathbb{1}$) in the Drinfeld center $\cZ_1(\cC_G)$ of a $G$-graded fusion category $\cC_G$. For example, $\cZ_1(\textbf{Vec}_{\Z_4})$ and $\cZ_1(\textbf{Vec}_{\Z_8})$ both do not contain $\Z_2$-conserved fermion. Therefore, general $G$-graded super fusion categories cannot be constructed by a fermion condensation from a $G$-graded fusion category, as the fermion condensation construction only covers fermionic global symmetries with trivial $\omega_2$.
\end{remark}

\begin{example}[Super Tambara-Yamagami category as a $\Z_2$-graded super fusion category]\label{eg.nonano4}
    Tambara-Yamagami category $\text{TY}_{\Z_{2N}}$ as a unitary fusion category is a generalization of Ising fusion category, which can be viewed as a $\Z_2$-graded unitary fusion category:
    \begin{align}
        \cC_{\Z_2}:=\text{TY}_{\Z_{2N}}=\cC_0\oplus\cC_1=\{0,1,...,2N-1\}\oplus\{\sigma\},
    \end{align}
    Let $a,b\in 2N$. Fusion rules are $a\ot b=[a+b]_{2N}$ ($[a]_{2N}$ means $a$ (mod $2N$)) and $\sigma\ot\sigma=\oplus_{a\in\Z_{2N}}a$.

    When $N$ is odd, the simple object $N$ (together with the half-braiding) is a fermion in $\cZ_1(\text{TY}_{\Z_{2N}})$\cite{fSN}. Then condensing $N$, we obtain the super Tambara-Yamagami category as a $\Z_2$-graded super fusion category:
    \begin{align}
        {^\svec\cC_{\Z_2}}={^\svec\cC_0}\oplus{^\svec\cC_1}=\{0,...,N-1\}\oplus\{\sigma\},
    \end{align}
    where $\sigma$ is a q-type object. Let $a,b\in\{0,...,N-1\}$. Quantum dimensions and fusion rules are listed in Example \ref{eg.TY}.
\end{example}

\begin{example}[${^\svec \text{SU}(2)_6}$ as a $\Z_2$-graded super fusion category]\label{eg.nonano5}
    $\text{SU}(2)_6$ can be viewed as a $\Z_2$-graded unitary fusion category. Denote simple objects in $\text{SU}(2)_6$ as $j=0,\frac{1}{2},1,\frac{3}{2},2,\frac{5}{2},3$.
    \begin{align}
        \cC_{\Z_2}:=\text{SU}(2)_6
        =\cC_0\oplus\cC_1=\{0,1,2,3\}\oplus\{\frac{1}{2},\frac{3}{2},\frac{5}{2}\},
    \end{align}
    where $f:=3$ is the fermion, and $\cC_0\cong \text{SO}(3)_6$. $F$-symbols in $\text{SU}(2)_6$ are rather complicated and listed in Ref. \onlinecite{gils2013anyonic}. Its Drinfeld center is
    \begin{align}
        \cZ_1(\text{SU}(2)_6)=
        &\text{SU}(2)_6\boxtimes\overline{\text{SU}(2)_6}
        \nonumber\\
        =&\{(j_1,j_2)|j_1,j_2=0,\frac{1}{2},1,\frac{3}{2},2,\frac{5}{2},3\},
    \end{align}
    where $(0,3)$ and $(3,0)$ are two fermions. By condensing their bound state $(3,3)$, or say, trivializing $j=3$ in $\text{SU}(2)_6$, we obtain a $\Z_2$-graded super fusion category:
    \begin{align}
        {^\svec\cC_{\Z_2}}={^\svec\cC_0}\oplus{^\svec\cC_1}=\{0,1\}\oplus\{\frac{1}{2},\frac{3}{2}\},
    \end{align}
    where $\frac{3}{2}$ is a q-type object and all other simple objects are m-type. Quantum dimensions and fusion rules are listed in Example \ref{eg.SU2}.
\end{example}

\begin{remark}\label{rmk.diff}
    {Here we discuss the difficulty on defining general $G$-graded super fusion category. Our definition of $G$-graded super fusion category in Definition~\ref{def.gsfc} only applies for the case that $\omega_2$ is trivial. This definition cannot be directly generalized to the general case with nontrivial $\omega_2$, which is due to that this definition is obtained by performing a fermion condensation for a $G$-graded unitary fusion category (equivalently, by enriching over $\svec$ for a $G$-graded unitary fusion category). However, in the general cases where $\omega_2$ is nontrivial, the structure of fermion condensation is unclear (more specifically, the usual definition of fermion condensation, as illustrated in Remark \ref{rmk.fc}, does not hold for $\omega_2$ nontrivial case in general). We show that why our definition of $G$-graded super fusion category fails in $\omega_2$ nontrivial case by a simple example. Since $\Z_4^f$ is the simplest symmetry group with nontrivial $\omega_2$, let us consider the $\Z_4^f$ trivial fSPT phase, which can be treated as a special fSET phase where the intrinsic topological order is trivial. Gauging the $\Z_4^f$ symmetry, we obtain a $\Z_4$ gauge theory, described by the category of $\Z_4$-graded vector spaces $\text{Vec}_{\Z_4}$, which can be treated as a special $G$-graded unitary fusion category $\cC_{\Z_4}$ where each $\cC_\Bg=\C$. According to our way to define $G$-graded super fusion category, we should start from the $G$-graded unitary fusion category $\text{Vec}_{\Z_4}$ and perform a fermion condensation. While its Drinfeld center (physically, anyons of $\Z_4$ gauge theory) $\cZ_1(\text{Vec}_{\Z_4})=\{(a_m,a_e)|a_m,a_e\in\Z_4\}$ contains no fermion (recall Definition \ref{def.fermion} and the $R$-symbol $R^{(a_m,a_e)(b_m,b_e)}=e^{2\pi i \frac{a_m b_e}{4}}$, the anyons labeled by $(0,0),(0,2),(2,0),(2,2)$ are all bosons). Therefore, fermion condensation cannot be performed on $\text{Vec}_{\Z_4}$. A possible way to solve this problem is to generalize the concept of fermion condensation, e.g., we can condense $\text{Rep}(\Z_4^f)$ ($\Z_4$ super-Tannakian category, or $\text{Rep}(\Z_4,z)$) instead of $\textbf{SVec}=\{1,f\}$. However, there does not exist a braided monoidal functor from $\text{Rep}(\Z_4^f)$ to $\cZ_1(\text{Vec}_{\Z_4})$ (an anyon condensation can be performed iff. there exists a braided monoidal functor from the category of condensed anyons to the category of all anyons, which is equivalent to the existence of a Lagrangian algebra for the category of all anyons), this naive generalization does not work. Therefore, how to define $G$-graded super fusion category in $\omega_2$ nontrivial case remains as an unsolved problem.}
\end{remark}



\begin{widetext}
\subsection{Drinfeld center of super fusion category}
\label{subsec.dcsfc}
We review the Drinfeld center of a super fusion category\cite{fc} in this Subsection. Let ${^\svec\cC}$ be the super fusion category obtained by fermion condensation or the canonical construction in Definition \ref{def.superfusion} from a unitary fusion category $\cC$. The Drinfeld center of ${^\svec\cC}$ is
\begin{align}\label{eq.dcsfc}
    \cZ_1({^\svec\cC})\cong{^\svec \cZ_2(\cZ_1(\cC),\svec)}\cong{^\svec \cZ_1(\cC)}_0,
\end{align}
where $\cZ_2(\cZ_1(\cC),\svec)$ is the relative M\"{u}ger center of $\cZ_1(\cC)$ relative to $\svec$ (Definition \ref{def.muger}), and
\begin{equation}
    {\cZ_1(\cC)}\cong \cZ_1(\cC)_0\oplus \cZ_1(\cC)_1
\end{equation}
is a spin modular category (Definition \ref{def.spinmodular}, and $0,1$ are group elements in $\Z_2^f$), and ${\cZ_1(\cC)}_0$ is a super modular category (Definition \ref{def.supermodular}). Then ${^\svec \cZ_1(\cC)}_0$ is defined as ${\cZ_1(\cC)}_0$ enriched in $\svec$ by a canonical construction of enriched braided monoidal category (inputted by a symmetric monoidal functor $\svec\to\cZ_2[{\cZ_1(\cC)}_0]$)\cite{kong2021enriched}. We clarify the relations among all mentioned categories by the following commutative diagram:
\begin{equation}\label{diag.centersfc}
    \begin{tikzcd}
	\cC &&&& {{^\svec \cC}} \\
	&& {[{^\svec \cZ_1(\cC)}_0]_{\Z_2^f}^\times} \\
	{\cZ_1(\cC)} &&&& {{^\svec \cZ_1(\cC)}_0} \\
	&& {\cZ_1(\cC)_0}
	\arrow["{\text{Fermion condensation}}", from=1-1, to=1-5]
	\arrow["{\text{Drinfeld center}}"', from=1-1, to=3-1]
	\arrow["{\text{Drinfeld center}}", from=1-5, to=3-5]
	\arrow["{\Z_2^f-\text{equivariantization}}"', from=2-3, to=3-1]
	\arrow["{\text{Fermion condensation}}", shift left, from=3-1, to=3-5]
	\arrow["{\Z_2^f\text{-crossed extension}}"', from=3-5, to=2-3]
	\arrow["{\text{Gauging }\Z_2^f}", shift left, from=3-5, to=3-1]
	\arrow["{\Z_2^f-\text{equivariantization}}", from=3-5, to=4-3]
	\arrow["{\Z_2^f\text{-modular extension}}", from=4-3, to=3-1]
    \end{tikzcd},
\end{equation}
where the $\Z_2^f$-modular extension is introduced in Definition \ref{def.mmm}, $G$-equivariantization is introduced in Definition \ref{def.gequiva}, and the commutativity of the lower-half diagram is introduced in Remark \ref{rmk.commu}. {Further, the $\Z_2^f$-modular extended category}
\begin{align}
    {[{^\svec \cZ_1(\cC)}_0]_{\Z_2^f}^\times
    \cong
    {^\svec \cZ_1(\cC)}_0\oplus
    {^\svec \cZ_1(\cC)}_1,}
\end{align}
{where ${^\svec \cZ_1(\cC)}_0$ is called the \textit{bounding sector} (the sector of anyons), and ${^\svec \cZ_1(\cC)}_1$ is called the \textit{non-bounding sector} (the sector of fermion-parity vortices) in Ref.~\onlinecite{fc}.}

\begin{remark}
    A fermion condensation on a spin modular category (Definition \ref{def.spinmodular}) $\cZ_1(\cC)$ consists of two steps:
    \begin{enumerate}
        \item Anyons with nontrivial braiding with the fermion will be confined, which is the inverse step of the $\Z_2^f$-modular extension.
        \item Two anyons differ by the fermion will be identified, which in the inverse step of the $\Z_2^f$-equivariantization.
    \end{enumerate}
\end{remark}

\begin{remark}\label{rmk.GSDtorus}
    {According to Ref.~\onlinecite{fc}, there is a way to obtain the ground state degeneracy (GSD) on torus $\mathbb{T}^2$. Denote the four spin structures on torus as $BB,NB,BN,NN$, where $B$ stands for bounding (anti-periodic) and $NB$ for non-bounding (periodic). We have}
    \begin{align}
        {\text{GSD}(\mathbb{T}^2_{BB})=\text{GSD}(\mathbb{T}^2_{BN})=
        \text{Number of simple objects in }^\svec {\cZ_1(\cC)}_0},
    \end{align}
    {or say, the number of anyon types in the bounding sector, and}
    \begin{align}
        {\text{GSD}(\mathbb{T}^2_{NB})=\text{GSD}(\mathbb{T}^2_{NN})=
        \text{Number of simple objects in }^\svec {\cZ_1(\cC)}_1},
    \end{align}
    {or say, the number of fermion-parity vortex types in the non-bounding sector.}
\end{remark}

\begin{example}[Anyons of Majorana toric code]
    Recall the notations in Example \ref{eg.mtc}. Denote (isomorphism classes of ) simple objects in Ising fusion category as $\textbf{Ising}=\{1,f,\sigma\}$, and simple objects in its Drinfeld center as $\cZ_1(\textbf{Ising})\cong \textbf{Ising}\boxtimes \overline{\textbf{Ising}}=\{\mathbb{1},f,\sigma\}\boxtimes\{\mathbb{1},\overline{f},\overline{\sigma}\}$, which describes anyons in the Ising string-net model. Then anyons in the Majorana toric code topological order is described by the Drinfeld center of a super fusion category $\cZ_1({^\svec \textbf{Ising}})\cong {^\svec \cZ_1(\textbf{Ising})}_0=\{m_\mathbb{1},m_f,m_\sigma\}$ (full data please refer to Ref.~\onlinecite{fc}), where $d_{m_\mathbb{1}}=d_{m_f}=1$, $d_{m_\sigma}=\sqrt{2}$. Then we have two different ways of gauging $\Z_2^f$:
    \begin{itemize}
        \item Performing a $\Z_2^f$-equivariantization on ${^\svec \cZ_1(\textbf{Ising})}_0$, we obtain $\cZ_1({\textbf{Ising}})_0=\{\mathbb{1},f,\bar{f},f\bar{f},\bar{\sigma},f\bar{\sigma}\}$ (and $\cZ_1({\textbf{Ising}})_1=\{\sigma,\bar{f}\sigma,\sigma\bar{\sigma}\}$). Then performing a $\Z_2^f$-modular extension on $\cZ_1({\textbf{Ising}})_0$, we obtain $\cZ_1(\textbf{Ising})$.
        \item Performing a $\Z_2^f$-crossed extension on ${^\svec \cZ_1(\textbf{Ising})}_0$, we obtain $[{^\svec \cZ_1(\textbf{Ising})}_0]_{\Z_2^f}^\times=\{m_\mathbb{1},m_f,m_\sigma\}\oplus\{q_\mathbb{1},q_f,q_\sigma\}$\cite{fc}, where $d_{q_\mathbb{1}}=d_{q_f}=\sqrt{2}, d_{q_\sigma}=2$. 
        Then performing a $\Z_2^f$-equivariantization, we obtain $\cZ_1(\textbf{Ising})$.
    \end{itemize}
     We conclude these relations in the following commutative diagram:
    \begin{equation}
    \begin{tikzcd}
	{\textbf{Ising}} && {\textbf{MTC}:={^\svec \textbf{Ising}}} \\
	& {[{^\svec \cZ_1(\textbf{Ising})}_0]_{\Z_2^f}^\times} \\
	{\cZ_1(\textbf{Ising})} && {{^\svec \cZ_1(\textbf{Ising})}_0} \\
	& {\cZ_1(\textbf{Ising})_0}
	\arrow["{\text{Fermion condensation}}", from=1-1, to=1-3]
	\arrow["{\text{Drinfeld center}}"', from=1-1, to=3-1]
	\arrow["{\text{Drinfeld center}}", from=1-3, to=3-3]
	\arrow["{\Z_2^f-\text{equivariantization}}"', from=2-2, to=3-1]
	\arrow["{\text{Fermion condensation}}", shift left, from=3-1, to=3-3]
	\arrow["{\Z_2^f\text{-crossed extension}}"', from=3-3, to=2-2]
	\arrow["{\text{Gauging }\Z_2^f}", shift left, from=3-3, to=3-1]
	\arrow["{\Z_2^f-\text{equivariantization}}", from=3-3, to=4-2]
	\arrow["{\Z_2^f\text{ modular extension}}", from=4-2, to=3-1]
    \end{tikzcd}.
    \end{equation}
\end{example}

\begin{example}[Drinfeld center of ${^\svec \frac{1}{2}\text{E}_6}$]
    $\frac{1}{2}\text{E}_6=\{\mathbb{1},x,y\}$ is a unitary fusion category, where the quantum dimensions are $d_\mathbb{1}=d_y=1, d_x=1+\sqrt{3}$, and the fusion rules are
    \begin{align}
        x\ot x=\mathbb{1}\oplus 2x\oplus y,\ \ 
        x\ot y=y\ot x=x,\ \ y\ot y=\mathbb{1}.
    \end{align}
    $F$-symbols (or associators) are listed in Ref.~\onlinecite{hong2008exotic}. Its Drinfeld center $\cZ_1(\frac{1}{2}\text{E}_6)=\{\mathbb{1},Y,X_1,X_2,X_3,X_4,X_5,U,V,W\}$ has 10 simple objects as introduced in Ref.~\onlinecite{hong2008exotic}, where $Y$ is a fermion (Definition \ref{def.fermion}). Since $y$ in $\frac{1}{2}\text{E}_6$ (together with half-braiding) is promoted to be the fermion $Y$ in its Drinfeld center, a fermion condensation can be performed on $\frac{1}{2}\text{E}_6$, and the resultant super fusion category is denoted as $\frac{1}{2}\text{E}_6/y\equiv{^\svec \frac{1}{2}\text{E}_6}$. Fusion rules in $^\svec \frac{1}{2}\text{E}_6$ are
    \begin{align}
    \mathbb{1}\otimes x=x\otimes \mathbb{1}=\mathbb{C}^{1|1}x, \ \ 
    x\otimes x=\mathbb{C}^{1|1}\mathbb{1}\oplus\mathbb{C}^{2|2}x.
    \end{align}
    $F$-symbols are listed in Ref.~\onlinecite{fSN}. Its Drinfeld center $\cZ_1(^\svec \frac{1}{2}\text{E}_6)={^\svec\cZ_1(\frac{1}{2}\text{E}_6)}_0=\{m_1,m_2,m_3\}$ as introduced in Ref~\onlinecite{fc}, where $d_{m_1}=1, d_{m_2}=2+\sqrt{3}, d_{m_3}=1+\sqrt{3}$. Then we have two different ways of gauging $\Z_2^f$:
    \begin{itemize}
        \item Performing a $\Z_2^f$-equivariantization on ${^\svec\cZ_1(\frac{1}{2}\text{E}_6)}_0$, we obtain ${\cZ_1(\frac{1}{2}\text{E}_6)}_0=\{\mathbb{1},Y,U,V,X_4,X_5\}$ (and ${\cZ_1(\frac{1}{2}\text{E}_6)}_1=\{W,X_1,X_2,X_3\}$). Then performing a $\Z_2^f$-modular extension on ${\cZ_1(\frac{1}{2}\text{E}_6)}_0$, we obtain ${\cZ_1(\frac{1}{2}\text{E}_6)}$.
        \item Performing a $\Z_2^f$-crossed extension on ${^\svec\cZ_1(\frac{1}{2}\text{E}_6)}_0$, we obtain $[{^\svec\cZ_1(\frac{1}{2}\text{E}_6)}_0]_{\Z_2^f}^\times=\{m_1,m_2,m_3\}\oplus\{q_1,q_2,m_4\}$\cite{fc}, where $d_{q_1}=3+\sqrt{3}, d_{q_2}= d_{m_4}=1+\sqrt{3}$. Then performing a $\Z_2^f$-equivariantization, we obtain ${\cZ_1(\frac{1}{2}\text{E}_6)}$.
    \end{itemize}
    We conclude these relations in the following commutative diagram:
    \begin{equation}
        \begin{tikzcd}
	{\frac{1}{2}\text{E}_6} &&&& {\frac{1}{2}\text{E}_6/y\equiv{^\svec \frac{1}{2}\text{E}_6}} \\
	&& {[^\svec\cZ_1(\frac{1}{2}\text{E}_6)_0]^\times_{\Z_2^f}} \\
	{\cZ_1(\frac{1}{2}\text{E}_6)} &&&& {^\svec\cZ_1(\frac{1}{2}\text{E}_6)_0} \\
	&& {\cZ_1(\frac{1}{2}\text{E}_6)_0}
	\arrow["{\text{Fermion condensation}}", from=1-1, to=1-5]
	\arrow["{\text{Drinfeld center}}"', from=1-1, to=3-1]
	\arrow["{\text{Drinfeld center}}", from=1-5, to=3-5]
	\arrow["{\Z_2^f-\text{equivariantization}}"', from=2-3, to=3-1]
	\arrow["{\text{Fermion condensation}}", shift left, from=3-1, to=3-5]
	\arrow["{\Z_2^f\text{-crossed extension}}"', from=3-5, to=2-3]
	\arrow["{\text{Gauging }\Z_2^f}", shift left, from=3-5, to=3-1]
	\arrow["{\Z_2^f-\text{equivariantization}}", from=3-5, to=4-3]
	\arrow["{\Z_2^f\text{-modular extension}}", from=4-3, to=3-1]
\end{tikzcd}.
    \end{equation}
\end{example}

\begin{example}[Drinfeld center of ${^\svec \text{SU}(2)_6}$]
    The Drinfeld center $\cZ_1(^\svec \text{SU}(2)_6)\cong{^\svec (\text{SO}(3)_6\boxtimes\overline{\text{SO}(3)_6}})$. Its $\Z_2^f$-crossed extension is denoted as [{${^\svec\cZ_1(\text{SU}(2)_6)}_0]_{\Z_2^f}^\times \cong{^\svec (\text{SU}(2)_6\boxtimes\overline{\text{SU}(2)_6}})
    $\cite{unpublished,fc}}. We conclude these relations in the following commutative diagram:
    \begin{equation}
        \begin{tikzcd}
	{\text{SU}(2)_6} && {{^\svec \text{SU}(2)_6}} \\
	& {{^\svec (\text{SU}(2)_6\boxtimes\overline{\text{SU}(2)_6})}} \\
	{\text{SU}(2)_6\boxtimes\overline{\text{SU}(2)_6}} && {{^\svec (\text{SO}(3)_6\boxtimes\overline{\text{SO}(3)_6})}} \\
	& {\text{SO}(3)_6\boxtimes\overline{\text{SO}(3)_6}}
	\arrow["{\text{Fermion condensation}}", from=1-1, to=1-3]
	\arrow["{\text{Drinfeld center}}"', from=1-1, to=3-1]
	\arrow["{\text{Drinfeld center}}", from=1-3, to=3-3]
	\arrow["{\Z_2^f-\text{equivariantization}}"', from=2-2, to=3-1]
	\arrow["{\text{Fermion condensation}}", shift left, from=3-1, to=3-3]
	\arrow["{\Z_2^f\text{-crossed extension}}"', from=3-3, to=2-2]
	\arrow["{\text{Gauging }\Z_2^f}", shift left, from=3-3, to=3-1]
	\arrow["{\Z_2^f-\text{equivariantization}}", from=3-3, to=4-2]
	\arrow["{\Z_2^f\text{-modular extension}}", from=4-2, to=3-1]
    \end{tikzcd}.
    \end{equation}
\end{example}

Recall Definition \ref{def.gsfc}, when $\omega_2$ is trivial, a $G$-graded super fusion category $\cS_G:={^\svec\cC_G}$, where $\cS_\one:={^\svec\cC_\one}$ is always a super fusion category. Therefore, $\cS_\one:={^\svec\cC_\one}$ always satisfies the commutative diagram (\ref{diag.centersfc}). Then recall Conjecture \ref{conj.anyon}, the Drinfeld center $\cZ_1(\cS_\one)$ of the trivial grading sector $\cS_\one$ describes the anyon/quasiparticle excitations of a fermionic symmetry-enriched string-net model inputted by $\cS_G$.

\subsection{Drinfeld center of \texorpdfstring{$G$}{}-graded super fusion category when \texorpdfstring{$\omega_2$}{} is trivial}

\begin{conjecture}
    The Drinfeld center of a $G$-graded super fusion category ${^\svec\cC_G}$ ($\omega_2$ is trivial) is
    \begin{align}
        \cZ_1(^\svec\cC_G)\cong{^\svec \cZ_2(\cZ_1(\cC_G),\svec)}\cong{^\svec \cZ_1(\cC_G)}_0,
    \end{align}
    where $\cZ_2(\cZ_1(\cC_G),\svec)$ is the relative M\"{u}ger center of spin modular category $\cZ_1(\cC_G)$ relative to $\svec$, and ${^\svec \cZ_1(\cC_G)}_0$ is defined as ${\cZ_1(\cC_G)}_0$ enriched in $\svec$ by canonical construction (inputted by a symmetric monoidal functor $\svec\to\cZ_2[{\cZ_1(\cC_G)}_0]$)\cite{kong2021enriched}.
\end{conjecture}

\begin{conjecture}
There exists a $G^f$-action (i.e., anyon permutation by $G^f$, see Subsection \ref{subsec.gaction}) on $\cZ_1(\cS_\one)$. We conjecture that after gauging $G$, i.e., performing a $G$-crossed extension and then a $G$-equivariantization, we obtain $\mathcal{Z}_1(^\svec\mathcal{C}_G)\cong{^\svec\cZ_1(\cC_G)_0}$, and the following diagram commutes:
\begin{equation}\label{diag.conj}
    \begin{tikzcd}
	\begin{array}{c} \mathcal{Z}_1(^\svec\mathcal{C}_\one)\cong{^\svec\cZ_1(\cC_\one)_0}\\ \text{ with }G^f\text{-action} \end{array} && \begin{array}{c} G\text{-crossed super-modular}\\ \text{category }\mathcal{Z}_1(^\svec\mathcal{C}_\one)_{G}^\times \end{array} && \begin{array}{c} \begin{pmatrix}\mathcal{Z}_1(^\svec\mathcal{C}_\one)_{G}^\times\end{pmatrix}^{G}\cong\mathcal{Z}_1(^\svec\mathcal{C}_G)\\ \cong{^\svec\cZ_1(\cC_G)_0} \end{array} \\
	\\
	{\mathcal{Z}_1(\cC_\one) \text{ with }G\text{-action}} && \begin{array}{c} G\text{-crossed spin-modular}\\ \text{category }\mathcal{Z}_1(\mathcal{C}_\one)_{G}^\times \end{array} && {\begin{pmatrix}\mathcal{Z}_1(\mathcal{C}_\one)_{G}^\times\end{pmatrix}^{G}\cong\mathcal{Z}_1(\mathcal{C}_G)}
	\arrow["{G\text{-crossed}}", shift left, from=1-1, to=1-3]
	\arrow["{\text{extension}}"', shift left, from=1-1, to=1-3]
	\arrow["{\text{Gauging }G}", curve={height=-30pt}, from=1-1, to=1-5]
	\arrow["{\text{Gauging }\mathbb{Z}_2^f}"', from=1-1, to=3-1]
	\arrow["{\text{Gauging }G^f}"{description}, from=1-1, to=3-5]
	\arrow["{G\text{-equivariantization}}"', shift left, from=1-3, to=1-5]
	\arrow["{\text{Gauging }\mathbb{Z}_2^f}", from=1-5, to=3-5]
	\arrow["{G\text{-crossed}}", shift left, from=3-1, to=3-3]
	\arrow["{\text{extension}}"', shift left, from=3-1, to=3-3]
	\arrow["{\text{Gauging }G}"', curve={height=30pt}, from=3-1, to=3-5]
	\arrow["{G\text{-equivariantization}}"', shift right, from=3-3, to=3-5]
    \end{tikzcd},
\end{equation}
where we note that in Ref.~\onlinecite{FSET}, a $G^f$-crossed extension is performed on the super modular category $\mathcal{Z}_1(^\svec\mathcal{C}_\one)$, while here we perform a $G$-crossed extension on it in order that the diagram looks ``symmetric''.
\end{conjecture}

Combining Diagrams (\ref{diag.centersfc}) and (\ref{diag.conj}), we have the following commutative diagram that reveals the spirit of the fermionic symmetry-enriched string-net model construction of 2+1D fSET phases when $\omega_2$ is trivial, i.e., inputting a $G$-graded super fusion category ${^\svec\cC_G}$, taking its Drinfeld center and then ungauging $G$ (the inverse process of gauging $G$), we obtain anyon excitations in this exactly solvable model:
\begin{equation}
    \begin{tikzcd}
	{^\svec\cC_G} && {\mathcal{Z}_1({^\svec\cC_G})} && {\mathcal{Z}_1({^\svec\cC_\one}) \text{ with }G^f\text{-action}} \\
	{\cC_G} && {\mathcal{Z}_1(\cC_G)} && {\mathcal{Z}_1(\cC_\one) \text{ with }G\text{-action}}
	\arrow["{\text{Drinfeld center}}", from=1-1, to=1-3]
	\arrow["{\text{Ungauging }G}", from=1-3, to=1-5]
	\arrow["{\text{Drinfeld center}}"', from=2-1, to=2-3]
	\arrow["{\text{Ungauging }G}"', from=2-3, to=2-5]
	\arrow["{\text{F.c.}}"', from=2-1, to=1-1]
	\arrow["{\text{F.c.}}"', from=2-3, to=1-3]
	\arrow["{\text{F.c.}}"', from=2-5, to=1-5]
    \end{tikzcd},
\end{equation}
where F.c. is short for fermion condensation, and the bosonic case in the second row is stated in Refs.~\onlinecite{Cheng2017} and \onlinecite{heinrich2016symmetry}.

\begin{example}[Fermionic $\Z_2$-enriched string-net model inputted by ${^\svec \text{SU}(2)_6}$]
    Let us consider the fermionic $\Z_2$-enriched string-net model inputted by ${^\svec \text{SU}(2)_6}$ as a $\Z_2$-graded super fusion category. The Drinfeld center $\cZ_1(^\svec \text{SU}(2)_6)$ describes the anyon excitations in this exactly solvable model after gauging $G$. Then by ungauging $G$, we obtain anyons with $\Z_2$ symmetry action described by $\cZ_1(^\svec\cC_\one)\cong\cZ_1({^\svec\text{SO}(3)_6})$. These relations are concluded in the following commutative diagram:
    \begin{equation}
        \begin{tikzcd}
	{{^\svec\text{SU}(2)_6}} && {{^\svec(\text{SO}(3)_6\boxtimes\overline{\text{SO}(3)_6})}} && {\cZ_1({^\svec\text{SO}(3)_6})} \\
	{\text{SU}(2)_6} && {\text{SU}(2)_6\boxtimes\overline{\text{SU}(2)_6}} && {\text{SO}(3)_6\boxtimes\overline{\text{SO}(3)_6}}
	\arrow["{\text{Drinfeld center}}", from=1-1, to=1-3]
	\arrow["{\text{Ungauging}}", from=1-3, to=1-5]
	\arrow["{G=\Z_2}"', from=1-3, to=1-5]
	\arrow["{\text{F.c.}}"', from=2-1, to=1-1]
	\arrow["{\text{Drinfeld center}}"', from=2-1, to=2-3]
	\arrow["{\text{F.c.}}"', from=2-3, to=1-3]
	\arrow["{\text{Ungauging}}", from=2-3, to=2-5]
	\arrow["{G=\Z_2}"', from=2-3, to=2-5]
	\arrow["{\text{F.c.}}"', from=2-5, to=1-5]
        \end{tikzcd}.
    \end{equation}
\end{example}
\end{widetext}

\section{Summary and discussion}\label{sec.sum}

In this paper, we first construct exactly solvable models for 2+1D non-chiral non-anomalous fSET phases. Recall that the total symmetry is $G^f=\Z_2^f\times_{\omega_2}G$. A new feature of our constructed fermionic symmetry-enriched string-net models is that $\omega_2$ appears as a phase factor in the symmetry action Eq.~(\ref{sc}), and hence also involved in the fermionic pentagon equation in Eq.~(\ref{fpenta2}).
For the fermionic symmetry-enriched string-net models defined in Section \ref{sec.fsesn}, the input data is a $G$-graded super fusion category, which is only partially defined in Subsection \ref{subsec.gsuper} when $\omega_2$ is trivial. Therefore, a complete definition of $G$-graded super fusion category is desired.
Next, we construct exactly solvable models for non-chiral surface fSET phases with $H^3(G,\Z_2)$ fermionic 't Hooft anomaly when $\omega_2$ is trivial. Each such anomalous phase is the surface of a bulk 3+1D fSPT phase described by a pair $(n_3,\nu_4)$, where $n_3\in H^3(G,\Z_2)$ and $\nu_4$ is a 4-cochain that satisfies $\dd \nu_4=(-1)^{n_3\smile_1 n_3}$. A generalization to $\omega_2$ nontrivial cases is expected in our future work. There are two key features in our constructed surface fermionic symmetry-enriched string-net models:
\begin{itemize}
    \item The fermion-parity conservation of a surface $\cF$-move is broken by $n_3\in H^3(G,\Z_2)$ in Eq.~(\ref{eq.n3anomaly}), which is due to that bulk fermions can run onto the surface, and we have $[n_3]=[o_3]$ (Remark \ref{rmk.n3o3}). While the fermion-parity is conserved if we perform the renormalization move for both bulk and boundary together, i.e., the bulk-boundary $\cF$-move defined in Eq.~(\ref{F1}).
    
    \item The symmetry action is anomalous (this is exactly the characteristics of a 't Hooft anomaly) in a way that a symmetry action involves a nontrivial phase factor $\Theta\nu_4$ as shown in Eq.~(\ref{eq.symaction}), where the phase factor $\Theta$ appears as a contribution of fermion creation and annihilation operators from the bulk fSPT phase, as defined in Eq.~(\ref{eq.omega}). Hence the surface fermionic pentagon equation in Eq.~(\ref{eq.anopenta}) is also obstructed by the phase factor $\Theta\nu_4$.
\end{itemize}


{One open question is how to compute the full anyon data in general fermionic symmetry-enriched string-net models. The \textit{tube algebra for fermionic string-net} models has been developed to calculate the anyon types, fusion rules, ground state degeneracy and modular $S,T$ transformations on torus\cite{fc}. However, the full anyon data also includes the $G$-action (anyon permutation) and symmetry fractionalization data, the computation of which remains unknown. Up to now, we only know how to calculate the anyon data in the simple case where the intrinsic topological order of a (fermionic) SET phase is an Abelian gauge theory and the $G$-action does not permute any gauge flux, with or without 't Hooft anomaly (illustrated in Subsection \ref{rmk.surfaceabeliangauge}). Therefore, a complete construction of \textit{tube algebra for (fermionic) symmetry-enriched string-net models} is desired, where the $G$-action and symmetry fractionalization data can be extracted.}

Bulk-boundary gauging is also a very interesting future direction. Fully gauging the symmetry for a bulk 3+1D symmetry-protected topological (SPT) phase together with an anomalous SET phase on its surface simultaneously, we obtain a bulk 3+1D topological order together with its gapped boundary (a surface anomalous 2+1D topological order). This has been studied for a particular example of bulk 3+1D $\Z_2\times\Z_2$-SPT phase with boundary projective semions\cite{wang2016twisted}. A complete framework of such bulk-boundary gauging process defined on lattice models is desired, which can also be generalized to fermionic systems and related to fermion condensation. For the example we studied in Subsection \ref{subsec.eg}, i.e., a bulk 3+1D $\Z_2^f\times\Z_2\times\Z_4$-SPT phase with boundary $\Z_4$ gauge theory anyons stacking with the physical fermion, we expect that after gauging the $\Z_2\times\Z_4$ symmetry for both boundary and bulk, we will obtain a bulk fermionic 3+1D $\Z_2\times\Z_4$ twisted gauge theory, whose gapped boundary is an anomalous fermionic $\Z_8\times\Z_4$ gauge theory. {With the fact that almost all 3+1D bosonic/fermionic topological orders can be obtained by gauging 3+1D bosonic/fermionic SPT phases \cite{lan2019classification}, we conjecture that the categorical symmetry correspondence\cite{ji2020categorical,kong2020algebraic,lan2024quantum,chatterjee2023symmetry,chatterjee2023holographic} between 3+1D and 2+1D is a simple case of such bulk-boundary gauging when the bulk SPT phase is trivial. A systematic study on such a topic also provides lattice model descriptions for the 2+1D gapped boundaries of a given 3+1D symmetry topological field theory (SymTFT)\cite{freed2024topological,kaidi2023symmetry} as a byproduct.}



{Finally, constructing exactly solvable models for the $H^2(G,\Z_2)$ fermionic 't Hooft anomaly is another future direction. Surface fermionic symmetry-enriched string-net models for fSET phases with $H^2(G,\Z_2)$ fermionic 't Hooft anomaly are expected to be technically involved. However, once the bulk-boundary gauging defined on lattice models illustrated above is fully understood, bulk-boundary ungauging certain 3+1D topological order together with one of its gapped boundary may provide a systematic way to construct a lattice model for an fSET phases with the $H^2(G,\Z_2)$ fermionic 't Hooft anomaly.}






\section{Acknowledgements}
We would like to thank Tian Lan, Chenjie Wang, Chenqi Meng and Qingrui Wang for enlightening discussions. This work is supported by funding from Hong Kong’s Research Grants Council (GRF No. 14306420,  GRF No. 14308223 and RGC Research Fellow Scheme (RFS) 2023/24,  No. RFS2324-4S02).

\appendix

\section{Super fusion category and super modular category}\label{appen.sfc}
Assuming basic knowledge on unitary (braided) fusion categories, this Section is a brief review of some definitions in Ref.~\onlinecite{chenqi,lan2016theory,usher2018fermionic,rowell2008unitarizability,bruillard2019classification,bruillard2020classification,bulmash2022fermionic,bulmash2022anomaly}.

\subsection{Super fusion category}
\begin{definition}[The category of super vector spaces with grade-even linear maps]\label{def.supervector}
    The category of super vector spaces over $\C$ with only grade-even linear maps, denoted as $\textbf{SVec}$, is a symmetric fusion category consists of
    \begin{itemize}
        \item Objects: $\Z_2$-graded vector spaces $V=V_0\oplus V_1$. We denote a super vector space as $\C^{p|q}$, where $p=\text{dim}(V_0)$, $q=\text{dim}(V_1)$. A non-zero element $v\in V_k$, $k=0,1$ is called homogeneous and we define the parity function $|v|=k$.
        \item Morphisms: Grade-even linear maps that preserve the gradings $\svec(V,W)=(W_0\ot V_0^*)\oplus (W_1\ot V_1^*)$.
        \item Tensor product: $V\ot W=(V_0\ot W_0\oplus V_1\ot W_1)\oplus(V_0\ot W_1\oplus V_1\ot W_0)$.
        \item Braidings: 
        \begin{align}
            c_{V,W}:V\ot W &\to W\ot V
            \nonumber\\
            v\ot w &\mapsto (-1)^{|v||w|}w\ot v,
        \end{align}
        where $v\in V$ and $w\in W$ are homogeneous.
    \end{itemize}
\end{definition}

\begin{remark}
    In some mathematical contexts, $\textbf{SVec}$ denotes for the category of $\Z_2$-graded vector spaces with both grade-even and grade-odd linear maps. We use this $\textbf{SVec}$ notation by a different meaning from these contexts.
\end{remark}

\begin{remark}
    $\textbf{SVec}$ and $\textbf{Vec}_{\Z_2}$ are equivalent as fusion categories, where $\textbf{Vec}_{\Z_2}$ is the category of $\Z_2$-graded vector spaces, while $\textbf{Vec}_{\Z_2}$ does not emphasize on the braiding structure.
\end{remark}

\begin{definition}[Super category]\label{def.supercat}
    A super category is an $\textbf{SVec}$-enriched category, denoted as ${^\svec\cC}$. (Please see the definition of enriched category in, e.g., Ref. \onlinecite{kong2021enriched}.) 

    Roughly speaking, important data in an $\textbf{SVec}$-enriched category ${^\svec\cC}$ are
    \begin{itemize}
        \item Objects: $X\in{^\svec\cC}$.
        \item Hom-objects: ${^\svec\cC}(X,Y)\in\textbf{SVec}$ for every pair of $X,Y\in\cS$.
    \end{itemize}
    $\textbf{SVec}$ is called the background category. $\cC$ is called the underlying category, defined by
    \begin{itemize}
        \item Objects: $\text{Ob}(\cC)=\text{Ob}({^\svec\cC})$.
        \item Morphisms: $\cC(X,Y):=\svec(\C,{^\svec\cC}(X,Y))$ for every $X,Y\in\cS$.
        \item Identity morphism and composition of morphisms: Omitted and please refer to Ref. \onlinecite{kong2021enriched}.
    \end{itemize}
    Intuitively, the underlying category $\cC$ takes the grade-0 hom-object of ${^\svec\cC}$ as its hom-space.
\end{definition}



\begin{prop}\label{prop.canonenrich}
    Let $\cA$ be a monoidal category. Given a left $\cA$-module category $\cC$, we obtain an $\cA$-enriched category ${^\cA\cC}$. This is called a \textit{canonical construction of enriched category}.
\end{prop}
\begin{proof}
    See Ref. \onlinecite{kong2021enriched}.
\end{proof}

\begin{prop}\label{prop.canon}
    Let $\cA$ be a braided monoidal category. Given a monoidal left $\cA$-module category $\cC$, we obtain an $\cA$-enriched monoidal category ${^\cA\cC}$. (Definition of enriched monoidal category please also refer to Ref. \onlinecite{kong2021enriched}.) This is called a \textit{canonical construction of enriched monoidal category}.
\end{prop}
\begin{proof}
    See Ref. \onlinecite{kong2021enriched}.
\end{proof}

\begin{definition}[Super fusion category]\label{def.superfusion}
    Let $\cC$ be a (unitary) fusion category and $\phi:\svec\to\cZ_1(\cC)$ be a braided monoidal functor, where $\cZ_1(\cC)$ is the Drinfeld center of $\cC$. Then $\cC$ equipped with the following $\svec$-module action
    \begin{align}\label{eq.moduleaction}
        \svec\times\cC \overset{\phi\times\id}{\longrightarrow}
        \cZ_1(\cC)\times\cC \overset{I\times\id}{\longrightarrow}
        \cC\times\cC \longrightarrow \cC
    \end{align}
    is a monoidal left $\svec$-module, where $I$ is the forgetful functor. Then according to Proposition \ref{prop.canon}, the $\svec$-enriched monoidal category ${^\svec\cC}$ obtained by canonical construction is called a (unitary) super fusion category.
    
    A simple object $X\in{^\svec\cC}$ is called m-type if ${^\svec\cC}(X,X)\cong \C^{1|0}$, and q-type if ${^\svec\cC}(X,X)\cong \C^{1|1}$.
\end{definition}

\begin{remark}\label{rmk.fc}
    Let $\cC$ be a unitary fusion category. The existence of the braided monoidal functor $\phi:\svec\cong\{\mathbb{1},f\}\to\cZ_1(\cC)$ implies that $\cZ_1(\cC)$ contains a fermion $f$ (Definition \ref{def.fermion}), i.e., $\cZ_1(\cC)$ is a spin modular category (Definition \ref{def.spinmodular}).
    
    The above definition of super fusion category by canonical construction physically corresponds to a \textit{fermion condensation}\cite{wan2017fermion,fc} in the string-net model with input $\cC$, and the resultant fermionic string-net model is inputted by a super fusion category, denoted as $\cC/f\equiv {^\svec\cC}$ in Ref.~\onlinecite{fc}.
\end{remark}

\begin{remark}
    $F$-symbols (or associators) in a super fusion category satisfy the fermionic/super pentagon equation\cite{2dtopo,usher2018fermionic}.
\end{remark}


\begin{example}[Fermion toric code]\label{eg.ftc}
    A fermionic toric code model is a fermionic string-net model. Input of a fermionic toric code is a (unitary) super fusion category $\textbf{FTC}:={^\svec\textbf{Vec}_{\Z_4}}=\{\mathbb{1},s\}$ with only m-type objects, 
    where $\textbf{Vec}_{\Z_4}=\{\mathbb{1},s,f,sf\}$ is a unitary fusion category. Fusion rules and $F$-symbols please refer to Ref. \onlinecite{gu2014lattice}.
\end{example}

\begin{example}[Majorana toric code]\label{eg.mtc}
    A Majorana toric code model is a fermionic string-net model. Input of a Majorana toric code is a super fusion category $\textbf{MTC}:={^\svec\textbf{Ising}}=\{\mathbb{1},\sigma\}$ with q-type object $\sigma$, where $\textbf{Ising}=\{1,f,\sigma\}$ is the Ising unitary fusion category. Fusion rules and $F$-symbols please refer to Ref. \onlinecite{fSN}.
\end{example}

\subsection{Super modular category and spin modular category}
\begin{definition}[pre-modular category]
    A pre-modular category is a spherical braided fusion category.
\end{definition}

\begin{definition}[(Relative) M\"{u}ger center]\label{def.muger}
    Let $\cD$ be a pre-modular category. The M\"{u}ger center $\cZ_2(\cD)$ of $\cD$ is the full subcategory
    \begin{align}
        \cZ_2(\cD)=\{X\in\cD|c_{X,Y}c_{Y,X}=\text{id}_{Y\otimes X},\forall Y\in\cD\},
    \end{align}
    where $c_{X,Y}:X\otimes Y\to Y\otimes X$ is the braiding natural isomorphism in $\cD$.

    Let $\cB$ be a subcategory of $\cD$. The relative M\"{u}ger center $\cZ_2(\cD,\cB)$ of $\cD$ relative to $\cB$ is the full subcategory
    \begin{align}
        \cZ_2(\cD,\cB)=\{X\in\cD|c_{X,Y}c_{Y,X}=\text{id}_{Y\otimes X},\forall Y\in\cB\}.
    \end{align}
\end{definition}

\begin{definition}[Symmetric, modular and super modular category]\label{def.supermodular}
    Let $\cD$ be a pre-modular category and $\textbf{Vec}$ be the category of vector spaces. $\cD$ is called symmetric if $\cZ_2(\cD)\cong \cD$. $\cD$ is called modular if $\cZ_2(\cD)\cong \textbf{Vec}$. $\cD$ is called super modular if $\cD$ is unitary and $\cZ_2(\cD)\cong \textbf{SVec}$.
\end{definition}

\begin{definition}[Fermion]\label{def.fermion}
    Let $\cD$ be a pre-modular category. A \textbf{fermion} $f$, if exists, is a distinguished object in $\cD$ satisfying $$f\otimes f\cong \mathbb{1}, \ \ \theta_f=-1,$$ where $\mathbb{1}$ is the tensor unit in $\cD$, and $\theta_a$ is the topological spin/twist of object $a$.
\end{definition}

\begin{remark}
    In other words, a pre-modular category $\cD$ is a super modular if its M\"{u}ger center is generated by a fermion $f$, where $\{\mathbb{1},f\}\cong\textbf{SVec}$.
\end{remark}

\begin{definition}[Split]
    A super modular category $\cD$ is called \textbf{split} if there exists a modular category $\cB$ such that $\cD\cong\cB\boxtimes\textbf{SVec}$, where $\boxtimes$ is the Deligne tensor product.
\end{definition}

\begin{definition}[Spin modular category]
\label{def.spinmodular}
    A spin modular category $(\widecheck{\cD},f)$ is a modular category $\widecheck{\cD}$ together with a distinguished fermion $f$.
\end{definition}

\begin{remark}\label{rmk.spinmodular}
    There is a canonical $\Z_2$-grading on a spin modular category
    \begin{align}
        \widecheck{\cD}\cong
        \widecheck{\cD}_0\oplus\widecheck{\cD}_1,
    \end{align}
    where $M_{a,f}=1$ if $a\in\widecheck{\cD}_0$ and $M_{a,f}=-1$ if $a\in\widecheck{\cD}_1$ ($M_{a,b}$ is the monodromy operator of objects $a$ and $b$). The grade-0 sector $\widecheck{\cD}_0$ is a super modular category. Physically, anyons in the grade-1 sector $\widecheck{\cD}_1$ are fermion-parity vortices (i.e., symmetry defects of $\Z_2^f$).
\end{remark}

\begin{example}\label{ex.toric}
    The toric code anyons $\mathbf{K}^{(0)}=\{\mathbb{1},e,m,\psi\}$ form a spin modular category, where $\psi$ is the fermion. We relabel
    $$\mathbb{1}\to I_0,\ \psi\to\psi_0,\ e\to I_1,\ m\to\psi_1,$$
    i.e., $\mathbf{K}^{(0)}\cong\mathbf{K}^{(0)}_0\oplus\mathbf{K}^{(0)}_1$, and $\mathbf{K}^{(0)}_0=\{1,\psi\}\cong\textbf{SVec}$, $\mathbf{K}^{(0)}_1=\{e,m\}$.
\end{example}

\begin{definition}[Minimal modular extension]\label{def.mmm}
    Let $\cD$ be a pre-modular category. A minimal modular extension of $\cD$ is a modular category $\cE$ such that $\cD\subset\cE$ and $\text{FPdim}(\cE)=\text{FPdim}(\cD)\text{FPdim}(\cZ_2(\cD))$, where $\text{FPdim}(\cD)$ is the Frobenius–Perron dimension of $\cD$.

    If $\cZ_2(\cD)\cong\svec$, this is also called a $\Z_2^f$-modular extension.
\end{definition}

\begin{remark}
    For any unitary (fusion) category $\cC$, the Frobenius–Perron dimension coincides with categorical dimension, i.e., $\text{FPdim}(X)=d_X:=\tilde{S}_{\mathbb{1},X}$, $\forall X\in\cC$ ($\tilde{S}$ is the unnormalized $S$-matrix in $\cC$), and $\text{FPdim}(\cC)=\text{dim}(\cC):=\sum_{X\in\cC} d_X^2$.
\end{remark}

\begin{remark}
    If $\cD$ is super modular, a minimal modular extension of $\cD$ is a spin modular category $(\widecheck{D},f)$, where $\widecheck{D}_0\cong\cD$ and $\text{dim}(\widecheck{D})=2\text{dim}(\cD)$. Physically, let the super modular category $\cD$ describe the anyon types (including the fermion $f$) of a fermionic topological phase, a $Z_2^f$-equivariantization together with a $\Z_2^f$-modular extension of $\cD$ corresponds to gauging the $\Z_2^f$ fermion-parity symmetry (see Diagram (\ref{diag.centersfc}), and gauging $Z_2^f$ is the inverse process of a fermion condensation).
\end{remark}

\begin{theorem}
    Every super modular category admits minimal modular extensions, and there are precisely 16 inequivalent such extensions\cite{bruillard2017fermionic,lan2017modular,johnson2024minimal}.
\end{theorem}

\begin{example}
    There are 16 minimal modular extensions for the simplest super modular category $\textbf{SVec}$, also known as the Kitaev's 16-fold way\cite{Kitaev2006anyons}. The 16 minimal modular extensions are denoted as $\mathbf{K}^{(\nu)}$, where $\nu\in\Z_{16}$, and the $\nu=0$ case is exactly the toric code anyon in Example \ref{ex.toric}. The fermion-parity vortex sector always has dim$(\mathbf{K}^{(\nu)}_1)=2$, corresponding to either Abelian vortices $\mathbf{K}^{(\nu\text{ is even})}_1=\{I_1,\psi_1\}$, or non-Abelian Ising vortices $\mathbf{K}^{(\nu\text{ is odd})}_1=\{\sigma_1\}$ ($d_{\sigma_1}=\sqrt{2}$). Topological spin of any vortex $a_1$ is $\theta_{a_1}=e^{i\frac{\pi}{8}\nu}$, which is related to the chiral central charge $c_{-}=\frac{\nu}{2}$ (mod 8).
\end{example}

\begin{remark}
    Anyons and symmetry defects of 2+1D fSET phases (including anomalous ones) are described by $G^f$-crossed super modular categories. Anyons and defects of non-anomalous 2+1D fSET phases are described by $G$-crossed spin modular categories\cite{FSET}. 2+1D fSET phases with fermionic 't Hooft anomalies correspond to those $G^f$-crossed super modular categories where there are obstructions on modular extension or gauging the $\Z^f_2$ symmetry.
\end{remark}

\section{Review of classification of anyons and symmetry defects in 2+1D bosonic SET phases}
\label{appen.anyonbset}
This Section is a brief review on Ref.~\onlinecite{BSET,galindo2017categorical}. Let $\cM$ be an unitary modular tensor category (UMTC) describing anyons of a 2+1D bosonic topological order, and let $G$ be a group describing the symmetry.

An UMTC $\cM$ is characterized by a set of gauge-invariant data $\{d_a,\theta_a,S_{ab}|a,b\in\cM\}$, where $d_a$ is the quantum dimension of anyon $a$, $\theta_a$ is the topological spin of $a$, and $S_{ab}$ is the topological $S$-matrix of anyons $a$ and $b$.

\begin{remark}
    A spin modular category (Definition \ref{def.spinmodular}) is a special UMTC that contains a fermion $f$.
\end{remark}

\begin{remark}\label{thm.phase}
    If we can assign a phase factor $e^{i\phi_a}$ for each anyon $a\in\cM$ such that
    \begin{equation}
        e^{i\phi_a}e^{i\phi_b}=e^{i\phi_c},\ \ 
        \text{whenever }N^{ab}_c \neq 0
    \end{equation}
     ($N^{ab}_c=\underset{x\in\cM}{\sum}\frac{S_{ax}S_{bx}S^*_{cx}}{S_{\mathbb{1} x}}$ is the fusion multiplicity, and $\mathbb{1}$ is the tensor unit in $\cM$), then we have
    \begin{align}
        e^{i\phi_a}=M^*_{ae},\ \ \forall a\in\cM,
    \end{align}
    for some Abelian anyon $e\in\cA\subseteq \cM$ ($\cA$ denotes for the subcategory of Abelian anyons in $\cM$), where $M_{ab}=\frac{S^*_{ab}S_{\mathbb{1}\mathbb{1}}}{S_{\mathbb{1} a}S_{\mathbb{1} b}}$ is the monodromy operator between anyons $a$ and $b$. Such phase factors form an Abelian group, denoted as $K(\cM)$, i.e. $K(\cM):=\{e^{i\phi_a}|e^{i\phi_a}e^{i\phi_b}=e^{i\phi_c}
        \text{ whenever }N^{ab}_c \neq 0, a,b,c\in\cM\}$.
\end{remark}

\subsection{\texorpdfstring{$G$}{}-action on \texorpdfstring{$\cM$}{}}
\label{subsec.gaction}
\begin{definition}[Braided auto-equivalence]
    A braided auto-equivalence in $\cM$ is a braided monoidal functor $\varphi:\cM\to\cM$.

    Given two braided auto-equivalences $\varphi$ and $\varphi'$, a natural isomorphism $\gamma:\varphi\to\varphi'$ consists of a collection of isomorphisms $\{\gamma_a:\varphi(a)\to\varphi'(a)|a\in\cM\}$ such that they satisfy the naturality condition and preserve both the monoidal and braiding structures.
\end{definition}

\begin{remark}\label{thm.anyonphase}
    We define the following map $\Upsilon$ as a "natural isomorphism" on a fusion vector:
    \begin{equation}
        \Upsilon|a,b;c,\mu\ra =\frac{\gamma_a\gamma_b}{\gamma_c} |a,b;c,\mu\ra,
    \end{equation}
    where $a,b,c\in\cM$, $|a,b;c,\mu\ra\in\Hom(a\ot b,c)$ is a fusion vector and $\mu$ denotes for different orthogonal fusion vectors. Here $\gamma_a,\gamma_b,\gamma_c$ are phase factors, which are also morphisms for $a,b,c$ of natural isomorphism $\gamma$ (we abuse notations to denote a morphism and the phase factor produced by this morphism).
\end{remark}

\begin{remark}
    Equivalence classes of braided auto-equivalences in $\cM$ (up to natural isomorphisms) form a group, denoted as $\Aut(\cM)$.
\end{remark}

\begin{definition}[The category of braided auto-equivalences in $\cM$]
    The category of braided auto-equivalences in $\cM$, denoted as $\underline{\Aut(\cM)}$, is a monoidal category consists of
    \begin{itemize}
        \item Object: A braided auto-equivalence $\varphi:\cM\to\cM$;
        \item Morphism: A monoidal natural isomorphism $\gamma:\varphi\to\varphi'$;
        \item Tensor product: Composition of functors.
    \end{itemize}
\end{definition}

\begin{definition}[Symmetry/$G$-action on $\cM$]
\label{def.symaction}
    Let $\underline{G}$ be the monoidal category whose objects are group elements in $G$ and tensor product is the group multiplication. A group action of $G$ on $\cM$ is a monoidal functor $\underline{\rho}:\underline{G}\to\underline{\Aut(\cM)}$. Spelling it out, the monoidal functor $\underline{\rho}$ consists of
    \begin{itemize}
        \item Braided monoidal functors $\underline{\rho_\Bg}:\cM\to\cM$, $\forall \Bg\in G$;
        \item Monoidal natural isomorphisms $\underline{\kappa_{\Bg,\Bh}}:\underline{\rho_{\Bg \Bh }}\to\underline{\rho_\Bg}\circ\underline{\rho_\Bh}$, $\forall \Bg,\Bh\in G$,
    \end{itemize}
    such that $\forall \Bg,\Bh,\Bk\in G$, $\forall a\in\cM$, the following diagram commutes:
    \begin{equation}\label{diag.groupact}
        \begin{tikzcd}
	{\underline{\rho_{\Bg \Bh \Bk}}(a)} && {\underline{\rho_{\Bg \Bh }}\circ\underline{\rho_{k}}(a)} \\
	\\
	{\underline{\rho_{\Bg}}\circ\underline{\rho_{\Bh \Bk}}(a)} && {\underline{\rho_{\Bg}}\circ\underline{\rho_{h}}\circ\underline{\rho_{k}}(a)}
	\arrow["{(\underline{\kappa_{\Bg\Bh, \Bk}})_a}", from=1-1, to=1-3]
	\arrow["{(\underline{\kappa_{\Bg,\Bh \Bk}})_a}"', from=1-1, to=3-1]
	\arrow["{\underline{\rho_{\Bg}}\circ(\underline{\kappa_{\Bh, \Bk}})_a}"', from=3-1, to=3-3]
	\arrow["{(\underline{\kappa_{\Bg,\Bh}})_{\underline{\rho_\Bk}(a)}}", from=1-3, to=3-3]
    \end{tikzcd}.
    \end{equation}
\end{definition}

\begin{remark}
    $\underline{\kappa_{\Bg,\Bh}}$ is the natural transformation that preserves monoidal structure in the definition of monoidal functor, and Diagram (\ref{diag.groupact}) is exactly the coherence relation of a monoidal functor. $\{(\underline{\kappa_{\Bg,\Bh}})_a:\underline{\rho_{\Bg \Bh }}(a)\to\underline{\rho_\Bg}\circ\underline{\rho_\Bh}(a)|a\in\cM\}$ is the collection of morphisms in the definition of a natural transformation.
\end{remark}

\begin{definition}[Obstruction to define a $G$-action on $\cM$] \label{def.ob}
Let $\rho:G\to \Aut(\cM)$ be a group homomorphism equipped with 
\begin{itemize}
     \item Braided monoidal functors $\rho_\Bg:\cM\to\cM$, $\forall \Bg\in G$;
        \item Monoidal natural isomorphisms ${\kappa_{\Bg,\Bh}}:{\rho_{\Bg \Bh }}\to{\rho_\Bg}\circ{\rho_\Bh}$, $\forall \Bg,\Bh\in G$,
\end{itemize}
where $\kappa_{\Bg,\Bh}|a,b;c,\mu\ra =\frac{\beta_a (\Bg,\Bh)\beta_b (\Bg,\Bh)}{\beta_c (\Bg,\Bh)}|a,b;c,\mu\ra$ (recall Theorem. \ref{thm.anyonphase}), i.e., $(\kappa_{\Bg,\Bh})_a=\beta_a(\Bg,\Bh)$ (we abuse notations to denote the morphism and the phase factor). Then the obstruction to define a $G$-action on $\cM$ is the obstruction to lift the group homomorphism $\rho:G\to \Aut(\cM)$ to a monoidal functor $\underline{\rho}:\underline{G}\to\underline{\Aut(\cM)}$, which is defined by the following diagram:
    \begin{equation}
        \begin{tikzcd}
	& {\rho_{\Bg \Bh \Bk}(a)} \\
	{\rho_\Bg \circ \rho_{\Bh \Bk}(a)} && {\rho_{\Bg \Bh } \circ \rho_{\Bk}(a)} \\
	{\rho_\Bg\circ\rho_\Bh\circ\rho_\Bk(a)} && {\rho_\Bg\circ\rho_\Bh\circ\rho_\Bk(a)}
	\arrow["{(\kappa_{\Bg,\Bh\Bk})_a=\beta_a(\Bg,\Bh\Bk)}"', from=1-2, to=2-1]
	\arrow["{\beta_{\rho_\Bg(a)}(\Bh,\Bk)}"', from=2-1, to=3-1]
	\arrow["{\beta_a(\Bg\Bh,\Bk)}", from=1-2, to=2-3]
	\arrow["{\beta_a(\Bg,\Bh)}", from=2-3, to=3-3]
	\arrow["{\Omega_a(\Bg,\Bh,\Bk)}"', from=3-1, to=3-3]
\end{tikzcd},
    \end{equation}
i.e., the obstruction is the phase factor
\begin{equation}
    \Omega_a(\Bg,\Bh,\Bk)=\beta_{\rho_\Bg(a)}(\Bh, \Bk)\beta^{-1}_a(\Bg \Bh ,\Bk)\beta_a(\Bg,\Bh \Bk)\beta^{-1}_a(\Bg,\Bh).
    \label{eq.obs}
\end{equation}
\end{definition}

\begin{remark}
    $\{\beta_a(\Bg,\Bh),\forall\Bg,\Bh\in G,a\in\cM\}$ is not a set of cochains, as $\beta_a(\Bg,\Bh)$ has a dependence on anyon $a$. However, fixing $a$, $\{\beta_a(\Bg,\Bh),\forall\Bg,\Bh\in G\}$ is a set of cochains. Thereby, the $\Omega_a(\Bg,\Bh,\Bk)$ defined above is in general not a cochain or further coboundary.
\end{remark}

\begin{prop}\label{prop.cochain}
    The obstruction phase factor $\Omega_a(\Bg,\Bh,\Bk)$ satisfies
    \begin{equation}
        \Omega_a(\Bg,\Bh,\Bk)\Omega_b(\Bg,\Bh,\Bk)=\Omega_c(\Bg,\Bh,\Bk),
    \end{equation}
    whenever $N^c_{ab}\neq 0$ (this can be checked directly by Eq.~(\ref{eq.obs})). Then recalling Remark \ref{thm.phase},
    \begin{equation}\label{eq.oanyon}
        \Omega_a(\Bg,\Bh,\Bk)=M^*_{a \textswab{O}(\Bg,\Bh,\Bk)},
    \end{equation}
    where $\textswab{O}(\Bg,\Bh,\Bk)\in C^3(G,\cA)$ is an anyon-valued 3-cochain.
\end{prop}

\begin{remark}
    Proposition \ref{prop.cochain} transforms a set of anyon-dependent phases $\{\Omega_a(\Bg,\Bh,\Bk),\forall\Bg,\Bh,\Bk\in G,a\in\cM\}$ to a set of cochains $\{\textswab{O}(\Bg,\Bh,\Bk),\forall\Bg,\Bh,\Bk\in G\}$.
\end{remark}

\begin{remark}
    We can check that $\textswab{O}(\Bg,\Bh,\Bk)$ satisfies the twisted 3-cocycle condition (depending on $\rho$), i.e., $\textswab{O}(\Bg,\Bh,\Bk)\in Z^3_{\rho}(G,\cA)$, and we can define the coboundary transformation on it (omitted). Denoting the equivalence class of group homomorphism $\rho$ as $[\rho]$ (up to natural isomorphism), and equivalence class of $\textswab{O}$ as $[\textswab{O}]$ (up to coboundary transformation), we have
    \begin{equation}
        [\textswab{O}]\in H^3_{[\rho]}(G,\cA),
    \end{equation}
    where $H^3_{[\rho]}(G,\cA)$ is the anyon-valued 3rd-cohomology group that depends on $[\rho]$.
\end{remark}

\subsection{Symmetry fractionalization}
\label{subsec.sf}
Let $U(\Bg)$ be the global symmetry operator of $\Bg\in G$, which forms linear represention of $G$, i.e., \begin{equation}\label{eq.linear}
    U(\Bg)U(\Bh)=U(\Bg\Bh).
\end{equation} 
Let $|\Psi_{a_1,...,a_n}\ra$ be a state with $n$ separated anyons $a_1,...,a_n$ such that all the anyons can fuse to vacuum, i.e., 
\begin{equation}\label{eq.vac}
    a_1\otimes ...\otimes a_n=\mathbb{1} \oplus....
\end{equation} 
Then we assume that any symmetry operator $U(\Bg)$ can be expressed in the following way:
\begin{equation}\label{eq.local}
    U(\Bg) |\Psi_{a_1,...,a_n}\ra
    =
    \prod^n_{j=1}U_{a_j}(\Bg) \rho_\Bg |\Psi_{a_1,...,a_n}\ra,
\end{equation}
where $U_{a_j}(\Bg)$ is a local unitary operator of $\Bg$-action localized near anyon $a_j$, and $\rho_\Bg$ is the anyon permutation by $\Bg\in G$ introduced in Definition \ref{def.ob}, which is called \emph{symmetry localization}.

Further, combining Eq.~(\ref{eq.linear}) and Eq.~(\ref{eq.local}), the local unitary operators form a "twisted" projective representation of the symmetry group:
\begin{equation}\label{eq.sf}
    U_{a_j}(\Bg) \rho_\Bg U_{a_j}(\Bh) \rho^{-1}_\Bg |\Psi_{a_1,...,a_n}\ra
    =\eta_{a_j}(\Bg,\Bh) U_{a_j}(\Bg\Bh) |\Psi_{a_1,...,a_n}\ra,
\end{equation}
which is called \emph{symmetry fractionalization}, and $\eta_{a_j}(\Bg,\Bh)$ is a phase factor satisfying (recall Definition \ref{def.ob})
\begin{equation}
    \prod_{j=1}^n \eta_{a_j}(\Bg,\Bh)=\kappa_{\Bg,\Bh}=\prod_{j=1}^n \beta_{a_j}(\Bg,\Bh).
\end{equation}
Here $\eta_{a}(\Bg,\Bh)$ in general differs from $\beta_{a}(\Bg,\Bh)$ by another phase factor
\begin{equation}\label{eq.somega}
    \omega_{a}(\Bg,\Bh)=\frac{\beta_{a}(\Bg,\Bh)}{\eta_{a}(\Bg,\Bh)},
\end{equation}
which satisfies
\begin{equation}
    \prod_{j=1}^n \omega_{a_j}(\Bg,\Bh)=1.
\end{equation}
Then recalling Theorem \ref{thm.phase} and Eq.~(\ref{eq.vac}), $\omega_{a}(\Bg,\Bh)$ can be written as the braiding phase with some Abelian anyon:
\begin{equation}\label{eq.omegaanyon}
    \omega_{a}(\Bg,\Bh)=M^*_{a \textswab{w}(\Bg,\Bh)},
\end{equation}
where $\textswab{w}(\Bg,\Bh)\in\cA\subset \cM$ is an Abelian anyon valued 2-cochain. 

Further, requiring associativity of the global symmetry operator, i.e., $(R_\Bg R_\Bh)R_\Bk=R_\Bg(R_\Bh R_\Bk)$, leads to the following condition on $\eta_{a}(\Bg,\Bh)$:
\begin{equation}
    \frac{\eta_{\rho_\Bg(a)}(\Bh, \Bk)\eta_{a}(\Bg,\Bh\Bk)}{\eta_{a}(\Bg\Bh,\Bk)\eta_{a}(\Bg,\Bh)}=1.
    \label{eq.eta}
\end{equation}
Combining Eq.~(\ref{eq.obs}), Eq.~(\ref{eq.somega}) and Eq.~(\ref{eq.eta}), we have
\begin{equation}\label{eq.sfob}
    \Omega_a(\Bg,\Bh,\Bk)=\omega_{\rho_\Bg(a)}(\Bh, \Bk)\omega^{-1}_a(\Bg \Bh ,\Bk)\omega_a(\Bg,\Bh \Bk)\omega^{-1}_a(\Bg,\Bh).
\end{equation}
Recalling Eq.~(\ref{eq.oanyon}) and Eq.~(\ref{eq.omegaanyon}),
\begin{align}\label{eq.cob}
    \textswab{O}(\Bg,\Bh,\Bk)=&\rho_\Bg(\textswab{w}(\Bh, \Bk))\overline{\textswab{w}(\Bg \Bh ,\Bk)}\textswab{w}(\Bg,\Bh \Bk)\overline{\textswab{w}(\Bg,\Bh)}
    \nonumber\\
    =&d\textswab{w}(\Bg,\Bh,\Bk),
\end{align}
where $\overline{a}$ denotes for the anti-particle of anyon $a\in\cM$. This means that a well-defined symmetry fractionalization requires $\textswab{O}(\Bg,\Bh,\Bk)$ to be a 3-coboundary. In other words, \emph{any nontrivial $[\textswab{O}]\in H^3_{[\rho]}(G,\cA)$ is an obstruction to define symmetry fractionalization}.

Then when $\textswab{O}$ is a 3-coboundary, except from $\textswab{w}$, 
\begin{align}
    \textswab{w}'(\Bg,\Bh)=\textswab{w}(\Bg,\Bh)\ot \textswab{t}(\Bg,\Bh)
\end{align}
is always another solution to Eq.~(\ref{eq.cob}), where $\textswab{t}(\Bg,\Bh)\in H^2_\rho(G,\cA)$ torsorially classify different symmetry fractionalization classes.

\subsection{\texorpdfstring{$G$}{}-crossed extension and \texorpdfstring{$G$}{}-equivariantization}

Omitted. Crucial points of a $G$-crossed extension are the definition of $G$-crossed braiding and the three consistency equations in FIG. 9 and FIG. 10 in Ref.~\onlinecite{BSET}. The $G$-crossed extended category of $\cM$ is denoted as $\cM_G^\times$. Physically, a $G$-crossed extension of $\cM$ corresponds to adding $G$-symmetry defects to the system such that they couple with the anyons in a consistent way.

By a $G$-equivariantization in $\cM_G^\times$, the symmetry-defects become anyons, and the resultant theory is a UMTC, denoted by $(\cM_G^\times)^G$.

\begin{definition}[$G$-equivariantization]\label{def.gequiva}
    A $G$-equivariant object in a fusion category $\cC$ is a pair $(X,\{u_\Bg\}_{\Bg\in G})$ consisting an $X\in\cC$ and a collection of isomorphisms $u_\Bg:\underline{\rho_\Bg}(X)\cong X$ ($\underline{\rho_\Bg}$ is a $G$-action on $\cC$ defined in Definition \ref{def.symaction}) such that the following diagram commutes:
    \begin{equation}
        \begin{tikzcd}
        	{\underline{\rho_\Bg}\circ
                \underline{\rho_\Bh}(X)} && {\underline{\rho_\Bg}(X)} \\
        	{\underline{\rho_{\Bg\Bh}}(X)} && X
        	\arrow["{\underline{\rho_\Bg}\circ u_\Bg}", from=1-1, to=1-3]
        	\arrow["{u_\Bg}", from=1-3, to=2-3]
        	\arrow["{\underline{\kappa_{\Bg,\Bh}}}", from=2-1, to=1-1]
        	\arrow["{u_{\Bg\Bh}}"', from=2-1, to=2-3]
        \end{tikzcd},
    \end{equation}
    where $\underline{\kappa_{\Bg,\Bh}}:\underline{\rho_{\Bg \Bh }}\to\underline{\rho_\Bg}\circ\underline{\rho_\Bh}$ is the monoidal natural isomorphism.

    Equivariant-objects in $\cC$ form a fusion category, called the equivariantization, denoted by $\cC^G$. Further, we have $\text{FPdim}(\cC^G)=|G|\text{FPdim}(\cC)$\cite{gelaki2009centers}.
\end{definition}

\begin{definition}[Orbit under $G$]
    The orbit of $a\in \cM_G^\times$ under $G$ is defined to be the following set:
    \begin{equation}
        [a]=\{\rho_\Bg(a),\forall \Bg\in G\}.
    \end{equation}
\end{definition}

\begin{definition}[Stablizer subgroup]
    Let $a$ be a representative in orbit $[a]. $The stablizer subgroup of $a$ is defined to be
    \begin{equation}
        G_a=\{\Bg\in G|\rho_\Bg(a)=a\}.
    \end{equation}
\end{definition}

\begin{remark}
    Simple objects in $(\cM_G^\times)^G$ are pairs
    \begin{equation}
        ([a],\pi_a),
    \end{equation}
    where $\pi_a$ is an irreducible projective representation of $G_a$, i.e.,
    \begin{equation}
        \pi_a(\Bg)\pi_a(\Bh)=\eta_a(\Bg,\Bh)\pi_a(\Bg \Bh),\ \ \Bg,\Bh\in G_a,
    \end{equation}
    where $\eta_a(\Bg,\Bh)$ is exactly the projective phase in symmetry fractionalization in the $G$-crossed theory in Eq.~(\ref{eq.sf}).
\end{remark}

\begin{widetext}
\begin{remark}\label{rmk.commu}
    If $\cM$ describes anyons of a non-chiral topological order, i.e., $\cM\cong\cZ_1(\cC)$ for some unitary fusion category $\cC$ (this means that this non-chiral topological order can be constructed as a string-net model with input $\cC$), and given that there is a $G$-action on $\mathcal{Z}_1(\mathcal{C})$, the following diagram commutes:
    \begin{equation}\label{diag.gauging}
    \begin{tikzcd}
	{\mathcal{Z}_1(\mathcal{C}) \text{ with }G\text{-action}} &&& {G\text{-crossed MTC }\mathcal{Z}_1(\mathcal{C})_{G}^\times} \\
	{\mathcal{Z}_2\begin{pmatrix}\mathcal{Z}_1(\mathcal{C}_G),\text{Rep}(G)\end{pmatrix}} &&& {\begin{pmatrix}\mathcal{Z}_1(\mathcal{C})_{G}^\times\end{pmatrix}^{G}\cong\mathcal{Z}_1(\mathcal{C}_G)}
	\arrow["{G\text{-crossed extension}}", from=1-1, to=1-4]
	\arrow["{G\text{-modular extension}}"', from=2-1, to=2-4]
	\arrow["{G\text{-equivariantization}}"', from=1-1, to=2-1]
	\arrow["{G\text{-equivariantization}}", from=1-4, to=2-4]
	\arrow["{\text{Gauging }G}"{description}, from=1-1, to=2-4]
    \end{tikzcd},
    \end{equation}
    where $\cC_G$ is a $G$-graded unitary fusion category such that $\cC_\one\cong\cC$ ($\one$ is trivial element in $G$). We note that our definition of gauging $G$\cite{cui2016gauging,lan2023gauging} is a $G$-crossed extension followed by a $G$-equivariantization, which is slightly different from that in Ref.~\onlinecite{BSET}. This commutative diagram implies that this non-chiral topological order $\mathcal{Z}_1(\mathcal{C})$ with symmetry $G$, i.e., a non-chiral SET phase, can be constructed as a symmetry-enriched string-net model with input $G$-graded fusion category $\cC_G$.
\end{remark}
\end{widetext}

\section{Relating symmetry-enriched string-net model data with anyon symmetry fractionalization data for Abelian gauge theory}\label{appen.relating}

\subsection{Non-anomalous cases}
Partially-gauging a 2+1D bosonic SPT phase to obatin an SET phase is discussed in Ref.~\onlinecite{Cheng2017}. Here we breifly review an relation between symmetry-enriched string-net model data and symmetry fractionalization data for anyons in Abelian gauge theory. 

Using the same notation as in Section \ref{subsec.partialgauge} (the short exact sequence is $1\to N \to \mathcal{G} \to G \to 1$, where $N$ is an Abelian gauge group), given a 2+1D SPT phase classied by a 3-cocycle $\nu_3\in H^3(\cG,U_T(1))$, and given a group homomorphism $\varphi:G\to\text{Aut}(N)$ and a cohomology class $[\mu]\in H^2_\varphi(G,N)$ that specify 
the partial-gauging, the $F$-move or $F$-symbol in the symmetry-enriched string-net model exactly equals to the 3-cocycle of the original SPT phase:
\begin{align}
    {^\one F^{a_\Bg b_\Bh c_\Bk}}=\nu_3(a_\Bg,b_\Bh,c_\Bk),
\end{align}
where $a,b,...\in N$, $\Bg,\Bh,...\in G$ and $a_\Bg,b_\Bh,...\in\cG$. In the following, we consider 3-cocycles that taking the following form:
\begin{align}
    \nu_3(a_\Bg,b_\Bh,c_\Bk)=
    \chi_c(\Bg,\Bh)\alpha_3(\Bg,\Bh,\Bk),
\end{align}
where the 3-cocycle condition $\dd \nu_3=1$ implies that $\chi$ and $\alpha_3$ should satisfy the following properties:
\begin{itemize}
    \item $\chi$ is a character on $N$:
    \begin{align}\label{eq.charac}
        \chi_a(\Bg,\Bh)\chi_b(\Bg,\Bh)=\chi_{a+b}(\Bg,\Bh),
    \end{align}
    where $+$ denotes for the group multiplication in $N$.
    \item Fixing $a$, $\chi_a\in Z^2(G,U(1))$ is a 2-cocycle satisfying $\chi_a(\Bg,\Bh)\chi_a(\Bg\Bh,\Bk)=\chi_a^{S(\Bg)}(\Bh,\Bk)\chi_a(\Bg,\Bh\Bk)$.
    \item $\alpha_3\in C^3(G,U(1))$ is a 3-cochain satisfying 
    \begin{align}
        \dd \alpha_3(\Bg,\Bh,\Bk,\Bl)=\chi_{\mu(\Bk,\Bl)}(\Bg,\Bh).
    \end{align}
\end{itemize}
The input of this symmetry-enriched string-net model is the $G$-graded unitary fusion category $\cC_G:=(\text{Vec}_N)_G$.

\subsection{Cases with \texorpdfstring{$H^3(G,Z_2)$}{} fermionic 't Hooft anomaly}
\label{subsec.relating}
Recall Eq.~(\ref{eq.anopenta}) for surface topological order being Abelian gauge theory:
\begin{align}
    \frac{{^\one F^{(a_\Bg\ot      
        b_\Bh)c_{\textbf{k}}d_{\textbf{l}}}}
        {^\one F^{a_{\textbf{g}}b_{\textbf{h}}(c_\Bk\ot d_\Bl)}}}
        {{^\one F^{a_{\textbf{g}}b_{\textbf{h}}c_{\textbf{k}}}}
        {^\one F^{a_{\textbf{g}}(b_\Bh\ot c_\Bk)d_{\textbf{l}}}}
        {^\one F^{b_{\textbf{h}}c_{\textbf{k}}d_{\textbf{l}}}}}
    ={^{\Bg}\Theta(b_\Bh,c_\Bk,d_\Bl)}
    {\nu_4 (\textbf{g},\textbf{h},\textbf{k},\textbf{l})}.
\end{align}
Plug in the special form in Eq. \eqref{eq.specialform}:
\begin{multline}
	\frac{\chi_d(\mb{gh,k})\chi_{c+d+\mu(\mb{k,l})}(\mb{g,h})}{\chi_c(\mb{g,h})\chi_{d}(\mb{g,hk})\chi_d^{S(\Bg)}(\mb{h,k})}
    =
    {^{\Bg}\Theta(b_\Bh,c_\Bk,d_\Bl)}
    {\nu_4 (\textbf{g},\textbf{h},\textbf{k},\textbf{l})}
    \\
	\frac{\alpha_3^{S(\Bg)}(\Bh, \Bk, \Bl)\alpha_3(\Bg, \Bh, \Bk)\alpha_3(\Bg,\mb{hk}, \Bl)}{\alpha_3(\mb{gh}, \Bk, \Bl)\alpha_3(\Bg,\Bh,\mb{kl})}.
	\label{eqn:xx1}
\end{multline}
First set $\Bl=\one$. Under the normalization $\alpha_3(\one, -, -)=\alpha_3(-,\one,-)=\alpha_3(-,-,\one)=1$ and $\nu_4(-,-,-,\one)=1$, The right-hand side only lefts with ${^{\Bg}\Theta(b_\Bh,c_\Bk,d_\one)}$, and the equation becomes
\begin{multline}
    {\chi_d(\mb{gh,k})\chi_{c+d}(\mb{g,h})}=\\
    {^{\Bg}\Theta(b_\Bh,c_\Bk,d_\one)}
    {\chi_c(\mb{g,h})\chi_{d}(\mb{g,hk})\chi_d^{S(\Bg)}(\mb{h,k})}.
    \label{eqn:xx2}
\end{multline}
If we further set $c=\mathbb{1}$ and use the normalization $\chi_\mathbb{1}=1$, we find $\chi_d$ satisfies the following condition:
\begin{equation}
    \frac{{\chi_d(\mb{gh,k})\chi_{d}(\mb{g,h})}}{{\chi_{d}(\mb{g,hk})\chi_d^{S(\Bg)}(\mb{h,k})}}
    ={^{\Bg}\Theta(b_\Bh,c_\Bk,d_\one)}.
\end{equation}
Eq. \eqref{eqn:xx2} then implies $\chi_{c+d}(\mb{g,h})=\chi_c(\mb{g,h})\chi_d(\mb{g,h})$. Finally, Eq. \eqref{eqn:xx1} is reduced to
\begin{equation}
    \chi_{\mu(\mb{k,l})}(\mb{g,h})=
    \nu_4(\Bg,\Bh,\Bk,\Bl)
    \dd \alpha_3(\Bg,\Bh,\Bk,\Bl).
\end{equation}

\section{\texorpdfstring{$H^2(G,\Z_2)$}{} and \texorpdfstring{$H^3(G,\Z_2)$}{} fermionic 't Hooft anomalies in the context of anyons}
\label{appen.h2h3}
This Section is a brief review on Ref. \onlinecite{bulmash2022anomaly}. Let $\cD$ be a super modular category that describe the anyons of a 2+1D fSET phase, and $(\widecheck{\cD},f)$ be a spin modular category that is the minimal modular extension of $\cD$, i.e., $\widecheck{\cD}=\widecheck{\cD}_0\oplus \widecheck{\cD}_1$ and $\widecheck{\cD}_0\cong\cD$.

Given a $G$-action on $\cD$, i.e., a group homomorphism $\rho:G\to\text{Aut}(\cD)$, there might be obstruction to extend the $G$-action to $\widecheck{\cD}$. Let us define a map $\widecheck{\rho}:G\to\text{Aut}(\widecheck{\cD})$, and a restriction map
\begin{align}\label{eq.restri}
    r:\text{Aut}(\widecheck{\cD})\to& \text{Aut}(\cD)
    \nonumber\\
    [\widecheck{\rho}_\Bg]\mapsto&[\rho_\Bg],
\end{align}
where $[\ ]$ is the equivalence class up to natural isomorphisms (recall Subsection \ref{subsec.gaction}), and $\ker r=\Z_2\cong\{\mathbb{1},f\}$ (we mean the group formed by the fusion rules in $\{\mathbb{1},f\}$ is isomorphic to $\Z_2$) as the vacuum in $\cD$ is $\cZ_2(\cD)=\{\mathbb{1},f\}$ (braids trivially with all other anyons). If $\widecheck{\rho}$ fails to be a group homomorphism, i.e., the group multiplication rule of $G$ is not preserved in $\text{Aut}(\widecheck{\cD})$, we say there is an \textit{obstruction to extend the $G$-action from $\cD$ to $\widecheck{\cD}$}, measured by
\begin{align}
    o_2(\Bg,\Bh):=\widecheck{\rho}_{\Bg\Bh}
    \widecheck{\rho}_{\Bh}^{-1}
    \widecheck{\rho}_{\Bg}^{-1},
\end{align}
where $o_2$ restricts to a trivial map in $\text{Aut}(\cD)$, i.e., $o_2\in\ker r=\Z_2$. So that $o_2$ is a 2-cochain in $C^2(G,\Z_2)$. Requiring associativity on $\widecheck{\rho_\Bg}$ gives the 2-cocycle condition of $o_2$, and after defining the 2-coboundary transformation properly, we conclude that $o_2\in H^2(G,\Z_2)$.

If the above $H^2(G,\Z_2)$ obstruction vanishes, i.e., $o_2$ is a coboundary, there might be obstruction to extend the symmetry fractionalization from $\cD$ to $\widecheck{\cD}$. Given a well-defined symmetry fractionalization on $\cD$, i.e., the obstruction function $\Omega_a$ (for all $a\in\cD$) in Eq.~(\ref{eq.sfob}) is trivial, or say, the Abelian anyon valued cochain $\textswab{O}$ defined in Eq.~(\ref{eq.oanyon}) is a coboundary. Denote the obstruction function to defined symmetry fractionalization in $\widecheck{\cD}$ as $\widecheck{\Omega}_a$, and the corresponding Abelian anyon valued 3-cochain $\widecheck{\textswab{O}}$ is defined by
\begin{align}
    \widecheck{\Omega}_a(\Bg,\Bh,\Bk)=M^*_{a \widecheck{\textswab{O}}(\Bg,\Bh,\Bk)},
\end{align}
for all $a\in\widecheck{\cD}$. Recall the phase factor in Eq.~(\ref{eq.somega}) and its corresponding Abelian anyon valued cochain in Eq.~(\ref{eq.omegaanyon}) defined in $\cD$. We extend them to $\widecheck{\cD}$, denoted as $\widecheck{\omega}_a$ and $\widecheck{\textswab{w}}$, where
\begin{equation}
    \widecheck{\omega}_{a}(\Bg,\Bh)=M^*_{a \widecheck{\textswab{w}}(\Bg,\Bh)},
\end{equation}
for all $a\in\widecheck{\cD}$, such that $r(\widecheck{\omega}_a)=\omega_a$. We say there is an \textit{obstruction to extend the symmetry fractionalization from $\cD$ to $\widecheck{\cD}$} if $\widecheck{\textswab{O}}$ fails to be a 3-coboundary, measured by the following mismatch
\begin{align}\label{eq.o3mismatch}
    O_a(\Bg,\Bh,\Bk)=
    \widecheck{\Omega}_a(\Bg,\Bh,\Bk)^{-1}
    \dd\widecheck{\omega}_a(\Bg,\Bh,\Bk),
\end{align}
for all $a\in\widecheck{\cD}$. We note that $O_a=1$ for all $a\in\cD$. Let us define
\begin{align}\label{eq.o3anyon}
    O_a(\Bg,\Bh,\Bk)=M^*_{a o_3(\Bg,\Bh,\Bk)},
\end{align}
for all $a\in\widecheck{\cD}$. Then the mismatch between $\widecheck{\Omega}_a$ and $\dd\widecheck{\omega}_a$ in Eq.~(\ref{eq.o3mismatch}), such that $\widecheck{\textswab{w}}$ fails to be a 3-coboundary, is transformed into the following anyon valued 3-cochain
\begin{align}
    o_3(\Bg,\Bh,\Bk)=\overline{\widecheck{\textswab{O}}(\Bg,\Bh,\Bk)}\ot 
    \dd\widecheck{\textswab{w}}(\Bg,\Bh,\Bk)\in\{\mathbb{1},f\}.
\end{align}
The obstruction $o_3$ is valued in $\cZ_2(\cD)=\{\mathbb{1},f\}$ because changing $\widecheck{\textswab{w}}\to\widecheck{\textswab{w}}\ot f$ does not affect the restriction map $r(\widecheck{\omega}_a)=\omega_a$, or say, $o_3$ should braid trivially with anyons in $\cD$ in order to preserve $r(\widecheck{\omega}_a)=\omega_a$. We see $\dd O_a=1$, which implies the 3-cocycle condition of $o_3$, and after defining 3-coboundary transformation on $o_3$ properly, we conclude that $o_3\in H^3(G,\Z_2)$ ($\{\mathbb{1},f\}\cong\Z_2$ as groups).

\begin{remark}\label{rmk.o3restri}
    If restricting the definition of $o_3$ in $a\in\cD$ in Eq.~(\ref{eq.o3anyon}), and recall that $O_a=1$ for $a\in\cD$, we have
    \begin{align}
        o_3(\Bg,\Bh,\Bk)=&\overline{\textswab{O}(\Bg,\Bh,\Bk)}\ot 
        \dd\textswab{w}(\Bg,\Bh,\Bk)
        \nonumber\\
        =&\dd\textswab{w}(\Bg,\Bh,\Bk)
        \in\{\mathbb{1},f\},
\end{align}
    where $\textswab{O}$ is trivial as symmetry fractionalization is well-defined in $\cD$.
\end{remark}

\section{Anyon data of surface \texorpdfstring{$\cZ_1(\textbf{Vec}_{\mathbb{Z}_4})\boxtimes\{\mathbb{1},f\}$}{} topological order with \texorpdfstring{$\mathbb{Z}_2^f\times \mathbb{Z}_2\times \mathbb{Z}_4$}{} symmetry}\label{eg.Z4anyon}

This Section is a brief review on Ref. \onlinecite{Fid2018}. The surface fSET phase of the $\nu=1$ 3+1D bulk $\mathbb{Z}_2^f\times \mathbb{Z}_2\times \mathbb{Z}_4$ fSPT phase (recall Section \ref{subsec.bulkfspt}) have quasiparticle/anyon types 
$\cZ_1(\textbf{Vec}_{\mathbb{Z}_4})\boxtimes \{\mathbb{1},f \},
$ which is a super modular category, where $\cZ_1(\textbf{Vec}_{\mathbb{Z}_4})$ describes anyons in the $\mathbb{Z}_4$ gauge theory. We choose the trivial solution that all $F$-symbols are $1$, whose $K$-matrix is $\begin{pmatrix}
0&4\\
4&0\\
\end{pmatrix}$. Here $\{1,f \}$ is the trivial fermionic theory, whose $K$-matrix is $\begin{pmatrix}
1&0\\
0&-1\\
\end{pmatrix}$.
We label the anyon types by
\begin{align}
    \{a=(a_m,a_e,a_f)| a_m, a_e=0,1,2,3, \ a_f=0,1 \},
\end{align}
where $a_m$ and $a_e$ label gauge fluxes and gauge charges of the $\mathbb{Z}_4$ gauge theory respectively, and $a_f=0,1$ labels whether the excitation is bosonic or fermionic. The $R$-symbols are
\begin{align}\label{eq.exrsymbol}
    R^{a,b}=e^{2\pi i (\frac{a_m b_e}{4}+\frac{a_f b_f}{2})}.
\end{align}

Recall Subsection \ref{subsec.gaction}, the symmetry action on anyons is characterized by a group homomorphism\cite{BSET}:
\begin{align}
    \rho: G\to \text{Aut}(\cZ_1(\textbf{Vec}_{\Z_4}\boxtimes \{\mathbb{1},f \}).
\end{align}
Denote the group elements as $\textbf{g}=\{(\textbf{g}_1,\textbf{g}_2)| \textbf{g}_1=0,1 \in \mathbb{Z}_2, \ \textbf{g}_2=0,1,2,3 \in \mathbb{Z}_4 \}$. For each group element,
\begin{align}
    \rho_\textbf{g} :\cZ_1(\textbf{Vec}_{\Z_4})\boxtimes \{\mathbb{1},f \} &\to \cZ_1(\textbf{Vec}_{\Z_4})\boxtimes \{\mathbb{1},f \}
    \nonumber\\
    a &\mapsto \rho_\textbf{g}(a)=(a_m,a_e+2\textbf{g}_2 a_m,a_f+\textbf{g}_2 a_m).
\end{align}
The group multiplication law of $\rho_\textbf{g}$ is satisfied up to a phase factor $\kappa_{\textbf{g},\textbf{h}}$: 
\begin{align}
    \kappa_{\textbf{g},\textbf{h}}
    \rho_\textbf{g}\rho_\textbf{h}
    =
    \rho_\textbf{gh}.
\end{align}
In this example, we can simply take $\kappa_{\textbf{g},\textbf{h}}=1$. The anyon valued 2-cochain characterizing the symmetry fractionalization class defined in Eq.~(\ref{eq.omegaanyon}) is
\begin{align}\label{eq.exsf}
    \textswab{w}(\textbf{g},\textbf{h})=
    (\textbf{g}_1 \textbf{h}_1,\textbf{g}_2 \textbf{h}_1,0).
\end{align}
We can read all symmetry fractionalization phase factors $\eta_a$ from Eq.~(\ref{eq.somega}) and Eq.~(\ref{eq.omegaanyon}), e.g., we see that under the $\mathbb{Z}_2$ symmetry, $e:=(0,1,0)$ carries $\frac{1}{4}$ charge and $m:=(1,0,0)$ carries no fractional charge. Recall Remark \ref{rmk.o3restri} for the expression of $o_3$ when restricting to the super modular category, we have
\begin{align}\label{eq.exo3}
    o_3(\textbf{g},\textbf{h},\textbf{k})
    &=
    d\textswab{w}(\textbf{g},\textbf{h},\textbf{k})
    \nonumber\\
    &=
    \rho_{\textbf{g}}(\textswab{w}(\textbf{h},\textbf{k}))\times
    \overline{\textswab{w}(\textbf{gh},\textbf{k})}\times
    \textswab{w}(\textbf{g},\textbf{hk})\times
    \overline{\textswab{w}(\textbf{g},\textbf{h})}
    \nonumber\\
    &=
    (0,0,[\textbf{g}_2 \textbf{h}_1 \textbf{k}_1]_2),
\end{align}
which is a nontrivial 3-cocycle in $H^3(G,\Z_2)$, where $[n]_2=n$ (mod 2). This $o_3\in H^3(G,\mathbb{Z}_2)$ fermionic 't Hooft anomaly can be compensated by the complex fermion layer data $n_3\in H^3(G,\mathbb{Z}_2)$ of a certain bulk 3+1D fSPT phase with the same symmetry group\cite{bulmash2022anomaly}, where $n_3$ is in the same cohomology class as $o_3$, making the whole bulk-boundary system consistent.

\section{Other renormalization moves in fermionic symmetry-enriched string-net models}\label{appen.othermove}
\subsection{\texorpdfstring{$\cO$}{}-move and \texorpdfstring{$\cY$}{}-move}
Let us consider the following local patch of a fermionic symmetry-enriched string-net model:
\begin{equation}
    \includegraphics[scale=.42]{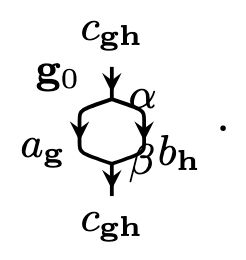}
\end{equation}
We define renormalization moves on such local patch as
\begin{itemize}
    \item When $c_{\Bg\Bh}$ is m-type, the $\cO$-move as a gfSLU transformation is defined as
\begin{equation}
    \includegraphics[scale=.42]{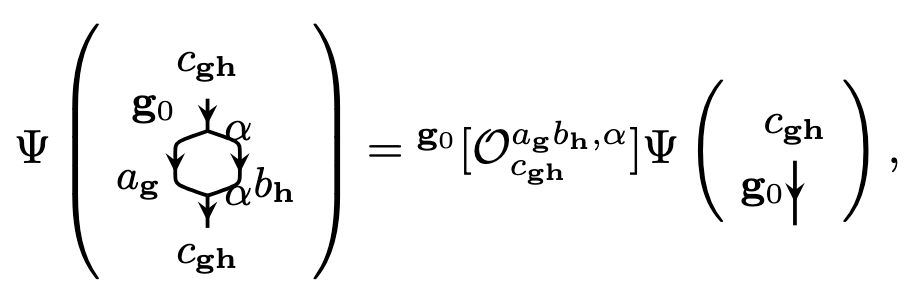}
\end{equation}
where we line up the basis choice in $\text{Hom}(a_\Bg\ot b_\Bh,c_{\Bg\Bh})$ on vertex $\underline{\beta}$ and $\text{Hom}(c_{\Bg\Bh},a_\Bg\ot b_\Bh)=\text{Hom}(a_\Bg\ot b_\Bh,c_{\Bg\Bh})^*$ on vertex $\underline{\alpha}$ (fusion states in the dual space are also denoted as $\alpha,\beta,...$) by the gauge transformation in Eq.~(\ref{eq.gauge}), and
\begin{equation}\label{eq.omove}
    {}^{\mb g_0}[\cO_{c_{\mb g \mb h}}^{a_{\mb g}b_{\mb h},\alpha}] =
    (c^{\Bg_0\dagger}_{\underline{\alpha}})^{s(\alpha)}
    (c^{\Bg_0\dagger}_{\underline{\beta}})^{s(\alpha)}
    {}^{\mb g_0}[O_{c_{\mb g \mb h}}^{a_{\mb g}b_{\mb h},\alpha}].
\end{equation}
The $\cO$-move satisfy the unitary condition:
\begin{align}\label{eq.o1}
    \sum_{a_{\mb g} b_{\mb h} \alpha}
    {^{\mb g_0}}[O_{c_{\mb g \mb h}}^{a_{\mb g}b_{\mb h},\alpha}](^{\mb g_0}[O_{c_{\mb g \mb h}}^{a_{\mb g}b_{\mb h},\alpha}])^*=1
\end{align}

\item When $c_{\Bg\Bh}$ is q-type, the three-vertex $\widetilde{\cO}$-move as a gfSLU transformation is defined similarly as in Ref.~\onlinecite{fSN} and omitted here. While we can still define two-vertex $\cO$-moves as in Eq.~(\ref{eq.omove}) when $c_{\Bg\Bh}$ is q-type, the value of which are related to three-vertex $\widetilde{\cO}$-moves. The unitary condition of $\widetilde{\cO}$-move implies the following condition on two-vertex $\cO$-move:
\begin{align}\label{eq.o2}
    2\sum_{a_{\mb g} b_{\mb h} \alpha}
    {^{\mb g_0}}[O_{c_{\mb g \mb h}}^{a_{\mb g}b_{\mb h},\alpha}](^{\mb g_0}[O_{c_{\mb g \mb h}}^{a_{\mb g}b_{\mb h},\alpha}])^*=1.
\end{align}
\end{itemize}
Combining Eqs. (\ref{eq.o1}) and (\ref{eq.o2}), we obtain
\begin{align}\label{eq.o3}
    n_{c_{\Bg\Bh}}\sum_{a_{\mb g} b_{\mb h} \alpha}
    {^{\mb g_0}}[O_{c_{\mb g \mb h}}^{a_{\mb g}b_{\mb h},\alpha}](^{\mb g_0}[O_{c_{\mb g \mb h}}^{a_{\mb g}b_{\mb h},\alpha}])^*=1.
\end{align}

The $\cY$-move as a gfSLU transformation is defined as
\begin{equation}
    \includegraphics[scale=.42]{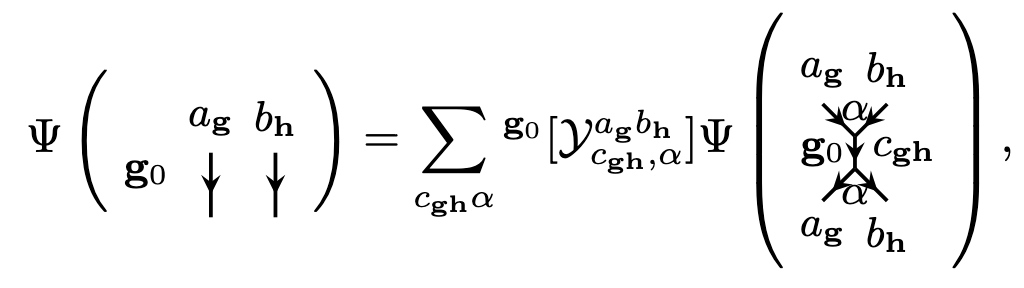}
\end{equation}
where 
\begin{equation}
    {}^{\mb g_0}[\cY_{c_{\mb g \mb h},\alpha}^{a_{\mb g}b_{\mb h}]}
    =
    (c^{\Bg_0}_{\underline{\beta}})^{s(\alpha)}
    (c^{\Bg_0}_{\underline{\alpha}})^{s(\alpha)}
    {}^{\mb g_0}[Y_{c_{\mb g \mb h},\alpha}^{a_{\mb g}b_{\mb h}}].
\end{equation}
We can show that the $\cY$-move is related to $\cO$-move by\cite{fSN}
\begin{align}\label{eq.ymove}
    {}^{\mb g_0}[Y_{c_{\mb g \mb h},\alpha}^{a_{\mb g}b_{\mb h}}]
    =
    \frac{1}
    {n_{c_{\Bg\Bh}}{}^{\mb g_0}[O_{c_{\mb g \mb h}}^{a_{\mb g}b_{\mb h},\alpha}]}.
\end{align}

\subsection{Dual \texorpdfstring{$\cF$}{}-move, \texorpdfstring{$\cH$}{}-move and dual \texorpdfstring{$\cH$}{}-move}\label{appen.othermove2}
The dual $\cF$-move as a gfSLU transformation is defined as
\begin{equation}
    \includegraphics[scale=.42]{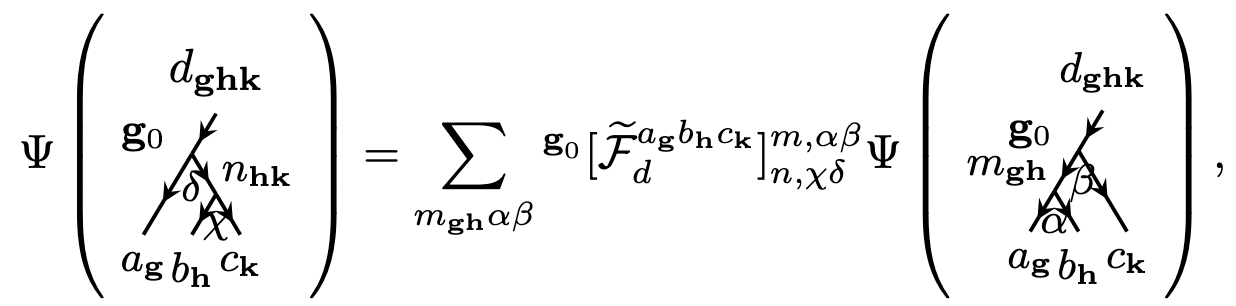}
\end{equation}
where
\begin{align}
    &{^{\textbf{g}_0}[{\widetilde{\cF}^{a_{\textbf{g}}b_{\textbf{h}}c_{\textbf{k}}}
    _{d}]
    ^{m,\alpha\beta}
    _{n,\chi\delta}}}
    \nonumber\\
    =&
     (c^{\textbf{g}_0\dagger}
     _{\underline{\delta}})^{s(\delta)}
    {(c^{\textbf{g}_0\textbf{g}\dagger}
    _{\underline{\chi}})^{s(\chi)}}
    (c^{\textbf{g}_0}_{\underline{\alpha}})^{s(\alpha)}
    (c^{\textbf{g}_0}_{\underline{\beta}})^{s(\beta)}
    {^{\textbf{g}_0}[{\widetilde{F}^{a_{\textbf{g}}b_{\textbf{h}}c_{\textbf{k}}}
    _{d}]
    ^{m,\alpha\beta}
    _{n,\chi\delta}}}.
\end{align}
The dual $\cF$-move is related to the $\cF$-move by\cite{fSN}
\begin{align}
    {^{\mb g_0}}[\widetilde{F}^{a_{\mb g }b_{\mb h}c_{\mb k}}_{d}]
    ^{m_{\mb g \mb h},\alpha\beta}
    _{n_{\mb h \mb k},\chi\delta}
    = 
    &\frac{n_{n_{\Bh\Bk}}}{n_{m_{\Bg\Bh}}}
    {^{\mb g_0}}[F^{a_{\mb g }b_{\mb h}c_{\mb k}}_{d}]
    ^{m_{\mb g \mb h},\alpha\beta}
    _{n_{\mb h \mb k},\chi\delta}
    {^{\mb g_0}}[O_{n_{\mb h \mb k}}^{b_{\mb h}c_{\mb k},\chi}]
    \nonumber\\&
    {^{\mb g_0}}[O_{d_{\mb g \mb h \mb k}}^{a_{\mb g}n_{\mb h \mb k},\delta}]
    {^{\mb g_0}}[O_{m_{\mb g \mb h}}^{a_{\mb g}b_{\mb h},\alpha}]^{-1}
    {^{\mb g_0}}[O_{d_{\mb g\mb h \mb k}}^{m_{\mb g \mb h}c_{\mb k},\beta}]^{-1}.
\end{align}
Then the two projective-unitary conditions of dual $\cF$-move should reduce to the two projective-unitary conditions of $\cF$-move in Eqs.~(\ref{eq.fprojuni1}) and (\ref{eq.fprojuni2}), which can be satisfied by the following ansatz:
\begin{align}\label{eq.o}
    ^{\mb g_0}[O_{c_{\mb g \mb h}}^{a_{\mb g}b_{\mb h},\alpha}] 
    =
    {}^{\mb 1}[O_{c_{\mb g \mb h}}^{a_{\mb g}b_{\mb h},\alpha}]
    =
    \Phi^{a_\Bg b_\Bh,\alpha}_{c_\Bk}
    \frac{1}{D}\sqrt{\frac{d_{a_{\mb g}}d_{b_{\mb h}}}
    { n_{a_\Bg} n_{b_\Bh} n_{c_{\Bg\Bh}} d_{c_{\mb g \mb h}}
    }}\delta_{c_{\mb g \mb h}}^{a_{\mb g}b_{\mb h}},
\end{align}
where $\Phi^{a_\Bg b_\Bh,\alpha}_{c_\Bk}$ is a phase factor\cite{fSN}, $\delta_{c_{\mb g \mb h}}^{a_{\mb g}b_{\mb h}} =1$ for $N_{c_{\mb g \mb h}}^{a_{\mb g}b_{\mb h}} >0$ and $\delta_{c_{\mb g \mb h}}^{a_{\mb g}b_{\mb h}} =0$ for
$N_{c_{\mb g \mb h}}^{a_{\mb g}b_{\mb h}} =0$, and $D=\sqrt{\sum_{\ag ag \in \mathcal{S}_G}d_{\ag ag}^2}$.

The $\cH$-move as a gfSLU transformation is defined as
\begin{equation}
    \includegraphics[scale=.42]{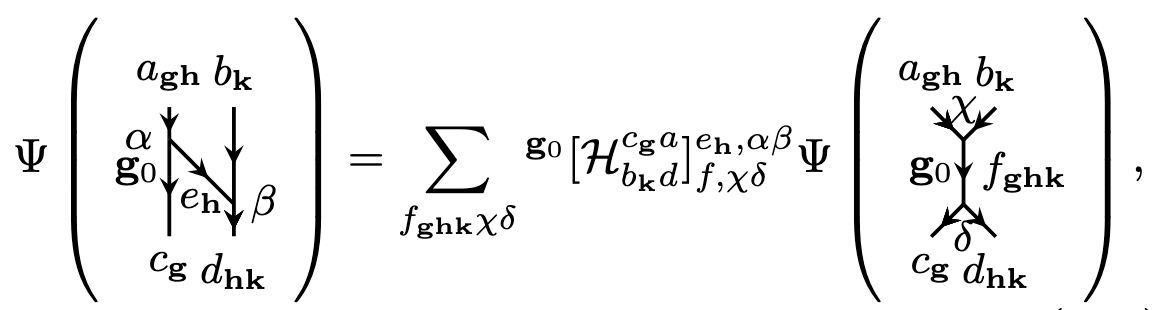}
\end{equation}
where
\begin{align}
    {^{\mb g_0 }}[\mathcal{H}^{c_{\mb g}a}_{b_{\mb k}d}]^{e_{\mb h},\alpha\beta}_{f,\chi\delta}
    =&
    (c^{\textbf{g}_0 \dagger}_{\underline{\alpha}})^{s(\alpha)}
    (c^{\textbf{g}_0 \dagger}_{\underline{\beta}})^{s(\beta)}
    (c^{\textbf{g}_0}_{\underline{\delta}})^{s(\delta)}
    {(c^{\textbf{g}_0\textbf{g}}
    _{\underline{\chi}})^{s(\chi)}}
    \nonumber\\&
    {^{\mb g_0 }}[H^{c_{\mb g}a}_{b_{\mb k}d}]^{e_{\mb h},\alpha\beta}_{f,\chi\delta}.
\end{align}
The $\cH$-move is related to the $\cF$-move by\cite{fSN}
\begin{align}\label{eq.hmove}
    {^{\mb g_0 }[{H}^{c_{\mb g}a}_{b_{\mb k}d}]^{e_{\mb h},\alpha\beta}_{f,\chi\delta}}
    =
    n_{a_{\Bg\Bh}}
    {^{\mb g_0 }Y^{c_\Bg d}_{f,\delta}}\cdot
    ({^{\mb g_0 }[F^{c_\mb g e_\Bh b_\mb k}_f]^{a,\alpha\chi}_{d,\beta\delta}})^*\cdot
    {^{\mb g_0 }O^{c_\mb g e_\mb h,\alpha}_{a}}.
\end{align}

The dual $\cH$-move as a gfSLU transformation is defined as
\begin{equation}
    \includegraphics[scale=.42]{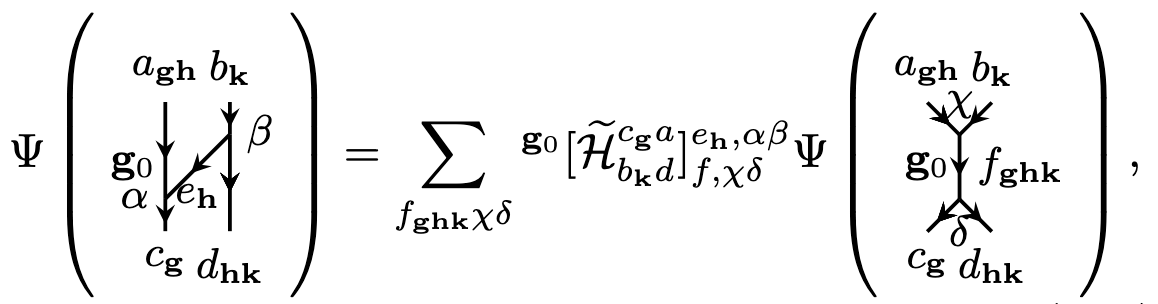}
\end{equation}
where
\begin{align}
   {^{\mb g_0 }}[\widetilde{\cH}^{c_{\mb g}a}_{b_{\mb k}d}]^{e_{\mb h},\alpha\beta}_{f,\chi\delta}
    =&
    (c^{\textbf{g}_0 \dagger}_{\underline{\beta}})^{s(\beta)}
    (c^{\textbf{g}_0 \dagger}_{\underline{\alpha}})^{s(\alpha)}
    (c^{\textbf{g}_0}_{\underline{\delta}})^{s(\delta)}
    {(c^{\textbf{g}_0\textbf{g}}
    _{\underline{\chi}})^{s(\chi)}}
    \nonumber\\&
    {^{\mb g_0 }}[\widetilde{H}^{c_{\mb g}a}_{b_{\mb k}d}]^{e_{\mb h},\alpha\beta}_{f,\chi\delta}.
\end{align}
The dual $\cH$-move is related to the $\cF$-move by\cite{fSN}
\begin{align}\label{eq.dualhmove}
    {^{\mb g_0 }[\widetilde{H}^{c_{\mb g}a}_{b_{\mb k}d}]^{e_{\mb h},\alpha\beta}_{f,\chi\delta}}
    =
    n_{b_\Bk}
    {^{\mb g_0 }Y^{c_\Bg d}_{f,\delta}}\cdot
    {^{\mb g_0 }[F^{a e_\Bh d}_f]^{c_\Bg,\alpha\delta}_{b_\Bk,\beta\chi}}\cdot
    {^{\mb g_0 }O^{e_\Bh d,\beta}_{b_\Bk}}.
\end{align}
Then the two projective-unitary conditions of dual $\cH$-move implies the two projective-unitary conditions of $\cF$-move in Eqs.~(\ref{eq.hproj1}) and (\ref{eq.hproj2}), proof of which is similar to that in Ref.~\onlinecite{fSN}. We note that the two projective-unitary conditions of $\cH$-move also implies two projective-unitary conditions of $\cF$-move, which are equivalent to (i.e., can be derived from) Eqs.~(\ref{eq.hproj1}) and (\ref{eq.hproj2}).


\begin{widetext}
\section{Partition function approach on the \texorpdfstring{$H^3(G,\Z_2)$}{} fermionic 't Hooft anomaly}
\label{appen.part}
Here we study the gapped symmetry-preserving boundary theory of 3+1D fSPT phases via the partition function approach. Since we do not know how to construct the fermionic partition function with Kitaev chain decoration in the bulk yet. Our method here can only deal with the simple case that the bulk fSPT phases can contain at most complex fermion decoration. The 3+1D bulk fermionic partition function in this simple case is given by\cite{Gu2014}
\begin{equation}
\cZ=
\sum_{\{\textbf{g}_i\}}
\int
\prod_{\text{interior }\langle ijkl \rangle}
d\theta^{n_3(ijkl)}_{ijkl}
d\overline{\theta}^{n_3(ijkl)}_{ijkl}
\prod_{\text{interior }\langle ijk \rangle} 
(-1)^{m_2(ijk)}
\prod_{\langle ijklm \rangle}
\mathcal{V}_4^{\sigma(ijklm)}(ijklm),
\label{partition}
\end{equation}
where $\sigma(ijklm)=\pm$ is the orientation of the 4-simplex $\langle ijklm \rangle$, and $m_2$ is a $\mathbb{Z}_2$-valued function satisfying $m_2(ijk)+m_2(ikl)+m_2(ijl)+m_2(jkl)=dm_2(ijkl)=n_3(ijkl)$ (mod 2), which contributes as the spin structure term. (Recall Convention \ref{conv.cochain}, $m_2(ijk):=m_2(\Bg_i,\Bg_j,\Bg_k)$ and so on, where $\Bg_i,\Bg_j,\Bg_k\in G$.) Further we define
\begin{align}
\mathcal{V}_4^{+}(01234)=&
(-1)^{m_2(013)+m_2(134)+m_2(123)}
\theta_{1234}^{n_3(1234)}
\theta_{0134}^{n_3(0134)}
\theta_{0123}^{n_3(0123)}
\overline{\theta}_{0234}^{n_3(0234)}
\overline{\theta}_{0124}^{n_3(0124)}
\nu_4(01234),
\end{align}
\begin{align}
\mathcal{V}_4^{-}(01234)=&
(-1)^{m_2(024)}
\theta_{0124}^{n_3(0124)}
\theta_{0234}^{n_3(0234)}
\overline{\theta}_{0123}^{n_3(0123)}
\overline{\theta}_{0134}^{n_3(0134)}
\overline{\theta}_{1234}^{n_3(1234)}
\nu_4(01234)^{-1}.
\end{align}

We relate the bulk 3+1D fSPT states with the 2+1D surface by adding a single vertex to the bulk and connecting the bulk vertex to all surface vertices, and we denote the single vertex in the bulk as $\textbf{g}_*$. Each surface $\cF$-move is associated with a bulk $\widetilde{\cF}$-move. In order to construct the wavefunction of the bulk move, we add a trivial vertex $\textbf{g}_{e}=\textbf{1}$ in one higher dimension and write down the partition function. Let $\textbf{g}=\textbf{g}_{0}^{-1}\textbf{g}_1$, $\textbf{h}=\textbf{g}_{1}^{-1}\textbf{g}_2$ and $\textbf{k}=\textbf{g}_{2}^{-1}\textbf{g}_3$. A surface $\mathcal{F}$-move $^{\textbf{g}_0}[{\mathcal{F}^{a_{\textbf{g}}b_{\textbf{h}}c_{\textbf{k}}}
_{d}]
^{m,\alpha\beta}
_{n,\chi\delta}}$ and its corresponding bulk move $^{\textbf{g}_0}[\widetilde{\mathcal{F}}^{\textbf{g}\textbf{h}\textbf{k}}]
(\textbf{g}_{*})$ together have expression:
\begin{align}
&{^{\textbf{g}_0}[{\mathcal{F}^{a_{\textbf{g}}b_{\textbf{h}}c_{\textbf{k}}}
_{d}]
^{m,\alpha\beta}
_{n,\chi\delta}}}
{^{\textbf{g}_0}[\widetilde{\mathcal{F}}^{\textbf{g}\textbf{h}\textbf{k}}]
(\textbf{g}_{*})}
\nonumber\\
=&
c^{\dagger s(\alpha)}_{\underline{\alpha}}
c^{\dagger s(\beta)}_{\underline{\beta}}
c_{\underline{\delta}}^{s(\delta)}
c_{\underline{\chi}}^{s(\chi)}
{^{\textbf{g}_0}[{F^{a_{\textbf{g}}b_{\textbf{h}}c_{\textbf{k}}}
_{d}]^{m,\alpha\beta}_{n,\chi\delta}}}
\nonumber\\
&
\int
d\theta^{n_3(e01*)}_{e01*}
d\overline{\theta}^{n_3(e01*)}_{e01*}
d\theta^{n_3(e02*)}_{e02*}
d\overline{\theta}^{n_3(e02*)}_{e02*}
d\theta^{n_3(e03*)}_{e03*}
d\overline{\theta}^{n_3(e03*)}_{e03*}
d\theta^{n_3(e12*)}_{e12*}
d\overline{\theta}^{n_3(12*)}_{e12*}
d\theta^{n_3(e13*)}_{e13*}
d\overline{\theta}^{n_3(e13*)}_{e13*}
d\theta^{n_3(e23*)}_{e23*}
d\overline{\theta}^{n_3(e23*)}_{e23*}
\nonumber\\
&\ \ \ \ 
(-1)^{m_2(e0*)+m_2(e1*)+m_2(e2*)+m_2(e3*)}
\mathcal{V}^{-}_4(e123*)\mathcal{V}^{-}_4(e013*)
\mathcal{V}^{+}_4(e023*)\mathcal{V}^{+}_4(e012*)
\nonumber\\
=&
c^{\dagger s(\alpha)}_{\underline{\alpha}}
c^{\dagger s(\beta)}_{\underline{\beta}}
c^{s(\delta)}_{\underline{\delta}}
c^{s(\chi)}_{\underline{\chi}}
{^{\textbf{g}_0}[{F^{a_{\textbf{g}}b_{\textbf{h}}c_{\textbf{k}}}
_{d}]^{m,\alpha\beta}_{n,\chi\delta}}}
\int
d\theta^{n_3(0123)}_{0123}
d\overline{\theta}^{n_3(0123)}_{0123}
\mathcal{V}^{-}_4(e0123)\mathcal{V}^{+}_4(0123*)
\nonumber\\
=&
(-1)^{n_3(0123)(n_3(123*)+n_3(013*))
}
c^{\dagger s(\alpha)}_{\underline{\alpha}}
c^{\dagger s(\beta)}_{\underline{\beta}}
c^{s(\delta)}_{\underline{\delta}}
c^{s(\chi)}_{\underline{\chi}}
{^{\textbf{g}_0}[{F^{a_{\textbf{g}}b_{\textbf{h}}c_{\textbf{k}}}
_{d}]^{m,\alpha\beta}_{n,\chi\delta}}}
\nonumber\\
&
\theta_{e013}^{n_3(e013)}\theta_{e123}^{n_3(e123)}\overline{\theta}_{e012}^{n_3(e012)}\overline{\theta}_{e023}^{n_3(e023)}
\theta_{123*}^{n_3(123*)}\theta_{013*}^{n_3(013*)}\overline{\theta}_{023*}^{n_3(023*)}\overline{\theta}_{012*}^{n_3(012*)}
\frac{\nu_4(0123*)}{\nu_4(e0123)}.
\end{align}
We should take care of which part belongs to the surface and which belongs to the bulk. Actually the complex fermion operators in tetrahedrons $\langle e012 \rangle$, $\langle e023 \rangle$, $\langle e013 \rangle$ and $\langle e123 \rangle$ all belong to surface, and they change the fermion parities on each vertex on surface. Therefore, for anomalous surface $\mathcal{F}$-move, we have
\begin{align}
^{\textbf{g}_0}[{\mathcal{F}^{a_{\textbf{g}}b_{\textbf{h}}c_{\textbf{k}}}
_{d}]
^{m,\alpha\beta}
_{n,\chi\delta}}
=
c^{\dagger s(\alpha)}_{\underline{\alpha}}
c^{\dagger s(\beta)}_{\underline{\beta}}
c^{s(\delta)}_{\underline{\delta}}
c^{s(\chi)}_{\underline{\chi}}
\theta_{e013}^{n_3(e013)}\theta_{e123}^{n_3(e123)}\overline{\theta}_{e012}^{n_3(e012)}\overline{\theta}_{e023}^{n_3(e023)}\cdot
{^{\textbf{g}_0}[{F^{a_{\textbf{g}}b_{\textbf{h}}c_{\textbf{k}}}
_{d}]^{m,\alpha\beta}_{n,\chi\delta}}},
\end{align}
which may violate fermion-parity conservation by $n_3(0123)=n_3(\Bg,\Bh,\Bk)$ (mod 2) in general. 
While for the bulk move, we have
\begin{align}
^{\textbf{g}_0}[\widetilde{\mathcal{F}}^{\textbf{g}\textbf{h}\textbf{k}}]
(\textbf{g}_{*})
=
(-1)^{n_3(0123)(n_3(123*)+n_3(013*))
}
\theta_{123*}^{n_3(123*)}\theta_{013*}^{n_3(013*)}\overline{\theta}_{023*}^{n_3(023*)}\overline{\theta}_{012*}^{n_3(012*)}
\frac{\nu_4(0123*)}{\nu_4(e0123)}.
\end{align}
The surface anomalous $\mathcal{F}$-move and the bulk $\widetilde{\mathcal{F}}$-move together constitute a gfSLU transformation.

\section{Consistency between the anomalous symmetric condition and the surface fermionic pentagon equation}
\label{appen.check}
\label{appen.d2}
In this Appendix, we check that the anomalous symmetry condition in Eq.~(\ref{eq.symaction}) is consistent with the surface fermionic pentagon equation in Eq.~(\ref{eq.ano}). By Eq.~(\ref{eq.symaction}) and recall the property ${^{\Bg_0}\Theta(a_\Bg,b_\Bh,c_\Bk)}{^{\Bg}\Theta(a_\Bg,b_\Bh,c_\Bk)}^{S(\Bg_0)}={^{\Bg_0\Bg}\Theta(a_\Bg,b_\Bh,c_\Bk)}$,
\begin{align}\label{eq.g0gtog}
    ^{\Bg_0\Bg}[{F^{b_{\textbf{h}}c_{\textbf{k}}d_{\textbf{l}}}_{p}]^{n,\eta\kappa}_{q,\delta\phi}}
    =&
    {^{\Bg_0\Bg}\Theta(b_{\textbf{h}},
    c_{\textbf{k}},d_{\textbf{l}}})
    \nu_4(\Bg_0\Bg,\Bh,\Bk,\Bl)
    (^{\one}[{F^{b_{\textbf{h}}c_{\textbf{k}}d_{\textbf{l}}}_{p}]^{n,\eta\kappa}_{q,\delta\phi}})^{S(\Bg_0\Bg)}
    \nonumber\\
    =&
    {^{\Bg_0}\Theta(b_{\textbf{h}},
    c_{\textbf{k}},d_{\textbf{l}}})
    \frac{\nu_4(\Bg_0\Bg,\Bh,\Bk,\Bl)}{\nu_4(\Bg,\Bh,\Bk,\Bl)^{S(\Bg_0)}}
    (^{\Bg}[{F^{b_{\textbf{h}}c_{\textbf{k}}d_{\textbf{l}}}_{p}]^{n,\eta\kappa}_{q,\delta\phi}})^{S(\Bg_0)},
\end{align}
wehre $S(\Bg_0)S(\Bg)=S(\Bg_0\Bg)$. Replacing the above equation into Eq.~(\ref{eq.ano}),
\begin{align}
    &\underset{\epsilon}{\sum} 
    {^{\Bg_0}\Theta(j_{\Bg\Bh},c_\Bk,d_\Bl)}
    \nu_4(\Bg_0,\Bg\Bh,\Bk,\Bl)
    ({^{\one}[{F^{j_{\textbf{gh}}c_{\textbf{k}}d_{\textbf{l}}}_{e}]^{m,\beta\chi}_{q,\delta\epsilon}}})^{S(\Bg_0)}\cdot
    {^{\Bg_0}\Theta(a_\Bg,b_\Bh,q_{\Bk\Bl})}
    \nu_4(\Bg_0,\Bg,\Bh,\Bk\Bl)
    ({^{\one}[{F^{a_{\textbf{g}}b_{\textbf{h}}q_{\textbf{kl}}}_{e}]^{j,\alpha\epsilon}_{p,\phi\gamma}}})^{S(\Bg_0)}
    \nonumber\\=&
    (-1)^{s(\alpha)s(\beta)}
    \underset{n_{\Bh\Bk}\eta\psi\kappa}{\sum} {^{\Bg_0}\Theta(a_{\textbf{g}},b_{\textbf{h}},c_{\textbf{k}}})
    \nu_4(\Bg_0,\Bg,\Bh,\Bk)
    ({^{\one}[{F^{a_{\textbf{g}}b_{\textbf{h}}c_{\textbf{k}}}_{m}]
    ^{j,\alpha\beta}_{n,\eta\psi}}})^{S(\Bg_0)}\cdot
    {^{\Bg_0}\Theta(a_{\textbf{g}},n_{\textbf{hk}},
    d_{\textbf{l}}})
    \nu_4(\Bg_0,\Bg,\Bh\Bk,\Bl)
    ({^{\one}[{F^{a_{\textbf{g}}n_{\textbf{hk}}d_{\textbf{l}}}_{e}]^{m,\psi\chi}_{p,\kappa\gamma}}})^{S(\Bg_0)}
    \nonumber\\&\quad\quad\quad\quad\quad\quad\quad
    {^{\Bg_0}\Theta(b_{\textbf{h}},
    c_{\textbf{k}},d_{\textbf{l}}})
    \frac{\nu_4(\Bg_0\Bg,\Bh,\Bk,\Bl)}{\nu_4(\Bg,\Bh,\Bk,\Bl)^{S(\Bg_0)}}
    (^{\Bg}[{F^{b_{\textbf{h}}c_{\textbf{k}}d_{\textbf{l}}}_{p}]^{n,\eta\kappa}_{q,\delta\phi}})^{S(\Bg_0)}
    .
\end{align}
Then by Eq.~(\ref{eq.omega}), and taking an $S(\Bg_0)$-conjugate for the whole equation, we obtain
\begin{align}
    \underset{\epsilon}{\sum} 
    {^{\one}[{F^{j_{\textbf{gh}}c_{\textbf{k}}d_{\textbf{l}}}_{e}]^{m,\beta\chi}_{q,\delta\epsilon}}}
    {^{\one}[{F^{a_{\textbf{g}}b_{\textbf{h}}q_{\textbf{kl}}}_{e}]^{j,\alpha\epsilon}_{p,\phi\gamma}}}
    =
    (-1)^{s(\alpha)s(\delta)}
    \underset{n_{\Bh\Bk}\eta\psi\kappa}{\sum} 
    {^{\one}[{F^{a_{\textbf{g}}b_{\textbf{h}}c_{\textbf{k}}}_{m}]^{j,\alpha\beta}_{n,\eta\psi}}}
    {^{\one}[{F^{a_{\textbf{g}}n_{\textbf{hk}}d_{\textbf{l}}}_{e}]^{m,\psi\chi}_{p,\kappa\gamma}}}
    ^{\textbf{g}}[{F^{b_{\textbf{h}}c_{\textbf{k}}d_{\textbf{l}}}_{p}]^{n,\eta\kappa}_{q,\delta\phi}}.
\end{align}

\section{Canonical choice of cochains}
\label{Appendix.cochain}

This appendix will be useful if we further consider the exactly solvable model for 2+1D surface SET phases with the $H^2(G,\Z_2)$ fermionic 't Hooft anomaly. It has been proved in Ref. \onlinecite{SPT} that we can choose a gauge for any n-cocycle $\omega_n$ such that
\begin{eqnarray}\label{General}
\omega_n(\underbrace{\bf{1,...,1}}_{m\text{ terms}},\textbf{g}_1,...,\textbf{g}_{n+1-m})=1,
\end{eqnarray}
where $m$ is the number of repeating index $\textbf{1}$ ($\textbf{1}\in G$ is the trivial element) with $2\leq m\leq n+1$ and $\textbf{g}_1\neq \textbf{1}$.

Here we show that we can also do such a canonical choice for the data in describing general fSPT phases\cite{Wang2020}, which are cochains $\nu_4$, $n_3$, $n_2$ and cocycle $n_1$, in the case that $\beta_4(01234) := s_1(01)n_2(134)n_2(123)=0\text{ (mod 2)}$. Recall that for any $n$-cochain $f_n(\one,\Bg_0,\Bg_1,\Bg_2,...)$, we write it in abbreviation as $f_n(e0123...)$. They satisfy
\begin{align}
&\dd n_1=0,\\
&\dd n_2 = \omega_2\smile n_1 + s_1\smile n_1\smile n_1 \text{ (mod 2)},\\
&\dd n_3 = \omega_2\smile n_2 + n_2\smile n_2 + s_1\smile(n_2\smile_1n_2) \text{ (mod 2)},\\
&\dd \nu_4 = \mathcal{O}_5[n_3],
\end{align}
where 
\begin{align}\nonumber
\mathcal O_5[n_3](012345)\big |_{\be_4=0} &= (-1)^{(\om_2\smile n_3 + n_3\smile_1 n_3 + n_3\smile_2 \dd n_3)(012345) + \om_2(013)\dd n_3(12345)}\\\nonumber
&\quad\times (-1)^{\dd n_3(02345)\dd n_3(01235) + \om_2(023) \left[\dd n_3(01245) + \dd n_3(01235) + \dd n_3(01234)\right]} \\
&\quad\times i^{\dd n_3(01245) \dd n_3(01234) \text{ (mod 2)}}
\times (-i)^{\left[\dd n_3(12345)+\dd n_3(02345)+\dd n_3(01345)\right] \dd n_3(01235) \text{ (mod 2)}}
\nonumber\\
&{=(-1)^{(\om_2\smile n_3 + n_3\smile_1 n_3 + n_3\smile_2 \dd n_3+ \dd n_3\smile_1 n_2+\zeta_5(n_2+\omega_2,n_2))(012345)}}
\nonumber\\
&{\quad\times i^{[n_2+\omega_2]_2\smile (n_2\smile_1 n_2)}
,}
\label{o5}
\end{align}
\end{widetext}
{where the above simplification is from Ref.~\onlinecite{barkeshli2024higher}, and $\zeta_5(a_2,b_2)(012345)=a_2(023)a_2(012)b_2(345)b_2(235)$ for any $a_2,b_2\in Z^2(G,\Z_2)$, and $s_1(\Bg_0,\Bg_1)\in\{0,1\}$ indicates $\Bg_0^{-1}\Bg_1$ is anti-unitary or not.}

Since $\omega_2$ and $s_1$ are cocycles, we can at first choose
\begin{align}
&\omega_2(\textbf{1},\textbf{1},\textbf{g}_1)=\omega_2(\textbf{1},\textbf{1},\textbf{1})=0\text{ (mod 2)},
\\
&s_1(\textbf{1},\textbf{1})=0\text{ (mod 2)}.
\end{align}

\subsection{\texorpdfstring{$\mathbb{Z}_2$}{}-valued cochain \texorpdfstring{$n_2$}{}}

First we show we can choose
\begin{align}
n_2(\textbf{1},\textbf{1},\textbf{1})=0 \text{ (mod 2)}.
\end{align}
Following the same argument in Ref. \onlinecite{SPT}, we introduce a coboundary $b_2(\textbf{1},\textbf{1},\textbf{1})=d\mu_1 (\textbf{1},\textbf{1},\textbf{1})=\mu_1 (\textbf{1},\textbf{1})$, where $\mu_1$ is a $\mathbb{Z}_2$-valued 1-cochain. If we require that $\mu_1 (\textbf{1},\textbf{1})=n'_2(\textbf{1},\textbf{1},\textbf{1})$, then after the gauge transformation $n_2=n'_2 +b_2 \text{ (mod 2)}$, we have $n_2(\textbf{1},\textbf{1},\textbf{1})=0 \text{ (mod 2)}$.

Second we show 
\begin{align}
n_2(\textbf{1},\textbf{1},\textbf{g}_1)=0 \text{ (mod 2)}.
\end{align}
We have 
\begin{align}
\dd n_2(\textbf{1},\textbf{1},\textbf{1},\textbf{g}_1)
=
n_2(\textbf{1},\textbf{1},\textbf{g}_1).
\end{align}
And we check
$
\dd n_2(eee1)
=
 \omega_2\smile n_1 (eee1)+ s_1\smile n_1\smile n_1 (eee1)=0 \text{ (mod 2)}
$.

\subsection{\texorpdfstring{$\mathbb{Z}_2$}{}-valued cochain \texorpdfstring{$n_3$}{}}

First we show 
\begin{align}
n_3(\textbf{1},\textbf{1},\textbf{1},\textbf{1})=0 \text{ (mod 2)}.
\end{align}
We have 
\begin{align}
\dd n_3(\textbf{1},\textbf{1},\textbf{1},\textbf{1},\textbf{1})
=
n_3(\textbf{1},\textbf{1},\textbf{1},\textbf{1}).
\end{align}
And we check
$
dn_3(eeeee)
=
\omega_2\smile n_2 (eeeee)+ n_2\smile n_2 (eeeee) + s_1\smile(n_2\smile_1n_2) (eeeee)=0 \text{ (mod 2)}
$.

Second we show we can choose 
\begin{align}
n_3(\textbf{1},\textbf{1},\textbf{1},\textbf{g}_1)=0 \text{ (mod 2)},
\end{align}
This is also done by the gauge transformation $n_3=n'_3 +b_3 \text{ (mod 2)}$, where $b_3 (\textbf{1},\textbf{1},\textbf{1},\textbf{g}_1)=d\mu_2 (\textbf{1},\textbf{1},\textbf{1},\textbf{g}_1)=\mu_2(\textbf{1},\textbf{1},\textbf{g}_1)+\mu_2(\textbf{1},\textbf{1},\textbf{1})$. By requiring $n'_3(\textbf{1},\textbf{1},\textbf{1},\textbf{g}_1)=\mu_2(\textbf{1},\textbf{1},\textbf{g}_1)+\mu_2(\textbf{1},\textbf{1},\textbf{1})$, we fall into this gauge.

Third we show in this gauge, 
\begin{align}
n_3(\textbf{1},\textbf{1},\textbf{g}_1,\textbf{g}_2)=0 \text{ (mod 2)}.
\end{align}
We have 
\begin{align}
&\dd n_3(\textbf{1},\textbf{1},\textbf{1},\textbf{g}_1,\textbf{g}_2)
\nonumber\\
=&
n_3(\textbf{1},\textbf{1},\textbf{g}_1,\textbf{g}_2)
+
n_3(\textbf{1},\textbf{1},\textbf{1},\textbf{g}_2)
+
n_3(\textbf{1},\textbf{1},\textbf{1},\textbf{g}_1)
\nonumber\\
=&
n_3(\textbf{1},\textbf{1},\textbf{g}_1,\textbf{g}_2).
\end{align}
And we check
$
dn_3(eee12)
=
\omega_2\smile n_2 (eee12)+ n_2\smile n_2 (eee12) + s_1\smile(n_2\smile_1n_2) (eee12)=0 \text{ (mod 2)}
$.

\subsection{\texorpdfstring{$U(1)$}{}-valued cochain \texorpdfstring{$\nu_4$}{}}

First we show we can choose
\begin{align}
\nu_4(\textbf{1},\textbf{1},\textbf{1},\textbf{1},\textbf{1})=1.
\end{align}
We introduce a coboundary $b_4(\textbf{1},\textbf{1},\textbf{1},\textbf{1},\textbf{1})=d\mu_3 (\textbf{1},\textbf{1},\textbf{1},\textbf{1},\textbf{1})=\mu_3 (\textbf{1},\textbf{1},\textbf{1},\textbf{1})$, where $\mu_3$ is a $U(1)$-valued 3-cochain. If we require that $\mu_3 (\textbf{1},\textbf{1},\textbf{1},\textbf{1})=\nu'_4(\textbf{1},\textbf{1},\textbf{1},\textbf{1},\textbf{1})$, then after the gauge transformation $\nu_4=\nu'_4 b^{-1}_4$, we have $\nu_4(\textbf{1},\textbf{1},\textbf{1},\textbf{1},\textbf{1})=1$.

And we can also choose
\begin{align}
\nu_4(\textbf{1},\textbf{1},\textbf{1},\textbf{g}_1,\textbf{g}_2)=1.
\end{align} 
We still consider gauge transformation $\nu_4=\nu'_4 b^{-1}_4$, where $b_4(\textbf{1},\textbf{1},\textbf{1},\textbf{g}_1,\textbf{g}_2)=d\mu_3 (\textbf{1},\textbf{1},\textbf{1},\textbf{g}_1,\textbf{g}_2)$\\$=\mu_3 (\textbf{1},\textbf{1},\textbf{g}_1,\textbf{g}_2)\mu^{-1}_3 (\textbf{1},\textbf{1},\textbf{1},\textbf{g}_2)\mu_3 (\textbf{1},\textbf{1},\textbf{1},\textbf{g}_1)$. By requiring $\nu'_4(\textbf{1},\textbf{1},\textbf{1},\textbf{g}_1,\textbf{g}_2)=$\\$ \mu_3 (\textbf{1},\textbf{1},\textbf{g}_1,\textbf{g}_2)
\mu^{-1}_3 (\textbf{1},\textbf{1},\textbf{1},\textbf{g}_2)\mu_3 (\textbf{1},\textbf{1},\textbf{1},\textbf{g}_1)$, we fall into this gauge.

Then we show in this gauge,
\begin{align}
&\nu_4(\textbf{1},\textbf{1},\textbf{1},\textbf{1},\textbf{g}_1)=1,
\\
&\nu_4(\textbf{1},\textbf{1},\textbf{g}_1,\textbf{g}_2,\textbf{g}_3)=1.
\end{align}
We have
\begin{align}
\dd \nu_4(\textbf{1},\textbf{1},\textbf{1},\textbf{1},\textbf{1},\textbf{g}_1)
=&
\nu_4(\textbf{1},\textbf{1},\textbf{1},\textbf{1},\textbf{g}_1),
\\
\dd \nu_4(\textbf{1},\textbf{1},\textbf{1},\textbf{g}_1,\textbf{g}_2,\textbf{g}_3)
=&
\frac{\nu_4(\textbf{1},\textbf{1},\textbf{g}_1,\textbf{g}_2,\textbf{g}_3)\nu_4(\textbf{1},\textbf{1},\textbf{1},\textbf{g}_1,\textbf{g}_3)}
{\nu_4(\textbf{1},\textbf{1},\textbf{1},\textbf{g}_2,\textbf{g}_3)
\nu_4(\textbf{1},\textbf{1},\textbf{1},\textbf{g}_1,\textbf{g}_2)}
\nonumber\\
=&
\nu_4(\textbf{1},\textbf{1},\textbf{g}_1,\textbf{g}_2,\textbf{g}_3)
.
\end{align}
And we check from Eq.~(\ref{o5}) that $\dd \nu_4(\textbf{1},\textbf{1},\textbf{1},\textbf{1},\textbf{1},\textbf{g}_1)=\dd \nu_4(\textbf{1},\textbf{1},\textbf{1},\textbf{g}_1,\textbf{g}_2,\textbf{g}_3)=1$.


\bibliography{SETSN.bib}

@article{Fid2018,
  title={Surface topological order and a new't Hooft anomaly of interaction enabled 3+ 1D fermion SPTs},
  author={Fidkowski, Lukasz and Vishwanath, Ashvin and Metlitski, Max A},
  journal={arXiv preprint arXiv:1804.08628},
  year={2018},
  url={https://arxiv.org/abs/1804.08628},
}

@article{Cheng2017,
   title={Exactly solvable models for symmetry-enriched topological phases},
   volume={96},
   ISSN={2469-9969},
   url={http://dx.doi.org/10.1103/PhysRevB.96.115107},
   DOI={10.1103/physrevb.96.115107},
   number={11},
   journal={Physical Review B},
   publisher={American Physical Society (APS)},
   author={Cheng, Meng and Gu, Zheng-Cheng and Jiang, Shenghan and Qi, Yang},
   year={2017},
   month={Sep}
}

@article{BSET,
   title={Symmetry fractionalization, defects, and gauging of topological phases},
   volume={100},
   ISSN={2469-9969},
   url={http://dx.doi.org/10.1103/PhysRevB.100.115147},
   DOI={10.1103/physrevb.100.115147},
   number={11},
   journal={Physical Review B},
   publisher={American Physical Society (APS)},
   author={Barkeshli, Maissam and Bonderson, Parsa and Cheng, Meng and Wang, Zhenghan},
   year={2019},
   month={Sep}
}

@article{FSET,
  title={Characterization and classification of fermionic symmetry enriched topological phases},
  author={Aasen, David and Bonderson, Parsa and Knapp, Christina},
  journal={arXiv preprint arXiv:2109.10911},
  year={2021},
   url={https://arxiv.org/abs/2109.10911},
}

@article{Wang2018,
   title={Towards a Complete Classification of Symmetry-Protected Topological Phases for Interacting Fermions in Three Dimensions and a General Group Supercohomology Theory},
   volume={8},
   ISSN={2160-3308},
   url={http://dx.doi.org/10.1103/PhysRevX.8.011055},
   DOI={10.1103/physrevx.8.011055},
   number={1},
   journal={Physical Review X},
   publisher={American Physical Society (APS)},
   author={Wang, Qing-Rui and Gu, Zheng-Cheng},
   year={2018},
   month={Mar}
}

@article{Gu2014,
   title={Symmetry-protected topological orders for interacting fermions: Fermionic topological nonlinear sigma models and a special group supercohomology theory},
   volume={90},
   ISSN={1550-235X},
   url={http://dx.doi.org/10.1103/PhysRevB.90.115141},
   DOI={10.1103/physrevb.90.115141},
   number={11},
   journal={Physical Review B},
   publisher={American Physical Society (APS)},
   author={Gu, Zheng-Cheng and Wen, Xiao-Gang},
   year={2014},
   month={Sep}
}

@article{Wang2020,
   title={Construction and Classification of Symmetry-Protected Topological Phases in Interacting Fermion Systems},
   volume={10},
   ISSN={2160-3308},
   url={http://dx.doi.org/10.1103/PhysRevX.10.031055},
   DOI={10.1103/physrevx.10.031055},
   number={3},
   journal={Physical Review X},
   publisher={American Physical Society (APS)},
   author={Wang, Qing-Rui and Gu, Zheng-Cheng},
   year={2020},
   month={Sep}
}

@article{Wen2002,
   title={Quantum orders and symmetric spin liquids},
   volume={65},
   ISSN={1095-3795},
   url={http://dx.doi.org/10.1103/PhysRevB.65.165113},
   DOI={10.1103/physrevb.65.165113},
   number={16},
   journal={Physical Review B},
   publisher={American Physical Society (APS)},
   author={Wen, Xiao-Gang},
   year={2002},
   month={Apr}
}

@article{2dtopo,
   title={Classification of two-dimensional fermionic and bosonic topological orders},
   volume={91},
   ISSN={1550-235X},
   url={http://dx.doi.org/10.1103/PhysRevB.91.125149},
   DOI={10.1103/physrevb.91.125149},
   number={12},
   journal={Physical Review B},
   publisher={American Physical Society (APS)},
   author={Gu, Zheng-Cheng and Wang, Zhenghan and Wen, Xiao-Gang},
   year={2015},
   month={Mar} }

@article{fSN,
  title = {Towards a complete classification of nonchiral topological phases in two-dimensional fermion systems},
  author = {Zhou, Jing-Ren and Wang, Qing-Rui and Gu, Zheng-Cheng},
  journal = {Phys. Rev. B},
  volume = {106},
  issue = {24},
  pages = {245120},
  numpages = {48},
  year = {2022},
  month = {Dec},
  publisher = {American Physical Society},
  doi = {10.1103/PhysRevB.106.245120},
  url = {https://link.aps.org/doi/10.1103/PhysRevB.106.245120}
}

@article{SPT,
	doi = {10.1103/physrevb.87.155114},
  
	url = {https://doi.org/10.1103%2Fphysrevb.87.155114},
  
	year = 2013,
	month = {apr},
  
	publisher = {American Physical Society ({APS})},
  
	volume = {87},
  
	number = {15},
  
	author = {Xie Chen and Zheng-Cheng Gu and Zheng-Xin Liu and Xiao-Gang Wen},
  
	title = {Symmetry protected topological orders and the group cohomology of their symmetry group},
  
	journal = {Physical Review B}
}

@article{fc,
  title={Fermion condensation and super pivotal categories},
  author={Aasen, David and Lake, Ethan and Walker, Kevin},
  journal={Journal of Mathematical Physics},
  volume={60},
  number={12},
  pages={121901},
  year={2019},
  publisher={AIP Publishing LLC},
  url={https://doi.org/10.1063/1.5045669},
}

@article{threeloop,
  title={Non-Abelian three-loop braiding statistics for 3D fermionic topological phases},
  author={Zhou, Jing-Ren and Wang, Qing-Rui and Wang, Chenjie and Gu, Zheng-Cheng},
  journal={Nature communications},
  volume={12},
  number={1},
  pages={1--10},
  year={2021},
  publisher={Nature Publishing Group},
  url={https://doi.org/10.1038/s41467-021-23309-3},
}

@article{galindo2017categorical,
  title={Categorical fermionic actions and minimal modular extensions},
  author={Galindo, C{\'e}sar and Venegas-Ram{\'\i}rez, C{\'e}sar F},
  journal={arXiv preprint arXiv:1712.07097},
  year={2017},
  url={https://arxiv.org/abs/1712.07097},
}

@article{rowell2008unitarizability,
  title={Unitarizability of premodular categories},
  author={Rowell, Eric C},
  journal={Journal of Pure and Applied Algebra},
  volume={212},
  number={8},
  pages={1878--1887},
  year={2008},
  publisher={Elsevier},
  url={https://doi.org/10.1016/j.jpaa.2007.11.004},
}

@article{bruillard2020classification,
  title={Classification of super-modular categories by rank},
  author={Bruillard, Paul and Galindo, C{\'e}sar and Ng, Siu-Hung and Plavnik, Julia Y and Rowell, Eric C and Wang, Zhenghan},
  journal={Algebras and Representation Theory},
  volume={23},
  pages={795--809},
  year={2020},
  publisher={Springer},
  url={https://doi.org/10.48550/arXiv.1705.05293},
}

@article{bruillard2019classification,
  title={Classification of super-modular categories},
  author={Bruillard, Paul and Plavnik, Julia Yael and Rowell, Eric C and Zhang, Qing},
  journal={arXiv preprint arXiv:1909.09843},
  year={2019},
  url={
https://doi.org/10.48550/arXiv.1705.05293
},
}

@article{bulmash2022fermionic,
  title={Fermionic symmetry fractionalization in (2+ 1) dimensions},
  author={Bulmash, Daniel and Barkeshli, Maissam},
  journal={Physical Review B},
  volume={105},
  number={12},
  pages={125114},
  year={2022},
  publisher={APS},
  url={
https://doi.org/10.1103/PhysRevB.105.125114
Focus to learn more
},
}

@article{bulmash2022anomaly,
  title={Anomaly cascade in (2+ 1)-dimensional fermionic topological phases},
  author={Bulmash, Daniel and Barkeshli, Maissam},
  journal={Physical Review B},
  volume={105},
  number={15},
  pages={155126},
  year={2022},
  publisher={APS},
  url={
https://doi.org/10.1103/PhysRevB.105.155126
},
}

@article{kitaev2006anyons,
  title={Anyons in an exactly solved model and beyond},
  author={Kitaev, Alexei},
  journal={Annals of Physics},
  volume={321},
  number={1},
  pages={2--111},
  year={2006},
  publisher={Elsevier},
  url={https://doi.org/10.1016/j.aop.2005.10.005
},
}

@article{lan2017modular,
  title={Modular extensions of unitary braided fusion categories and 2+ 1D topological/SPT orders with symmetries},
  author={Lan, Tian and Kong, Liang and Wen, Xiao-Gang},
  journal={Communications in Mathematical Physics},
  volume={351},
  pages={709--739},
  year={2017},
  publisher={Springer},
  url={
https://doi.org/10.1007/s00220-016-2748-y
},
}

@article{lan2016theory,
  title={Theory of (2+ 1)-dimensional fermionic topological orders and fermionic/bosonic topological orders with symmetries},
  author={Lan, Tian and Kong, Liang and Wen, Xiao-Gang},
  journal={Physical Review B},
  volume={94},
  number={15},
  pages={155113},
  year={2016},
  publisher={APS},
  url={
https://doi.org/10.1103/PhysRevB.94.155113
},
}

@article{heinrich2016symmetry,
  title={Symmetry-enriched string nets: Exactly solvable models for set phases},
  author={Heinrich, Chris and Burnell, Fiona and Fidkowski, Lukasz and Levin, Michael},
  journal={Physical Review B},
  volume={94},
  number={23},
  pages={235136},
  year={2016},
  publisher={APS},
  url={
https://doi.org/10.1103/PhysRevB.94.235136
},
}

@article{gu2014lattice,
  title={Lattice model for fermionic toric code},
  author={Gu, Zheng-Cheng and Wang, Zhenghan and Wen, Xiao-Gang},
  journal={Physical Review B},
  volume={90},
  number={8},
  pages={085140},
  year={2014},
  publisher={APS},
  url={
https://doi.org/10.1103/PhysRevB.90.085140
},
}

@article{chen2015anomalous,
  title={Anomalous symmetry fractionalization and surface topological order},
  author={Chen, Xie and Burnell, Fiona J and Vishwanath, Ashvin and Fidkowski, Lukasz},
  journal={Physical Review X},
  volume={5},
  number={4},
  pages={041013},
  year={2015},
  publisher={APS},
  url={
https://doi.org/10.1103/PhysRevX.5.041013
},
}

@article{essin2013classifying,
  title={Classifying fractionalization: Symmetry classification of gapped Z 2 spin liquids in two dimensions},
  author={Essin, Andrew M and Hermele, Michael},
  journal={Physical Review B},
  volume={87},
  number={10},
  pages={104406},
  year={2013},
  publisher={APS},
  url={
https://doi.org/10.1103/PhysRevB.87.104406
},
}

@article{chen2017symmetry,
  title={Symmetry fractionalization in two dimensional topological phases},
  author={Chen, Xie},
  journal={Reviews in Physics},
  volume={2},
  pages={3--18},
  year={2017},
  publisher={Elsevier},
  url={
https://doi.org/10.1016/j.revip.2017.02.002},
}

@article{etingof2010fusion,
  title={Fusion categories and homotopy theory},
  author={Etingof, Pavel and Nikshych, Dmitri and Ostrik, Victor},
  journal={Quantum topology},
  volume={1},
  number={3},
  pages={209--273},
  year={2010},
  url={
https://doi.org/10.48550/arXiv.0909.3140
},
}

@article{gelaki2009centers,
  title={Centers of graded fusion categories},
  author={Gelaki, Shlomo and Naidu, Deepak and Nikshych, Dmitri},
  journal={Algebra \& Number Theory},
  volume={3},
  number={8},
  pages={959--990},
  year={2009},
  publisher={Mathematical Sciences Publishers},
  url={
https://doi.org/10.48550/arXiv.0905.3117},
}

@article{cui2016gauging,
  title={On gauging symmetry of modular categories},
  author={Cui, Shawn X and Galindo, C{\'e}sar and Plavnik, Julia Yael and Wang, Zhenghan},
  journal={Communications in Mathematical Physics},
  volume={348},
  number={3},
  pages={1043--1064},
  year={2016},
  publisher={Springer},
  url={
https://doi.org/10.1007/s00220-016-2633-8
},
}

@article{lan2023gauging,
  title={Gauging of generalized symmetry},
  author={Lan, Tian and Yue, Gen and Wang, Longye},
  journal={arXiv preprint arXiv:2312.15958},
  year={2023},
  url={
https://doi.org/10.48550/arXiv.2312.15958
},
}

@article{wang2021domain,
  title={Domain wall decorations, anomalies and spectral sequences in bosonic topological phases},
  author={Wang, Qing-Rui and Ning, Shang-Qiang and Cheng, Meng},
  journal={arXiv preprint arXiv:2104.13233},
  year={2021},
  url={
https://doi.org/10.48550/arXiv.2104.13233},
}

@article{wang2022exactly,
  title={Exactly solvable models for U (1) symmetry-enriched topological phases},
  author={Wang, Qing-Rui and Cheng, Meng},
  journal={Physical Review B},
  volume={106},
  number={11},
  pages={115104},
  year={2022},
  publisher={APS}
}

@article{jiang2021generalized,
  title={Generalized Lieb-Schultz-Mattis theorem on bosonic symmetry protected topological phases},
  author={Jiang, Shenghan and Cheng, Meng and Qi, Yang and Lu, Yuan-Ming},
  journal={SciPost Physics},
  volume={11},
  number={2},
  pages={024},
  year={2021},
  url={
https://doi.org/10.21468/SciPostPhys.11.2.024
},
}

@article{cheng2019fermionic,
  title={Fermionic Lieb-Schultz-Mattis theorems and weak symmetry-protected phases},
  author={Cheng, Meng},
  journal={Physical Review B},
  volume={99},
  number={7},
  pages={075143},
  year={2019},
  publisher={APS},
  url={
https://doi.org/10.1103/PhysRevB.99.075143
},
}

@article{wang2018symmetric,
  title={Symmetric gapped interfaces of SPT and SET states: systematic constructions},
  author={Wang, Juven and Wen, Xiao-Gang and Witten, Edward},
  journal={Physical Review X},
  volume={8},
  number={3},
  pages={031048},
  year={2018},
  publisher={APS},
  url={
https://doi.org/10.1103/PhysRevX.8.031048},
}

@article{barkeshli2020relative,
  title={Relative anomalies in (2+ 1) D symmetry enriched topological states},
  author={Barkeshli, Maissam and Cheng, Meng},
  journal={SciPost Physics},
  volume={8},
  number={2},
  pages={028},
  year={2020},
  url={
https://doi.org/10.21468/SciPostPhys.8.2.028},
}

@article{cheng2023lieb,
  title={Lieb-Schultz-Mattis, Luttinger, and't Hooft-anomaly matching in lattice systems},
  author={Cheng, Meng and Seiberg, Nathan},
  journal={SciPost Physics},
  volume={15},
  number={2},
  pages={051},
  year={2023},
  url={
https://doi.org/10.21468/SciPostPhys.15.2.051
},
}

@article{burnell2014exactly,
  title={Exactly soluble model of a three-dimensional symmetry-protected topological phase of bosons with surface topological order},
  author={Burnell, Fiona J and Chen, Xie and Fidkowski, Lukasz and Vishwanath, Ashvin},
  journal={Physical Review B},
  volume={90},
  number={24},
  pages={245122},
  year={2014},
  publisher={APS},
  url={
https://doi.org/10.1103/PhysRevB.90.245122
},
}

@article{green2023enriched,
  title={Enriched string-net models and their excitations},
  author={Green, David and Huston, Peter and Kawagoe, Kyle and Penneys, David and Poudel, Anup and Sanford, Sean},
  journal={arXiv preprint arXiv:2305.14068},
  year={2023},
  url={
https://doi.org/10.22331/q-2024-03-28-1301
},
}

@article{fidkowski2013non,
  title = {Non-Abelian Topological Order on the Surface of a 3D Topological Superconductor from an Exactly Solved Model},
  author = {Fidkowski, Lukasz and Chen, Xie and Vishwanath, Ashvin},
  journal = {Phys. Rev. X},
  volume = {3},
  issue = {4},
  pages = {041016},
  numpages = {17},
  year = {2013},
  month = {Nov},
  publisher = {American Physical Society},
  doi = {10.1103/PhysRevX.3.041016},
  url = {https://link.aps.org/doi/10.1103/PhysRevX.3.041016}
}

@article{cheng2018microscopic,
  title={Microscopic theory of surface topological order for topological crystalline superconductors},
  author={Cheng, Meng},
  journal={Physical Review Letters},
  volume={120},
  number={3},
  pages={036801},
  year={2018},
  publisher={APS},
  url={
https://doi.org/10.1103/PhysRevLett.120.036801
},
}

@article{metlitski2014interaction,
  title={Interaction effects on 3D topological superconductors: surface topological order from vortex condensation, the 16 fold way and fermionic Kramers doublets},
  author={Metlitski, Max A and Fidkowski, Lukasz and Chen, Xie and Vishwanath, Ashvin},
  journal={arXiv preprint arXiv:1406.3032},
  year={2014},
  url={
https://doi.org/10.48550/arXiv.1406.3032},
}

@article{hong2017topological,
  title={Topological order and symmetry anomaly on the surface of topological crystalline insulators},
  author={Hong, Sungjoon and Fu, Liang},
  journal={arXiv preprint arXiv:1707.02594},
  year={2017},
  url={
https://doi.org/10.48550/arXiv.1707.02594
},
}

@article{wang2017anomaly,
  title={Anomaly indicators for time-reversal symmetric topological orders},
  author={Wang, Chenjie and Levin, Michael},
  journal={Physical review letters},
  volume={119},
  number={13},
  pages={136801},
  year={2017},
  publisher={APS},
  url={
https://doi.org/10.1103/PhysRevLett.119.136801
},
}

@article{wang2014interacting,
  title={Interacting fermionic topological insulators/superconductors in three dimensions},
  author={Wang, Chong and Senthil, T},
  journal={Physical Review B},
  volume={89},
  number={19},
  pages={195124},
  year={2014},
  publisher={APS},
  url={
https://doi.org/10.1103/PhysRevB.89.195124
},
}

@article{shirley2022three,
  title={Three-dimensional quantum cellular automata from chiral semion surface topological order and beyond},
  author={Shirley, Wilbur and Chen, Yu-An and Dua, Arpit and Ellison, Tyler D and Tantivasadakarn, Nathanan and Williamson, Dominic J},
  journal={PRX Quantum},
  volume={3},
  number={3},
  pages={030326},
  year={2022},
  publisher={APS},
  url={https://doi.org/10.1103/PRXQuantum.3.030326},
}

@article{delmastro2021global,
  title={Global anomalies on the Hilbert space},
  author={Delmastro, Diego and Gaiotto, Davide and Gomis, Jaume},
  journal={Journal of High Energy Physics},
  volume={2021},
  number={11},
  pages={1--68},
  year={2021},
  publisher={Springer},
  url={
https://doi.org/10.1007/JHEP11%282021%29142
},
}

@article{cho2014conflicting,
  title={Conflicting symmetries in topologically ordered surface states of three-dimensional bosonic symmetry protected topological phases},
  author={Cho, Gil Young and Teo, Jeffrey CY and Ryu, Shinsei},
  journal={Physical Review B},
  volume={89},
  number={23},
  pages={235103},
  year={2014},
  publisher={APS},
  url={
https://doi.org/10.1103/PhysRevB.89.235103},
}

@article{wang2016bulk,
  title={Bulk-boundary correspondence for three-dimensional symmetry-protected topological phases},
  author={Wang, Chenjie and Lin, Chien-Hung and Levin, Michael},
  journal={Physical Review X},
  volume={6},
  number={2},
  pages={021015},
  year={2016},
  publisher={APS},
  url={https://doi.org/10.1103/PhysRevX.6.021015},
}

@article{xu2013three,
  title={Three-dimensional z 2 topological phases enriched by time-reversal symmetry},
  author={Xu, Cenke},
  journal={Physical Review B},
  volume={88},
  number={20},
  pages={205137},
  year={2013},
  publisher={APS},
  url={
https://doi.org/10.1103/PhysRevB.88.205137
},
}

@article{wang2013boson,
  title={Boson topological insulators: A window into highly entangled quantum phases},
  author={Wang, Chong and Senthil, T},
  journal={Physical Review B},
  volume={87},
  number={23},
  pages={235122},
  year={2013},
  publisher={APS},
  url={
https://doi.org/10.1103/PhysRevB.87.235122
},
}

@article{vishwanath2013physics,
  title={Physics of three-dimensional bosonic topological insulators: Surface-deconfined criticality and quantized magnetoelectric effect},
  author={Vishwanath, Ashvin and Senthil, T},
  journal={Physical Review X},
  volume={3},
  number={1},
  pages={011016},
  year={2013},
  publisher={APS},
  url={
https://doi.org/10.1103/PhysRevX.3.011016
},
}

@article{bonderson2013time,
  title={A time-reversal invariant topological phase at the surface of a 3D topological insulator},
  author={Bonderson, Parsa and Nayak, Chetan and Qi, Xiao-Liang},
  journal={Journal of Statistical Mechanics: Theory and Experiment},
  volume={2013},
  number={09},
  pages={P09016},
  year={2013},
  publisher={IOP Publishing},
  url={
https://doi.org/10.1088/1742-5468/2013/09/P09016
},
}

@article{metlitski2015symmetry,
  title={Symmetry-respecting topologically ordered surface phase of three-dimensional electron topological insulators},
  author={Metlitski, Max A and Kane, CL and Fisher, Matthew PA},
  journal={Physical Review B},
  volume={92},
  number={12},
  pages={125111},
  year={2015},
  publisher={APS},
  url={
https://doi.org/10.1103/PhysRevB.92.125111
},
}

@article{tata2023anomalies,
  title={Anomalies in (2+ 1) D fermionic topological phases and (3+ 1) D path integral state sums for fermionic SPTs},
  author={Tata, Srivatsa and Kobayashi, Ryohei and Bulmash, Daniel and Barkeshli, Maissam},
  journal={Communications in Mathematical Physics},
  volume={397},
  number={1},
  pages={199--336},
  year={2023},
  publisher={Springer},
  url={
https://doi.org/10.1007/s00220-022-04484-w
},
}

@article{bruillard2017fermionic,
  title={Fermionic modular categories and the 16-fold way},
  author={Bruillard, Paul and Galindo, C{\'e}sar and Hagge, Tobias and Ng, Siu-Hung and Plavnik, Julia Yael and Rowell, Eric C and Wang, Zhenghan},
  journal={Journal of Mathematical Physics},
  volume={58},
  number={4},
  year={2017},
  publisher={AIP Publishing},
  url={
https://doi.org/10.1063/1.4982048
},
}

@article{cheng2016translational,
  title={Translational symmetry and microscopic constraints on symmetry-enriched topological phases: A view from the surface},
  author={Cheng, Meng and Zaletel, Michael and Barkeshli, Maissam and Vishwanath, Ashvin and Bonderson, Parsa},
  journal={Physical Review X},
  volume={6},
  number={4},
  pages={041068},
  year={2016},
  publisher={APS},
  url={
https://doi.org/10.1103/PhysRevX.6.041068},
}

@article{usher2018fermionic,
  title={Fermionic 6j-symbols in superfusion categories},
  author={Usher, Robert},
  journal={Journal of Algebra},
  volume={503},
  pages={453--473},
  year={2018},
  publisher={Elsevier},
  url={https://doi.org/10.48550/arXiv.1606.03466
},
}

@article{kong2021enriched,
  title={Enriched monoidal categories I: centers},
  author={Kong, Liang and Yuan, Wei and Zhang, Zhi-Hao and Zheng, Hao},
  journal={Quantum Topology},
  year={2024},
  url={
https://doi.org/10.4171/qt/217
},
}

@article{chenqi,
  title={Dictionary of sVec0-enriched category},
  author={Chenqi Meng},
  journal={Unpublished note},
  year={2023}
}

@article{wen1995topological,
  title={Topological orders and edge excitations in fractional quantum Hall states},
  author={Wen, Xiao-Gang},
  journal={Advances in Physics},
  volume={44},
  number={5},
  pages={405--473},
  year={1995},
  publisher={Taylor \& Francis},
  url={
https://doi.org/10.1080/00018739500101566},
}

@article{PhysRevB.90.115119,
  title = {Topological quasiparticles and the holographic bulk-edge relation in $(2+1)$-dimensional string-net models},
  author = {Lan, Tian and Wen, Xiao-Gang},
  journal = {Phys. Rev. B},
  volume = {90},
  issue = {11},
  pages = {115119},
  numpages = {27},
  year = {2014},
  month = {Sep},
  publisher = {American Physical Society},
  doi = {10.1103/PhysRevB.90.115119},
  url = {https://link.aps.org/doi/10.1103/PhysRevB.90.115119}
}

@article{wang2017interacting,
  title={Interacting fermionic symmetry-protected topological phases in two dimensions},
  author={Wang, Chenjie and Lin, Chien-Hung and Gu, Zheng-Cheng},
  journal={Physical Review B},
  volume={95},
  number={19},
  pages={195147},
  year={2017},
  publisher={APS},
  url={
https://doi.org/10.1103/PhysRevB.95.195147
},
}

@article{gils2013anyonic,
  title={Anyonic quantum spin chains: Spin-1 generalizations and topological stability},
  author={Gils, Charlotte and Ardonne, Eddy and Trebst, Simon and Huse, David A and Ludwig, Andreas WW and Troyer, Matthias and Wang, Zhenghan},
  journal={Physical Review B},
  volume={87},
  number={23},
  pages={235120},
  year={2013},
  publisher={APS},
  url={
https://doi.org/10.1103/PhysRevB.87.235120
},
}

@article{levin2005string,
  title={String-net condensation: A physical mechanism for topological phases},
  author={Levin, Michael A and Wen, Xiao-Gang},
  journal={Physical Review B},
  volume={71},
  number={4},
  pages={045110},
  year={2005},
  publisher={APS},
  url={
https://doi.org/10.1103/PhysRevB.71.045110
},
}

@article{wang2015topological,
  title={Topological invariants for gauge theories and symmetry-protected topological phases},
  author={Wang, Chenjie and Levin, Michael},
  journal={Physical Review B},
  volume={91},
  number={16},
  pages={165119},
  year={2015},
  publisher={APS},
  url={
https://doi.org/10.1103/PhysRevB.91.165119
},
}

@article{levin2012braiding,
  title={Braiding statistics approach to symmetry-protected topological phases},
  author={Levin, Michael and Gu, Zheng-Cheng},
  journal={Physical Review B},
  volume={86},
  number={11},
  pages={115109},
  year={2012},
  publisher={APS},
  url={
https://doi.org/10.1103/PhysRevB.86.115109},
}

@article{wan2017fermion,
  title={Fermion condensation and gapped domain walls in topological orders},
  author={Wan, Yidun and Wang, Chenjie},
  journal={Journal of High Energy Physics},
  volume={2017},
  number={3},
  pages={1--39},
  year={2017},
  publisher={Springer},
  url={
https://doi.org/10.1007/JHEP03%282017%29172
},
}

@article{unpublished,
  title={Exactly Solvable Models of Fermionic Non-Abelian Topological Phas},
  author={Zhen, Bi and Yi-Zhuang, You and Meng, Cheng},
  journal={Unpublished note},
  year={2016},
}

@article{chen2010local,
  title={Local unitary transformation, long-range quantum entanglement, wave function renormalization, and topological order},
  author={Chen, Xie and Gu, Zheng-Cheng and Wen, Xiao-Gang},
  journal={Physical Review B—Condensed Matter and Materials Physics},
  volume={82},
  number={15},
  pages={155138},
  year={2010},
  publisher={APS},
  url={
https://doi.org/10.1103/PhysRevB.82.155138
},
}

@article{chen2012symmetry,
  title={Symmetry-protected topological orders in interacting bosonic systems},
  author={Chen, Xie and Gu, Zheng-Cheng and Liu, Zheng-Xin and Wen, Xiao-Gang},
  journal={Science},
  volume={338},
  number={6114},
  pages={1604--1606},
  year={2012},
  publisher={American Association for the Advancement of Science},
  url={
https://doi.org/10.48550/arXiv.1301.0861
},
}

@article{chen2011complete,
  title={Complete classification of one-dimensional gapped quantum phases in interacting spin systems},
  author={Chen, Xie and Gu, Zheng-Cheng and Wen, Xiao-Gang},
  journal={Physical Review B—Condensed Matter and Materials Physics},
  volume={84},
  number={23},
  pages={235128},
  year={2011},
  publisher={APS},
  url={
https://doi.org/10.1103/PhysRevB.84.235128
},
}

@article{chen2011classification,
  title = {Classification of gapped symmetric phases in one-dimensional spin systems},
  author = {Chen, Xie and Gu, Zheng-Cheng and Wen, Xiao-Gang},
  journal = {Phys. Rev. B},
  volume = {83},
  issue = {3},
  pages = {035107},
  numpages = {19},
  year = {2011},
  month = {Jan},
  publisher = {American Physical Society},
  doi = {10.1103/PhysRevB.83.035107},
  url = {https://link.aps.org/doi/10.1103/PhysRevB.83.035107}
}

@article{gu2009tensor,
  title = {Tensor-entanglement-filtering renormalization approach and symmetry-protected topological order},
  author = {Gu, Zheng-Cheng and Wen, Xiao-Gang},
  journal = {Phys. Rev. B},
  volume = {80},
  issue = {15},
  pages = {155131},
  numpages = {23},
  year = {2009},
  month = {Oct},
  publisher = {American Physical Society},
  doi = {10.1103/PhysRevB.80.155131},
  url = {https://link.aps.org/doi/10.1103/PhysRevB.80.155131}
}

@article{cheng2018classification,
  title={Classification of symmetry-protected phases for interacting fermions in two dimensions},
  author={Cheng, Meng and Bi, Zhen and You, Yi-Zhuang and Gu, Zheng-Cheng},
  journal={Physical Review B},
  volume={97},
  number={20},
  pages={205109},
  year={2018},
  publisher={APS},
  url={
https://doi.org/10.1103/PhysRevB.97.205109},
}

@article{johnson2024minimal,
  title={Minimal nondegenerate extensions},
  author={Johnson-Freyd, Theo and Reutter, David},
  journal={Journal of the American Mathematical Society},
  volume={37},
  number={1},
  pages={81--150},
  year={2024},
  url={
https://doi.org/10.1090/jams/1023},
}

@article{lan2019classification,
  title={Classification of 3+ 1 d bosonic topological orders (ii): The case when some pointlike excitations are fermions},
  author={Lan, Tian and Wen, Xiao-Gang},
  journal={Physical Review X},
  volume={9},
  number={2},
  pages={021005},
  year={2019},
  publisher={APS},
  url={
https://doi.org/10.1103/PhysRevX.9.021005
},
}

@article{ji2020categorical,
  title={Categorical symmetry and noninvertible anomaly in symmetry-breaking and topological phase transitions},
  author={Ji, Wenjie and Wen, Xiao-Gang},
  journal={Physical Review Research},
  volume={2},
  number={3},
  pages={033417},
  year={2020},
  publisher={APS},
  url={
https://doi.org/10.1103/PhysRevResearch.2.033417
},
}

@article{chatterjee2023symmetry,
  title={Symmetry as a shadow of topological order and a derivation of topological holographic principle},
  author={Chatterjee, Arkya and Wen, Xiao-Gang},
  journal={Physical Review B},
  volume={107},
  number={15},
  pages={155136},
  year={2023},
  publisher={APS},
  url={
https://doi.org/10.1103/PhysRevB.107.155136},
}

@article{chatterjee2023holographic,
  title={Holographic theory for continuous phase transitions: Emergence and symmetry protection of gaplessness},
  author={Chatterjee, Arkya and Wen, Xiao-Gang},
  journal={Physical Review B},
  volume={108},
  number={7},
  pages={075105},
  year={2023},
  publisher={APS},
  url={
https://doi.org/10.1103/PhysRevB.108.075105
},
}

@article{kong2020algebraic,
  title={Algebraic higher symmetry and categorical symmetry: A holographic and entanglement view of symmetry},
  author={Kong, Liang and Lan, Tian and Wen, Xiao-Gang and Zhang, Zhi-Hao and Zheng, Hao},
  journal={Physical Review Research},
  volume={2},
  number={4},
  pages={043086},
  year={2020},
  publisher={APS},
  url={
https://doi.org/10.1103/PhysRevResearch.2.043086
},
}

@article{lan2024quantum,
  title={Quantum current and holographic categorical symmetry},
  author={Lan, Tian and Zhou, Jing-Ren},
  journal={SciPost Physics},
  volume={16},
  number={2},
  pages={053},
  year={2024},
  url={
https://doi.org/10.21468/SciPostPhys.16.2.053
},
}

@article{wang2016twisted,
  title = {Twisted gauge theories in three-dimensional Walker-Wang models},
  author = {Wang, Zitao and Chen, Xie},
  journal = {Phys. Rev. B},
  volume = {95},
  issue = {11},
  pages = {115142},
  numpages = {23},
  year = {2017},
  month = {Mar},
  publisher = {American Physical Society},
  doi = {10.1103/PhysRevB.95.115142},
  url = {https://link.aps.org/doi/10.1103/PhysRevB.95.115142}
}

@article{hong2008exotic,
  title={On exotic modular tensor categories},
  author={Hong, Seung-Moon and Rowell, Eric and Wang, Zhenghan},
  journal={Communications in Contemporary Mathematics},
  volume={10},
  number={supp01},
  pages={1049--1074},
  year={2008},
  publisher={World Scientific},
  url={
https://doi.org/10.48550/arXiv.0710.5761
},
}

@article{chen2021disentangling,
  title={Disentangling supercohomology symmetry-protected topological phases in three spatial dimensions},
  author={Chen, Yu-An and Ellison, Tyler D and Tantivasadakarn, Nathanan},
  journal={Physical Review Research},
  volume={3},
  number={1},
  pages={013056},
  year={2021},
  publisher={APS}
}

@article{barkeshli2024higher,
  title={Higher-group symmetry in finite gauge theory and stabilizer codes},
  author={Barkeshli, Maissam and Chen, Yu-An and Hsin, Po-Shen and Kobayashi, Ryohei},
  journal={SciPost Physics},
  volume={16},
  number={4},
  pages={089},
  year={2024}
}

@article{loo2025systematic,
  title={Systematic construction of interfaces and anomalous boundaries for fermionic symmetry-protected topological phases},
  author={Loo, Kevin and Wang, Qing-Rui},
  journal={Physical Review B},
  volume={111},
  number={20},
  pages={205102},
  year={2025},
  publisher={APS}
}

@article{ellison2019disentangling,
  title={Disentangling interacting symmetry-protected phases of fermions in two dimensions},
  author={Ellison, Tyler D and Fidkowski, Lukasz},
  journal={Physical Review X},
  volume={9},
  number={1},
  pages={011016},
  year={2019},
  publisher={APS}
}

@article{barkeshli2022classification,
  title={Classification of (2+ 1) D invertible fermionic topological phases with symmetry},
  author={Barkeshli, Maissam and Chen, Yu-An and Hsin, Po-Shen and Manjunath, Naren},
  journal={Physical Review B},
  volume={105},
  number={23},
  pages={235143},
  year={2022},
  publisher={APS}
}

@article{freed2024topological,
  title={Topological symmetry in quantum field theory},
  author={Freed, Daniel S and Moore, Gregory W and Teleman, Constantin},
  journal={Quantum Topology},
  volume={15},
  number={3},
  pages={779--869},
  year={2024}
}

@article{kaidi2023symmetry,
  title={Symmetry TFTs for non-invertible defects},
  author={Kaidi, Justin and Ohmori, Kantaro and Zheng, Yunqin},
  journal={Communications in Mathematical Physics},
  volume={404},
  number={2},
  pages={1021--1124},
  year={2023},
  publisher={Springer}
}

\end{document}